\documentclass[12pt,preprint]{aastex6}
\usepackage{amsmath}
\usepackage{natbib}
\usepackage{hyperref}
\usepackage{rotating}
\usepackage{color}

\begin{document} 
%plainnat
\bibliographystyle{plainnat}

\title{\Large{Gamma Ray Burst afterglow and prompt-afterglow relations: an overview}}

\author{Dainotti M. G.\altaffilmark{1,2,3}, Del Vecchio R.\altaffilmark{3}}

\altaffiltext{1}{Physics Department, Stanford University, Via Pueblo Mall 382, Stanford, CA, USA, E-mail: mdainott@stanford.edu}
\altaffiltext{2}{INAF-Istituto di Astrofisica Spaziale e Fisica cosmica, Via Gobetti 101, 40129, Bologna, Italy}
\altaffiltext{3}{Astronomical Observatory, Jagiellonian University, ul. Orla 171, 30-244 Krak{\'o}w, Poland E-mails: delvecchioroberta@hotmail.it, mariagiovannadainotti@yahoo.it}

\begin{abstract}
The mechanism responsible for the afterglow emission of Gamma Ray Bursts (GRBs) and its connection to the 
prompt $\gamma$-ray emission is still a debated issue.
Relations between intrinsic properties of the prompt
or afterglow emission can help to discriminate between plausible
theoretical models of GRB production. 
Here we present an overview of the afterglow and prompt-afterglow two parameter relations, 
their physical interpretations, their use as redshift estimators and as possible cosmological tools.
A similar task has already been correctly achieved for Supernovae (SNe) Ia by using the peak 
magnitude-stretch relation, known in the literature as the Phillips relation \citep{phillips93}.
The challenge today is to make GRBs, which are amongst the farthest objects ever observed, 
standardizable candles as the SNe Ia through well established and robust relations.
Thus, the study of relations amongst the observable and physical
properties of GRBs is highly relevant together with selection biases in their physical quantities.\\
\noindent Therefore, we describe the state of the art of the existing GRB relations, 
their possible and debated interpretations in view of the current theoretical models and how relations 
are corrected for selection biases. 
We conclude that only after an appropriate evaluation and correction for selection effects can GRB relations
be used to discriminate
among the theoretical models responsible for the prompt and 
afterglow emission and to estimate cosmological parameters.
\end{abstract}

\maketitle
\pagebreak
\tableofcontents
\pagebreak

\section{Introduction}
GRBs, amongst the farthest and the most powerful objects ever observed in the Universe, are still a mystery after
50 years from their 
discovery time by the Vela Satellites \citep{klebesadel73}.
Phenomenologically, GRBs are traditionally classified in short SGRBs ($T_{90}<2$s) and long LGRBs 
($T_{90}>2$s) \citep{mazets81,kouveliotou93}, depending on their duration, where $T_{90}$ is the time
in which the 90\% (between 5\% and 95\%) of radiation is emitted in the prompt emission. 
However, \cite{norris2006} discovered the existence 
of an intermediate class (IC), or SGRBs with Extended Emission
(SGRBsEE), that shows mixed properties between SGRBs and LGRBs. Another relevant classification related to the spectral
features distinguishing normal GRBs from X-ray Flashes (XRFs) appears. 
The XRFs \citep{Heise2001,Kippen2001} are extra-galactic transient X-ray sources with spatial distribution,
spectral and temporal characteristics similar to LGRBs. The remarkable property that distinguishes XRFs from GRBs is 
that their $\nu F_{\nu}$ prompt emission spectrum peaks at energies 
typically one order of magnitude lower than the 
observed peak energies of GRBs. XRFs are empirically defined by a greater fluence (time-integrated flux) in the X-ray 
band 
($2-30$ keV) than in the gamma-ray band ($30-400$ keV). This classification is also relevant for 
the investigation
of GRB relations since some of them become stronger or weaker by introducing different GRB categories, see sec. 
\ref{Dainotti}.\\
One of the historical models used to explain the GRB phenomenon is the ``fireball" model \citep{wijers97,meszaros1998,meszaros2006} in 
which a compact 
central engine (either the collapsed core of a massive star or the merger product of a neutron star binary) launches a highly 
relativistic, and jetted electron/positron/baryon plasma. Interactions of blobs within the jet are believed to produce the prompt emission,
which consists of high photon energies such as gamma rays and hard X-rays. Instead, the interaction of the jet with the ambient 
material causes the afterglow phase, namely a long lasting multi-wavelength emission (X-ray, optical and sometimes 
also radio), which follows the prompt.
However, problems in explaining the light curves within this model have been shown by \cite{Willingale2007}, hereafter W07. 
More specifically, for $\sim 50\%$ of GRBs, the observed afterglow is in agreement with the model, but for the rest, the temporal and 
spectral indices do not conform and are suggestive of continued late energy injection. The difficulty of the standard fireball models 
appeared when Swift\footnote{The Swift satellite was launched in 2004.
With the instruments on board, the Burst Alert Telescope (BAT, divided in four standard channels 15-25; 25-50; 50-100; 100-150 keV), 
the X-Ray Telescope (XRT, 0.3-10 keV), and
the Ultra-Violet/Optical Telescope (UVOT, 170-650 nm), Swift provides a rapid follow-up of the afterglows in several 
wavelengths with better coverage than previous missions.} observations had revealed a more complex behaviour of the light curves 
\citep{Obrien06,sakamoto07,zhang07c} than in the past and pointed out that GRBs often follow 
``canonical" light curves \citep{Nousek2006}. In fact, the light curves can be divided into two, three and even more 
segments. The second segment, when it is flat, is called plateau emission.
X-ray plateaus can be interpreted as occurring due to
an accreting black hole (BH) \citep{Cannizzo2009,cannizzo2011,Kumar2008}
or a top-heavy jet evolution \citep{duffell15}. In addition, the fact that a
newly born magnetar could be formed either via the collapse
of a massive star or during the merger of two neutron stars
motivated the interpretation of the X-ray plateaus as
resulting from the delayed injection of rotational energy
($\dot{E}_{rot}\sim 10^{50}-10^{51}$ erg s$^{-1}$) from a fast spinning
magnetar \citep{usov92,zhang2001,dallosso2011,metzger11,rowlinson12,rowlinson14,rea15}.
These models are summarized in sec. \ref{Dainotti2008interpretation}.

Therefore, in this context, the discovery of relations amongst relevant
physical parameters between prompt and plateau phases is very important so as to use them 
as possible model discriminators.
In fact, many theoretical models have 
been presented in the literature to explain the wide variety of observations, but each
model has some advantages and drawbacks. The use of the phenomenological relations corrected for selection biases 
can boost the understanding of the mechanism responsible for such emissions.
Moreover, being observed at much larger redshift range than the SNe, it has
long been tempting to consider GRBs as useful cosmological probes, extending the redshift range by almost
an order of a magnitude further than the available SNe Ia, observed up to $z=2.26$ \citep{rodney15}. 
Indeed, GRBs are observed up to redshift $z=9.4$ \citep{cucchiara11},
which is much more distant than SNe Ia, and, therefore, they can help to understand the nature of the dark energy (DE), 
which is the main goal of modern cosmology, and determine the evolution of the equation of state (EoS), $w$, at very 
high $z$.
So far, the most robust standard candles are the SNe Ia which, by being excellent distance indicators, provide a unique probe for measuring
the expansion history of the Universe whose discovery has been awarded the Nobel Prize in 2011 \citep{riess98,perlmutter98}.
Up-to-date, $w$ has been measured to be \textcolor{red}{$-1$} within $5\%$ of the Einstein's cosmological constant, $\Omega_{\Lambda}$, the pure vacuum
energy. Measurement of the Hubble constant, $H_0$, provides another constraint on $w$ when combined
with Cosmic Microwave Background Radiation (CMBR) and Baryon Acoustic Oscillation (BAO) measurements \citep{weinberg2013}.
Therefore, the use of other estimates provided by GRBs would be helpful to confirm further and/or constrain the ranges of values of 
$H_0$.
However, different from the SNe Ia, which originate from white dwarves reaching
the Chandrasekhar limit and always releasing the same amount of energy, 
GRBs cannot yet be considered standard candles with their isotropic energies spanning over $8$ orders of magnitude. Therefore, finding out 
universal relations among observable properties can help to standardize their energetics and/or luminosities. 
It is for this reason that the study of GRB relations is relevant for both understanding the GRB emission mechanism, for finding a good distance
indicator and for estimating the cosmological parameters at high $z$.\\
Until now, for cosmological purposes, the most used relations are the prompt emission relations: 
Amati \citep{AmatiEtal02} and Ghirlanda relations \citep{Ghirlanda2004}. The scatter of these relations is 
significantly reduced providing constraints on the cosmological parameters, see \cite{ghirlanda06} and 
\cite{ghirlanda09b} for details. 
By adopting a maximum likelihood approach which allows for correct quantification of 
the extrinsic scatter of the relation, \cite{Amati2008} constrained the matter density 
$\Omega_M$ (for a flat Universe) 
to 0.04-0.40 (68\% confidence level, CL), with a best-fit value of
$\Omega_M \sim 0.15$, and exclude $\Omega_M=1$ at $> 99.9$\% CL. Releasing the assumption
of a flat Universe, they found evidence for a low value of $\Omega_M$ (0.04-0.50
at 68\% CL) as well as a weak dependence of the dispersion of the relation between the prompt peak energy in 
the $\nu F_\nu$ spectrum and the total gamma isotropic energy, $\log E_{\gamma,peak}-\log E_{\gamma,iso}$,
on $\Omega_{\Lambda}$ (with an upper limit of $\Omega_{\Lambda} \sim 1.15$ at 90\% CL). This
approach makes no assumptions about the $\log E_{\gamma,peak}-\log E_{\gamma,iso}$ relation and it does not use
other calibrators to set the normalization of the relation. Therefore, the treatment of the
data is not affected by the so-called circularity problem (to calibrate the GRB luminosity relations
for constraining cosmological models a particular cosmological model 
has to be assumed a priori) and the results are independent of those derived
via SNe Ia (or other cosmological probes).
Nowadays, the values of the cosmological parameters confirmed by measurements from the Planck Collaboration 
for the $\Lambda$CDM model are $\Omega_M=0.3089\pm0.0062$, $\Omega_{\Lambda}=0.6911\pm 0.0062$, and 
$H_0=67.74\pm0.46$ Km s$^{-1}$ Mpc$^{-1}$. 
For the investigation of the properties of DE, \cite{amati13} showed the 68\% CL
contours in the $\Omega_M-\Omega_{\Lambda}$ plane obtained by assuming a sample of 250 GRBs
expected shortly compared to those from other cosmological probes such as SNe Ia, CMB and Galaxy Clusters.\\
They obtained the simulated data sets via Monte Carlo techniques
by taking into account the slope, normalization, and dispersion of the observed 
$\log E_{\gamma,peak}-\log E_{\gamma,iso}$ relation, 
the observed $z$ distribution of GRBs and the distribution
of the uncertainties in the measured values of $\log E_{\gamma,peak}$ and $\log E_{\gamma,iso}$. 
These simulations indicated that with a sample of 250 GRBs, the
accuracy in measuring $\Omega_M$ would be comparable to that currently provided by SNe
data. In addition, they reported the estimates of $\Omega_M$ and the parameter of the DE EoS, $w_0$, derived from
the present and expected future samples. They assumed that the $\log E_{\gamma,peak}-\log E_{\gamma,iso}$ 
relation
is calibrated with a 10\% accuracy by using, e.g., the luminosity distances provided
by SNe Ia and the self-calibration of the relation with a large enough number of GRBs
lying within a narrow range of z ($\Delta z \sim 0.1-0.2$).
Generally speaking, as the number of GRBs in each redshift bin increases,
also the feasibility and accuracy of the self-calibration of GRB relations will improve.
\textcolor{red}{For a review on GRB prompt relations, see \cite{dainotti2016b}}.\\
\textcolor{red}{Even though the errors on $\Omega_M$ obtained in \cite{amati13} may lead to GRBs as promising 
standard candles, because they are 
almost comparable with SNe (0.06 for GRBs versus 0.04 for SNe, as provided for the SNe sample by \citealt{betoule14} and \citealt{calcino17}), these 
results show that $\Omega_M$ has an error which is 20 times larger then the value obtained by Planck. Thus, GRBs in a 
near future can be comparable with SNe Ia, but not likely with Planck. On the other hand, there is discrepancy among 
the values of $H_0$ computed by CMB and 
SNe \citep{Planck2015} and thus adding a new effective cosmological probe as GRBs can help to cast light on 
this discrepancy and break the degeneracy among several cosmological parameters.}\\
It is clear from this context that selection biases play a major and crucial role even for the close-by probes such 
as SNe Ia in determining the correct cosmological parameters.
This problem is more relevant for GRBs, which are particularly affected by the Malmquist bias effect 
(Malmquist 1920, Eddington 1940) that favours the brightest objects against faint ones at large distances.
Therefore, it is necessary to investigate carefully the problem of selection effects and how to overcome them before 
using GRB relations as distance estimators, as cosmological probes, and as model discriminators. This is indeed the 
major aim of this review.
Besides, this work is useful, especially for those embarking on
the study of GRB relations, because it aims at constituting a brief, but a complete compendium of afterglow and 
prompt-afterglow relations.\\
The review is organized as follows: in section \ref{notations}, we explain the nomenclature and definitions in all 
review, in sections \ref{Afterglow correlations} and \ref{promptaftcor}, we analyze the relations between the afterglow 
parameters and between parameters of both the prompt and afterglow phases. In section \ref{Selection effects}, we 
describe how these relations can be affected by selection biases. In section \ref{redshiftestimator}, we present how to obtain a 
redshift estimator and in section \ref{cosmology}, we report the use of the Dainotti relation as an example of GRB application as a cosmological tool. 
Finally, \textcolor{red}{in section \ref{discussion}, we briefly summarize some findings about the physical models and the 
cosmological usage of the analyzed relations, while in the last section} we draw our conclusions.

\section{Notations}\label{notations}
\begin{figure}[htbp]
\centering
\includegraphics[width=0.37\hsize,angle=0]{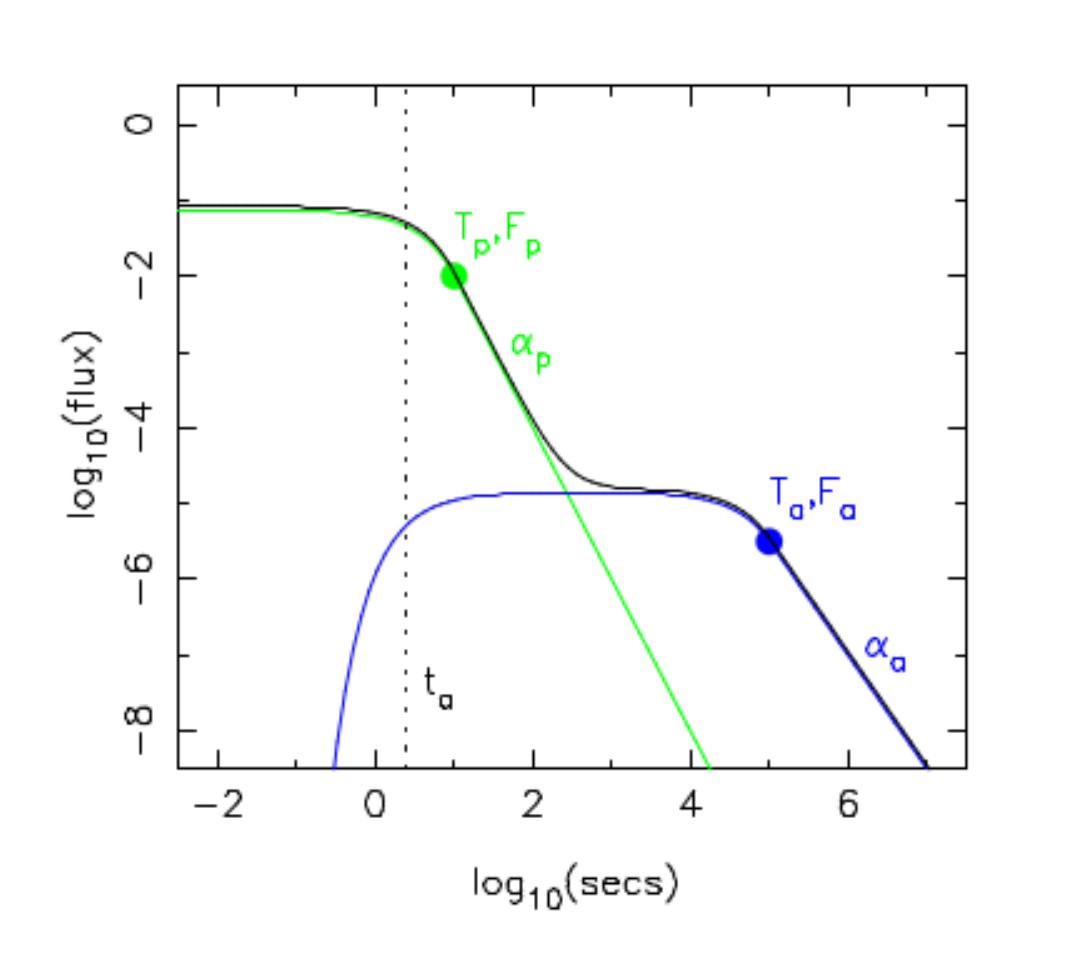}
\includegraphics[width=0.37\hsize,angle=0]{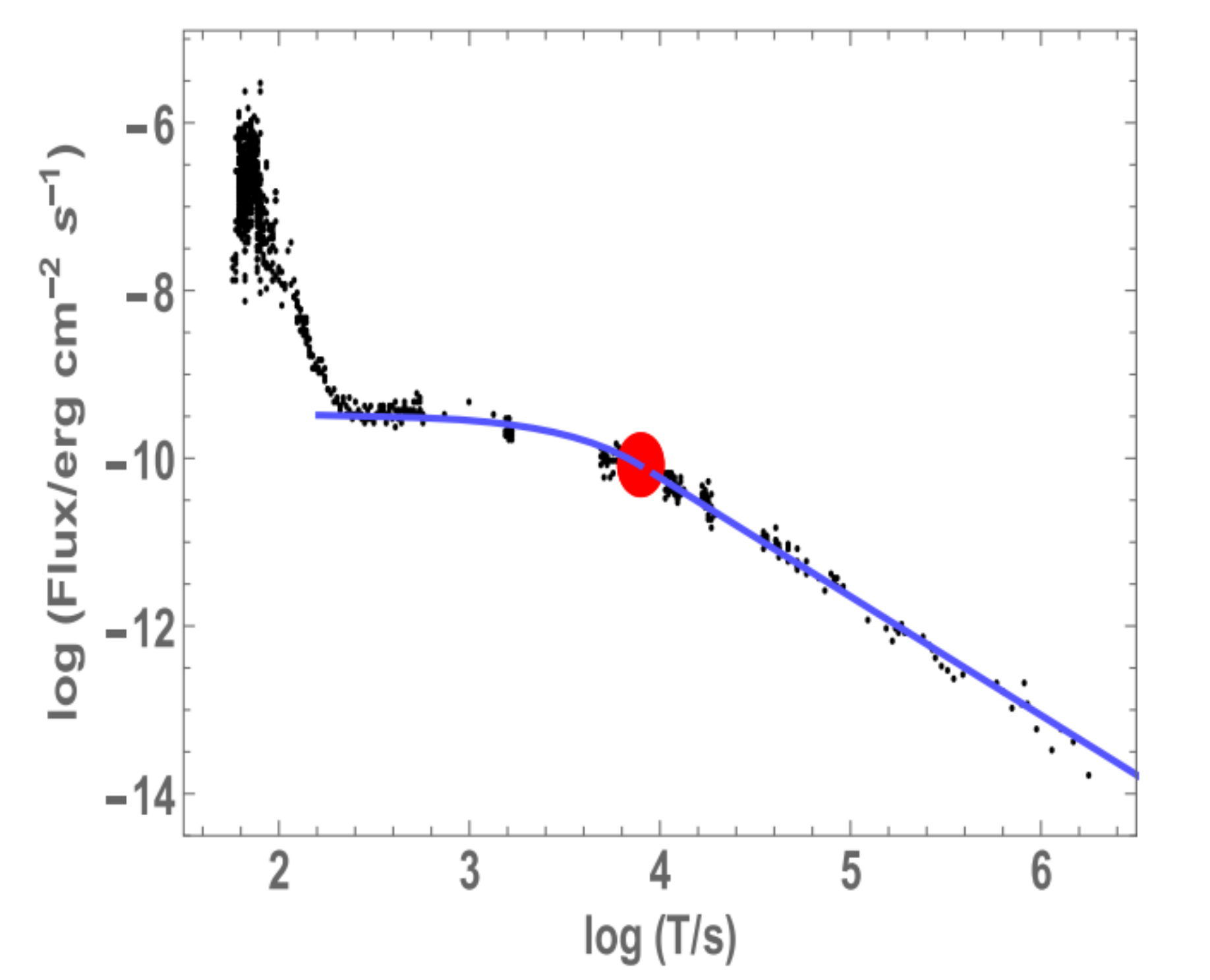}
 \caption{\footnotesize Left panel: the functional form of the fitting model from \cite{Willingale2007}. Right panel: 
 the observed light curve for GRB 061121 with the best-fit W07
model superimposed from \cite{dainotti16a}. The red dot marks the end of the flat plateau phase in the
X-ray afterglow ($T_{X,a}$, $F_{X,a}$). A similar configuration appears in the optical range.}
 \label{fig:notation}
\end{figure}

For clarity, we report a summary of the nomenclature adopted in the review.
\begin{itemize}
\item $L$, $E$, $F$, $S$, and $T$ indicate the luminosity, the energy, the flux, the fluence and the time
which can be observed in several
wavelengths, denoted with the first subscript, and at different times or part of the light curve, 
denoted instead with the second subscript. In addition, with
$\alpha$, $\beta$ and $\nu$, we represent the temporal and spectral decay indices and the frequencies.\\

More specifically:
\item $T_{X,a}$ and $T_{O,a}$ denote the time in the X-ray at the end of the plateau and the same time, but in the optical
wavelength respectively. $F_{X,a}$ are $F_{O,a}$ are their respective fluxes, while $L_{X,a}$ and $L_{O,a}$ are their
respective luminosities. An approximation of the energy of the plateau is $E_{X,plateau}=(L_{X,a}\times T^*_{X,a})$, 
see the left panel of Fig. \ref{fig:notation}.
\item $T_{O,peak}$ and $T_{X,f}$ are the peak time in the optical and the
time since ejection of the pulse.
$L_{O,peak}$ and $L_{X,f}$ are their respective luminosities. $F_{O,peak}$ is the respective flux of $T_{O,peak}$.
\item $T_{X,peak}$ is the peak time in the X-ray and $F_{X,peak}$ and $L_{X,peak}$ are its flux and luminosity
respectively.
\item $T_{X,p}$ and $T_{X,t}$ are the time at the end of the prompt emission
within the W07 model and the time at which the flat and the step decay behaviours of
the light curves join respectively. 
\item $T_{90}$ and $T_{45}$ are the times in which the 
90\% (between 5\% and 95\%) and 45\% (between 5\%-50\%) of radiation is emitted in the prompt emission respectively.
\item $\tau_{lag}$ and $\tau_{RT}$ are the differences in arrival time to the observer of the high 
energy photons and low energy photons and the shortest time over which the light curve increases by the $50\%$ of the 
peak flux of the pulse.
\item $L_{X,200\rm{s}}$, $L_{X,10}$, $L_{X,11}$, $L_{X,12}$, $L_{X,1\rm{d}}$ and $L_{O,200\rm{s}}$, $L_{O,10}$, $L_{O,11}$, 
$L_{O,12}$, $L_{O,1\rm{d}}$ 
are the X-ray and optical luminosities at 200 s, at 10, 11, 12 hours and at 1 day respectively; 
$L_{O,100s}$, $L_{O,1000s}$, $L_{O,10000s}$, $L_{O,7}$ are the optical luminosity at 100 s, 1000 s,
10000 s and 7 hours; $L_{\gamma,iso}$ and $L_L(\nu,T_{X,a})$ are the isotropic prompt emission mean 
luminosity and the optical 
or X-ray luminosity of the late prompt emission at the time $T_{X,a}$.
\item $F_{X,11}$, $F_{X,1\rm{d}}$ and $F_{O,11}$, $F_{O,1\rm{d}}$ are the X-ray and optical fluxes at 11 hours and at 1 day respectively;
$F_{\gamma,prompt}$, $F_{X,afterglow}$ are the gamma-ray flux in the prompt and the X-ray flux in the afterglow 
respectively. 
$E_{\gamma,prompt}$ and $E_{X,afterglow}$ are their respective isotropic energies and $L_{\gamma,prompt}$ and $L_{X,afterglow}$
are the respective luminosities. 
$S_{\gamma,prompt}$ indicates 
the prompt fluence in the gamma band correspondent to the rest frame isotropic prompt energy $E_{\gamma,prompt}$.
\item $E_{O,afterglow}$, $E_{\gamma,iso}$ and $E_{X,f}$ are the optical isotropic energy in the afterglow phase, 
the total gamma isotropic energy and the prompt emission energy of the pulse.
\item $E_{k,aft}$, $E_{\gamma,peak}$ and $E_{\gamma,cor}$ are the isotropic kinetic afterglow energy in X-ray, 
the prompt peak energy in 
the $\nu F_\nu$ spectrum and the isotropic energy corrected for the beaming factor.
\item $\alpha_{X,a}$, $\alpha_{O,>200\rm{s}}$, $\alpha_{X,>200\rm{s}}$, 
$\alpha_{\nu,fl}$ and $\alpha_{\nu,st}$ are the
X-ray temporal decay index in the afterglow phase, in the optical after $200$ s,
in the X-ray after $200$ s and the optical or X-ray flat and steep temporal decay indices respectively.
\item $\beta_{X,a}$, $\beta_{OX,a}$ and $\beta_{O,>200\rm{s}}$ are the spectral index of the 
plateau emission in X-ray, the optical-to-X-ray spectral index for 
the end time of the plateau and the optical spectral index after $200$ s.
\item $\nu_X$, $\nu_O$, $\nu_c$, $\nu_m$ are the X-ray and optical frequencies, and the cooling and the peak frequencies
of the synchrotron radiation.
\end{itemize}

All the time quantities described above are given in the observer frame, while with the upper index $*$ 
we denote in the text the observables in the GRB rest frame. The rest frame times are the observed times 
divided by the cosmic time expansion, for example, $T^*_{X,a}=T_{X,a}/(1+z)$ denotes the rest frame time 
at the end of the plateau emission. 
\pagebreak

In the following table we will give a list of the abbreviations/acronyms used through the text:

\begin{table}[htbp]
\footnotesize
\begin{center}{
\begin{tabular}{|c|c|}
\hline
Abbreviation& Meaning\\
\hline
DE& Dark Energy\\
EoS& Equation of State\\
CL& Confidence Level\\
IC& Intermediate Class GRB\\
SGRB& Short GRB\\
LGRB& Long GRBs\\
SGRBsEE& Short GRBs with extended emission\\
XRFs& X-ray Flashes\\
SNe&Supernovae\\
BH& Black Hole\\
z& redshift\\
FS& Forward Shock\\
RS& Reverse Shock\\
$H_0$&Hubble constant\\
$\Omega_M$& Matter density in $\Lambda$CDM model\\
$\Omega_{\Lambda}$ & Dark Energy density in $\Lambda$CDM model\\
$\Omega_k$ & curvature in $\Lambda$CDM model\\
$\sigma_{\log L_{X,a}}$ & error on the luminosity\\
$\sigma_{\log T^{*}_{X,a}}$& error on the time\\
E4& sample with $\sigma_E=(\sigma_{\log L_{X,a}}^2+\sigma_{\log T^{*}_{X,a}}^2)^{1/2}<4$\\
E0095& sample with $\sigma_E=(\sigma_{\log L_{X,a}}^2+\sigma_{\log T^{*}_{X,a}}^2)^{1/2}<0.095$ \\
W07& Willingale et al. (2007)\\
$\Gamma$& Lorentz Factor\\
$V$&Variability of the GRB light curve\\
$h$& Hubble constant divided by 100\\
$w_0$, $w_a$& coefficients of the DE EoS $w(z)=w_0+w_a z(1+z)^{-1}$ \\
HD & Hubble Diagram\\
a & normalization of the relation\\
b & slope of the relation\\
$\sigma_{int}$ & intrinsic scatter of the relation\\
$b_{int}$ & intrinsic slope of the relation\\
\hline
\end{tabular}}
\caption{\footnotesize Table with abbreviations.}
\label{abbreviations}
\end{center}
\end{table}

\section{The Afterglow Relations} 

\label{Afterglow correlations}
Several relations appeared in literature relating only parameters in the afterglow, such as the $L_X(T_{a})-T^{*}_{X,a}$ relation \citep{Dainotti2008} 
and similar ones in the optical and X-ray bands such as the unified $L_X(T_{a})$-$T_{X,a}^*$ and $L_{O,a}$\,-\,$T^*_{O,a}$ \citep{ghisellini09} and the $L_{O,200\rm{s}}$\,-\,$\alpha_{O,>200\rm{s}}$ relations \citep{oates2012}.

\subsection{The Dainotti relation (\texorpdfstring{$L_X(T_{a})$\,-\,$T^{*}_{X,a}$}{Lg})} \label{Dainotti}
The first relation to shed light on the plateau properties has been the $L_X(T_a)$\,-\,$T^{*}_{X,a}$ 
one, hereafter also referred as LT. The phenomenon is an anti-relation between the X-ray luminosity
at the end of the plateau, $L_X(T_a)$, and the time in the X-ray at the end of the plateau, $T^{*}_{X,a}$,
for simplicity of notation we will refer to $L_X(T_a)$ as $L_{X,a}$.\\ 
It was discovered by \cite{Dainotti2008} using
33 LGRBs detected by the Swift satellite in the X-ray energy band observed by XRT. 
Among the 107 GRBs fitted by W07 phenomenological model, shown in the left panel of Fig. \ref{fig:notation}, only the GRBs that 
have a good spectral 
fitting of the plateau and firm determination of $z$ have been chosen. The functional form of the LT 
relation obtained is the following:

\begin{equation}
\log L_{X,a} = a + b \times \log T^{*}_{X,a}, 
\end{equation}

with a normalization $a=48.54$, a slope $b=-0.74 ^{+0.20}_{-0.19}$, an intrinsic scatter, $\sigma_{int}=0.43$ and a 
Spearman correlation coefficient\footnote{A computation of statistical dependence 
between two variables stating how good the relation
between these variables can be represented employing a monotonic function. It assumes a value between $-1$ and $+1$.}
$\rho=-0.74$.
$L_{X,a}$ in the Swift XRT passband, $(E_{min}, E_{max})=(0.3,10)$ keV, has been 
computed from the following equation:

\begin{equation}
L_{X,a} (z)= 4 \pi D_L^2(z, \Omega_M, h) \, F_{X,a} \times K
\label{eq: lx}
\end{equation}

where $D_L(z, \Omega_M, h)$ represents the GRB luminosity distance for a given $z$, $F_{X,a}$ indicates the flux
in the X-ray at the end of the plateau,
and $K=\frac{1}{(1+z)^{(1-\beta_{X,a})}}$ denotes the K-correction for cosmic expansion \citep{Bloom2001}.
This anti-relation shows that the shorter the plateau duration, the more luminous the plateau. 
Since the ratio between the errors on both variables is close to unity, it means that both errors need to be considered 
and the Marquardt Levenberg algorithm is not the best fitting method to be applied in this circumstance. 
Therefore, a Bayesian 
approach \citep{Dagostini2005} needs to be considered. This method takes into account the errors of both variables and an intrinsic scatter, $\sigma_{int}$, of 
unknown nature. However, the results of both the D'Agostini method and the Marquardt Levenberg algorithm 
are comparable. Due to the higher accuracy of the first method from now on the authors prefer this technique in their papers.
Evidently, the tighter the relation, the better the chances to constrain
the cosmological parameters. With this specific challenge in mind, 
a subsample of bursts has been chosen with particular selection criteria both
on luminosity and time, namely $\log L_{X,a} > 45$ and $1 \le \log T_{X,a}^{*} \le 5$.
After this selection has been applied, a subsample of 28 LGRBs was obtained with 
$(a, b, \sigma_{int})=(48.09, -0.58 \pm 0.18, 0.33)$, thus reducing considerably the scatter.

\begin{figure}[htbp]
\centering
\includegraphics[width=8.1cm,height=7.4cm,angle=0]{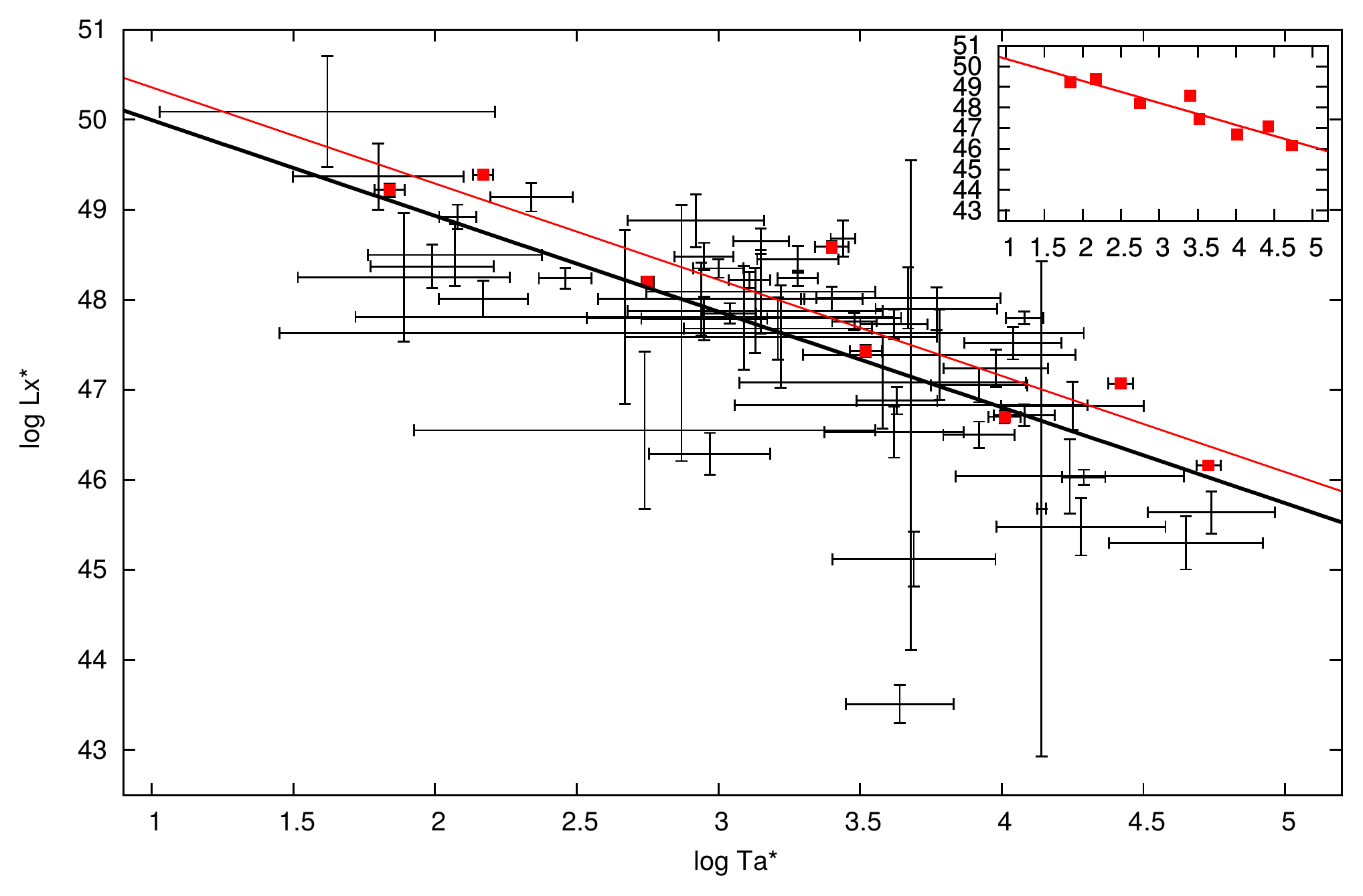}
\includegraphics[width=8.1cm,height=7.3cm,angle=0]{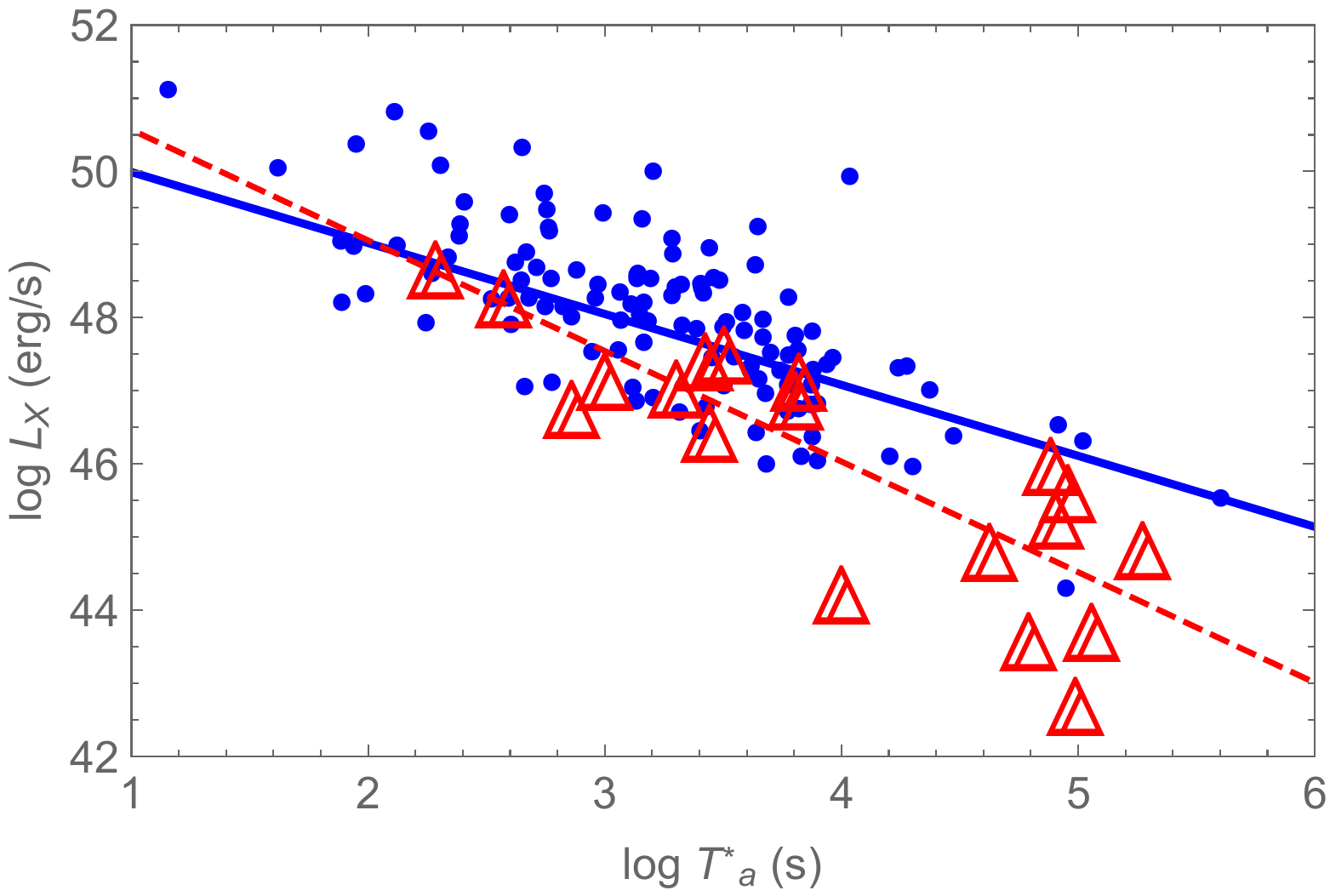}
\caption{\footnotesize Left panel: $\log L_{X,a}$ (equivalent to $\log L_X^*$ in this plot) vs. $\log T^{*}_{X,a}$ for $62$ 
long afterglows with the error energy parameter $\sigma_E < 4$, and the best 
fitted relation line in black, from \cite{Dainotti2010}. The red line fitted to the 8 lowest error
(red) points produces an upper envelope of the full data set. 
The upper envelope points with the best fitted line are separately presented in an inset panel. 
\textcolor{red}{Right panel: LONG-NO-SNe 128 GRBs (blue points fitted with
a solid blue line) and the 19 events from LONG-SNe (red
empty triangles) fitted with a red dashed line from \cite{dainotti16c}}}
\label{fig:Dainotti2010}
\end{figure}

\begin{figure}[htbp]
\centering
\includegraphics[width=7cm,height=6cm,angle=0]{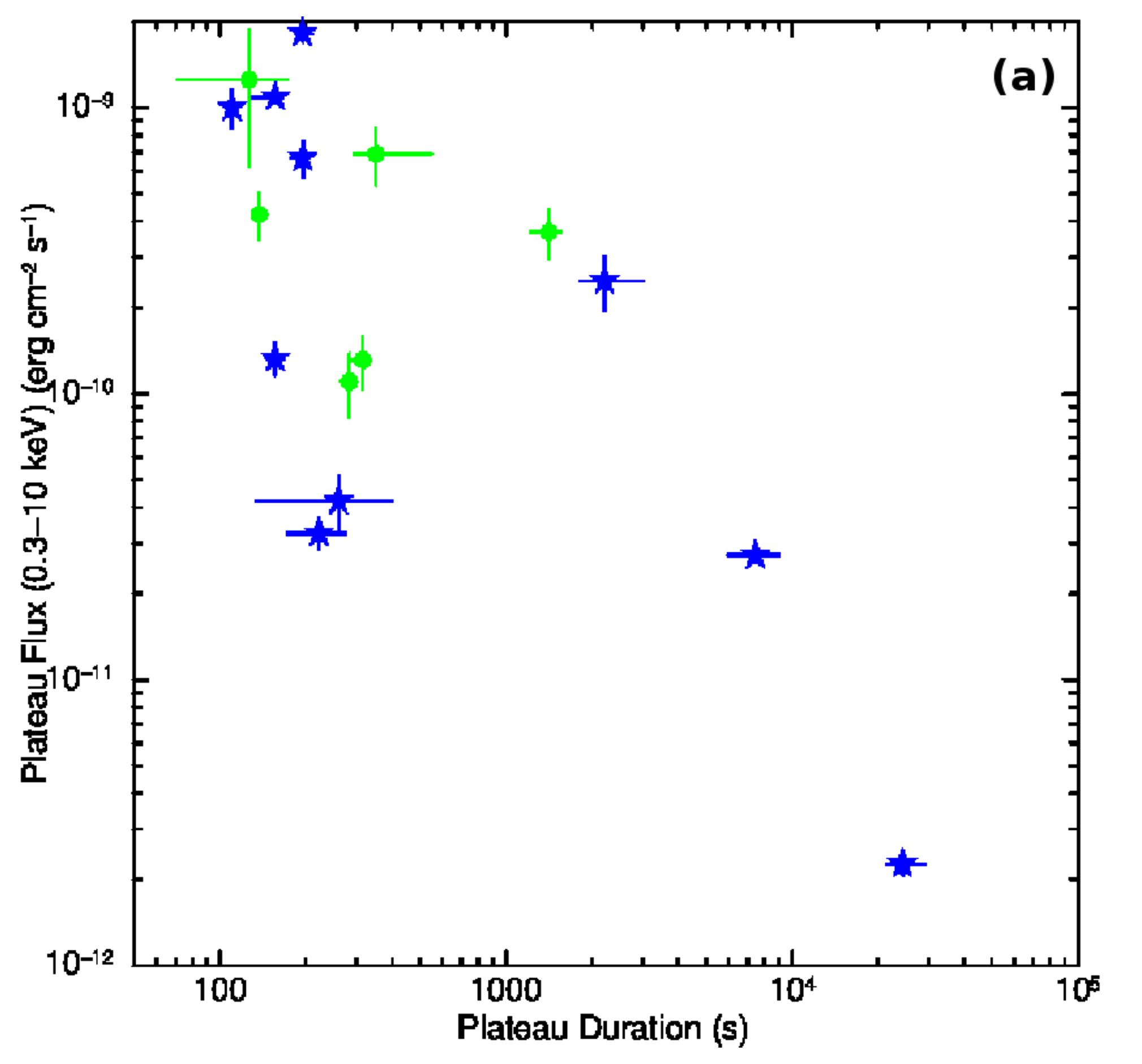}
\includegraphics[width=8.2cm,height=6.2cm,angle=0]{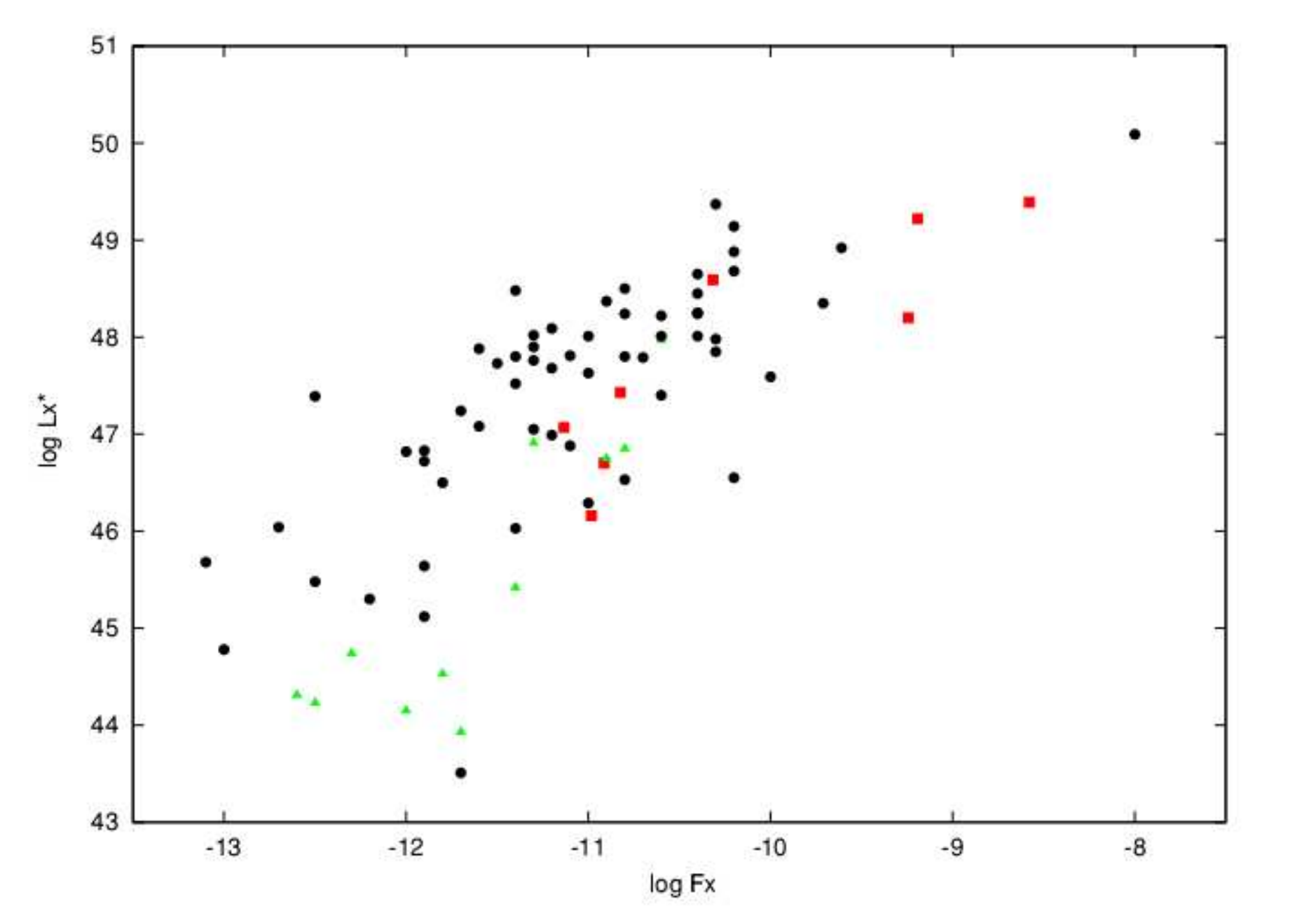}
\caption{\footnotesize Left panel: 
the plateau flux versus the plateau duration for a sample of 22 SGRBs from \cite{rowlinson2013}. Blue stars are GRBs with two or more 
breaks in their light curves, while green circles have one break.
Right panel: 
$\log L_{X,a}$ versus $\log F_{X,a}$ for the full GRB sample from \cite{Dainotti2010}. The 8 upper envelope 
points are shown as red squares, while the IC GRBs are represented by green triangles.}
 \label{fig:Dainotti2010b}
\end{figure}

In agreement with these results, through the analysis of the late prompt phase in optical and X-ray light curves 
of 33 LGRBs,
also \cite{ghisellini09} found a common observational model for optical and X-ray light curves
with the same value for the slope, $b=-0.58^{+0.18}_{-0.18}$, obtained by \cite{Dainotti2008} 
when the time is limited between $1 \le \log T_{X,a}^{*} \le 5$.\\
Instead, \cite{Dainotti2010} from a sample of 62 LGRBs found $b=-1.06^{+0.27}_{-0.28}$, while for
the 8 IC GRBs pointed out a much steeper relation ($b=-1.72^{+0.22}_{-0.21}$). Finally, taking into
account the errors on luminosity ($\sigma_{\log L_{X,a}}$) and time ($\sigma_{\log T^{*}_{X,a}}$), the 
8 GRBs with the smallest errors were defined as the ones with $\sigma_E=(\sigma_{\log L_{X,a}}^2+\sigma_{\log T^{*}_{X,a}}^2)^{1/2}<0.095$. 
For this subsample, \cite{Dainotti2010} found a slope $-1.05^{+0.19}_{-0.20}$, see Fig. \ref{fig:Dainotti2010}, the right panel of Fig. 
\ref{fig:Dainotti2010b} and Table \ref{tbl7}.\\
Similar to \cite{Dainotti2010}, also \cite{bernardini2012} and \cite{Sultana2012}, with a sample of 64 and 14 LGRBs 
respectively, found a slope $b\approx -1$, for details see Table \ref{tbl7}.\\
Expanding the sample again to 77 LGRBs, \cite{dainotti11a} discovered a relation with $b=-1.20^{+0.27}_{-0.30}$. 
Later, \cite{mangano12}, considering in their sample of 50 LGRBs those GRBs with no visible plateau phase and employing 
a broken power law as a fitting model, found a steeper slope ($b=-1.38^{+0.16}_{-0.16}$). 
Thus, from all these analyses it is clear that a steepening of the slope has been 
observed when the sample size is increased.\\
Therefore, before going further with additional analysis, 
\cite{Dainotti2013a} decided to show how selection biases can influence the slope of the relation. They showed that 
the steepening of the relation results from selection biases, while the intrinsic slope of the relation is 
$b=-1.07^{+0.09}_{-0.14}$, see section \ref{Selection effects}. Summarizing, \cite{Dainotti2013a} with a sample of 
101 GRBs, confirmed the previous results from \cite{Dainotti2010}, as well as \cite{rowlinson14}, with a data set 
of 159 GRBs. 

\cite{dainotti15} also confirmed previous results of \cite{Dainotti2013a} but with a larger sample of 123 LGRBs.
All the samples discussed are observed by SWIFT/XRT.

\begin{table}[htbp]
\footnotesize
\begin{center}{
\begin{tabular}{|c|c|c|c|c|c|c|c|}
\hline
Author & N& Type& Slope& Norm & Corr.coeff. & P \\
\hline
Dainotti et al. (2008)&28 & $1<T^{*}_{X,a}<5$&$-0.58^{+0.18}_{-0.18}$&48.09& -0.80&$1.6\times10^{-7}$\\
Dainotti et al. (2008)&33 &All GRBs&$-0.74^{+0.20}_{-0.19}$&48.54& -0.74&$10^{-9}$\\
Cardone et al. (2009)&28&L&$-0.58^{+0.18}_{-0.18}$& 48.09&-0.74&$10^{-9}$\\
Ghisellini et al. (2009)&33&L&$-0.58^{+0.18}_{-0.18}$&48.09& -0.74&$10^{-9}$\\
Cardone et al. (2010)&66&L&$-1.04^{+0.23}_{-0.22}$&$50.22^{+0.77}_{-0.76}$&-0.68&$7.6\times10^{-9}$\\
Dainotti et al. (2010) & 62 &L&$-1.06^{+0.27}_{-0.28}$&$51.06^{+1.02}_{-1.02}$&-0.76&$1.85\times10^{-11}$\\
Dainotti et al. (2010) & 8 &high luminosity&$-1.05^{+0.19}_{-0.20}$&$51.39^{+0.90}_{-0.90}$&-0.93&$1.7\times10^{-2}$\\
Dainotti et al. (2010) & 8 & IC&$-1.72^{+0.22}_{-0.21}$&$52.57^{+1.04}_{-1.04}$&-0.66&$7.4\times10^{-2}$\\
Dainotti et al. (2011a) &77 &L &$-1.20^{+0.27}_{-0.30}$&$51.04^{+0.27}_{-0.30}$&-0.69& $7.7\times10^{-8}$\\
Sultana et al. (2012)&14&L&$-1.10^{+0.03}_{-0.03}$&$51.57^{+0.10}_{-0.10}$&-0.88&$10^{-5}$\\
Bernardini et al. (2012)& 64 & L&$-1.06^{+0.06}_{-0.06}$&51.06&-0.68&$7.6\times10^{-9}$\\
Mangano et al. (2012) & 50 & L& $-1.38^{+0.16}_{-0.16}$&$52.2^{+0.06}_{-0.06}$&-0.81&$2.4\times10^{-10}$\\
Dainotti et al. (2013a)& 101&ALL intrinsic &$-1.07^{+0.09}_{-0.14}$&52.94&-0.74&$10^{-18}$  \\
Dainotti et al. (2013b) & 101&All GRBs&$-1.32^{+0.18}_{-0.17}$&$52.8^{+0.9}_{-0.3}$&-0.74&$10^{-18}$ \\
Dainotti et al. (2013b)& 101&without short&$-1.27^{+0.18}_{-0.26}$&52.94&-0.74&$10^{-18}$  \\
Dainotti et al. (2013b) & 101&simulated&$-1.52^{+0.04}_{-0.24}$&$53.27_{-0.48}^{+0.54}$&-0.74&$10^{-18}$  \\
Postnikov et al. (2014)&101&L ($z<1.4$)&$-1.51^{+0.26}_{-0.27}$&$53.27_{-0.48}^{+0.54}$&-0.74&$10^{-18}$\\
Rowlinson et al. (2014)&159 &intrinsic &$-1.07^{+0.09}_{-0.14}$&52.94&-0.74&$10^{-18}$ \\
Rowlinson et al. (2014)&159 &observed &$-1.40^{+0.19}_{-0.19}$&$52.73^{+0.52}_{-0.52}$&-0.74&$10^{-18}$\\
Rowlinson et al. (2014)&159 &simulated&$-1.30^{+0.03}_{-0.03}$&$52.73^{+0.52}_{-0.52}$&-0.74&$10^{-18}$\\
Dainotti et al (2015)&123&L&$-0.90^{+0.19}_{-0.17}$&$51.14^{+0.58}_{-0.58}$&-0.74&$10^{-15}$\\
\textcolor{red}{Dainotti et al. (2016c)}&\textcolor{red}{19}& \textcolor{red}{L-SNe}&\textcolor{red}{$-1.5^{+0.3}_{-0.3}$}&\textcolor{red}{$51.85^{+0.94}_{-0.94}$}&\textcolor{red}{-0.83}&\textcolor{red}{$5\times10^{-6}$}\\
\hline
\end{tabular}}
\caption{\footnotesize Summary of the LT relation. All the measurements are performed by the Swift XRT Telescope. 
The first column represents the authors, the second one the number of GRBs in the used sample, 
the third one the GRB type (S=Short, L=Long, IC=Intermediate), the fourth and the fifth ones are
the slope and normalization of the relation and the last two columns are the correlation coefficient and the chance 
probability, P.} 
\label{tbl7}
\end{center}
\end{table}

In the context of reducing the scatter of the LT relation, \cite{Delvecchio16} investigated the temporal decay indices 
$\alpha_{X,a}$ after the plateau phase for a sample of $176$ GRBs detected by Swift within two different models:
a simple power law, considering the decaying phase after the plateau phase, and the W07 one. It is pointed out that the results
are independent of the chosen model. It was checked if there are some common characteristics in GRBs phenomena that can
allow them to be used as standardizable candles like SNe Ia and to obtain some constraints revealing which is the best physical 
interpretation describing the plateau emission.
The interesting result is that the LT relation for the low and high luminosity GRBs seems to depend differently on the $\alpha_{X,a}$ 
parameter, thus possibly implying a diverse density medium.\\ 
\textcolor{red}{Continuing the search for a standard set of GRBs, \cite{dainotti16c} analyzed 176 GRB afterglow plateaus observed by 
Swift with known redshifts which revealed that the subsample of LGRBs associated with SNe (LONG-SNe) presents a very
high correlation coefficient for the LT relation. They investigated the category of LONG GRBs associated 
spectroscopically with SNe in order to compare the LT correlation for this sample with the one for LGRBs for which 
no associated SN has been observed (hereafter LONG-NO-SNe, 128 GRBs). They checked if there is a difference among these 
subsamples. They adopted first a non-parametric statistical method, the \cite{Efron1992} one, to take into 
account redshift evolution and check if and how this effect may steepen the slope for the
LONG-NO-SNe  sample. This procedure is necessary due to the fact that this sample is observed at much higher redshift 
than the GRB-SNe sample. Therefore, removing selection bias is the first step before any comparison among samples 
observed at different redshifts could be properly performed. They have demonstrated that there is no evolution for the 
slope of the LONG-NO-SNe sample and no evolution is expected for the LONG-SNe sample. The difference among the slopes 
is statistically significant with the probability P=0.005 for LONG-SNe. This possibly suggests that the LONG-SNe with 
firm spectroscopic features of the SNe associated might not require a standard energy reservoir in the plateau phase 
unlike the LONG-NO-SNe. Therefore, this analysis may open new perspectives in future theoretical investigations of the 
GRBs with plateau emission and associated with SNe. They also discuss how much this difference can be due to the jet 
opening angle effect. The difference between the SNe-LONG (A+B) and LONG-NO-SNe sample is only statistically significant 
at the 10\% level when we consider the beaming correction. Thus, one can argue that the difference in slopes can be 
partially due to the effect of the presence of low luminosity GRBs in the LONG-SNe sample that are not corrected for beaming. 
However, the beaming corrections are not very accurate due to indeterminate jet opening angles, so the debate remains 
open and it can only be resolved when we will gather more data.}\\
In Table \ref{tbl7}, we report a summary of the parameters $a$ and $b$ with $\rho$ and $P$ for the LT relation.
In conclusion, the most reliable parameters for this relation are those from 
\cite{Dainotti2013a}, because they have demonstrated that the intrinsic slope not affected by selection biases 
is determined to be $-1$ as computed through the Efron and Petrosian (EP) method.

\subsubsection{Physical interpretation of the Dainotti relation \texorpdfstring{($L_X(T_a)$\,-\,$T^{*}_{X,a}$)}{Lg}}\label{Dainotti2008interpretation}
\begin{figure}[htbp]
\includegraphics[width=8cm,height=6cm,angle=0]{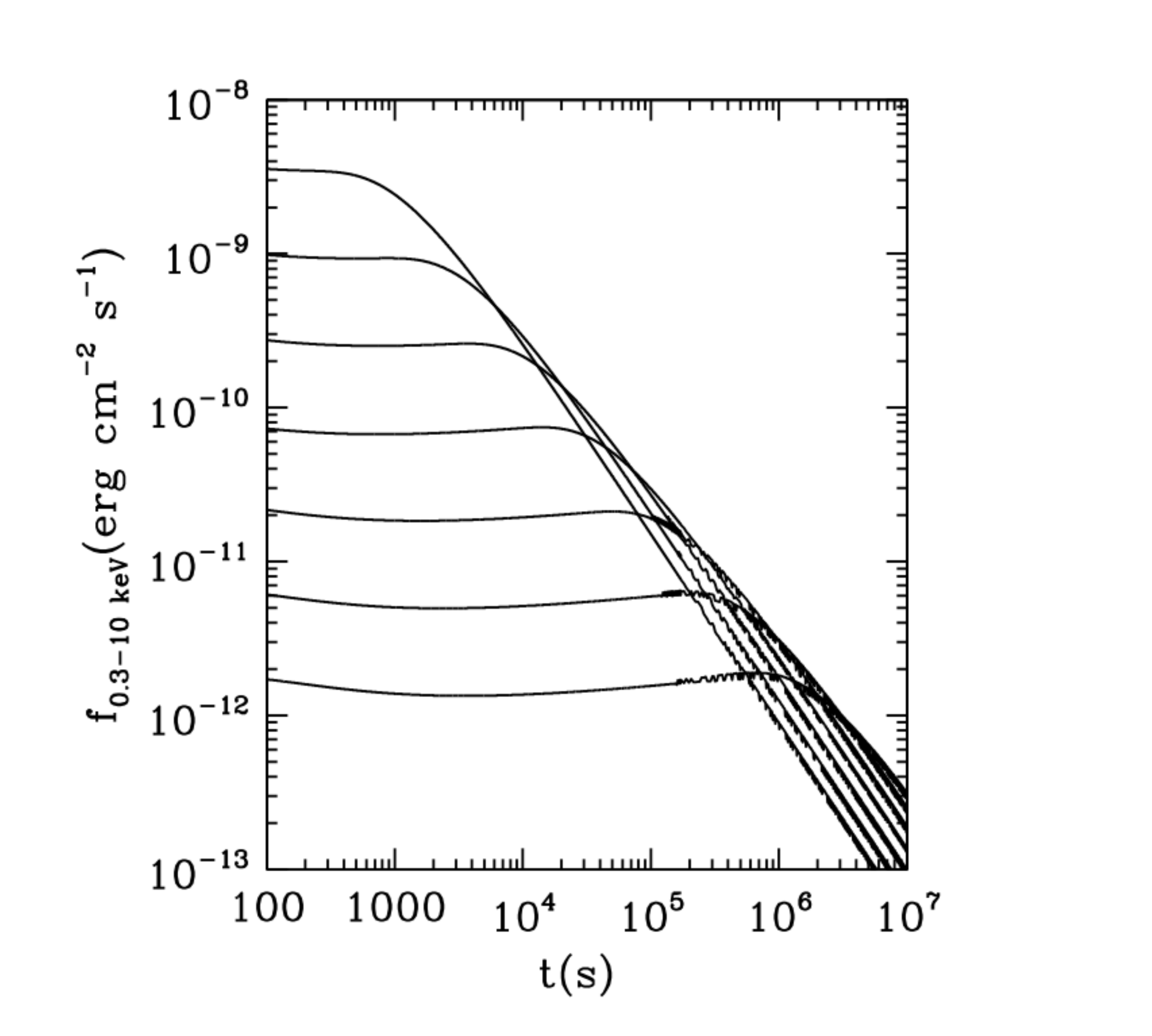}
\includegraphics[width=8cm,height=6cm,angle=0]{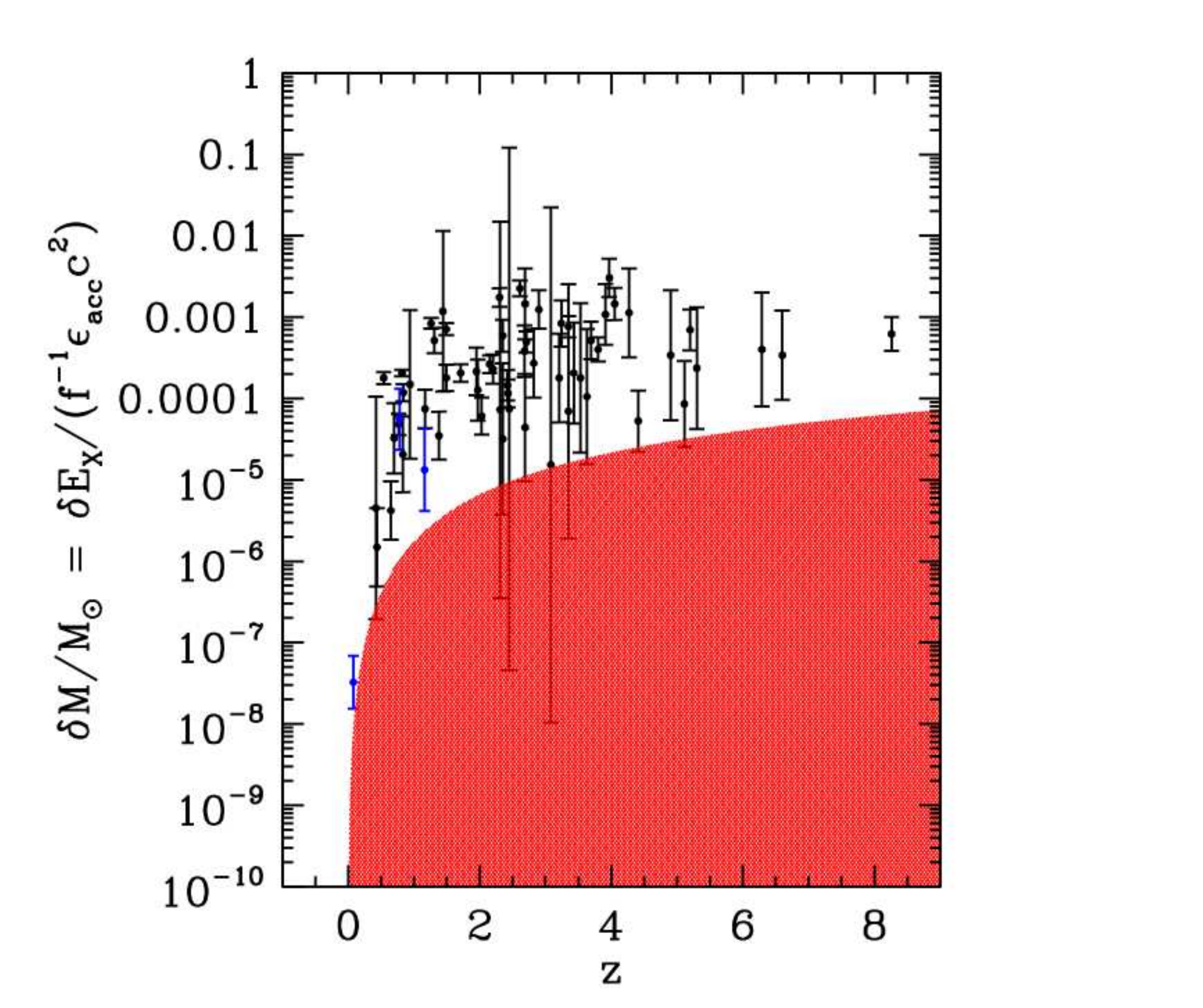}
\caption{\footnotesize Left panel: model light curves for LGRB parameters from \cite{cannizzo2011}, keeping
the starting fall-back disk mass constant at $10^{-4} M_{\odot}$ but changing the initial radius and normalization.
Right panel: total accretion mass for the plateau $+$ later decay phases
of GRBs from \cite{cannizzo2011}, considering 62 LGRBs from \cite{Dainotti2010}. The region in red 
represents a limiting XRT detection flux level $f_{\rm II} = 10^{-12}$ erg cm$^{-1}$ s$^{-1}$
(assuming a plateau duration $t_{\rm II} = 10^4$ s) in order to study a plateau to sufficient accuracy.
A beaming factor $f=1/300$ and a net efficiency for powering the X-ray flux 
$\epsilon_{\rm net}=\epsilon_{\rm acc}\epsilon_X=0.03$ were assumed.}
\label{fig:cannizzo11}
\end{figure}

Here, we revise the theoretical interpretation of the LT relation, which is based mainly
on the accretion \citep{Cannizzo2009,cannizzo2011} and the magnetar models \citep{zhang2001,dallosso2011,rowlinson12,rowlinson2013,rowlinson14}.\\
The first one states that an accretion disc is created from the motion of
the material around the GRB progenitor star collapsing towards its progenitor core. After it is compressed by
the gravitational forces, the GRB emission takes place. For LGRBs, the early rate of decline in the initial steep 
decay phase of the light curve 
may provide information about the radial density distribution within the progenitor \citep{Kumar2008}.\\
\cite{Cannizzo2009} predicted a steeper relation slope (\textcolor{red}{-3/2}) than the observed one 
(\textcolor{red}{$\sim -1$}), which on the other
hand is in good agreement with the prior emission model of \cite{yamazaki09}.\\
Later, \cite{cannizzo2011}, using a sample of $62$ LGRBs and few SGRBs simulated the fall-back disks 
surrounding the BH. They found that a circularization radius of the mass around the BH with value $10^{10}-10^{11}$ cm can give an 
estimate for the plateau 
duration of around $10^4$s for LGRBs maintaining the initial fall back mass at $10^{-4}$ solar masses ($M_{\odot}$), 
see the left panel 
of Fig. \ref{fig:cannizzo11}. For SGRBs the radius is estimated to be $10^8$ cm.
The LT relation provides a lower limit for the accreting mass estimates $\Delta M \approx 10^{-4}$ to $10^{-3} M_{\odot}$\footnote{This value can be 
derived considering the total inferred accretion mass $\Delta M/M=\Delta E_X/f^{-1}*\epsilon_{acc}*c^2$ where c 
is the light speed, f is the X-ray afterglow beaming factor, $\epsilon_{acc}$ is the efficiency of the accretion onto the BH and $E_X$ 
is the observed total energy of the plateau + later decay phases (the integral over time between $T_{X,t}$ and the end of afterglow, 
see Eq. 2 of W07).}. From their results, it was claimed that the LT relation could be obtained if a typical energy reservoir
in the fall-back mass is assumed, see the right panel of Fig. \ref{fig:cannizzo11}. However, in their analysis the very steep 
initial decay following the prompt emission, which have been modelled by \cite{lindner10} as fall-back of the 
progenitor core, is not considered.

\begin{figure}[htbp]
\includegraphics[width=0.495\hsize,angle=0]{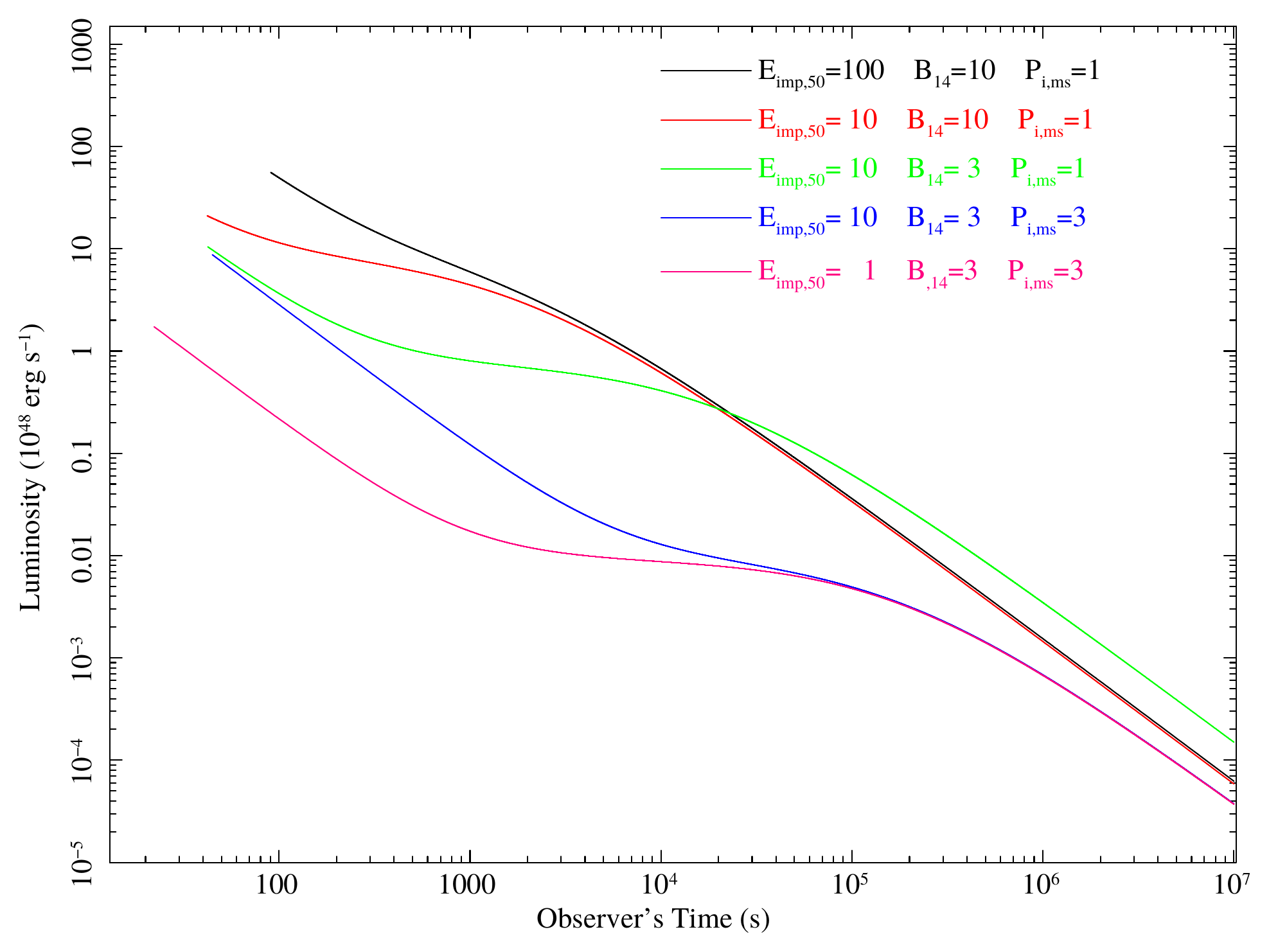}
\includegraphics[width=0.495\hsize,angle=0]{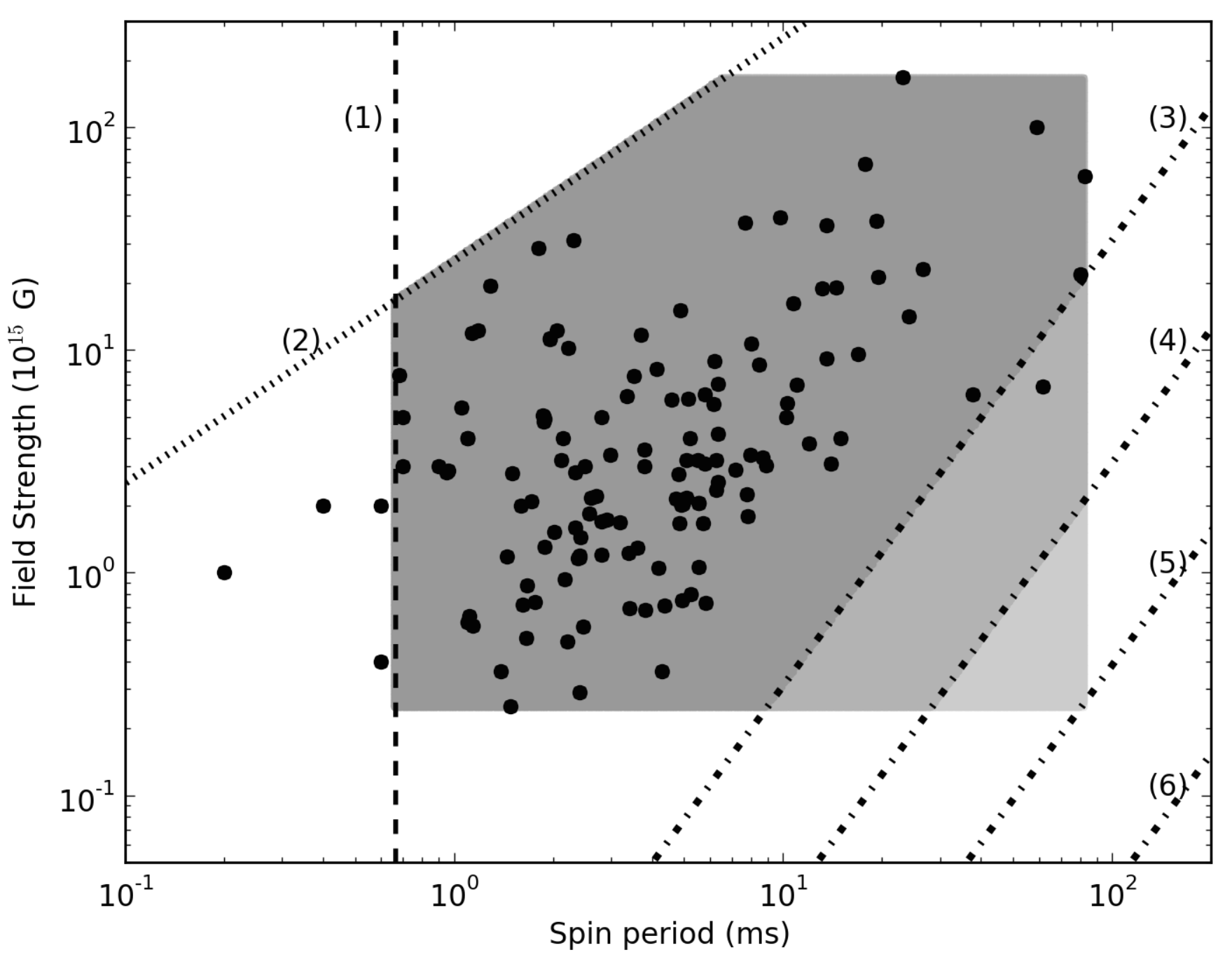}
\caption{\footnotesize Left panel: five theoretical light curves obtained by \cite{dallosso2011}, changing the initial energy of the afterglow, 
the dipole magnetic field, B, and the initial spin 
period of the NS, P. Right panel: the grey shaded areas are the homogeneous distribution of B and P employed to simulate the observable magnetar plateaus from \cite{rowlinson14}. The upper and lower limits on B and the upper
limit on P are computed considering the sample of GRBs fitted with the magnetar model 
\citep{lyons2010,dallosso2011,bernardini2012,gompertz2013,rowlinson2013,yi2014,lu2014}.
The dashed black vertical line (1) at 0.66 ms is the minimum P allowed. The dotted black line (2) indicates a limit on P and B strengths imposed by the fastest slew time of XRT in their sample in the rest frame of the highest $z$ GRB, as plateaus with durations shorter than the
slew time are unobservable. The black dash-dot lines (3-6) are the observational constraints for the dimmest XRT plateau
observable assuming the lowest $z$ in the GRB sample. These limits vary as a function of the beaming and efficiency of
the magnetar emission: (3) Minimum beaming angle and efficiency (1 degree and 1\% respectively), (4) Minimum efficiency (1\%) 
and maximum beaming angle (isotropic), (5) Maximum efficiency (100\%) and minimum beaming angle, (6) Maximum efficiency and beaming
angle. The observed distributions indicate that the samples have low efficiencies and
small beaming angles.}
\label{fig:rowlinson2014}
\end{figure}

Regarding the magnetar model, \cite {zhang2001} studied the effects of an injecting central
engine on the GRB afterglow radiation, concentrating on a strongly magnetized millisecond pulsar. 
For specific starting values of rotation period and magnetic field of the pulsar, the afterglow light curves should 
exhibit an achromatic bump lasting from minutes to months, and the observation of such characteristics could
set some limits on the progenitor models. 
More recently, \cite{dallosso2011} investigated the energy evolution in a relativistic
shock from a spinning down magnetar in spherical symmetry. With their fit of few observed Swift XRT light curves and 
the parameters of this model, namely a spin period of ($1-3$ ms), and high values of magnetic fields ($B\sim 10^{14}-10^{15}$ G), 
they managed to well reproduce the properties of the shallow decay phase and the LT relation, see the left panel of 
Fig. \ref{fig:rowlinson2014}.\\
Afterward, \cite{bernardini2012} with a sample of 64 LGRBs confirmed, as previously founded by \cite{dallosso2011}, that
the shallow decay phase of the GRB light curves and the LT relation can be well explained.\\
Then, \cite{rowlinson12} and \cite{rowlinson2013} pointed out that energy injection is a fundamental mechanism
for describing the plateau emission of both LGRBs and SGRBs. In fact, the remnant of NS-NS mergers can form a magnetar,
and indeed 
the origin of the majority of SGRBs is well explained through the energy injection by a magnetar.

Later, \cite{rowlinson14}, using 159 GRBs from Swift catalogue, analytically demonstrated that the central 
engine model accounts for the LT relation assuming that the compact object is
injecting energy into the forward shock (FS), a shock driven out
into the surrounding circumstellar medium. The luminosity and plateau duration can be computed
as follows:

\begin{equation}
\log L_{X,a} \sim \log (B^2_{p}P^{-4}_{0}R^6)
\label{eq10}
\end{equation}
and
\begin{equation}
\log T^*_{X,a} = \log (2.05 \times IB^{-2}_{p}P^2_{0}R^{-6}),
\label{eq11}
\end{equation}

where $T^*_{X,a}$ is in units of $10^3$ s, $L_{X,a}$ is in units of $10^{49}$ erg s$^{-1}$, $I$ is the moment of inertia in
units of $10^{45}$ g cm$^2$, $B_{p}$ is the magnetic field strength at the poles
in units of $10^{15}$ G, $R$ is the radius of the NS in units of $10^6$ cm and
$P_{0}$ is the initial period of the compact object in milliseconds.
Then, substituting the radius from eq. \ref{eq11} into eq. \ref{eq10}, it was derived that:

\begin{equation}
\log \ (L_{X,a}) \sim \log \ (10^{52}I^{-1}P^{-2}_{0})-\log \ (T^*_{X,a}).
\end{equation}

Therefore, an intrinsic relation $\log L_{X,a} \sim - \log T_{X,a}^{{*}}$ is confirmed directly from this formulation.
Although some magnetar plateaus are inconsistent with energy injection into the FS,
\cite{rowlinson14} showed that this emission is narrowly beamed 
and has $\leq 20$\% efficiency in conversion of rotational energy from the compact object into the
observed plateau luminosity. In addition, the intrinsic LT relation slope, namely the one where
the selection biases are
appropriately removed, is explained within the spin-down of a newly formed magnetar at 1 $\sigma$ level, 
see right panel of Fig. \ref{fig:rowlinson2014}. 
The scatter in the relation is mainly due to the range of the initial spin periods.\\
After several papers discussing the origin of the LT relation within the context of the magnetar model, 
very recently a debate has been opened by \cite{rea15} on the reliability of this model as the correct 
interpretation for the X-ray plateaus.
Using GRBs with known $z$ detected by Swift from its launch to August 2014, \cite{rea15} concluded that the initial 
magnetic field distribution,
used to interpret the GRB X-ray plateaus within the magnetar model does not match the features
of GRB-magnetars with the Galactic magnetar
population. Therefore, even though there are large uncertainties in these estimates due to GRB rates, metallicity and
star 
formation, the GRB-magnetar model in its present
form is safe only if two kinds of magnetar progenitors are allowed. Namely, the GRB should be different from Galactic 
magnetar ones 
(for example for different metallicities) and should be considered supermagnetars (magnetars with an initial magnetic
field significantly large). 
Finally, they set a limit of about $ \leq 16$ on the number of stable magnetars produced in the Milky Way via a GRB in
the past Myr. \textcolor{red}{However, it can be argued that since the rates of Galactic magnetars and GRBs are 
really different, the number of Galactic magnetars cannot fully describe the origin of GRBs. 
In fact the Galactic magnetar rate is likely to be greater than 10\% than the core collapse SNe rate, while GRB rate is 
much lower than that. In addition, the number of magnetars in the Milky Way may not be used as a constraint on the GRB 
rate because the spin-down rates of GRB magnetars should be very rapid. Due to the low GRB rate it would not be easy to 
detect these supermagnetars.} 
Thus, it can be claimed that no conflict stands among this paper and the previous ones.

\begin{figure}[htbp]
\centering
\includegraphics[width=0.495\hsize,angle=0]{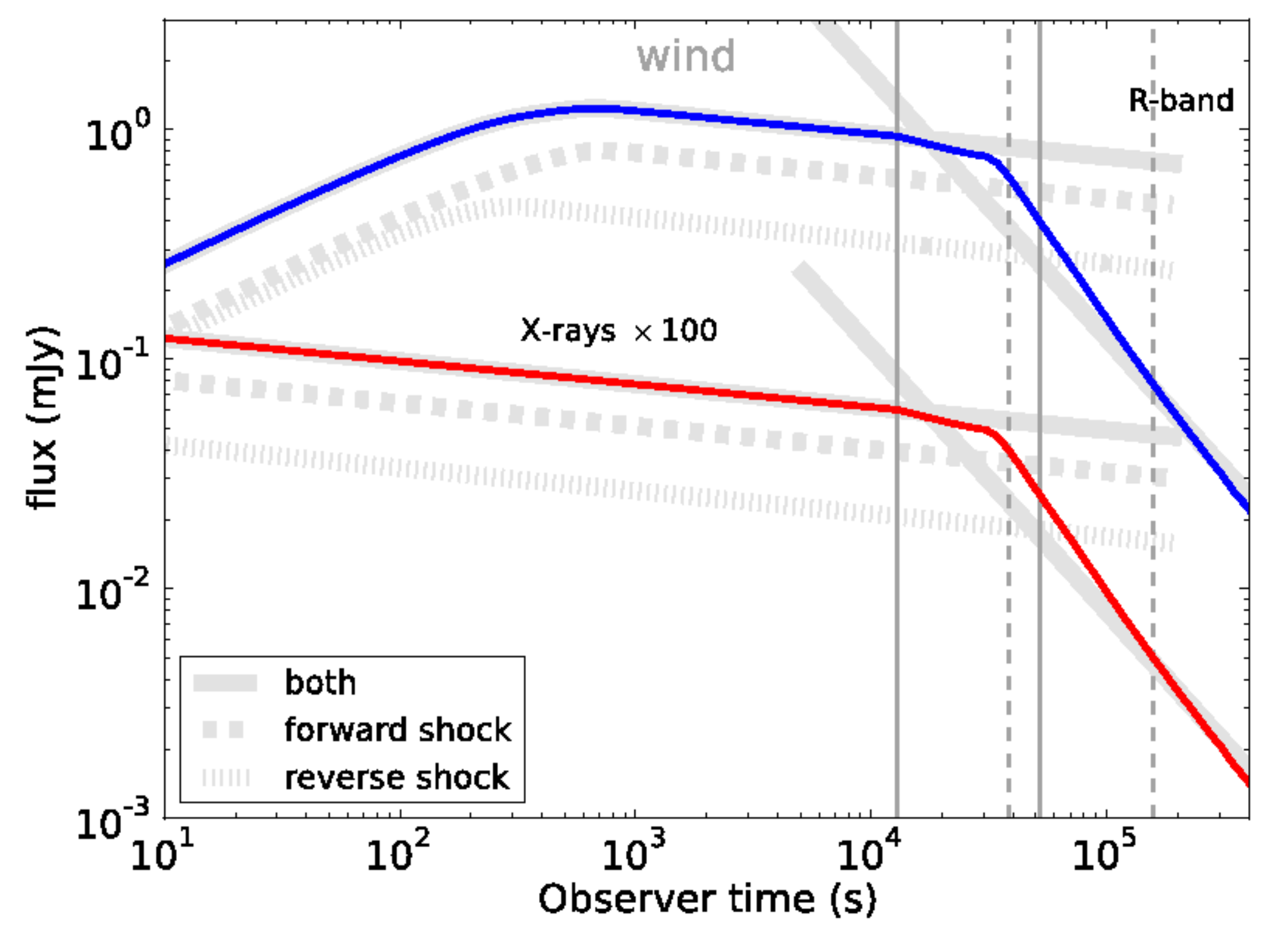}
\includegraphics[width=0.495\hsize,angle=0]{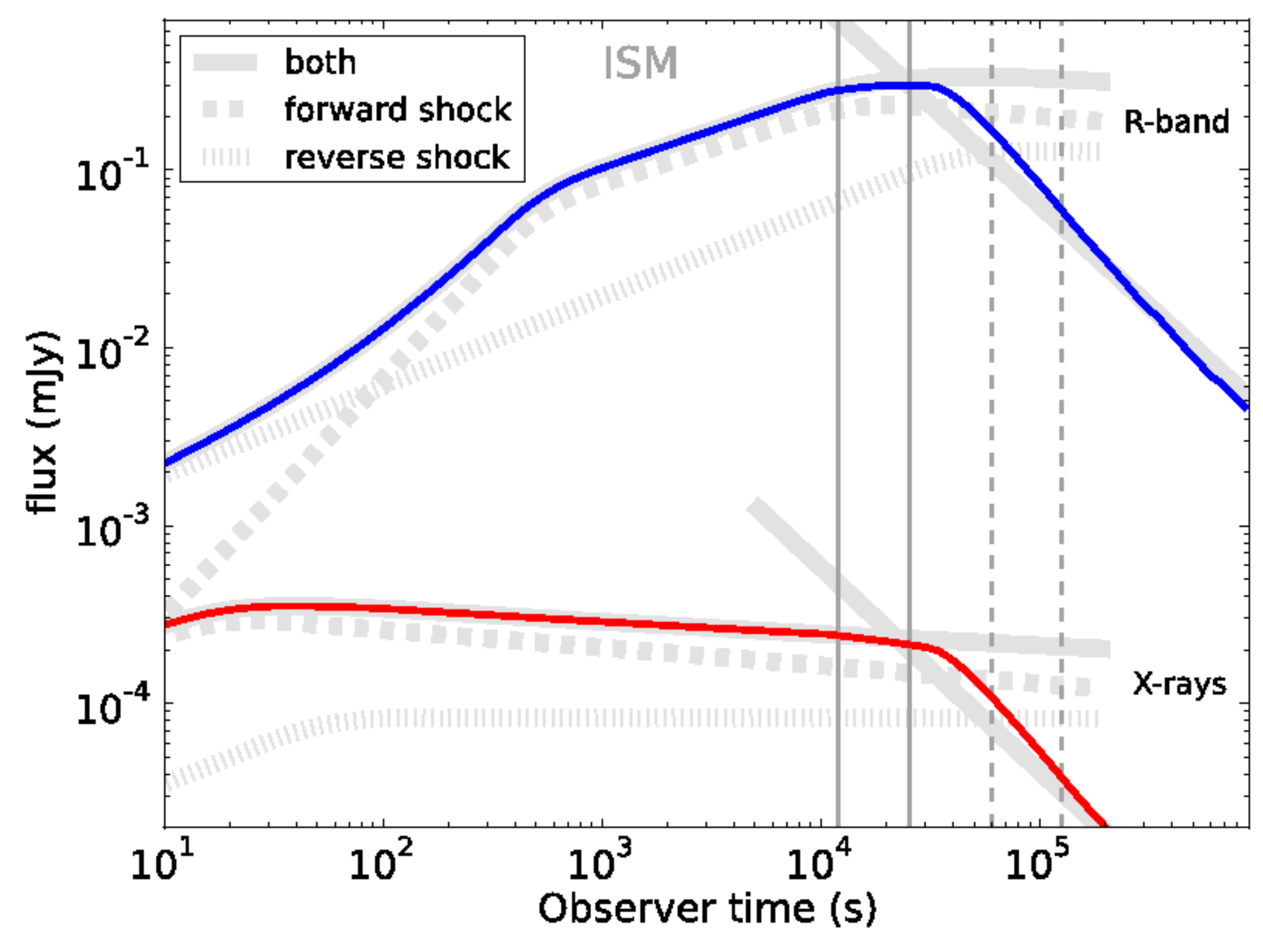}
\caption{\footnotesize Optical and X-ray light curves for wind (left panel) and ISM (right plot) 
scenario's from \cite{vaneerten14a}.
Thick light grey curves represent the analytical solutions for
prolonged and impulsive energy injection. Thick dashed light
grey and the thick dotted light grey curves indicate the forward shock region emission only and the reverse shock region only respectively. The grey vertical lines show
(1) the arrival time of emission
from the jet back and (2) the arrival time of emission
from the jet front. The solid vertical lines indicate arrival times of emission along the jet axis
for these two events; the dashed vertical lines express the arrival
times of emission from an angle $\theta = 1/\gamma$.}
\label{fig:vaneerten14a}
\end{figure}

Still in the context of the energy injection models, \cite{vaneerten14a} found a relation between the optical flux
at the end of the plateau and the time at the end of the plateau itself $F_{O,a} \sim T_{O,a}^{-0.78 \pm 0.08}$ 
\citep{Panaitescu2011,Li2012} for which observed frame variables were 
considered. The range of all parameters describing the emission
($E_{\gamma,iso}$, the fraction of the magnetic energy, $\epsilon_B$, the initial density, $n_0$) is the principal cause of the scatter in the
relation, but it does not affect the slope. Finally, it was claimed that both the wind ($\propto A/r^2$, where A is a constant)
and the interstellar medium can reproduce the observed relation within both the reverse shock (RS, a shock driven back into the 
expanding bubble of the ejecta) and FS scenarios, see Fig. \ref{fig:vaneerten14a}.

\begin{figure}[htbp]
\centering
\includegraphics[width=8cm,height=6cm,angle=0]{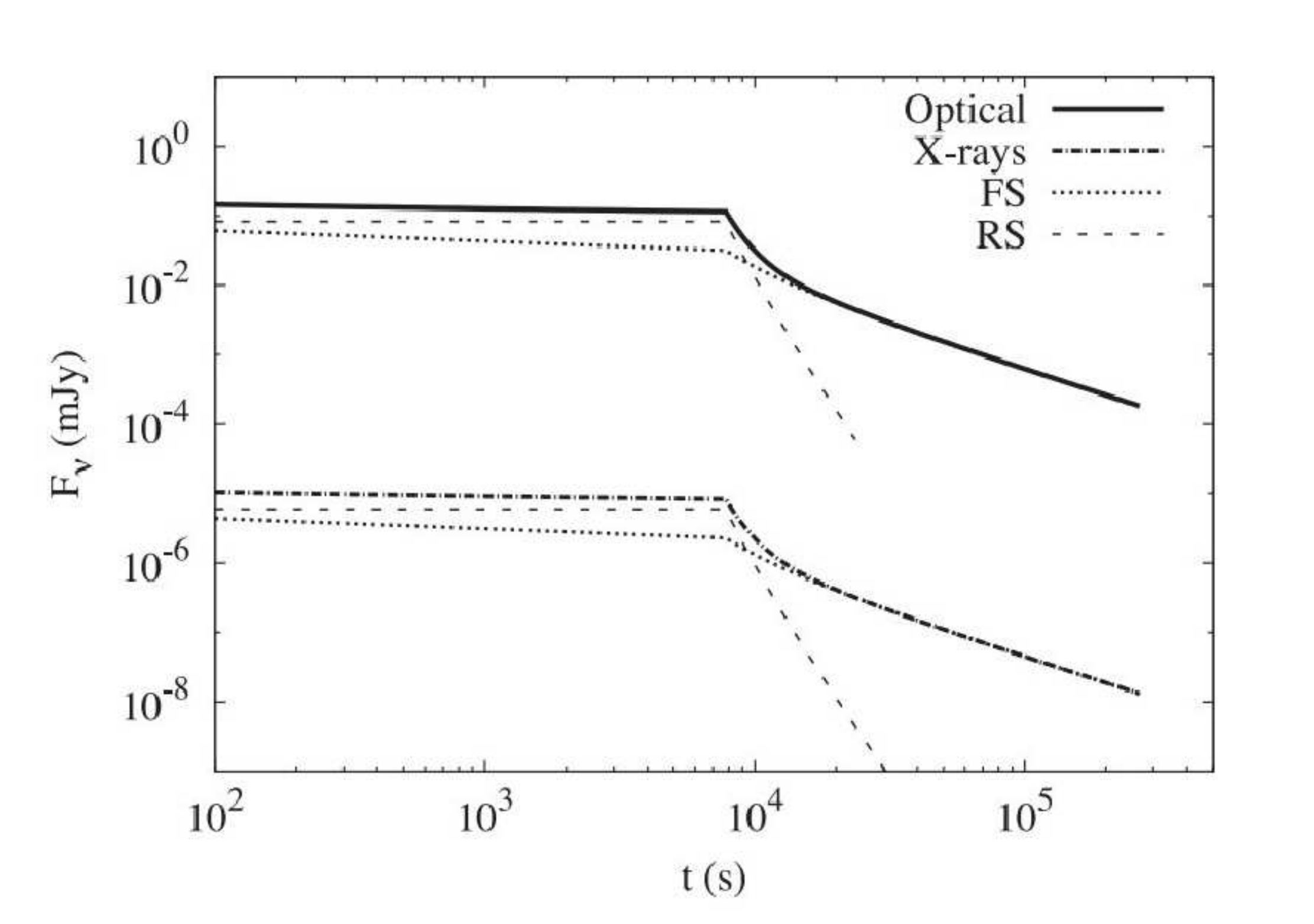}
\includegraphics[width=8cm,height=6.2cm,angle=0]{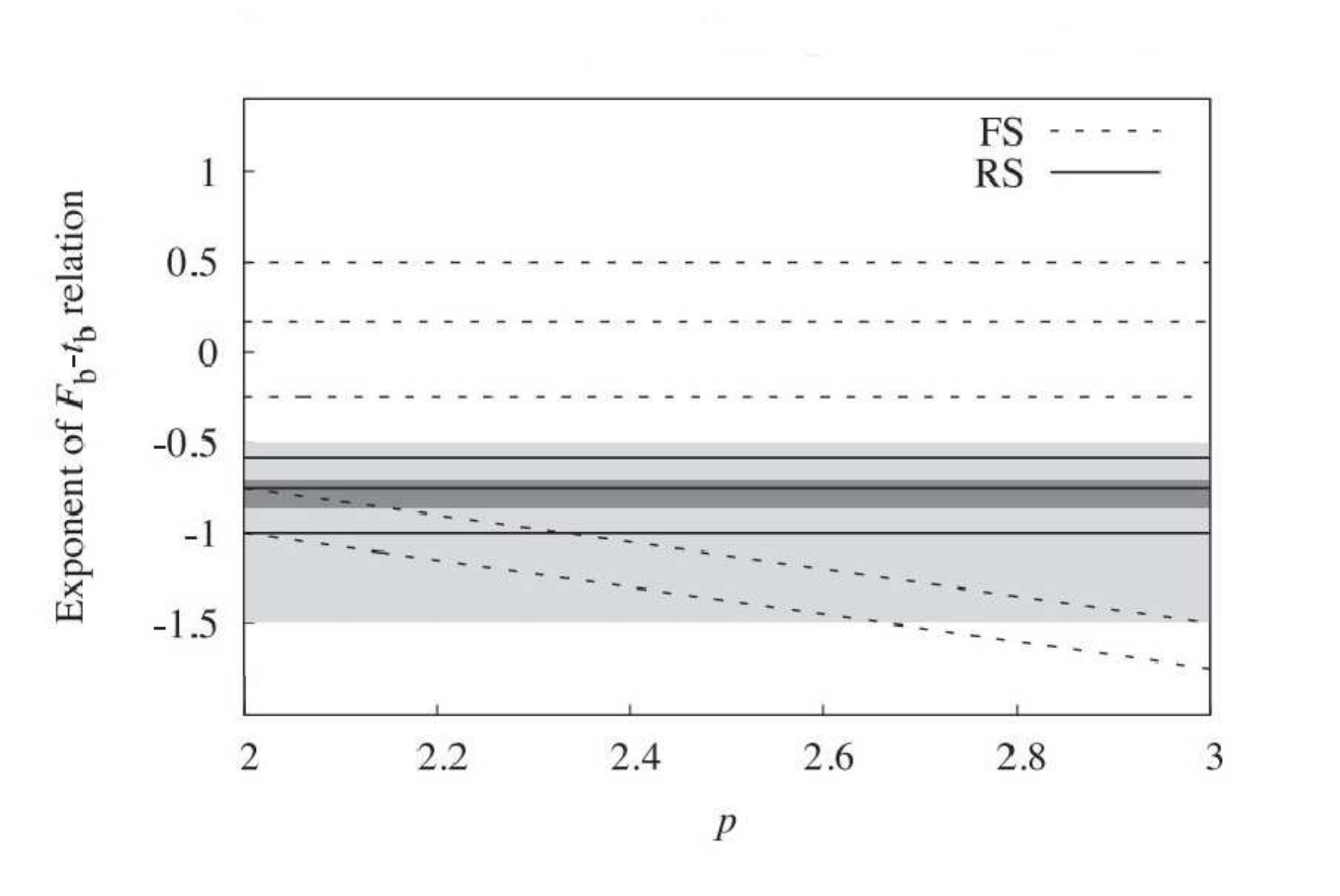}
\caption{\footnotesize Left panel: optical and X-ray light curves before and after
the injection break from \cite{Leventis2014}. The contributions
of the FS (dotted line) and RS (dashed line) are shown for both. The considered parameters are 
$E = 10^{51}$ erg, $n_1=50$ cm$^{-3}$, $\Delta t=5 \times 10^3$ s, $\eta=600$,
q=0, $\epsilon_e=\epsilon_B = 0.1$, p=2.3, $\theta_j=90^o$, $d=10^{28}$ cm and $z=0.56$. 
Right panel: index of the $F_{X,a}-T_{X,a}$ relation as a function of the electron distribution index, p, for the FS
and the RS from \cite{Leventis2014}. The lightly shaded region includes values allowed
by the scaling from \cite{Panaitescu2011}, while the darker region
indicates the scaling from \cite{Li2012}. The five dashed lines show 
the five possible indices for the FS, while the three solid lines display the three possible 
(independent of p) indices for the RS.}
\label{fig:leventis14}
\end{figure}

Considering alternative models explaining the LT relation, \cite{sultana2013} studied the evolution of 
the Lorentz gamma factor, $\Gamma=1/ \sqrt{1-v^2/c^2}$ (where $v$ is the relative velocity between the inertial reference 
frames and $c$ is the light speed), during the whole duration of the light curves within the context of the Supercritical Pile Model. 
This model provides an explanation for both the steep-decline and the plateau or the steep-decline and the power law decay phase 
of the GRB afterglow, as observed in a large number of light curves, and for the 
LT relation. One of their most important results is that the duration of the 
plateau in the evolution of $\Gamma$ becomes shorter with a decreasing value of $M_0c^2$, where $M_0$ is the initial rest mass of 
the flow. This occurrence means that the more luminous the plateau, the shorter its duration and the 
smaller the $M_0c^2$, namely the energy.

\begin{figure}[htbp]
\centering 
\includegraphics[width=0.495\hsize,angle=0]{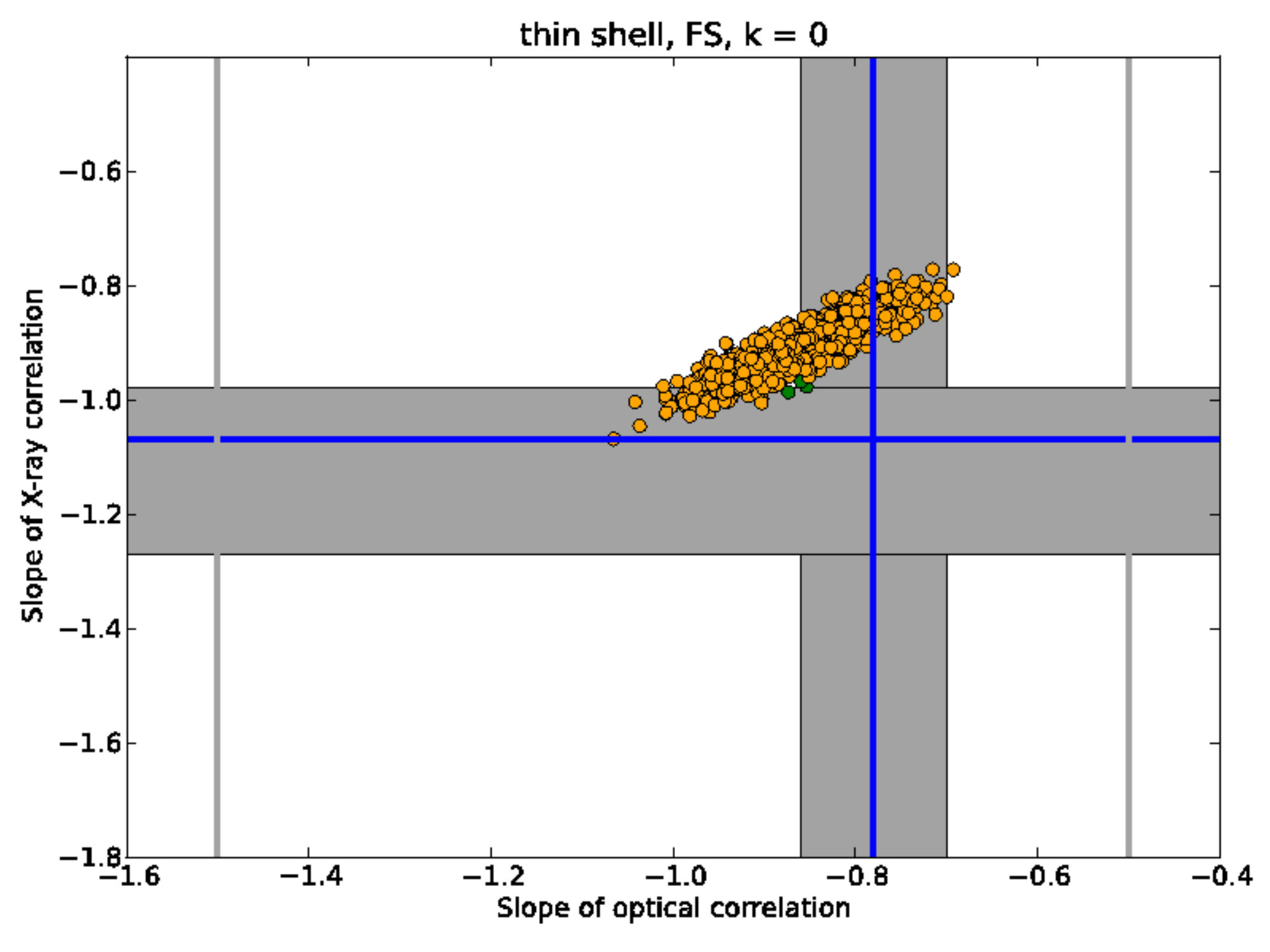}
\includegraphics[width=0.495\hsize,angle=0]{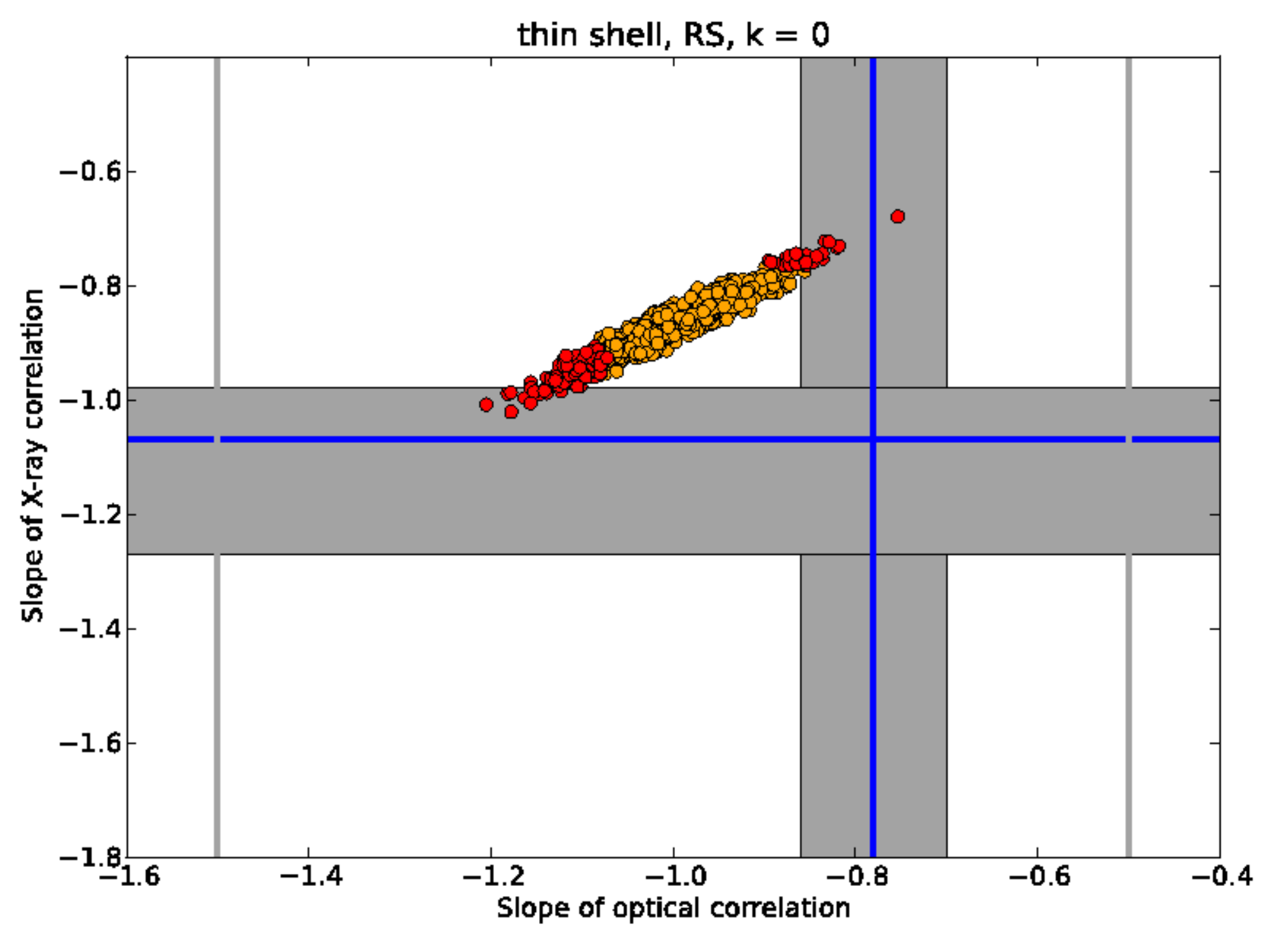}
\caption{\footnotesize Comparison of the slopes for 1000 thin shell data set runs and slopes of the observed LT relation in optical 
(horizontal direction) and the LT relation in X-ray (vertical direction) from \cite{vaneerten14b}
for the FS (left panel) and the RS (right panel) cases. 
Grey band expresses 1 $\sigma$ errors on the relations, while green dots represent runs consistent at 1 $\sigma$ error bars
for both, orange dots are compatible at 3 $\sigma$, but not at 1 $\sigma$ and red dots pass neither test. 
Vertical grey lines show more scattered LT in optical error bars from \cite{Panaitescu2011}.}
\label{fig:vaneerten14b}
\end{figure}

Instead, in the context of the RS and FS emissions, \cite{Leventis2014}, investigating the synchrotron radiation 
in the thick shell scenario (i.e. when the RS is relativistic), found that this radiation
is compatible with the presence of the plateau phase, see the left panel of Fig. \ref{fig:leventis14}. 
In addition, analyzing the $\log F_{X,a}$\,-\,$\log T_{X,a}$ 
relation in the framework of this model, they arrived at the conclusion that smooth energy injection 
through the RS is favoured respect to the FS, see the right panel of Fig. \ref{fig:leventis14}.\\
\cite{vaneerten14b}, with a simulated sample of GRBs, found out that the observed LT relation 
rules out basic thin shell models, but not basic thick ones. In fact, in the thick model, the
plateau phase comes from the late central
source activity or from additional kinetic energy transfer from slower ejecta
which catches up with the blast wave. As a drawback, in this
context, it is difficult to distinguish between FS and RS
emissions, or homogeneous and stellar wind-type environments.\\
In the thin shell case, the plateau phase is given by the pre-deceleration emission from a
slower component in a two-component or jet-type model, but this scenario is not in agreement with the observed LT 
relation, see Fig. \ref{fig:vaneerten14b}. 
This, however, does not imply that acceptable fits using a thin shell model are not possible, but 
further analysis is needed to exclude without any doubts thin shell models.
\textcolor{red}{Another model which has not been tested yet on this correlation is the photospheric emission model 
from stratified jets \citep{Ito2014}.}
\pagebreak
\subsection{The unified \texorpdfstring{$L_X(T_a)$-$T_{X,a}^*$ and $L_{O,a}$\,-\,$T^*_{O,a}$}{Lg} relations}\label{ghisellini}
In order to describe the unified picture of the X-ray and optical afterglow, it is necessary to introduce relevant 
features regarding optical luminosities. To this end, \cite{boer2000} studied 
the afterglow decay index in 8 GRBs in both X-ray and optical wavelengths. In the X-ray, the brightest GRBs had decay
indices around $1.6$ and the dimmest GRBs had decay indices around $1.11$. 
Instead, they didn't observe this trend for the optical light curves, 
probably due to the host galaxy absorption.\\
Later, \cite{nardini06a} discovered that the 
monochromatic optical luminosities at 12 hours, $L_{O,12}$, of 24 LGRBs cluster at 
$\log L_{O,12} = 30.65$ erg s$^{-1}$ Hz$^{-1}$, 
with $\sigma_{int} = 0.28$. The distribution of $L_{O,12}$ is less scattered
than the one of $L_{X,12}$, the luminosity at 12 hours in the X-ray, and the one of the 
ratio $L_{O,12}/E_{\gamma,prompt}$, where $E_{\gamma,prompt}$ is the rest frame isotropic prompt 
energy. Three bursts are outliers because they have luminosity which is smaller by a
factor $\sim 15$. This result suggests the existence of a family of intrinsically optically underluminous dark 
GRBs, namely GRBs where the optical-to-X-ray
spectral index, $\beta_{OX,a}$, is shallower than the X-ray spectral index minus 0.5, $\beta_{X,a}-0.5$.\\ 
\cite{Liang2006} confirmed these results. 
They found a bimodal distribution of $L_{O,1\rm{d}}$ using 44 GRBs. 
\cite{nardini08} also confirmed these findings. They
analyzed selection effects present in their observations extending 
the sample to 55 LGRBs
with known $z$ and rest-frame optical extinction detected by the Swift satellite.\\
In contrast, \cite{melandri2008}, \cite{oates2009}, \cite{zaninoni13} and \cite{melandri14} found no bimodality in 
the distributions of $L_{O,12}$, $L_{O,1\rm{d}}$ and $L_{O,11}$, investigating samples 
of 44, 24, 40 and 47 GRBs respectively.\\
Instead, with the aim of finding a unifying representation of the GRB afterglow phase, 
\cite{ghisellini09} fitted the light curves assuming this functional form:

\begin{equation}
 L_L(\nu,t)=L_L(\nu,T_{X,a})\frac{(t/T_{X,t})^{-\alpha_{\nu,fl}}}{1+(t/T_{X,t})^{\alpha_{\nu,st}-\alpha_{\nu,fl}}}.
\end{equation}

\begin{figure}[htbp]
\centering
\includegraphics[width=0.3\hsize,angle=0]{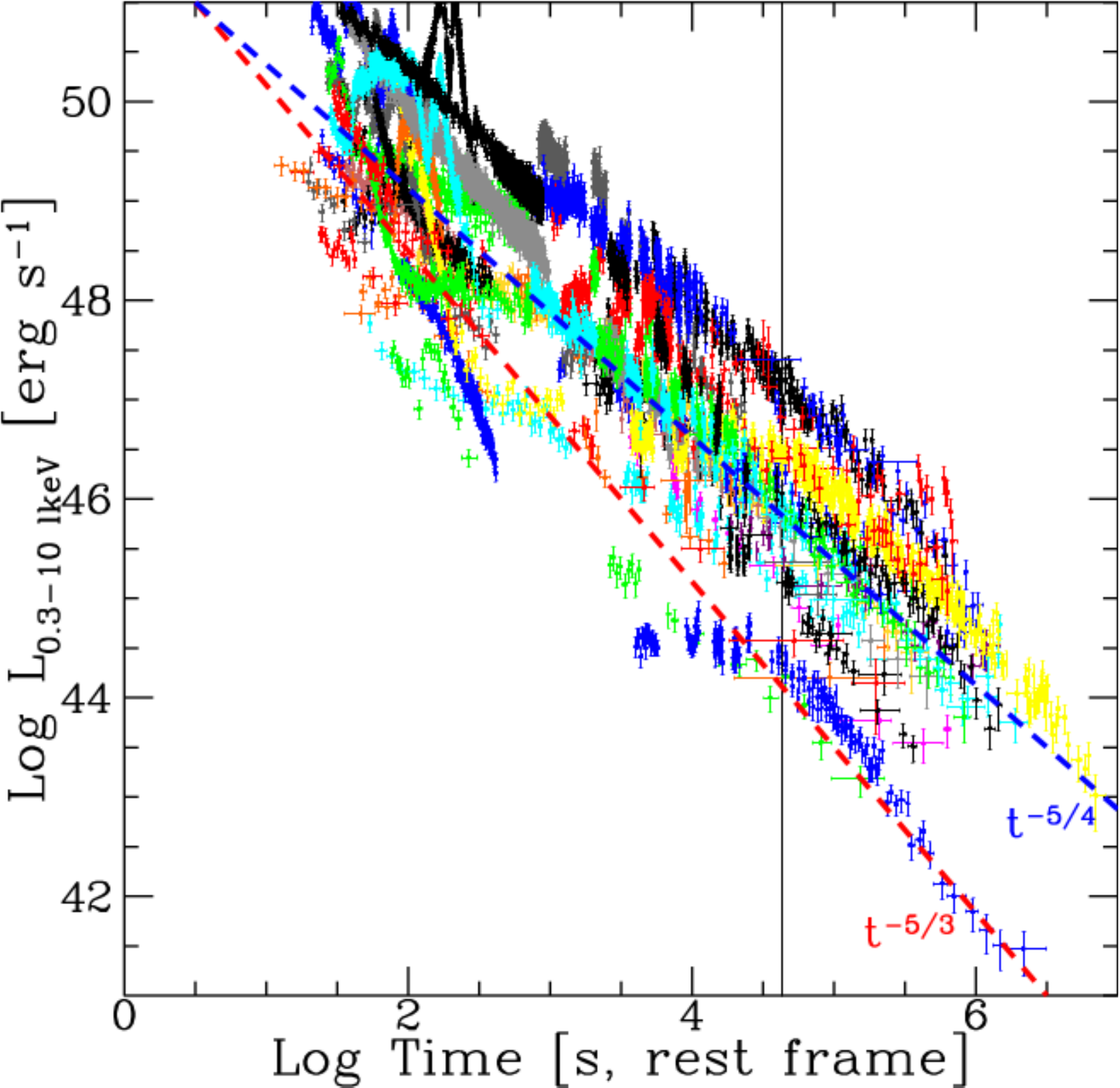}  
\includegraphics[width=0.3\hsize,angle=0]{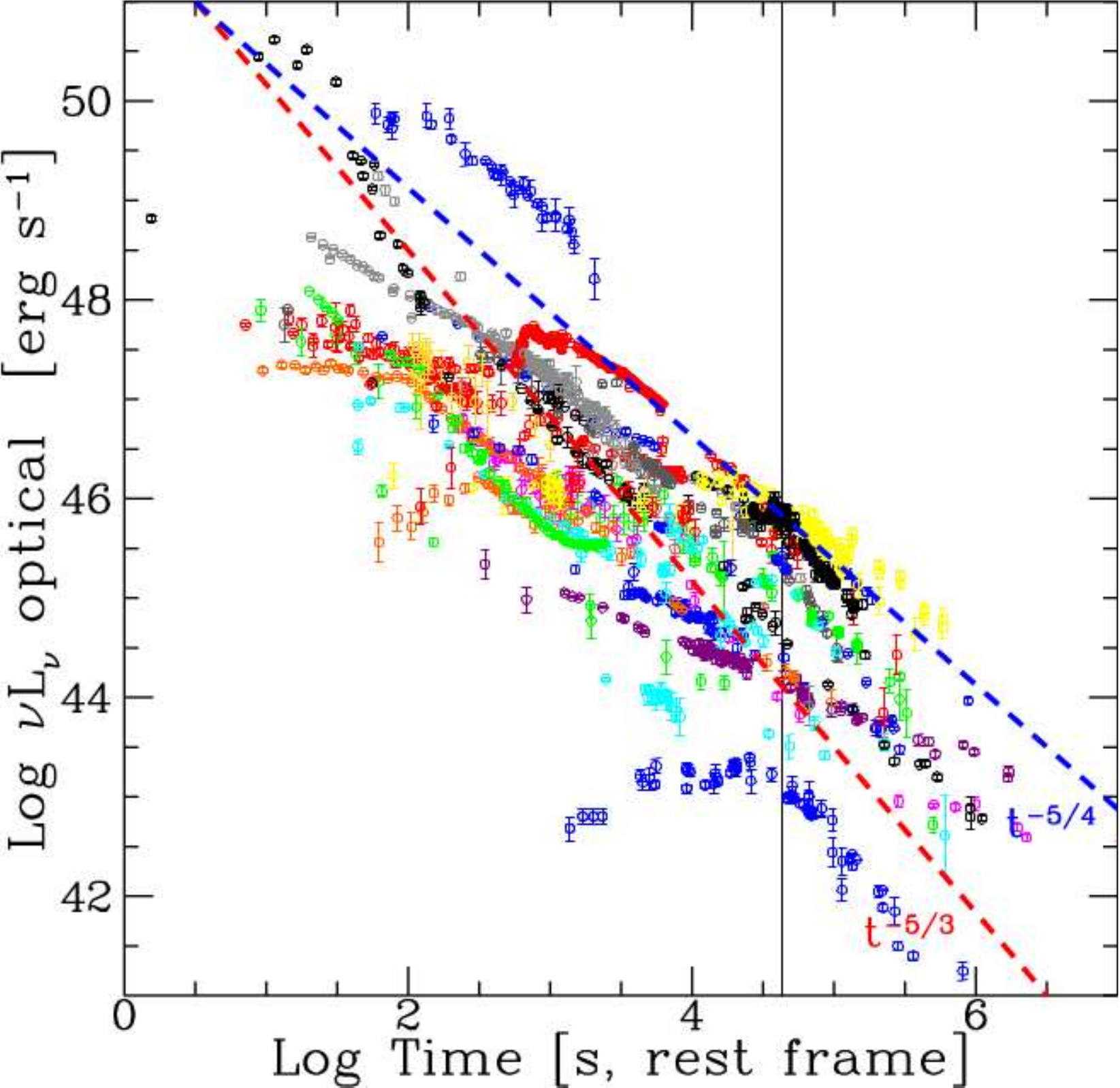}   
\includegraphics[width=0.385\hsize,angle=0]{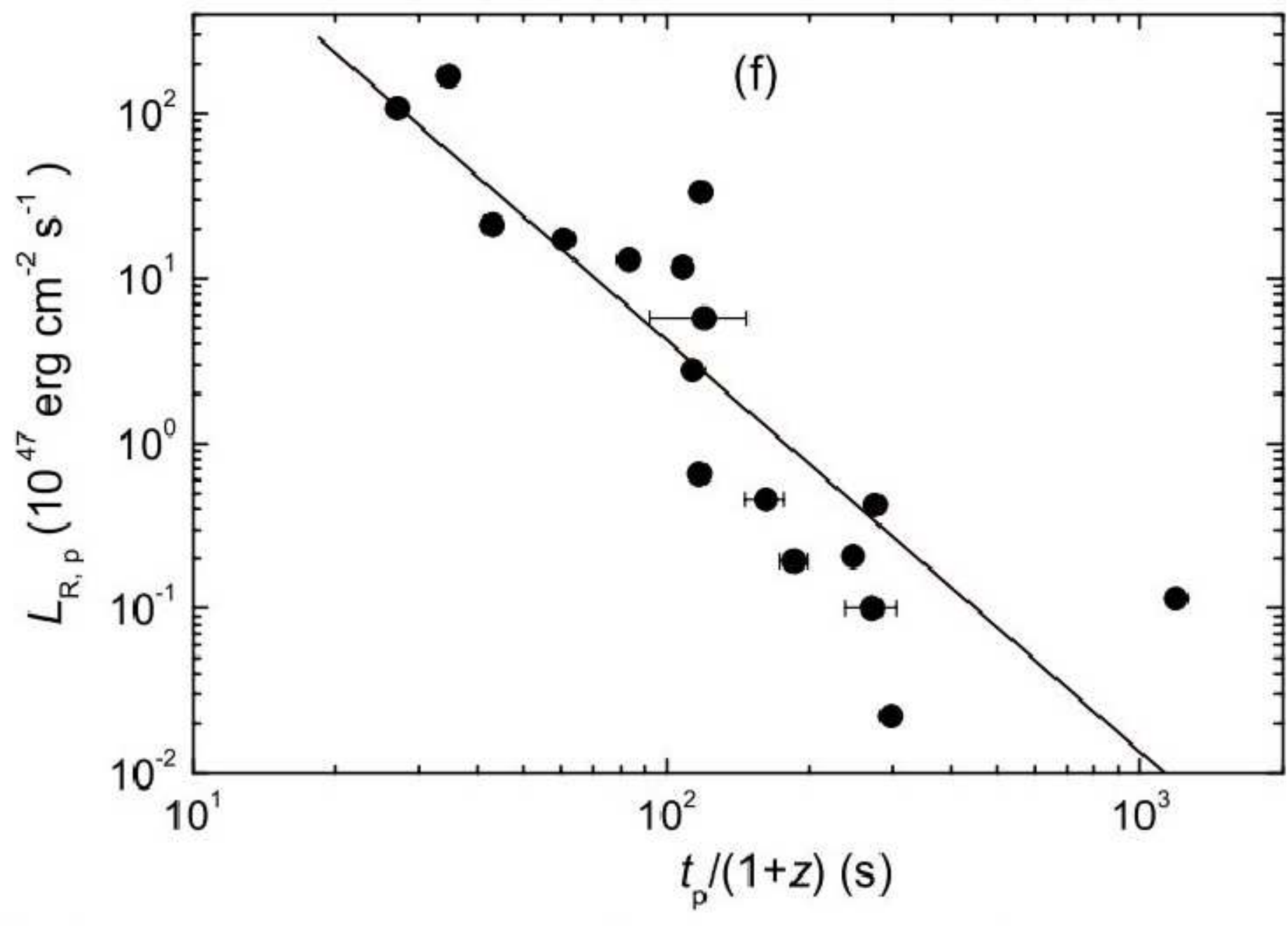}   
\caption{\footnotesize The light curves of the full sample from \cite{ghisellini09} in the X-rays (left
panel) and optical (middle panel). The vertical lines represent $\log L_{X,12}$ and $\log L_{O,12}$ in the rest frame time
respectively. Instead, the dashed lines indicate the $\log t^{-5/4}$ (blue) and the $\log t^{-5/3}$ (red) behaviours.
Right panel: relation between $ L_{O,peak}$ (equivalent to $L_{R,p}$ in the picture) and $T^*_{O,peak}$ of the data set from \cite{Liang2010}. Line represents the best fit.}
  \label{fig:ghisellini}
\end{figure}

They used a data sample of 33 LGRBs detected by Swift in X-ray (0.3-10 keV) and optical R 
bands (see the left and middle panels of Fig. \ref{fig:ghisellini}).
Within this approximation, the agreement with data is reasonably good, and they confirmed the X-ray 
LT relation.

Through their analysis using a data sample of 32 Swift GRBs, \cite{Liang2010} found that the
optical peak luminosity, $L_{O,peak}$, in the R band 
in units of $10^{47}$ erg s$^{-1}$ and the optical peak time, $T^*_{O,peak}$, are anti-correlated, see 
the right panel of Fig. \ref{fig:ghisellini}, with a slope $b=-2.49\pm 0.39$ and $\rho=-0.90$.
They deduced that a fainter bump has its maximum later than brighter ones and it also presents a 
longer duration.\\
\cite{Panaitescu2011} showed a similar relation to the one presented in \cite{Liang2010}. 
They found a $\log F_{O,a} \sim \log T^{-1}_{O,a}$ anti-relation using 37 Swift GRBs.
This result may indicate a unique mechanism for the optical 
afterglow even though the optical energy has a quite large scatter.

\begin{figure}[htbp]
\centering
\includegraphics[width=5.35cm,angle=0]{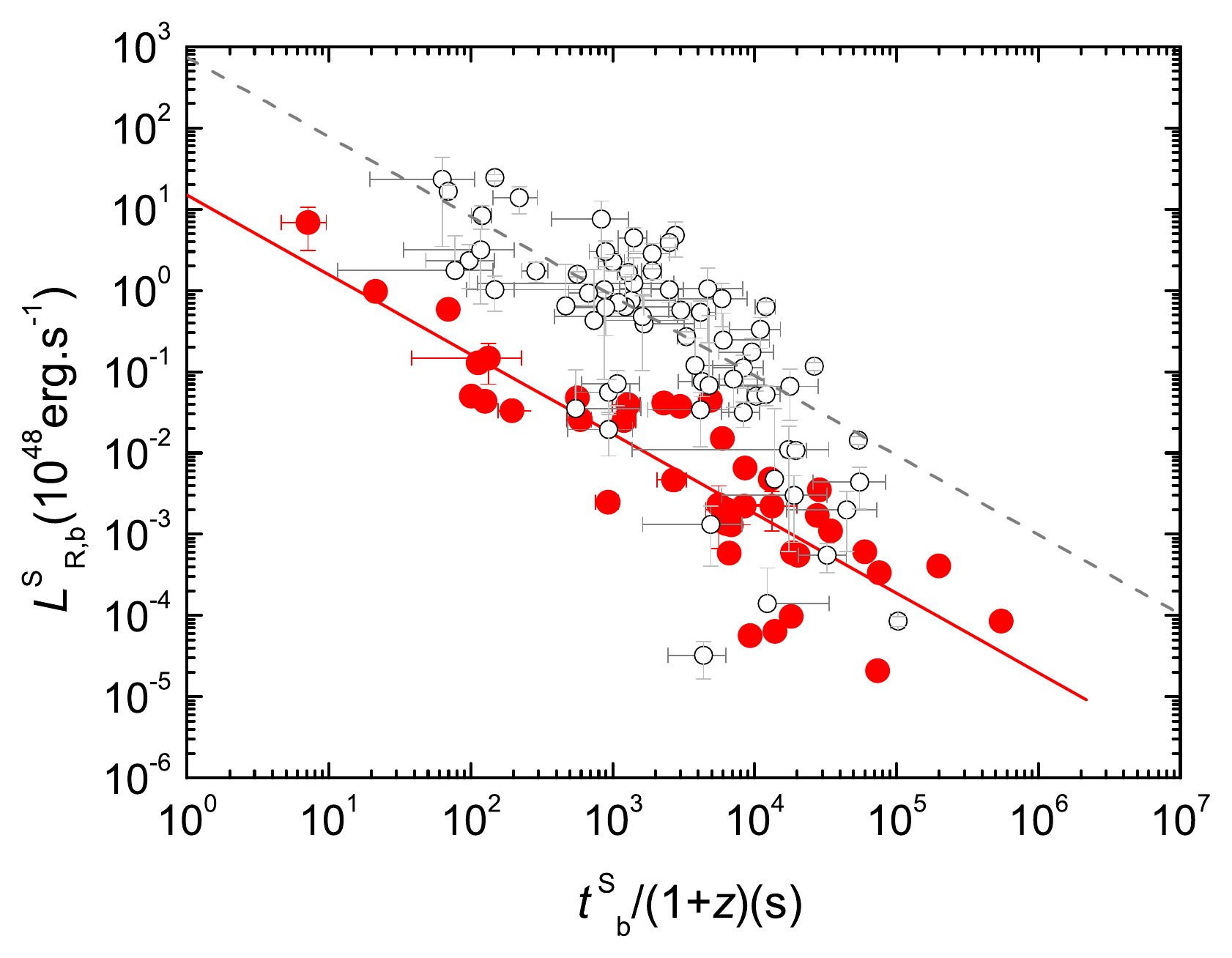}
\includegraphics[width=5.4cm,angle=0]{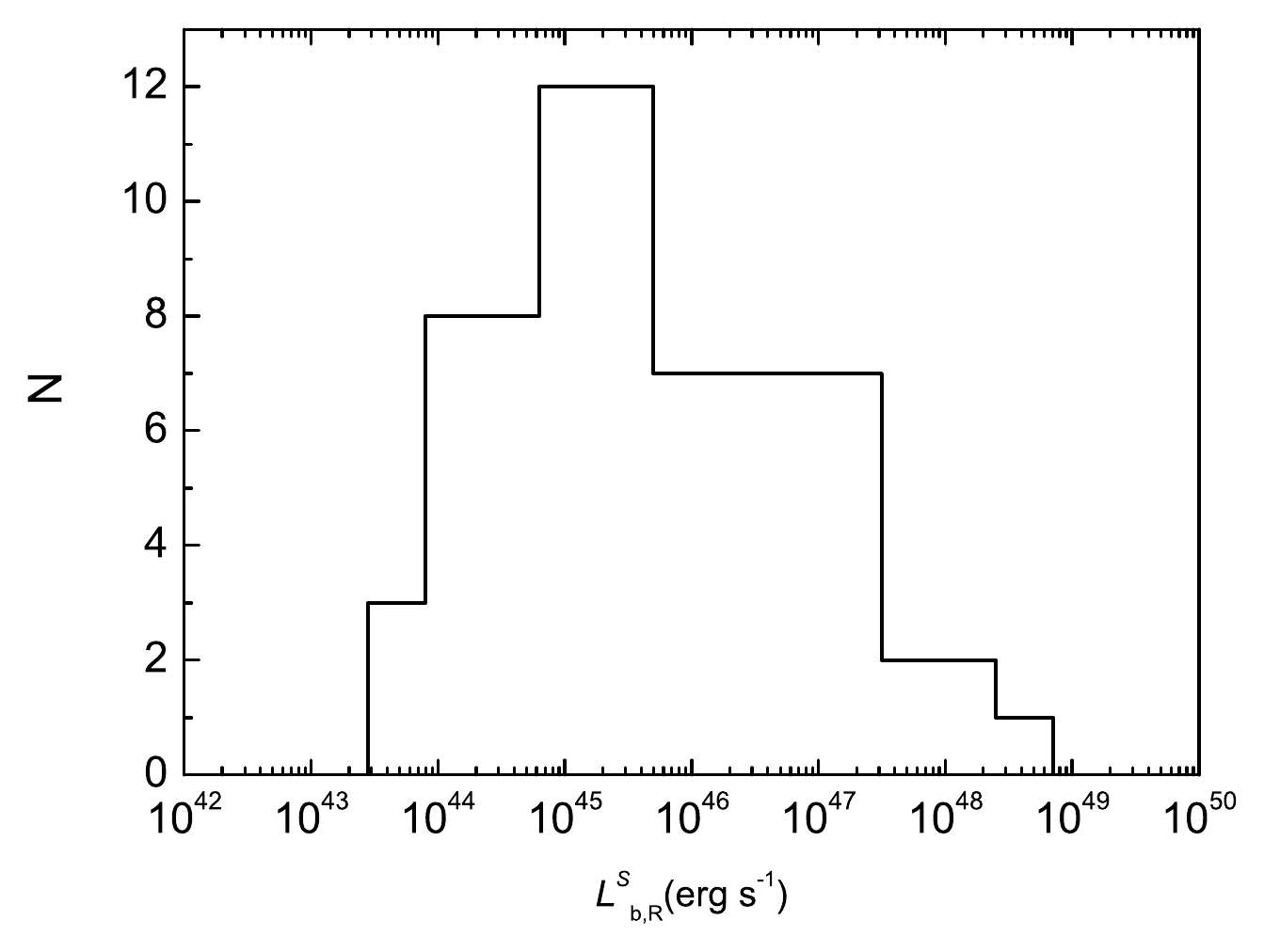}
\includegraphics[width=5.25cm,angle=0]{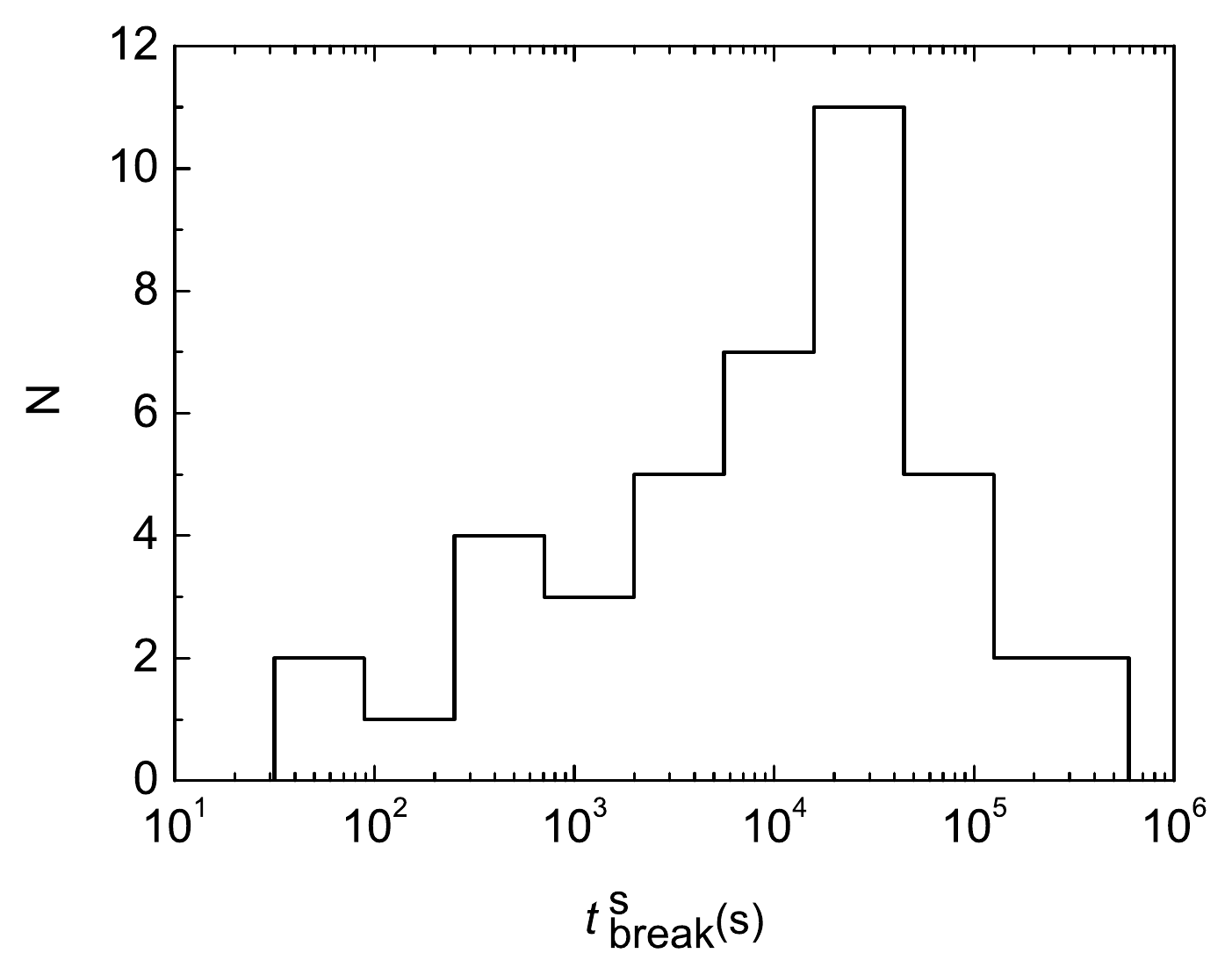}
\caption{\footnotesize Left panel: $ L^{S}_{O,a}$ (equivalent to $L^S_{R,p}$ in the picture) as a 
function of $T^{S,*}_{O,a}$
(equivalent to $t_{b}$ in the picture) from \cite{Li2012}. The grey circles represent 
the X-ray data from \cite{Dainotti2010}. Lines correspond to the best fit lines. Middle and Right panels: 
$L^{S}_{O,a}$ and $T^{S}_{O,a}$ distributions for the full GRB data set from \cite{Li2012}.}
  \label{fig:li2012a}
\end{figure}

Afterwards, \cite{Li2012} found a relation (see the left panel of Fig. \ref{fig:li2012a})
similar to the LT relation, but in the R band. They used 39 GRBs with optical data available in the literature. 
This relation is between the optical luminosity at the end of the plateau, $L_{O,a}$, in units of 
$10^{48}$ erg s$^{-1}$ and the optical end of the plateau time,
$\log T^*_{O,a}$, in the shallow decay phase of the GRB light 
curves, denoted with the index S. They found a slope $b=-0.78\pm 0.08$, $\rho=0.86$ and $P<10^{-4}$. 

\begin{table}[htbp]
\footnotesize
\begin{center}
\begin{tabular}{|c|c|c|c|c|c|}
\hline
Correlations & Author & N& Slope& Corr.coeff.& P \\
\hline
$L_{O,peak}-T_{O,peak}$ & Liang et al. (2010) &32 &$-2.49\pm 0.39$&-0.90& \\
$L_{O,a}-T_{O,a}$&Panaitescu \& Vestrand (2011)&37&$-1$&&\\
$L^{S}_{O,a}$-$T^{S}_{O,a}$&Li et al. (2012)&39&$-0.78\pm 0.08$&$0.86$&$<10^{-4}$ \\
\hline
\end{tabular}
\caption{\footnotesize Summary of the relations in this section. The first column represents the relation in 
log scale,
the second one the authors, and the third one the number of GRBs in the used sample. Afterwards, the fourth column is
the slope of the relation and the last two columns are the correlation coefficient and the chance 
probability, P.}
\label{tblopt}
\end{center}
\end{table}

$L^{S}_{O,a}$ varies from $10^{43}$ to $10^{47}$ erg s$^{-1}$,
and in some GRBs with an early break reaches $\sim 10^{49}$ erg s$^{-1}$, see the middle panel of Fig. 
\ref{fig:li2012a}. $T^S_{O,a}$ spans from tens of seconds to several days after the GRB trigger, 
with a typical shallow peak
time $T^{S}_{O,a}$ of $\sim 10^4$ seconds, see the right panel of Fig. \ref{fig:li2012a}. 
By plotting $L_{O,a}$ in units of $10^{48}$ erg s$^{-1}$ as a function of 
$T^*_{O,a}$ in the burst frame, they observed that optical data have a similar trend to the X-ray data. 
In fact, this power law relation, presented in the left panel of Fig. \ref{fig:li2012a},
with an index of $-0.78\pm 0.08$ is similar to that derived for the X-ray
flares (see sec. \ref{lisotp}). XRF phenomena are described in sec. 1. As a consequence, 
they recovered the LT relation. In Table \ref{tblopt} a summary of the relations described in this section is 
displayed.

\subsubsection{Physical interpretation of the unified \texorpdfstring{$L_X(T_a)$-$T_{X,a}^*$ and $L_{O,a}$\,-\,$T^*_{O,a}$}{Lg} relations}
In the unified $L_X(T_a)$-$T_{X,a}^*$ and $L_{O,a}$\,-\,$T^*_{O,a}$ relations 
\cite{ghisellini09} considered the flux as the sum of synchrotron radiation caused by
the standard FS due to the fireball impacting the
circumburst medium and of another component may be produced by a long-lived central engine, which resembles mechanisms
attributed to a ``late prompt". 
Even if based in part on a phenomenological model, \cite{ghisellini09} explained situations in which achromatic and 
chromatic
jet break are either present or not in the observed light curves.\\
In addition, from their analysis, the decay slope of the late prompt emission results to be
$\alpha_{X,a}=-5/4$ (see blue dashed line for X-ray and optical emission
in the left and middle panels of Fig. \ref{fig:ghisellini} respectively), really close within the errors 
to the value of the temporal accretion rate
of fall-back material (i.e. $\sim \log t^{-5/3}$,
see red dashed line for X-ray and for optical emission in the left and middle panels of 
Fig. \ref{fig:ghisellini} respectively). This explains the activity of the central engine for such a long duration.
For a similar interpretation within the context of the accretion onto the BH related to LT relation 
see sec. \ref{Dainotti2008interpretation}.\\
\cite{Liang2010} claimed that the external shock model explains well the anti-relation between
$L_{O,peak}$ and $T_{O,peak}$, because later deceleration time is equivalent to slower ejecta and thus to a less 
luminous emission.\\
Furthermore, \cite{panaitescu08} from the analysis of the $\log L_{O,a}-\log T^*_{O,a}$ relation explained the peaky 
afterglows (those with $L_{O,a}\propto T^{-1}_{O,a}$) as being a bit outside the cone of view, while the plateau as 
off-axis events and due to the angular structure of the jet. 
Later, \cite{Panaitescu2011} asserted that 
the double-jet structure of the ejecta is problematic. To overcome this issue, they suggested a
model in which both the peaky and plateau afterglows depend on how much time the central engine allows for the energy
injection.
More specifically, impulsive ejecta with a narrow range of $\Gamma$ are responsible for the peaky afterglows, while the plateau 
afterglows are produced by a distribution of initial $\Gamma$ which keeps the energy injection till $10^5$ s.\\
Later, \cite{Li2012} pointed out that late
GRB central engine activities can account for both optical flares
and the optical shallow-decay segments. These activities
can be either erratic (for flares) or steady (for internal plateaus).
A normal decay follows the external plateaus with $\alpha_{X,a}$ typically around $-1$, thus possibly originated by an 
external shock with the shallow decay segment 
caused by continuous energy injection into the blast wave \citep{rees98,dai98,sari2000,zhang2001}. Instead, the internal 
plateaus, found first by \cite{troja07} in GRB 070110 and later studied statistically by \cite{liang2007}, are 
followed by a much steeper decay ($\alpha_{X,a}$ steeper than -3), which
needs to be powered by internal dissipation of a late outflow. In summary, the afterglow can be interpreted as a mix 
of internal and external components.

\subsection{The \texorpdfstring{$L_{O,200\rm{s}}$\,-\,$\alpha_{O,>200\rm{s}}$ relation and its physical interpretation}{Lg}} \label{Oates}
\begin{figure}[htbp]
\centering
\includegraphics[width=0.49\hsize,clip]{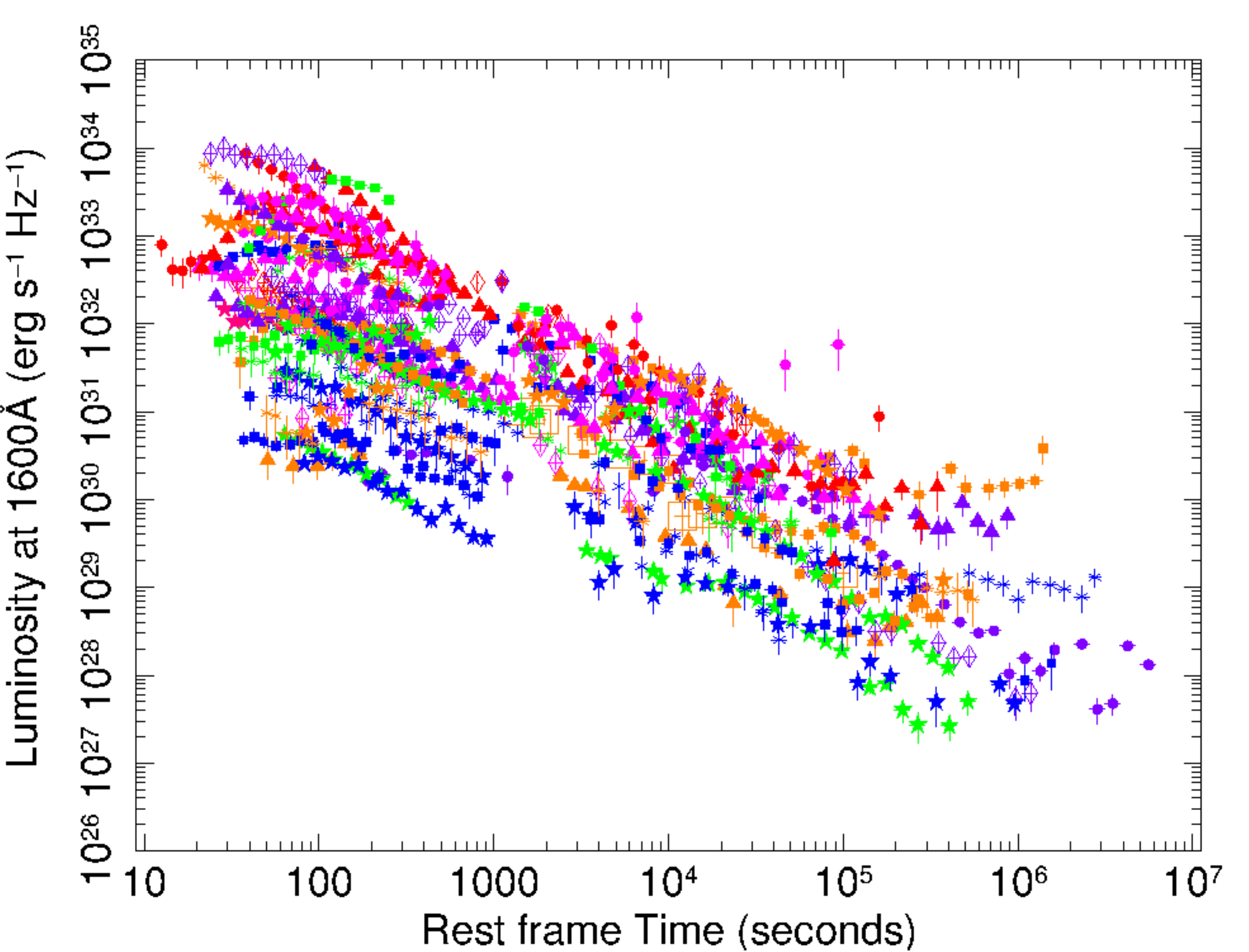}
  \includegraphics[width=0.49\hsize,angle=0,clip]{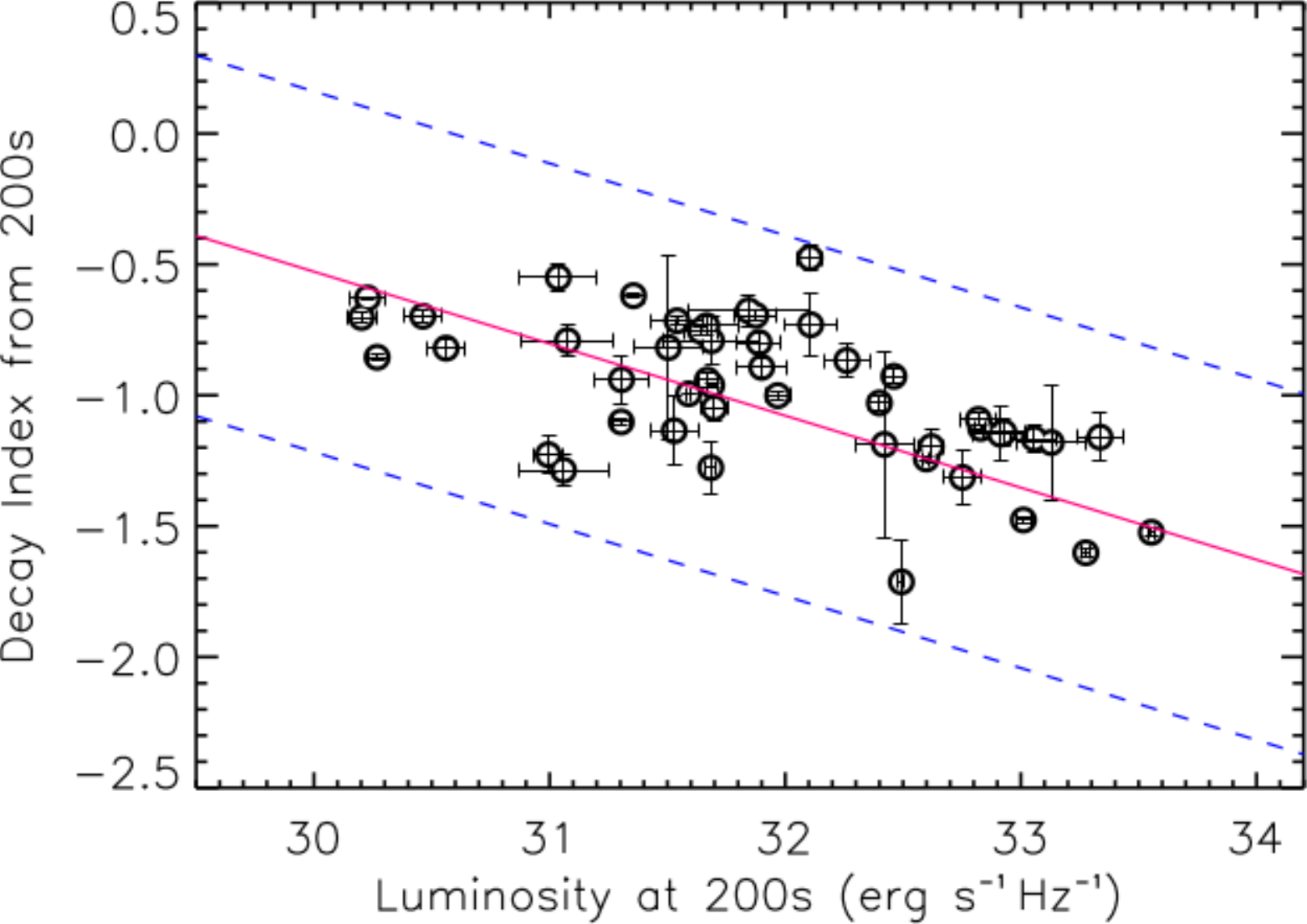}
    \caption{\footnotesize Left panel: ``optical light curves of 56 GRBs from \cite{oates2012}".
   Right panel: ``$\log L_{O,200\rm{s}}$ vs. $\alpha_{O,>200\rm{s}}$ from \cite{oates2012}. 
   The red solid line indicates the best fit line
   and the blue dashed lines show the 3 $\sigma$ variance".}
   \label{fig:14}
\end{figure}

\begin{figure}[htbp]
\centering
\includegraphics[width=0.495\hsize,angle=0,clip]{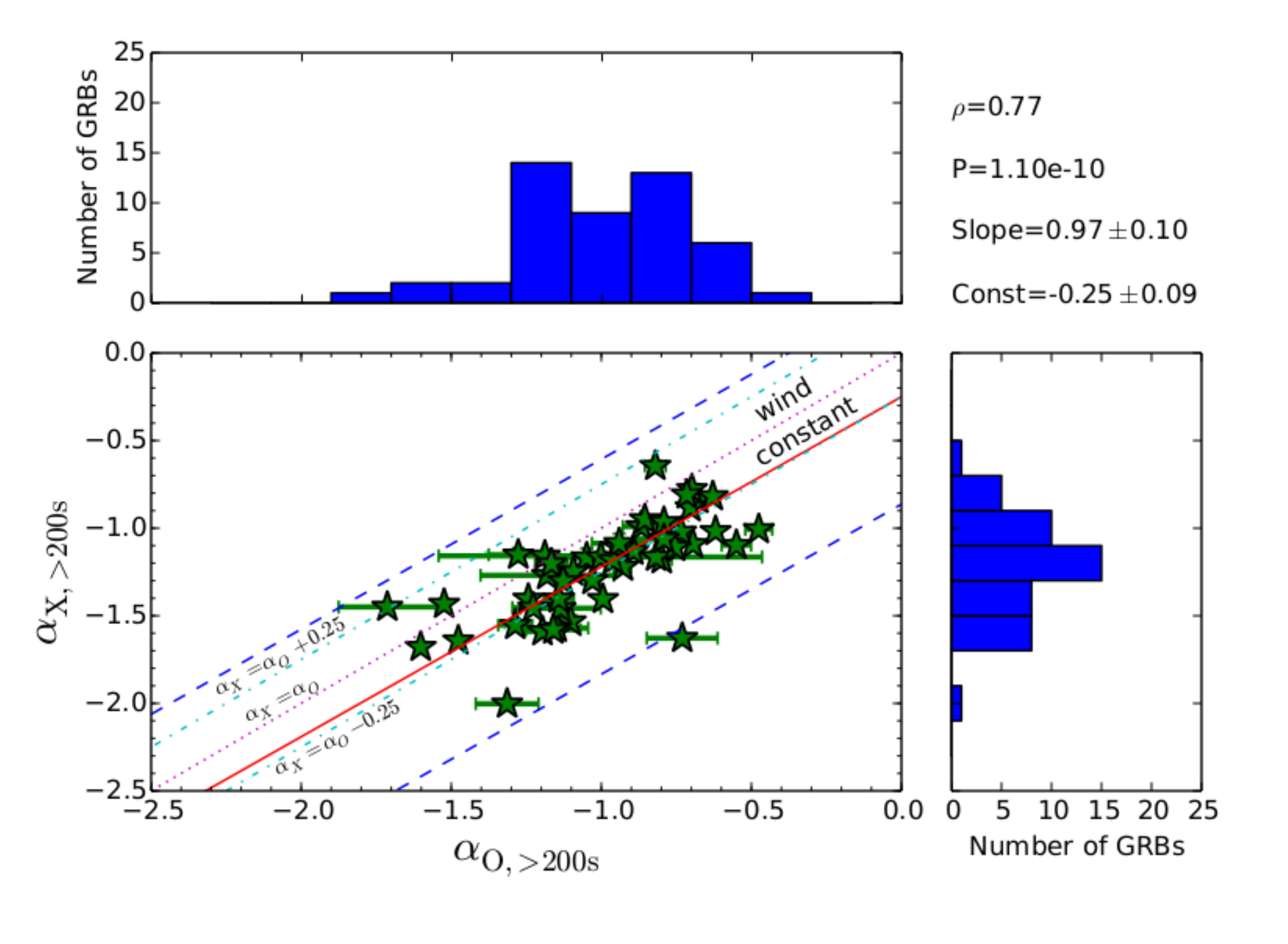}
\includegraphics[width=0.495\hsize,angle=0,clip]{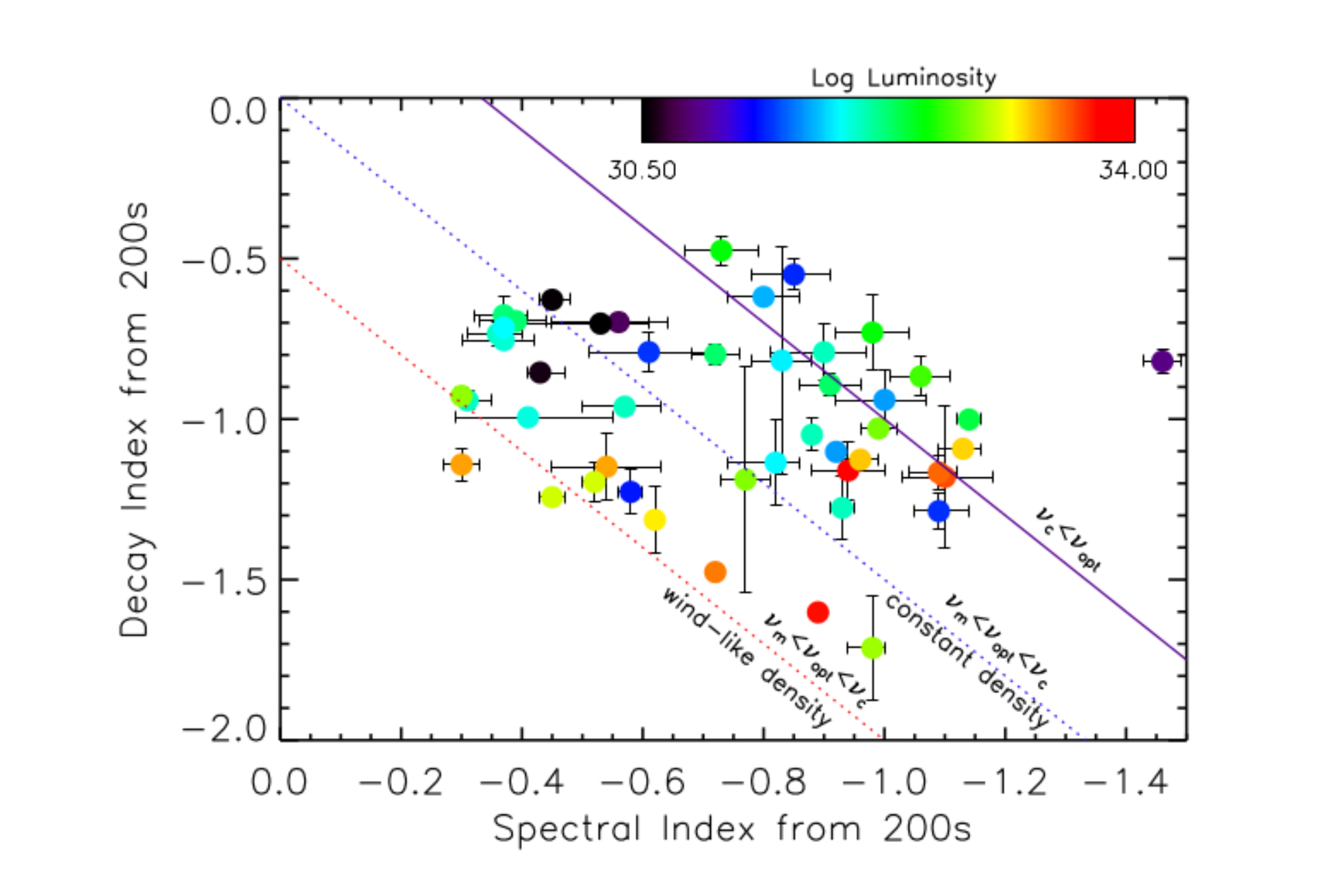}
\caption{\footnotesize Left panel: ``$\alpha_{O,>200\rm{s}}$ and $\alpha_{X,>200\rm{s}}$ 
from \cite{oates15}. 
The red solid line represents the best fit regression and the blue dashed lines 
represent 3 times the root mean square (RMS) deviation. The relationships expected between the optical/UV 
X-ray light curves from the GRB closure relations are also shown. 
The pink dotted line represents $\alpha_{O,>200\rm{s}}=\alpha_{X,>200\rm{s}}$. The light blue dotted-dashed 
lines represent $\alpha_{X,>200\rm{s}}=\alpha_{O,>200\rm{s}}\pm0.25$. In the top right corner it is given
the coefficient $\rho$ with $P$, and it is  
provided the best fit slope and constant determined by linear regression". Right panel: ``$\alpha_{O,>200\rm{s}}$ and $\beta_{O,>200\rm{s}}$
for the sample of 48 LGRBs from \cite{oates2012}. The lines represent 3 closure relations and a colour scale is used to
display the range in $\log L_{O,200\rm{s}}$".}
   \label{fig:oates15}
\end{figure}

\cite{oates2012} discovered a relation between the optical luminosity at 200 s, 
$\log L_{O,200\rm{s}}$, and the optical temporal decay index after 200 s, $\alpha_{O,>200\rm{s}}$, 
see the right panel of Fig. \ref{fig:14}. They used a sample of 48 LGRB afterglow light curves at $1600$ \AA{}
detected by UVOT on board of the Swift satellite, see the left panel of Fig. \ref{fig:14}.
The best fit line for this relation is given by:

\begin{equation}
\log \, L_{O,200 \rm{s}} = (28.08 \pm 0.13)-(3.636 \pm 0.004) \times \alpha_{O,>200\rm{s}},
\end{equation}

with $\rho=-0.58$ and a significance of 99.998\% (4.2 $\sigma$). 
This relation means that the brightest GRBs decay faster than the dimmest ones. To obtain
the light curves employed for building the relation, they used the criteria 
from \cite{oates2009} in order to guarantee that the entire UVOT light curve is not noisy, 
namely with a high signal to noise (S/N) ratio\textcolor{red}{: the optical/UV light curves must 
be observed in the V filter of the UVOT with a magnitude $\le 17.8$, UVOT observations must
have begun within the first 400 s after the BAT trigger and the afterglow must have been observed until at 
least $10^5$ s after the trigger.}
Their results pointed out the dependence of this relation is on the differences in the observing angle and on
the rate of the energy release from the central engine.\\
As a further step, \cite{oates15}, using the same data set, investigated the same
relation both in optical and in X-ray wavelengths in order to make a comparison, and they confirmed previous 
optical results finding a similar slope for both relations.
In addition, they analyzed the connection between 
the temporal decay indices after $200$ s (in X-ray and optical) obtaining as best fit relation
$\alpha_{X,>200\rm{s}} = \alpha_{O,>200\rm{s}} - 0.25$, see the left panel of Fig. \ref{fig:oates15}. 
They yielded some similarities between optical and 
X-ray components of GRBs from these studies. \textcolor{red}{Their results were in disagreement
with those previously found by \cite{urata07}, who investigated the relation between the optical and X-ray temporal 
decay 
indices in the normal decay phase derived from the external shock model. 
In fact, a good fraction of outliers was found in this previous work.}\\
\cite{racusin16} studied a similar relation using 237 Swift LGRBs, 
but in X-ray. For this relation, it was found that slope \textcolor{red}{$b=-0.27$}$\pm0.04$ and solid evidence 
for a strong connection between optical and X-ray components of GRBs was discovered as well. 
In conclusion, the Monte Carlo simulations and the statistical tests validated 
the relation between $\log L_{O,200\rm{s}}$ and $\alpha_{O,>200\rm{s}}$
by \cite{oates2012}. In addition,
it shows a possible connection with its equivalent,
the LT relation in X-ray, implying a common physical mechanism. 
In Table \ref{tbloates} a summary of the relations described in this section is reported.

Regarding the physical interpretation of the $\log L_{O,200\rm{s}}$\,-\,$\alpha_{O,>200\rm{s}}$ relation, 
\cite{oates2012} explored several scenarios. The first one implies that the relation can be due to the interaction
of the jet
with the external medium. In a 
straightforward scenario $\alpha_{O,>200\rm{s}}$ is not a fixed value and all optical afterglows stem from
only one closure relation where $\alpha_{O,>200\rm{s}}$ and $\beta_{O,>200\rm{s}}$ are related linearly. 
Thus a relation between 
$\log L_{O,200\rm{s}}$ and $\beta_{O,>200\rm{s}}$ should naturally appear.
Contrary to this expectation, $\alpha_{O,>200\rm{s}}$ and $\beta_{O,>200\rm{s}}$ are poorly correlated, see the right 
panel of Fig. \ref{fig:oates15}, and there
is no evidence for the existence of a relation between $\beta_{O,>200\rm{s}}$ and $\log L_{O,200\rm{s}}$. Therefore, this scenario cannot be 
ascribed as the cause of the $\log L_{O,200\rm{s}}$\,-\,$\alpha_{O,>200\rm{s}}$ relation.

In the second scenario, they assumed that the $\log L_{O,200\rm{s}}$\,-\,$\alpha_{O,>200\rm{s}}$
relation is produced by few closure relations indicated by lines in the right panel of Fig. \ref{fig:oates15}. 
However, from this picture, the $\alpha_{O,>200\rm{s}}$ and $\beta_{O,>200\rm{s}}$ values with similar luminosities do not gather around a particular 
closure relation, thus also the basic standard model is not a good explanation of the
$\log L_{O,200\rm{s}}-\alpha_{O,>200\rm{s}}$ relation. 
As a conclusion, the afterglow model is more complex than it was 
considered in the past. It is highly likely that there are physical properties that control the emission mechanism and the decay 
rate in the afterglow that still need to be investigated.\\
Therefore, \cite{oates2012} proposed two additional alternatives.
The first is related to some properties of the central engine
influencing the rate of energy release so that for fainter
afterglows, the energy is released more slowly. Otherwise, the relation can be due to different observing angles where 
observers at smaller viewing angles see brighter and faster decaying light curves. 

\begin{table}[htbp]
\footnotesize
\begin{center}
\begin{tabular}{|c|c|c|c|c|c|c|c|}
\hline
Correlations & Author & N& Slope& Norm & Corr.coeff.& P \\
\hline
$L_{O,200\rm{s}}$\,-\,$\alpha_{O,>200\rm{s}}$ & Oates et al. (2012) &48 &$-3.636 \pm 0.004$&$28.08 \pm 0.13$&$-0.58$&$2\times10^{-4}$ \\
&Oates et al. (2015)&48 &$-3.636 \pm 0.004$&$28.08 \pm 0.13$&$-0.58$&$2\times10^{-4}$ \\
$L_{X,200\rm{s}}$\,-\,$\alpha_{X,>200\rm{s}}$&Racusin et al. (2016)&237&\textcolor{red}{$-0.27$}$\pm0.04$&$-6.99\pm1.11$&$0.59$&$10^{-6}$\\
\hline
\end{tabular}
\caption{\footnotesize Summary of the relations in this section. The first column represents the relation in 
log scale,
the second one the authors, and the third one the number of GRBs in the used sample. Afterwards, the fourth and 
fifth columns are
the slope and normalization of the relation and the last two columns are the correlation coefficient and the chance 
probability, P.}
\label{tbloates}
\end{center}
\end{table}

As pointed out by \cite{Dainotti2013a}, the $\log L_{O,200\rm{s}}$\,-\,$\alpha_{O,>200\rm{s}}$ 
relation is related to the LT one since both show an anti-relation between luminosity and decay rate of the
light curve or time. The key point would be to understand how they relate to each other and the possible 
common physics that eventually drives both of them.
To this end, \cite{oates15} compared the observed relations with the ones obtained with the simulated sample. 
The luminosity-decay relationship in the optical/UV is in agreement with that in the X-ray, inferring a common 
mechanism.

\section{The Prompt-Afterglow Relations}\label{promptaftcor}
As we have discussed in the previous paragraphs, the nature of the plateau and the relations 
(e.g. the optical one) based on similar physics and directly related to the plateau are still under investigation.
For this reason, several models have been proposed. To further enhance its theoretical understanding, it is necessary 
to evaluate
the connection between plateaus and prompt phases. To this end, we hereby review the prompt-afterglow relations, 
thus helping to establish a more complete picture of the plateau GRB phenomenon.

\subsection{The \texorpdfstring{$E_{\gamma,afterglow}-E_{X,prompt}$ relation}{Lg} and its physical interpretation}
\begin{figure}[htbp]
\includegraphics[width=0.495\hsize,angle=0]{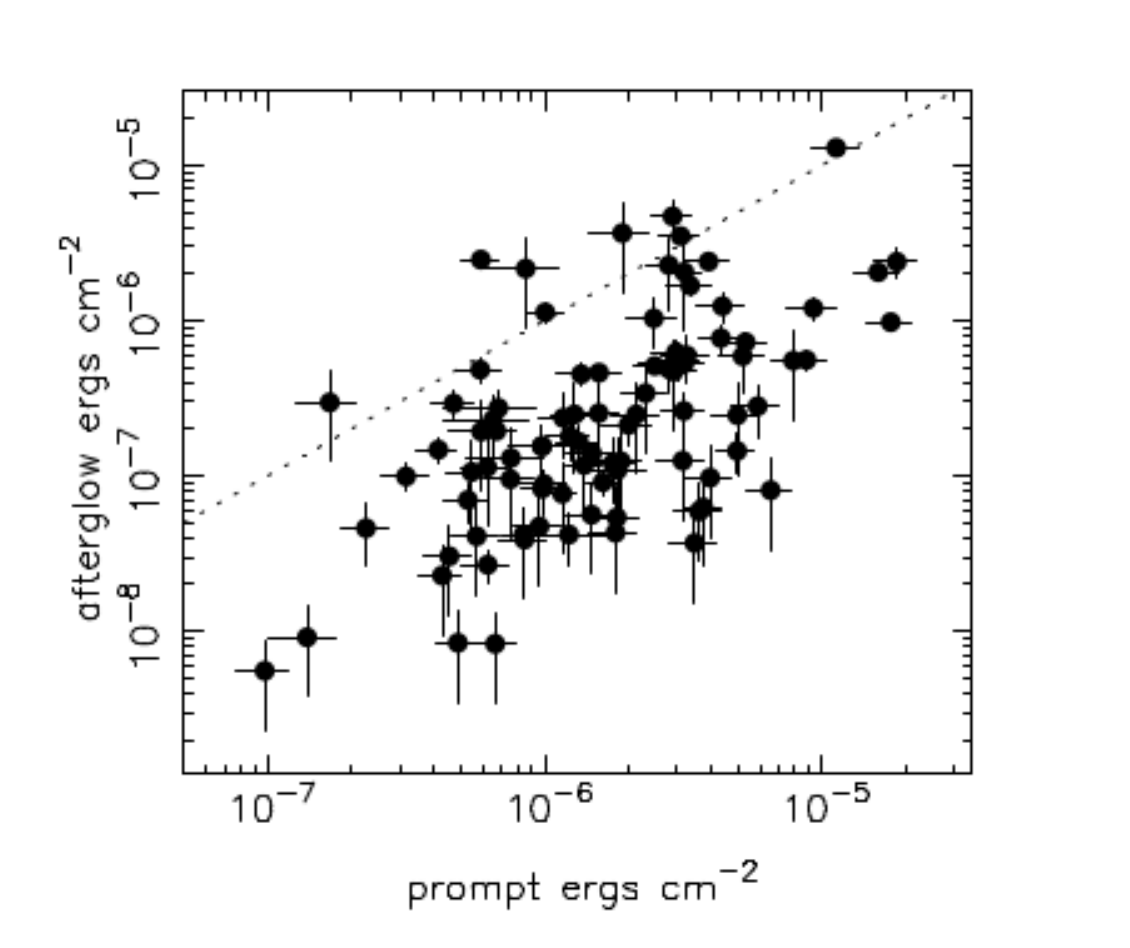}
\includegraphics[width=0.495\hsize,angle=0]{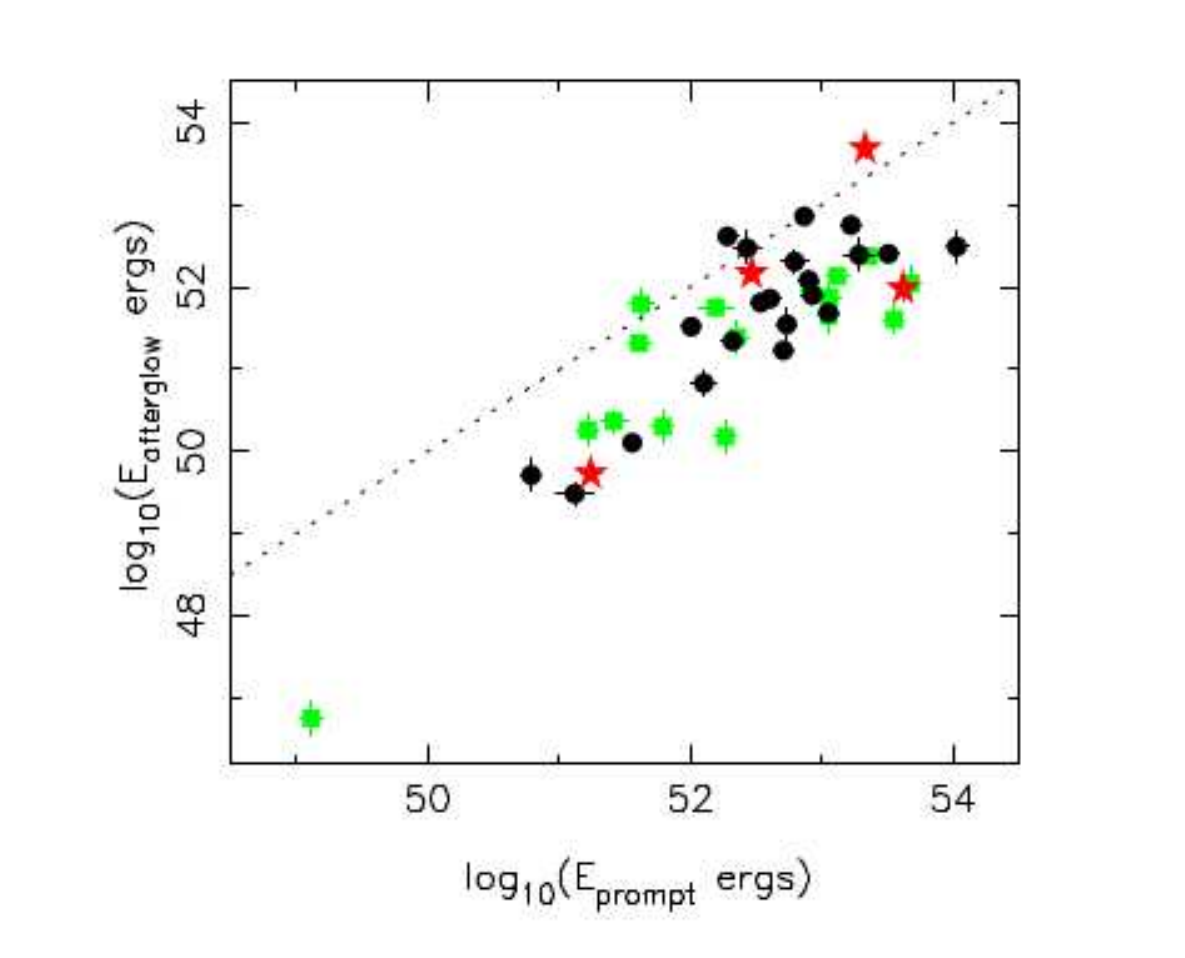}
\includegraphics[width=0.495\hsize,angle=0]{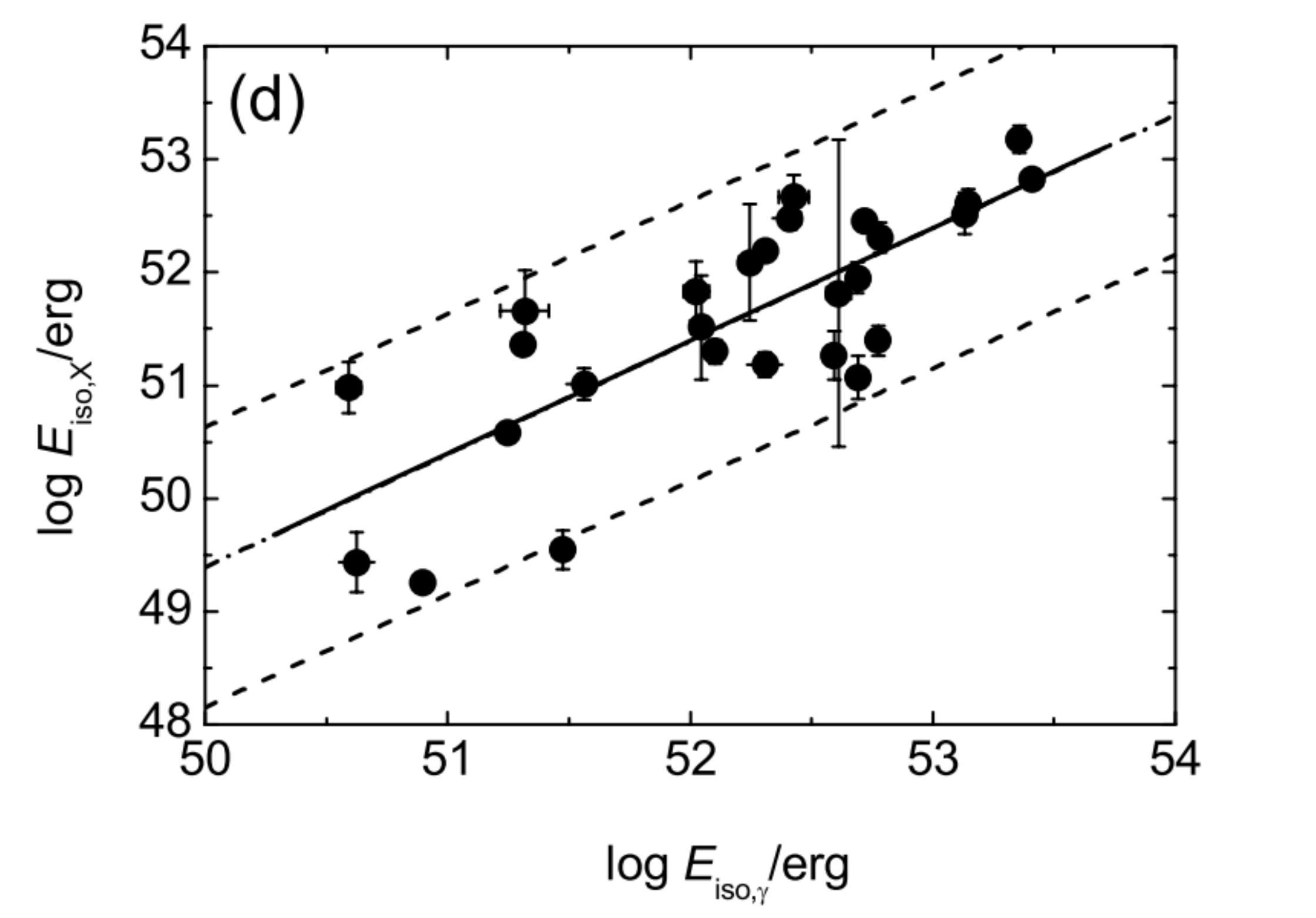}
\includegraphics[width=0.41\hsize,angle=0]{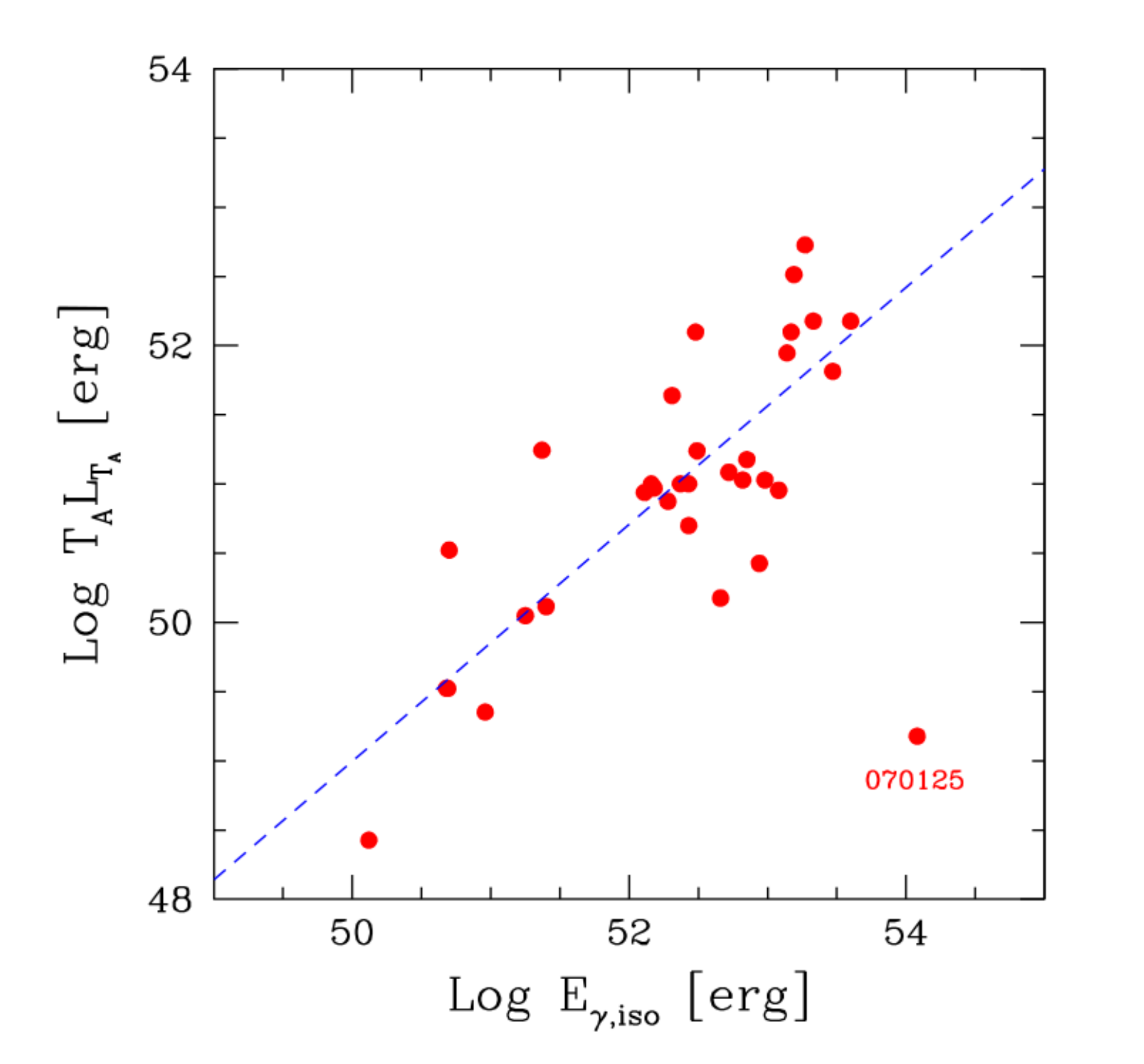}
\caption{\footnotesize Upper left panel: the $F_{X,afterglow}$ in the XRT
band (0.3-10 keV) vs. $F_{\gamma,prompt}$ computed from the BAT $T_{90}$ flux (15-150 keV) 
plus the XRT flux (0.3-10 keV)
from \cite{Willingale2007}. The dotted line represents where $F_{X,afterglow}$ and $F_{\gamma,prompt}$ are identical. 
Upper right panel: $\log E_{\gamma,prompt}$ vs. $\log E_{X,afterglow}$ 
from \cite{Willingale2007}. Symbols show the position of the afterglow in the $\beta_{X,a}$-$\alpha_{X,a}$
plane. GRBs that fall in the pre-jet-break region are plotted as dots, those that fall above this in the
post-jet-break region are plotted as stars, and those below the pre-jet-break band are plotted as squares. 
The dotted line represents equality between $\log E_{\gamma,prompt}$ and $\log E_{X,afterglow}$. 
Bottom left panel: the $\log E_{\gamma,prompt}-\log E_{X,afterglow}$ relation ($E_{\gamma,iso}$ and 
$E_{X,iso}$ respectively in the picture) from \cite{liang2007}. 
The solid line is the best fit. The dashed line indicates the 2 $\sigma$ area.
Bottom right panel: $\log E_{X,plateau} $ vs. $\log E_{\gamma,prompt}$ from \cite{ghisellini09}.  
The dashed line represents the least square fit with $ \log E_{X,plateau}$ ($T_{a}L_{T_a}$ in the picture) 
$ \sim 0.86\times \log E_{\gamma,prompt}$ 
($E_{\gamma,iso}$ in the picture) ($P= 2\times 10^{-7}$, without the outlier GRB 070125).}
\label{ee2} 
\end{figure}

W07 analyzed the relation between the gamma flux in the prompt phase, $F_{\gamma,prompt}$, and the X-ray flux
in the afterglow, $F_{X,afterglow}$ using 107 Swift GRBs, see the upper left panel of Fig. \ref{ee2}.
They calculated $F_{X,afterglow}$ in the XRT band (0.3-10 keV), while 
$F_{\gamma,prompt}$ in the BAT (15-150 keV) plus the XRT (0.3-10 keV) energy band.
For GRBs with known redshift, as shown in the upper right panel of Fig. \ref{ee2}, they investigated 
the prompt isotropic energy, $E_{\gamma,prompt}$, and the afterglow isotropic energy, $E_{X,afterglow}$, 
assuming a cosmology with $H_0 = 71$ km s$^{-1}$ Mpc$^{-1}$, $\Omega_{\Lambda}= 0.73$ and $\Omega_M= 0.27$.

\begin{figure}[htbp]
\includegraphics[width=5.2cm,height=4.7cm,angle=0]{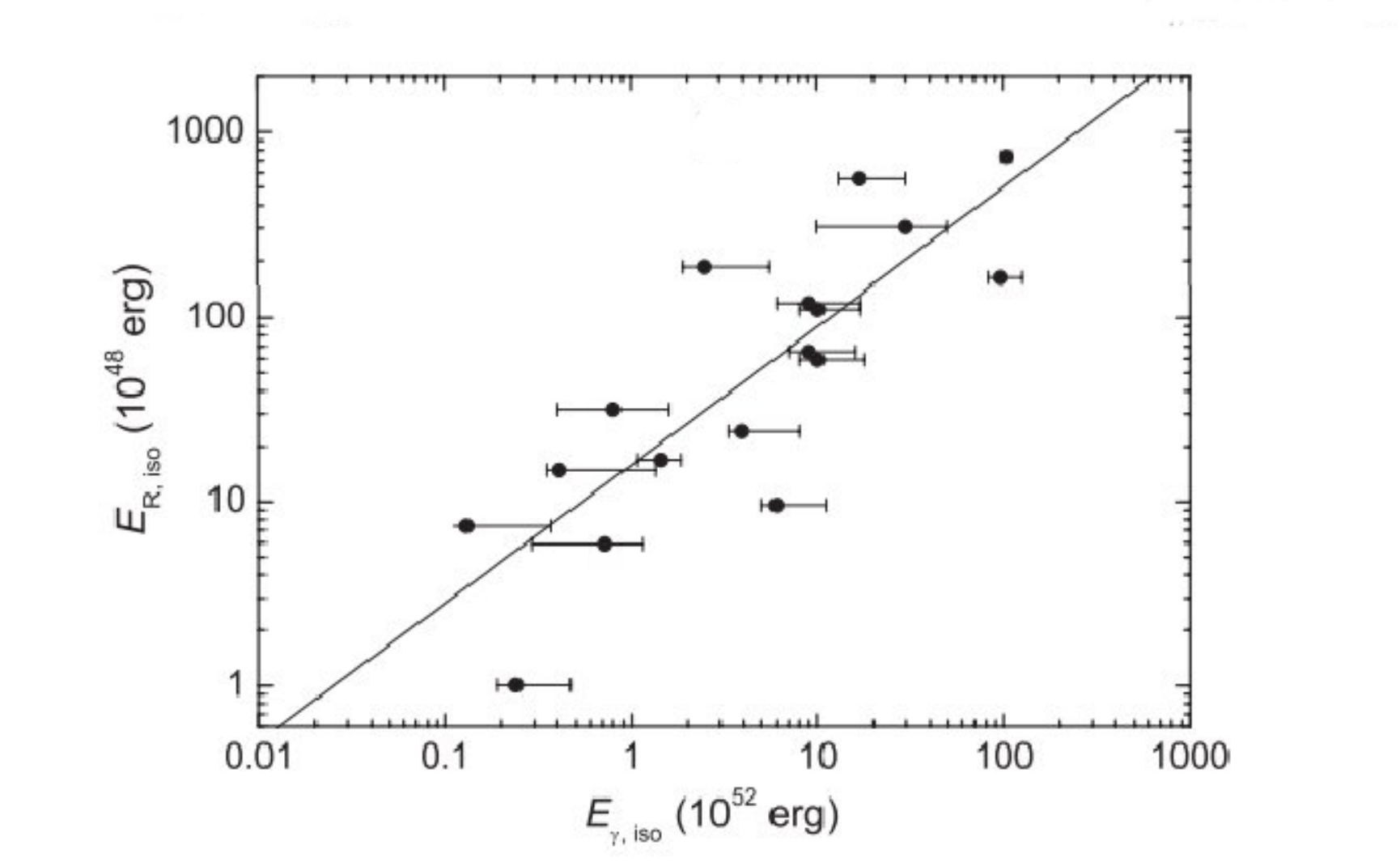}
\includegraphics[width=5.2cm,height=4.8cm,angle=0]{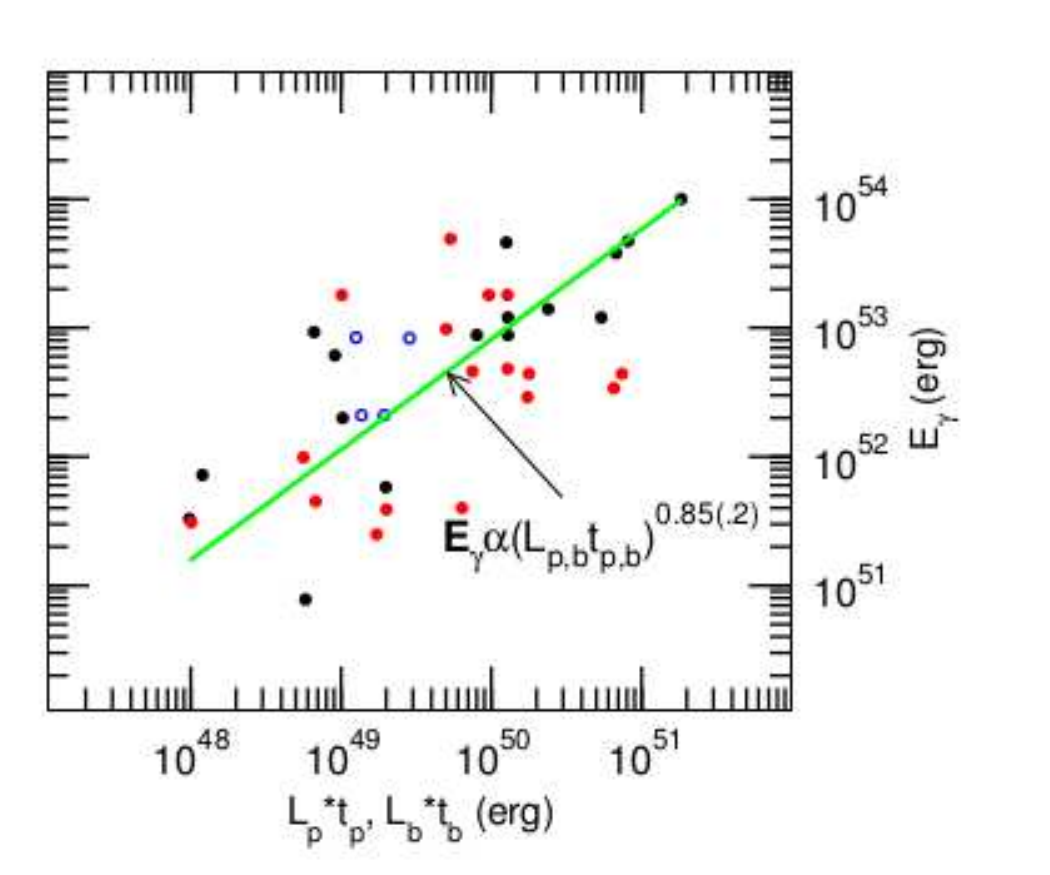}
\includegraphics[width=5.8cm,height=4.8cm,angle=0]{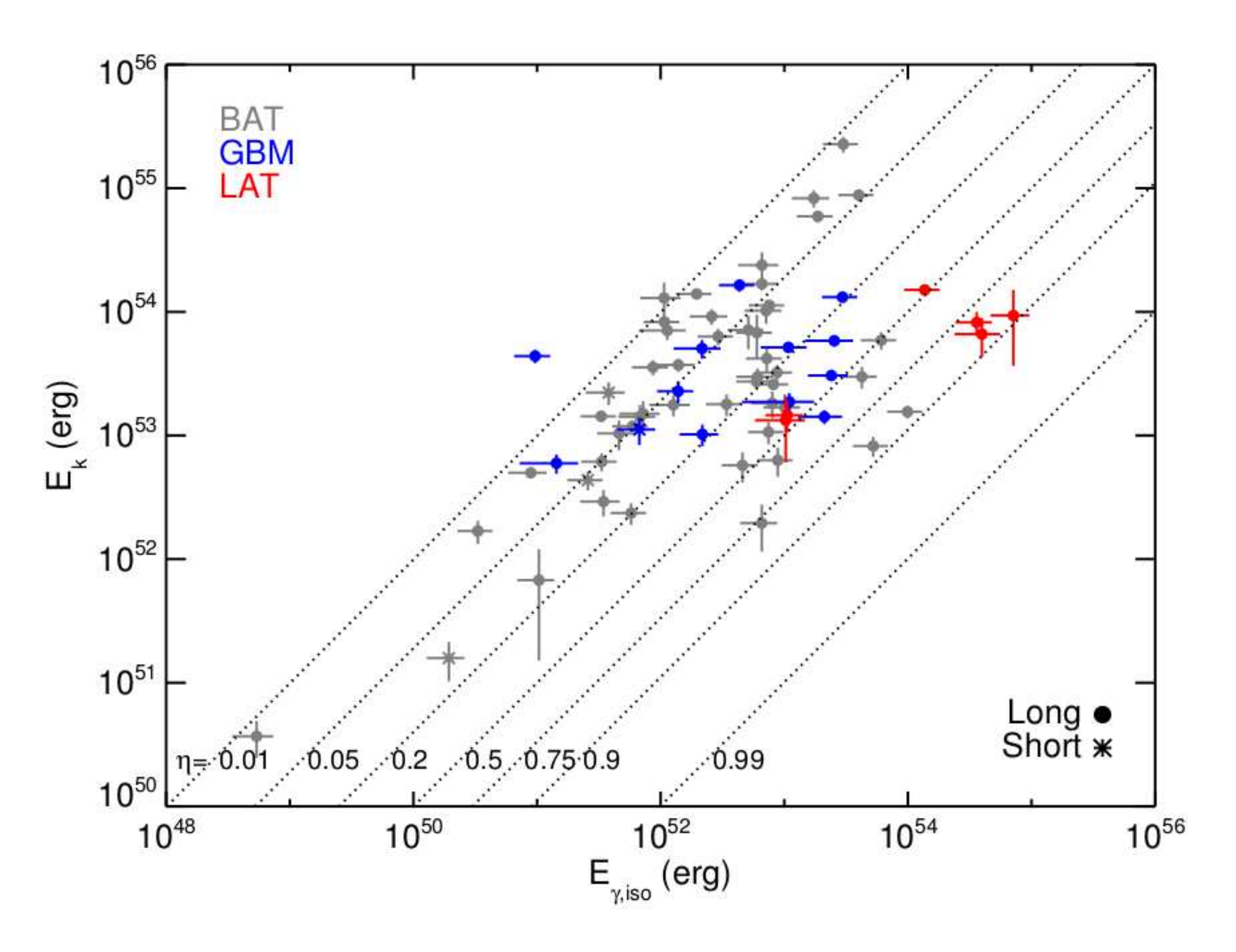}
\caption{\footnotesize Left panel: ``relation between 
$ E_{\gamma,prompt}$ and $ E_{O,afterglow}$ ($E_{\gamma,iso}$ and 
$E_{R,iso}$ respectively in the picture), for the optically selected sample, from \cite{Liang2010}. 
Line is the best fit".
Middle panel: ``relation between 
$\log E_{\gamma,prompt}$ 
and $\log E_{O,afterglow}$ ($E_{\gamma,iso}$ and 
$L_p \times t_p$ respectively in the picture) from \cite{Panaitescu2011}. Black symbols are for afterglows with optical 
peaks, 
red symbols
for optical plateaus, open circles for afterglows of uncertain type. $r(\log E_{X,afterglow},\log E_{\gamma,prompt}) = 0.66$ 
for all 37 afterglows.
This linear correlation coefficients correspond to a probability $P=10^{-5.3}$".
Right panel: ``$E_{k,aft}$ as a function of $E_{\gamma,prompt}$ from \cite{racusin11}. The dashed lines
indicated different values of $\eta$. The bursts detected by LAT on board of Fermi 
tend to have high $E_{\gamma,prompt}$ ($E_{\gamma,iso}$ in the picture), but average $E_{k,aft}$, and therefore
higher values of $\eta$ than the samples from BAT on board of Swift or GBM on board of Fermi".}
\label{ee3} 
\end{figure}

At the same time, \cite{liang2007} focused on the relation between 
$E_{\gamma,prompt}$ and $E_{X,afterglow}$ using a sample of 53 LGRBs. They pointed out a good relation with 
$b=1\pm0.16$, see the bottom left panel of Fig. \ref{ee2}.\\
In agreement with these results, \cite{Liang2010} and \cite{Panaitescu2011} analyzed this relation, 
using respectively 32 and 37 GRBs, but considering energy bands different from that used in \cite{liang2007}; 
they obtained the slopes $b=0.76\pm0.14$ and $b=1.18$ respectively (see the left and middle panels of Fig. \ref{ee3}).

\begin{figure}[htbp]
\includegraphics[width=4cm,height=3.85cm,angle=0]{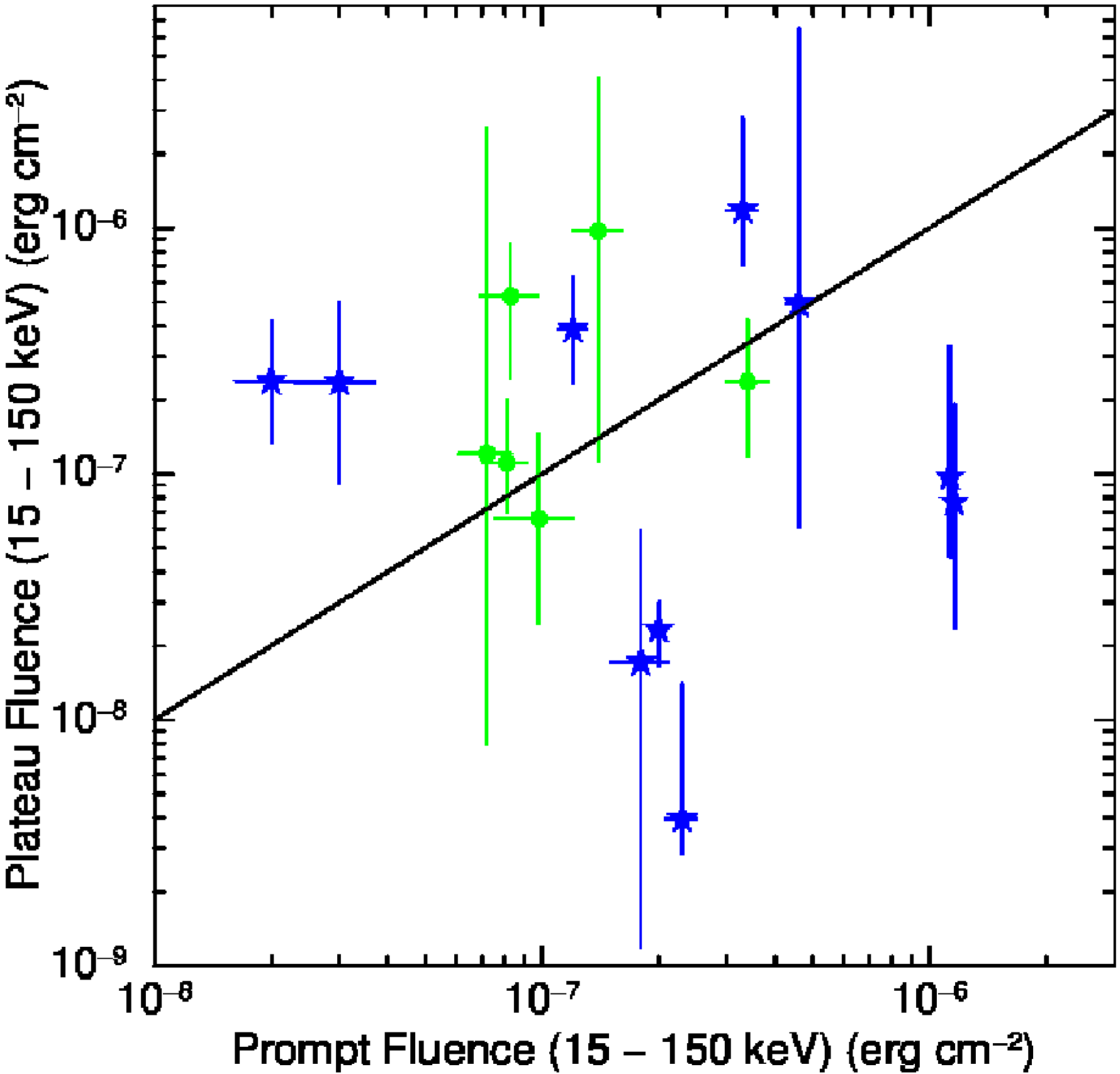}
\includegraphics[width=6.3cm,angle=0]{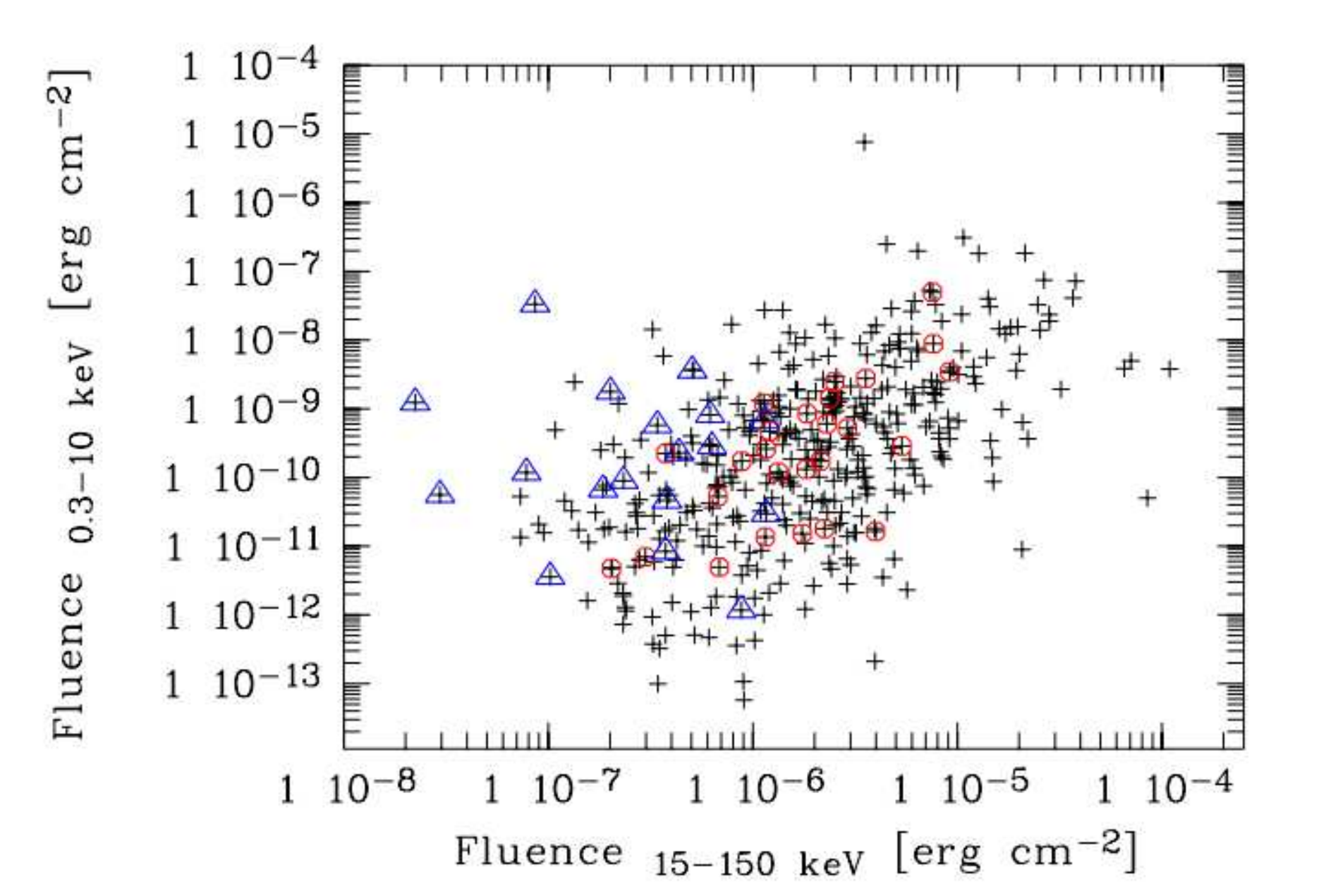}
\includegraphics[width=6cm,angle=0]{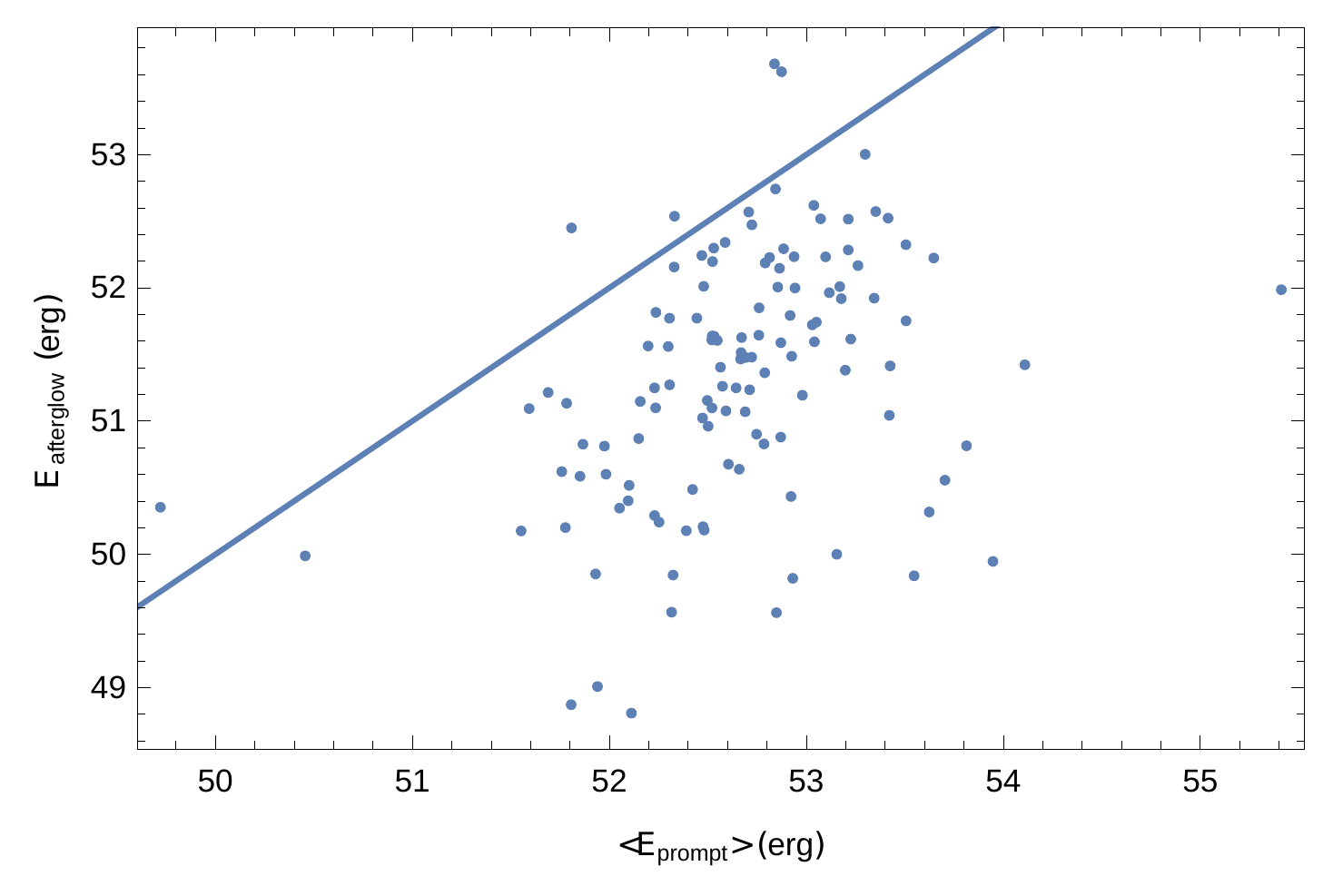}
\caption{\footnotesize Left panel: the prompt BAT 15-150 keV fluence vs. the X-ray fluence in the 15-150
keV energy band from \cite{rowlinson2013}. Blue stars are GRBs with 2 or more breaks in their light curves, 
green circles have 1 break and red triangles have
no significant breaks in their light curves. The black line
indicates where the shallow decay phase fluence is equal to the prompt fluence. Middle panel: relations of the 0.3-10 keV XRT 
fluence with fluence in the 15-150 keV BAT band from \cite{grupe2013}. Short bursts are represented with
triangles and high-redshift (z $> 3.5$) bursts with circles. 
Right panel: $<\log E_{\gamma,prompt}>$ vs. $\log E_{X,afterglow}$ relation from \cite{dainotti15} for 123 LGRBs. 
The solid line for equal $\log E_{\gamma,prompt}$ and $\log E_{X,afterglow}$ is given for reference.}
\label{fig:grupe13}
\end{figure}

\cite{rowlinson2013} and \cite{grupe2013} confirmed these results, see the left and middle panels of Fig. 
\ref{fig:grupe13}. In fact, they obtained a $E_{\gamma,prompt}-E_{X,afterglow}$ relation with slope 
$b\sim 1$ using 43 SGRBs and 232 GRBs with spectroscopic redshifts detected by Swift respectively.\\
Finally, \cite{dainotti15} analyzed this relation to find some constraints on the ratio of 
$E_{X,afterglow}$ to $E_{\gamma,prompt}$, considering a sample of 123 LGRBs, see the right panel of Fig. \ref{fig:grupe13}.\\
Instead, \cite{ghisellini09}, with a sample of 33 LGRBs, considered a similar relation, but assuming the X-ray plateau 
energy, $E_{X,plateau}$, as an estimation of $E_{X,afterglow}$, see the bottom right panel of Fig. \ref{ee2}; they found 
a slope $b=0.86$. \\
In addition, \cite{ghisellini09} also investigated the relation between $E_{\gamma,prompt}$
and the kinetic isotropic energy in the afterglow, $E_{k,aft}$, with the same sample, 
finding a relation with $b=0.42$. 
Similarly, \cite{racusin11} studied the same relation, using 69 GRBs and assuming different efficiencies to find some
limits between $E_{k,aft}$ and $E_{\gamma,prompt}$, see the right panel of Fig. \ref{ee3}.\\
This relation was most likely used to study the differences in detection
of several instruments and to analyze the transferring process of kinetic energy into the prompt emission in GRBs, 
making the relation by \cite{racusin11} the most reliable one.\\
\textcolor{red}{To summarize, for comparing the energies in the prompt and the afterglow phases, 
a $E_{\gamma,prompt}-E_{X,afterglow}$ relation was studied by \cite{liang2007} and 
confirmed by \cite{rowlinson2013}, \cite{grupe2013} and \cite{dainotti15}. The last study found also some limitations 
on the ratio among the prompt and the afterglow energies. Furthermore, instead of $E_{X,afterglow}$, $E_{X,plateau}$ 
was considered for the investigation, although this quantity provided similar results to the previous ones 
\citep{ghisellini09}. Finally, the relation between 
$E_{\gamma,prompt}$ and $E_{k,aft}$ was studied by \cite{ghisellini09} and confirmed by \cite{racusin11}, who examined 
the energy transfer in the prompt phase. These relations are relevant because of their usefulness for investigating 
the efficiency of the emission processes during the transition from the prompt phase to the afterglow one, and for 
explaining which the connection between these two emission phases is. As a main result, \cite{ghisellini09}
and \cite{racusin11} claimed that the fraction of kinetic energy transferred from the prompt phase to the afterglow one 
is around 10\%. In particular, \cite{racusin11} yielded that this value of the transferred kinetic 
energy, for BAT-detected GRBs, is in agreement with the analysis by \cite{zhang07b} for which the internal shock model 
well describes this value in the case of a late energy transfer from the fireball to the surrounding medium 
\citep{zhang2005}.}\\
In Table \ref{tbl3}, a summary of the relations described in this section is presented.\\
As regards the physical interpretation of the $E_{X,afterglow}-E_{\gamma,prompt}$ relation, 
\cite{racusin11}, estimating the efficiency parameter $\eta$ for the BAT sample, confirmed 
the \cite{zhang07b} result for which $\sim57\%$ of BAT bursts have $\eta<10\%$. 
However, for the samples from the Gamma Burst Monitor (GBM)
and the Large Area Telescope (LAT), on board the Fermi satellite\footnote{The Fermi Gamma ray
Space Telescope (FGST), launched in 2008 and still running, is a space observatory being
used to perform gamma ray astronomy observations from low Earth orbit. Its main instrument is the Large Area Telescope (LAT), an 
imaging gamma ray detector, (a pair-conversion instrument) which detects photons with energy from about 20 MeV to 300 GeV with a
field of view of about 20\% of the sky; it is a sequel to the EGRET instrument on the Compton gamma ray observatory (CGRO). Another 
instrument aboard Fermi is the Gamma Ray Burst Monitor (GBM), which is used to study prompt GRBs from $8$ keV to $30$ MeV.},
they found that only 25\% of the GBM bursts and none of the LAT bursts have
$\eta<10\%$. This implies that Fermi GRBs are more efficient at transferring kinetic energy
into prompt radiation.

\begin{table}[htbp]
\footnotesize
\begin{center}
\begin{tabular}{|c|c|c|c|c|c|c|c|}
\hline
Correlations & Author & N& Slope& Norm & Corr.coeff.& P \\
\hline
$E_{X,afterglow}-E_{\gamma,prompt}$ & Liang et al. (2007) & 53 &$1.00^{+0.16}_{-0.16}$&$-0.50^{+8.10}_{-8.10}$&0.79&$<10^{-4}$ \\
$E_{O,afterglow}-E_{\gamma,prompt}$ &Liang et al. (2010)&32&$0.76^{+0.14}_{-0.14}$&$1.30^{+0.14}_{-0.14}$&0.82&$<10^{-4}$\\
 &Panaitescu \& Vestrand (2011)&37&1.18&&0.66& $10^{-5.3}$\\
 \hline
$E_{X,plateau}-E_{\gamma,prompt}$ & Ghisellini et al. (2009)&33&0.86&&&$2\times10^{-7}$ \\
\hline
$ E_{k,aft}- E_{\gamma,prompt}$ & Ghisellini et al. (2009)&33&0.42&&&$10^{-3}$ \\
\hline
\end{tabular}
\caption{\footnotesize Summary of the relations in this section. The first column represents the relation in log scale,
the second one the authors, and the third one the number of GRBs in the used sample. Afterwards, the fourth and fifth columns are
the slope and normalization of the relation and the last two columns are the correlation coefficient and the chance probability, P.}
\label{tbl3}
\end{center}
\end{table}

\subsection{The \texorpdfstring{$L_{X,afterglow}-E_{\gamma,prompt}$}{Lg} relation and its physical interpretation}
\begin{figure}[htbp]
\centering
\includegraphics[width=0.85\hsize,clip]{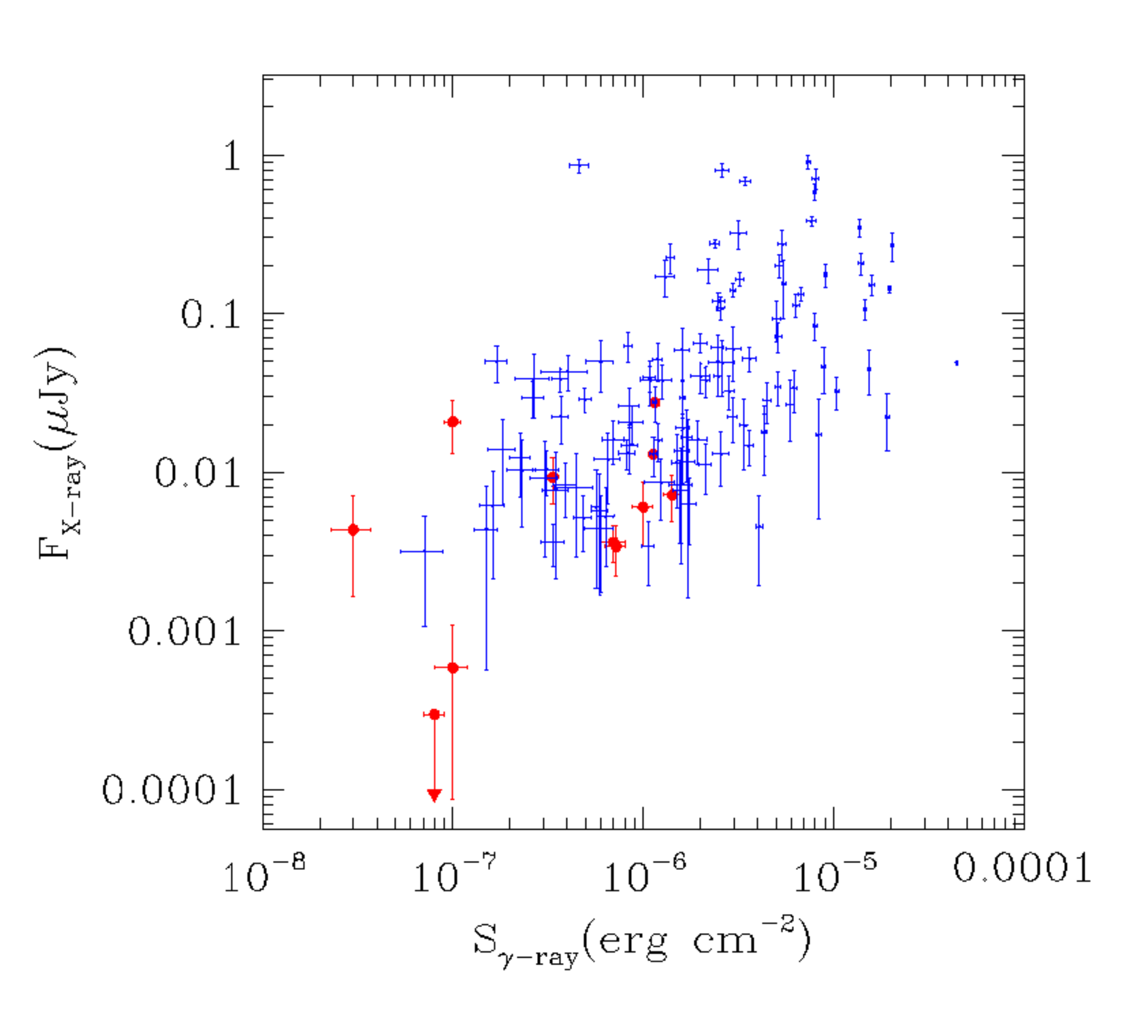}
\caption{\footnotesize $F_{X,11}$-$S_{\gamma,prompt}$ ($F_{X-ray}$ and 
$S_{\gamma-ray}$ respectively in the picture) relation for Swift
 SGRBs (in red) and LGRBs (in blue) from \cite{gehrels08} .
 The XRT $F_{X,11}$ are computed at 3 keV and the BAT
$S_{\gamma,prompt}$ are detected between 15 and 150 keV \citep{sakamoto08}.}
\label{fig:gehrels08b}
\end{figure}

\cite{berger07} investigated the prompt and afterglow energies in the observed frame of 16 SGRBs. 
A large fraction of them (80\%) follows a linear relation between 
the prompt fluence in the gamma band, $S_{\gamma,prompt}$, in the BAT range and 
the X-ray flux at 1 day, $F_{X,1\rm{d}}$, in the XRT band given by:

\begin{equation}
\log F_{X,1\rm{d}} \sim (1.01 \pm 0.09) \times \log S_{\gamma,prompt},
\end{equation}

with $\rho=0.86$ and $P=5.3\times10^{-5}$. 
\cite{gehrels08} confirmed his results investigating the same relation, but with X-ray 
fluxes at 11 hours, $F_{X,11}$, see Fig. \ref{fig:gehrels08b}.

\begin{figure}[htbp]
\centering
\includegraphics[width=0.495\hsize,angle=0,clip]{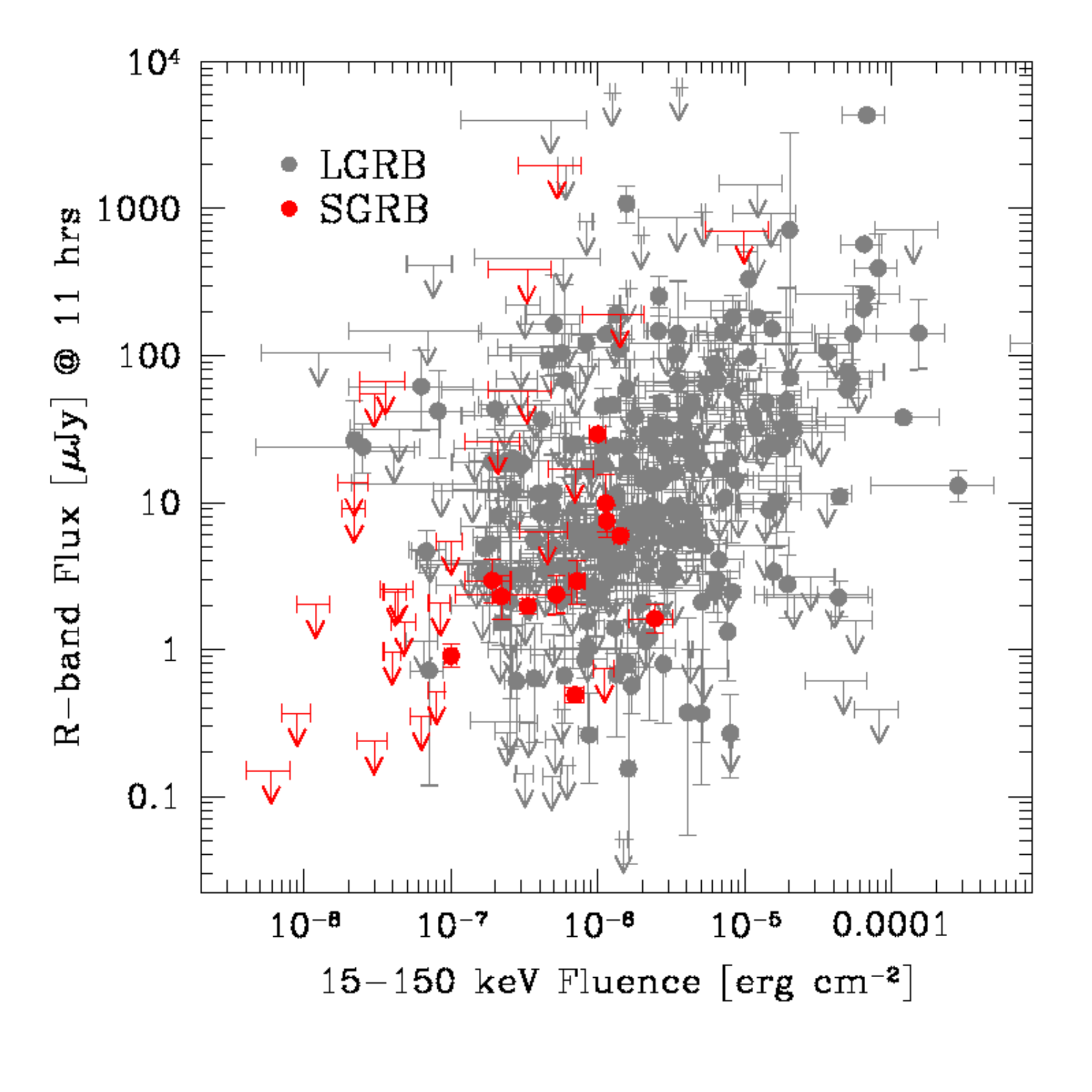}
\includegraphics[width=0.495\hsize,angle=0,clip]{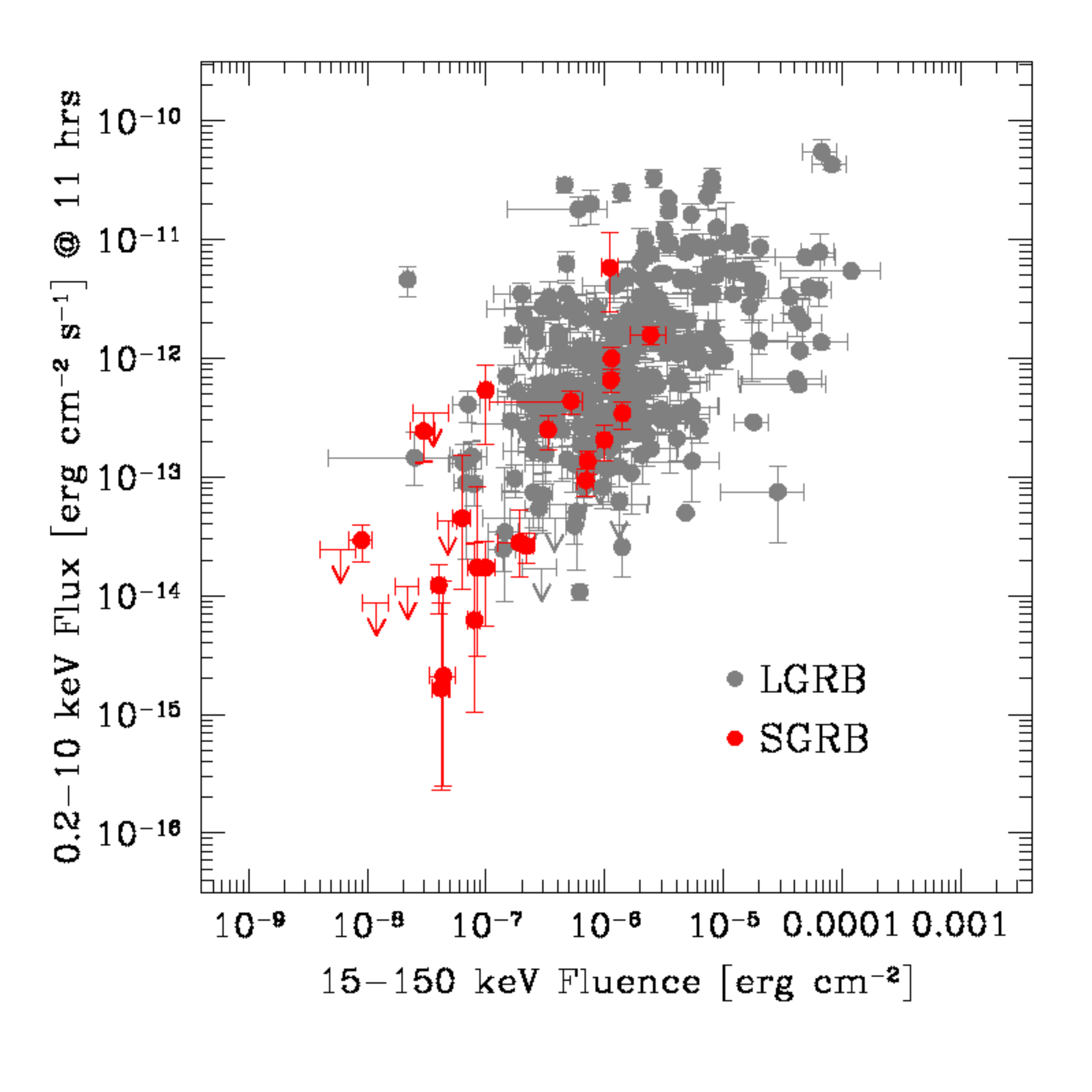}
\includegraphics[width=0.495\hsize,angle=0,clip]{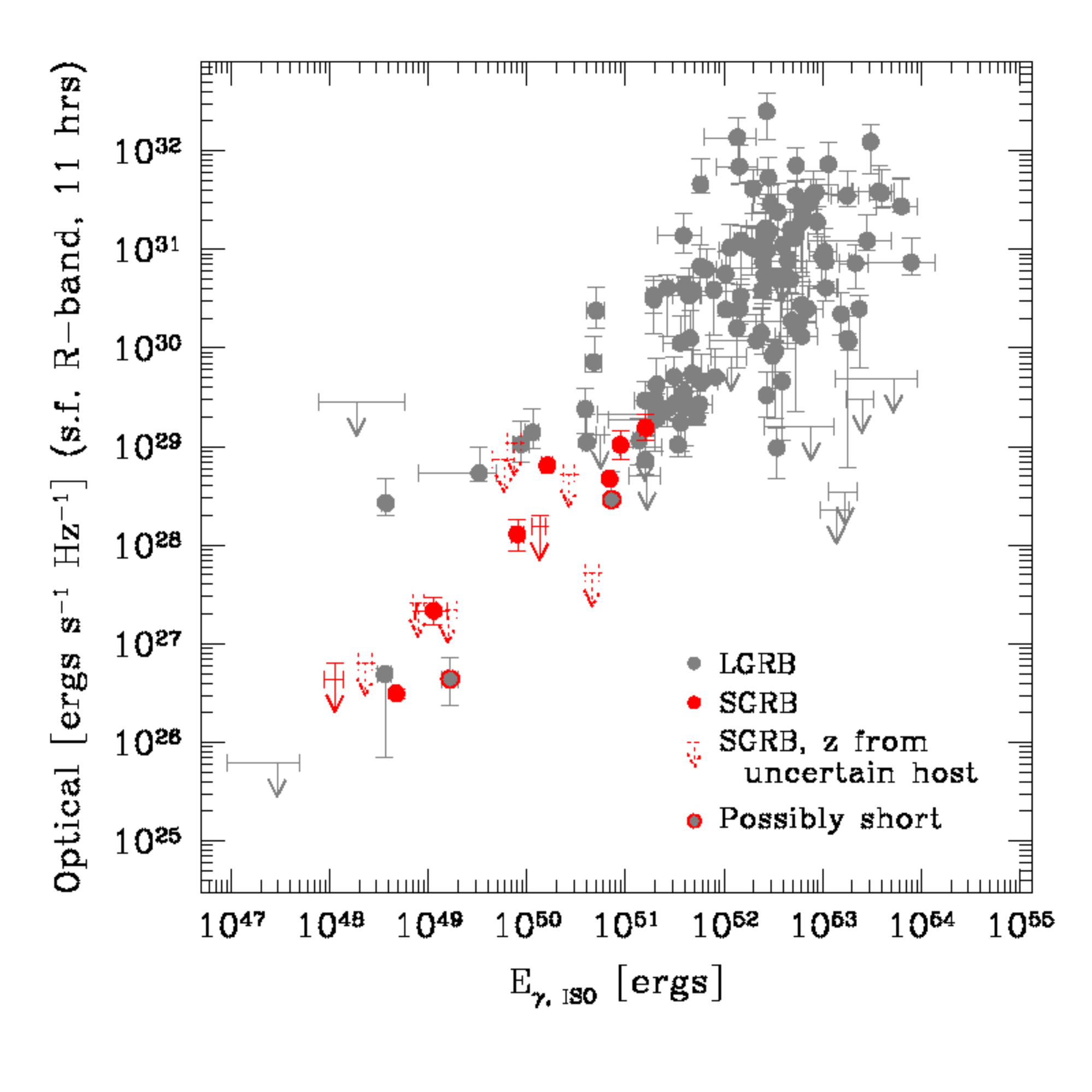}
\includegraphics[width=0.495\hsize,angle=0,clip]{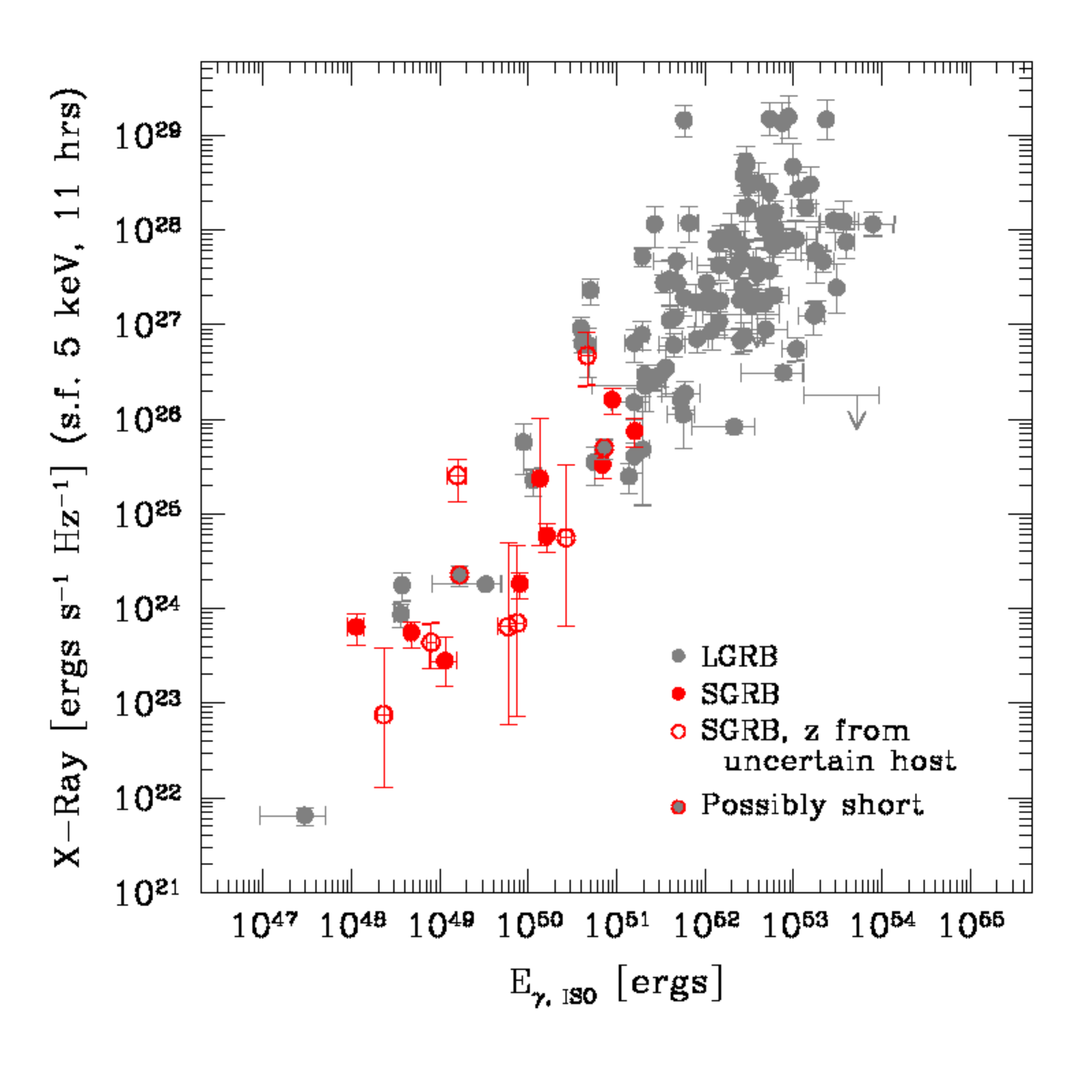}
   \caption{\footnotesize Upper left panel: ``a plot of $F_{O,11}$ (corrected for Galactic
  extinction) vs. 15-150 keV $S_{\gamma,prompt}$
  for both LGRBs (grey) and SGRBs (red) from \cite{nysewander09}.  Note that
  below a fluence of 10$^{-7}$ erg cm$^{-2}$, no optical afterglow of
  an SGRB has been discovered, while above 10$^{-7}$, all reasonably
  deep observing campaigns, but one (GRB 061210) have detected an
  optical afterglow". Upper right panel: ``a plot of $F_{X,11}$ 
  vs. 15-150 keV $S_{\gamma,prompt}$ ($E_{\gamma,iso}$ in the picture) for both 
  LGRBs (grey) and SGRBs (red) from \cite{nysewander09}". 
    Bottom left panel: ``a plot of $L_{O,11}$ (corrected for
  Galactic extinction) vs. $E_{\gamma,prompt}$ from \cite{nysewander09}.  Dashed upper limits represent SGRBs with a host
  galaxy determined by XRT error circle only.  The classification of
  GRB 060614 and GRB 060505 is uncertain, therefore, they are labelled
  as ``possibly short" ". Bottom right panel: ``a plot of $L_{X,11}$ vs. $E_{\gamma,prompt}$ from \cite{nysewander09}. The open circles
  represent SGRBs with a host galaxy determined by XRT error circle
  only.  The classification of GRB 060614 and GRB 060505 is uncertain,
  therefore, they are labelled as ``possibly short" ".}
\label{fig:nysawander}
\end{figure}

\begin{figure}[htbp]
\centering
\includegraphics[width=8.1cm,height=6.3cm,angle=0,clip]{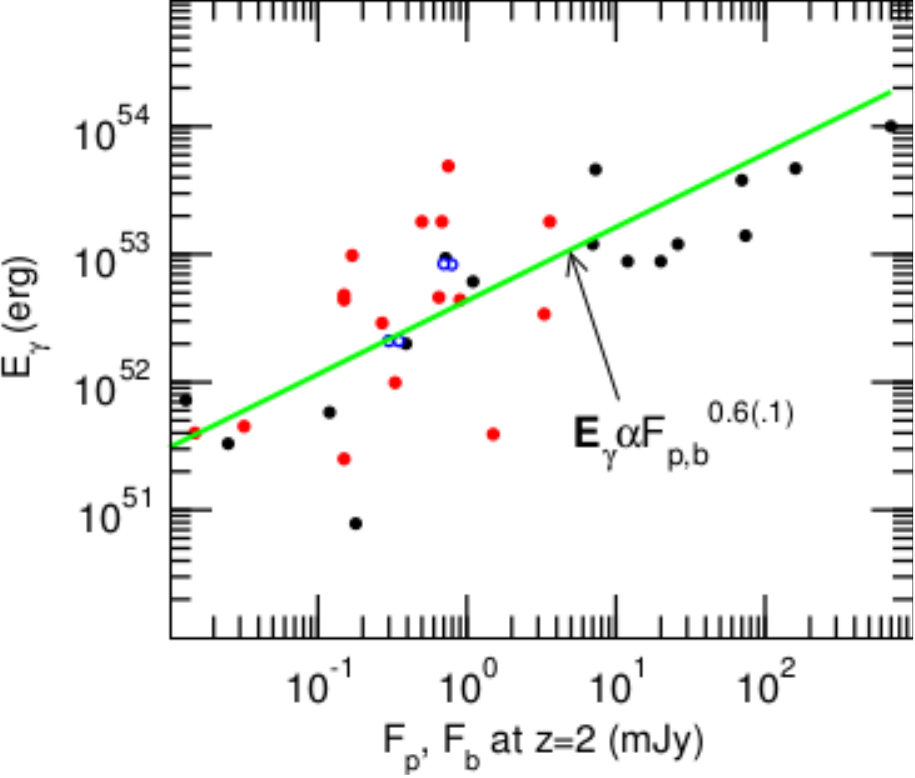}
\includegraphics[width=8.1cm,height=6.3cm,angle=0,clip]{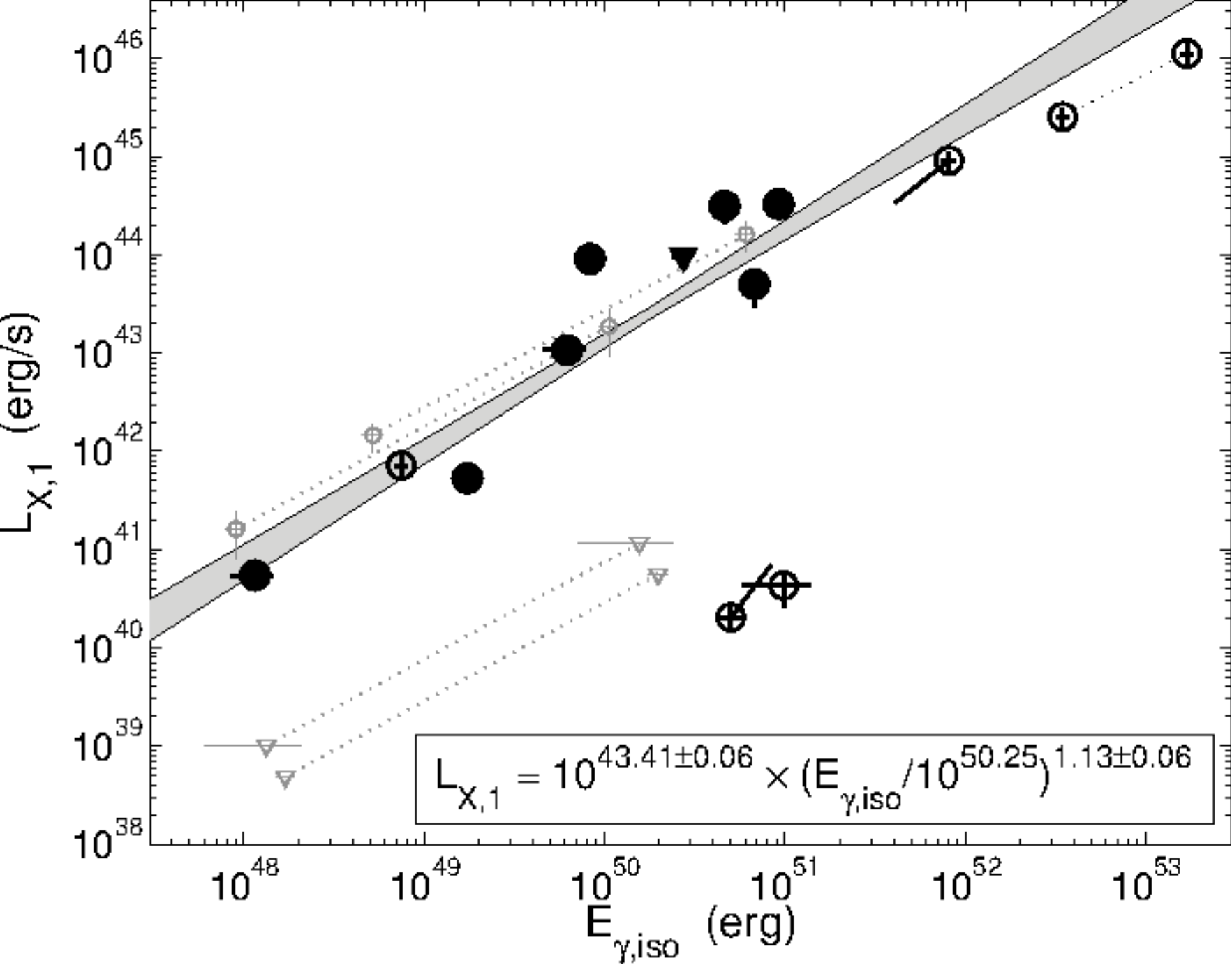}
   \caption{\footnotesize Left panel: $E_{\gamma,prompt}$-$F_{O,a}$ ($E_{\gamma,iso}$ and 
$F_p$ respectively in the picture) relation from \cite{Panaitescu2011}.
Black symbols are for afterglows with optical peaks, red symbols for optical plateaus,
   open circles for afterglows of unknown kind. Right panel: $L_{X,1\rm{d}}$ vs. $E_{\gamma,prompt}$ ($E_{\gamma,iso}$ 
   in the picture) for the SGRBs with a known $z$ (solid black circles),
redshift constraints (open black circles) and without any redshift
information (grey symbols connected by dotted lines) from \cite{berger07}.}
\label{fig:gehrels08}
\end{figure}

\begin{figure}[htbp]
 \centering
\includegraphics[width=5.5cm,height=5cm,angle=0,clip]{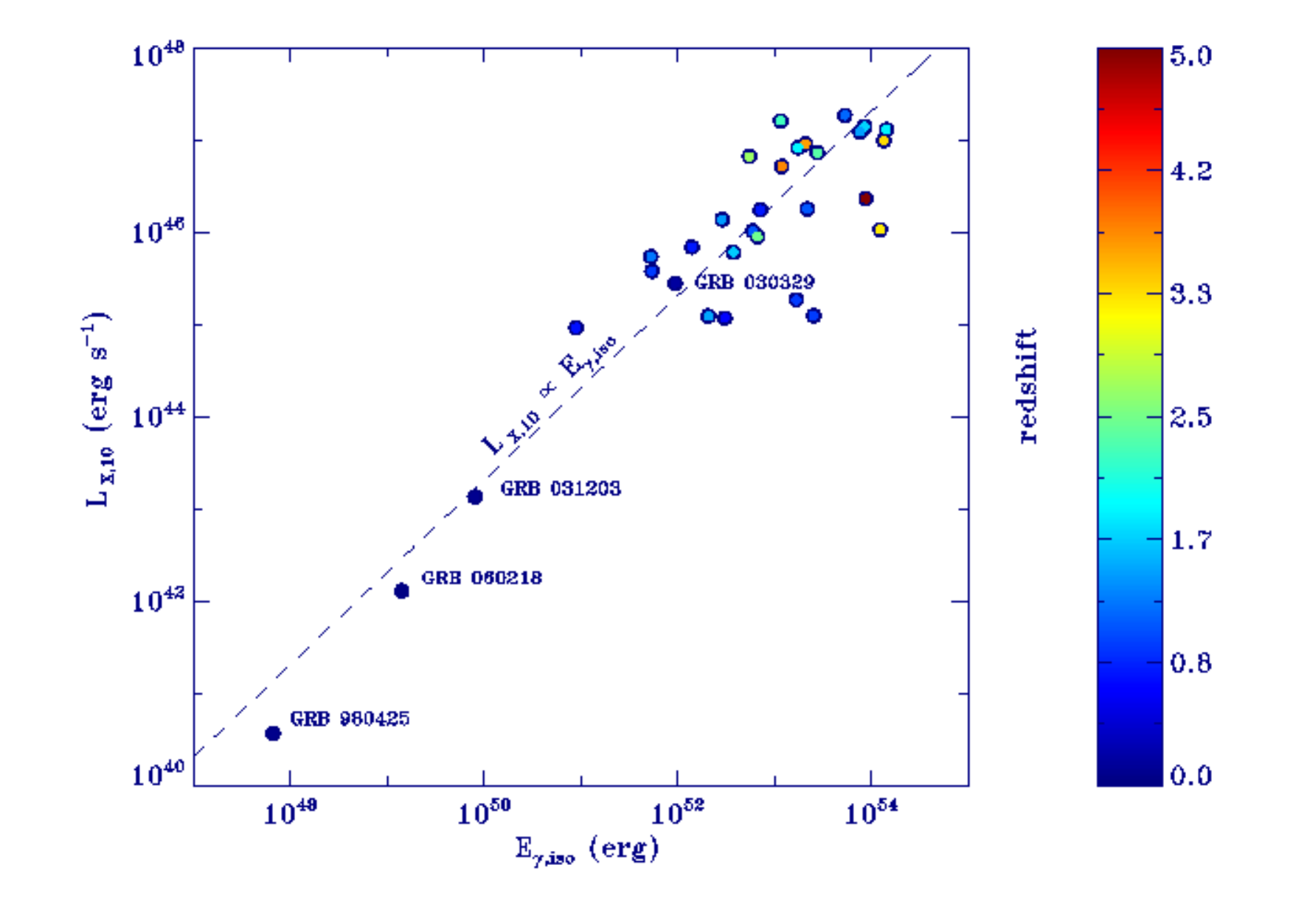}
\includegraphics[width=5cm,height=5cm,angle=0,clip]{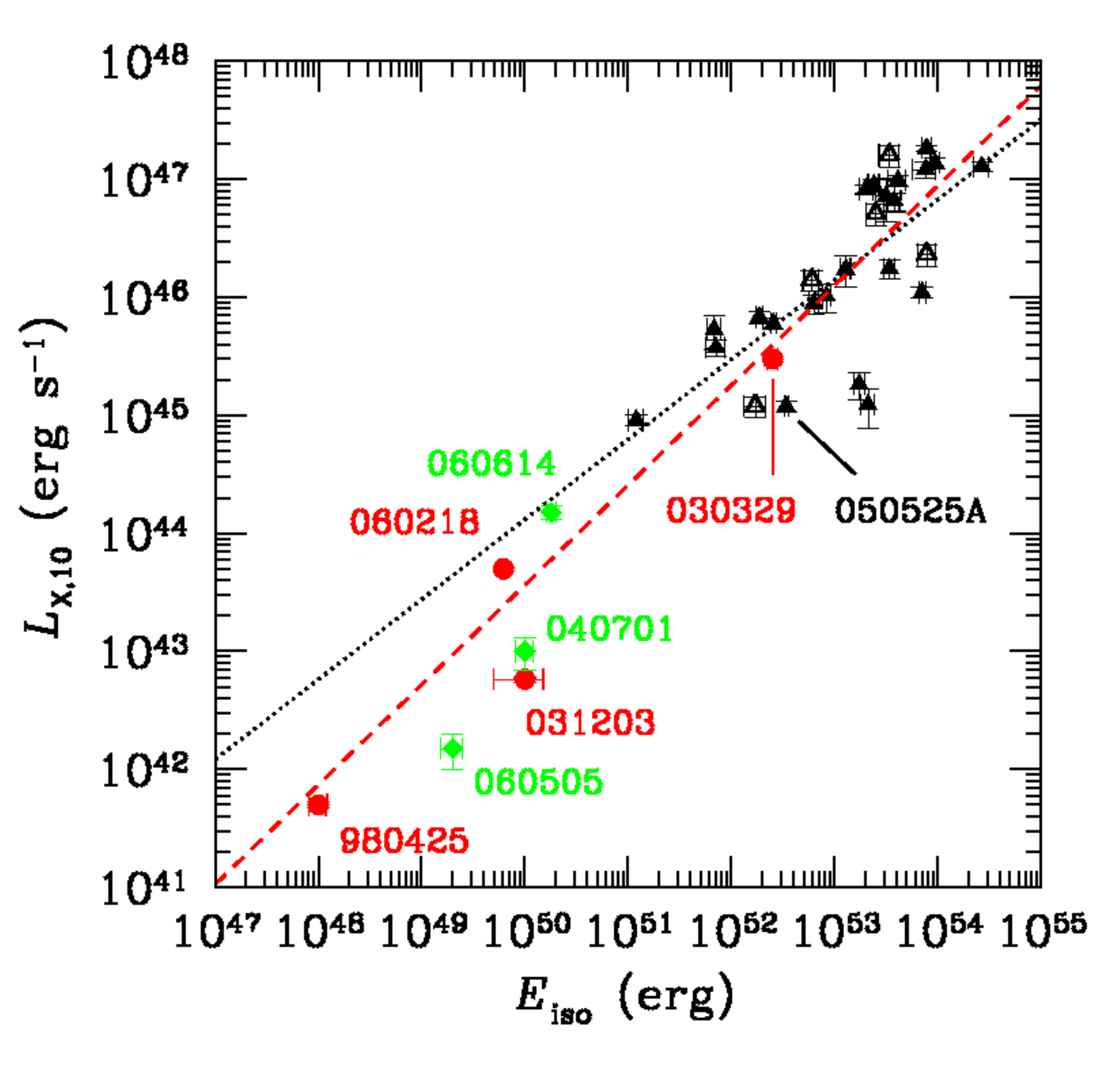}
\includegraphics[width=5.75cm,height=5cm,angle=0,clip]{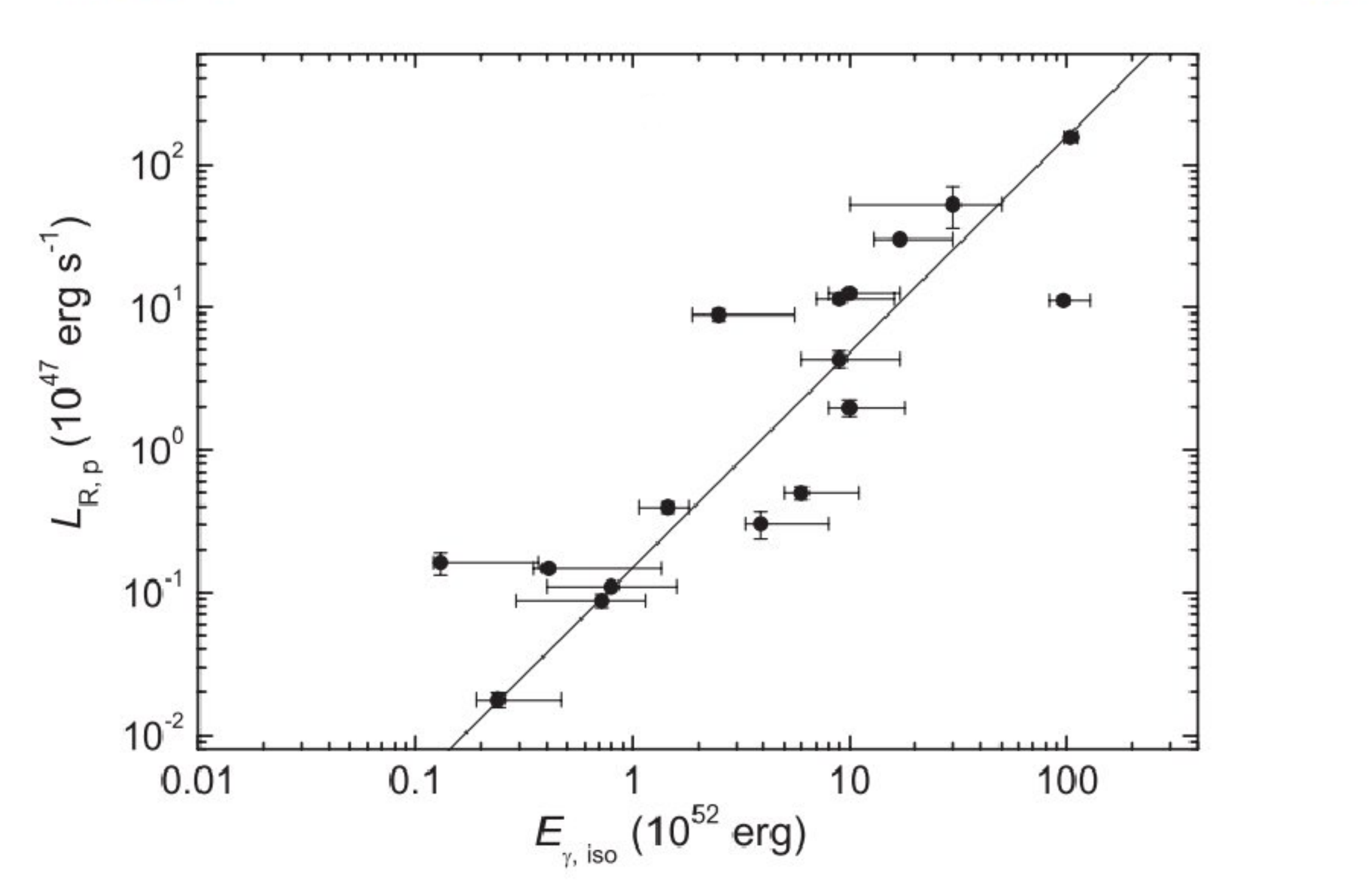}
\caption{\footnotesize Left panel: ``$L_{\rm X,10}$ of SN-GRBs (source frame: 2$-$10 keV) as a 
function of their $E_{\gamma,prompt}$, $E_{\gamma,iso}$ in the picture, (20$-$2000 keV) 
from \cite{kaneko07}.
$z$ for each event is also shown in colour". Middle panel: ``$L_{X,10}$ (in 2-10 keV range)
vs. $E_{\gamma,prompt}$ ($E_{iso}$ in the picture) for the events included in the sample of \cite{Nousek2006} (triangles) plus the 3 sub-energetic GRB\,980425, 
GRB\,031203, GRB\,060218,  the other GRB/SN event GRB\,030329 (circles), and
3 GRBs with known $z$ and deep limits to the peak magnitude
of associated SN, XRF\,040701, GRB\,060505 and GRB\,060614 (diamonds) from \cite{amati07}. Empty triangles
indicate those GRBs for which the 1-10000 keV $E_{\gamma,prompt}$ was computed based on 
the 100-500 keV $E_{\gamma,prompt}$ reported by \cite{Nousek2006} by assuming
an average spectral index. 
The plotted lines are the best-fit power laws obtained
without (dotted) and with (dashed) sub-energetic GRBs and GRB\,030329". Right panel: ``relation between $E_{\gamma,prompt}$ 
and $L_{O,peak}$ ($E_{\gamma,iso}$ and 
$L_{R,p}$ respectively in the picture) for the optically selected sample from \cite{Liang2010}. Line is the best fit".}
\label{fig:kaneko}
\end{figure}

Later, \cite{nysewander09} considered the relation between $F_{X,11}$ or the optical flux at 11 hours, $F_{O,11}$, 
and $E_{\gamma,prompt}$, finding an almost linear relation, see Fig. \ref{fig:nysawander}. They used a data set 
of 37 SGRBs and 421 LGRBs detected by Swift.
\cite{Panaitescu2011} confirmed, in part, these results. They found a similar relation between 
$E_{\gamma,prompt}$ and $F_{O,a}$ using 37 GRBs, but with a higher slope ($b=1.67$), see the left panel of Fig. 
\ref{fig:gehrels08}.\\
\cite{kaneko07} showed a linear relation $L_{X,10}\propto E_{\gamma,prompt}$, where $L_{X,10}$ is the X-ray
luminosity at 10 hours calculated in the 
2-10 keV energy range, while $E_{\gamma,prompt}$ in the 20-2000 keV energy range, see the left panel of 
Fig. \ref{fig:kaneko}.
This relation compares four long events spectroscopically associated with SNe with ``regular" energetic LGRBs 
($E_{\gamma,prompt}\sim 10^{52}-10^{54}$ erg). The results possibly indicate a common
efficiency $\eta$ for transforming kinetic energy into gamma rays
in the prompt phase for both these four events and for ``regular" energetic LGRBs.\\
The same relation has been studied in the context of the low luminosity versus normal luminosity GRBs.
Indeed, \cite{amati07} found that the relation between $L_{X,10}$, in the 2-10 keV band,
and $E_{\gamma,prompt}$, in the 1-10000 keV band, becomes stronger ($P\sim 10^{-11}$) including sub-energetic GRBs 
as GRB 060218, GRB 980425 
and GRB 031203, see the middle panel of Fig. \ref{fig:kaneko}. Therefore, it is claimed 
that sub-energetic GRBs are intrinsically faint and are considered to some extent normal cosmological GRBs.\\
Finally, \cite{berger07} also analyzed the relation between the X-ray luminosity at one day, 
$L_{X,1\rm{d}}$, and $E_{\gamma,prompt}$, using 13 SGRBs with measured $z$. They found a slope $b=1.13\pm0.16$ 
(see the right panel of Fig. \ref{fig:gehrels08}). 

\begin{figure}[htbp]
\centering
\includegraphics[width=5.5cm,height=4.4cm]{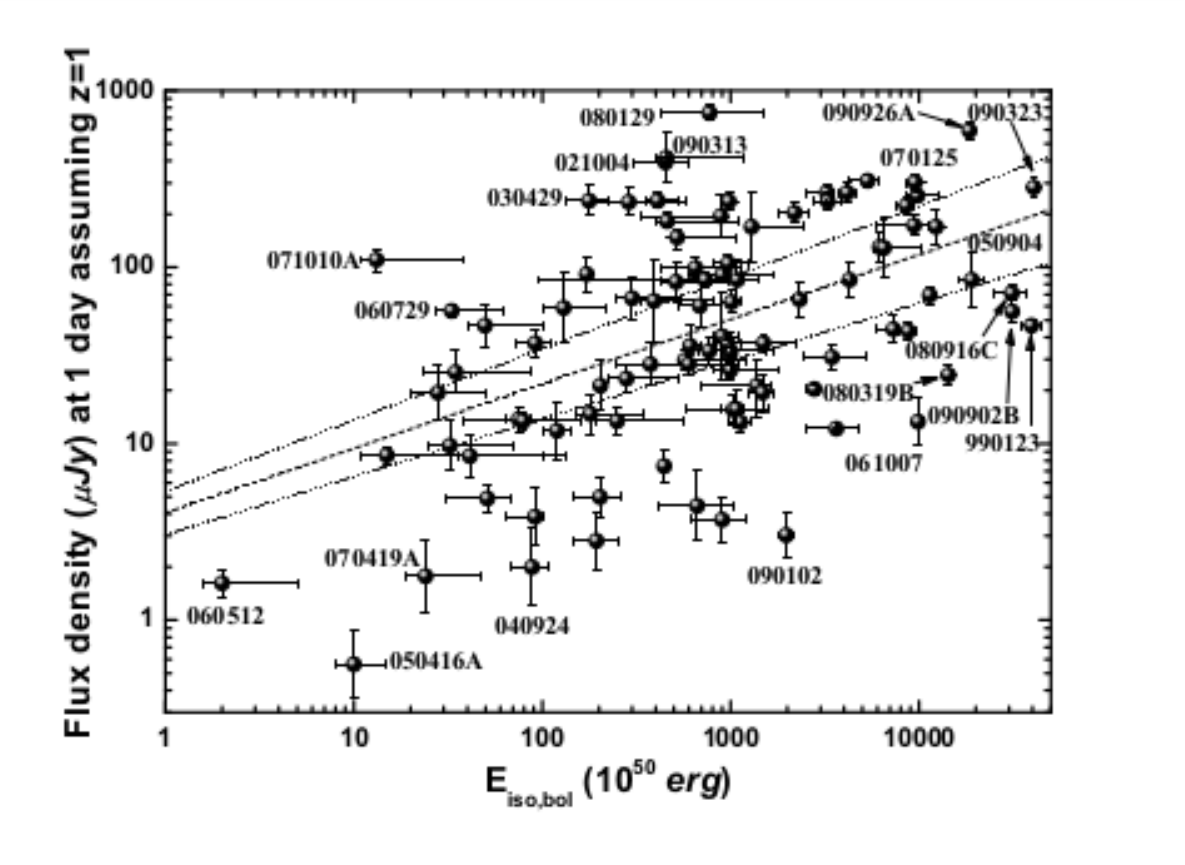}
\includegraphics[width=0.325\hsize,angle=0,clip]{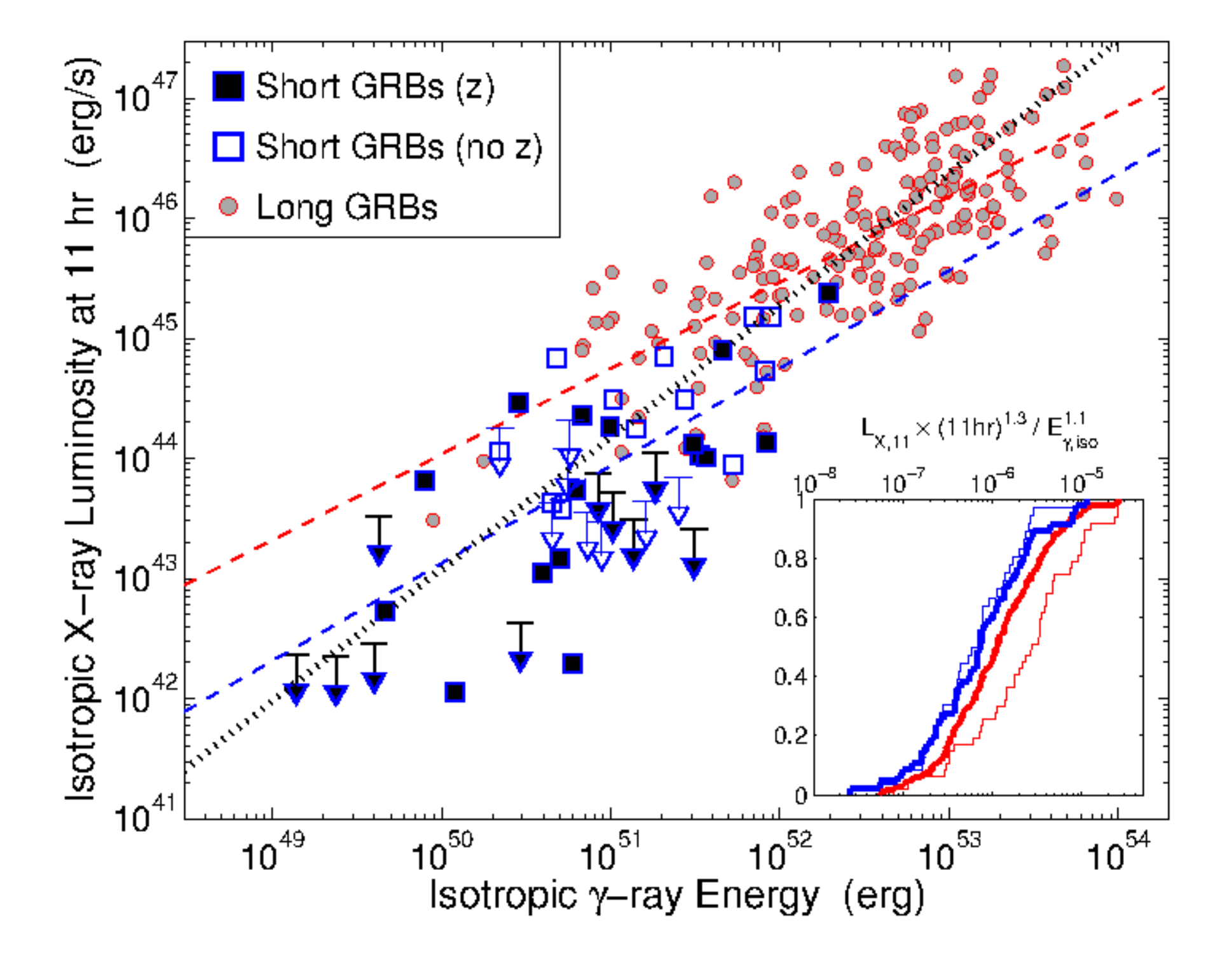}  
\includegraphics[width=0.325\hsize,angle=0,clip]{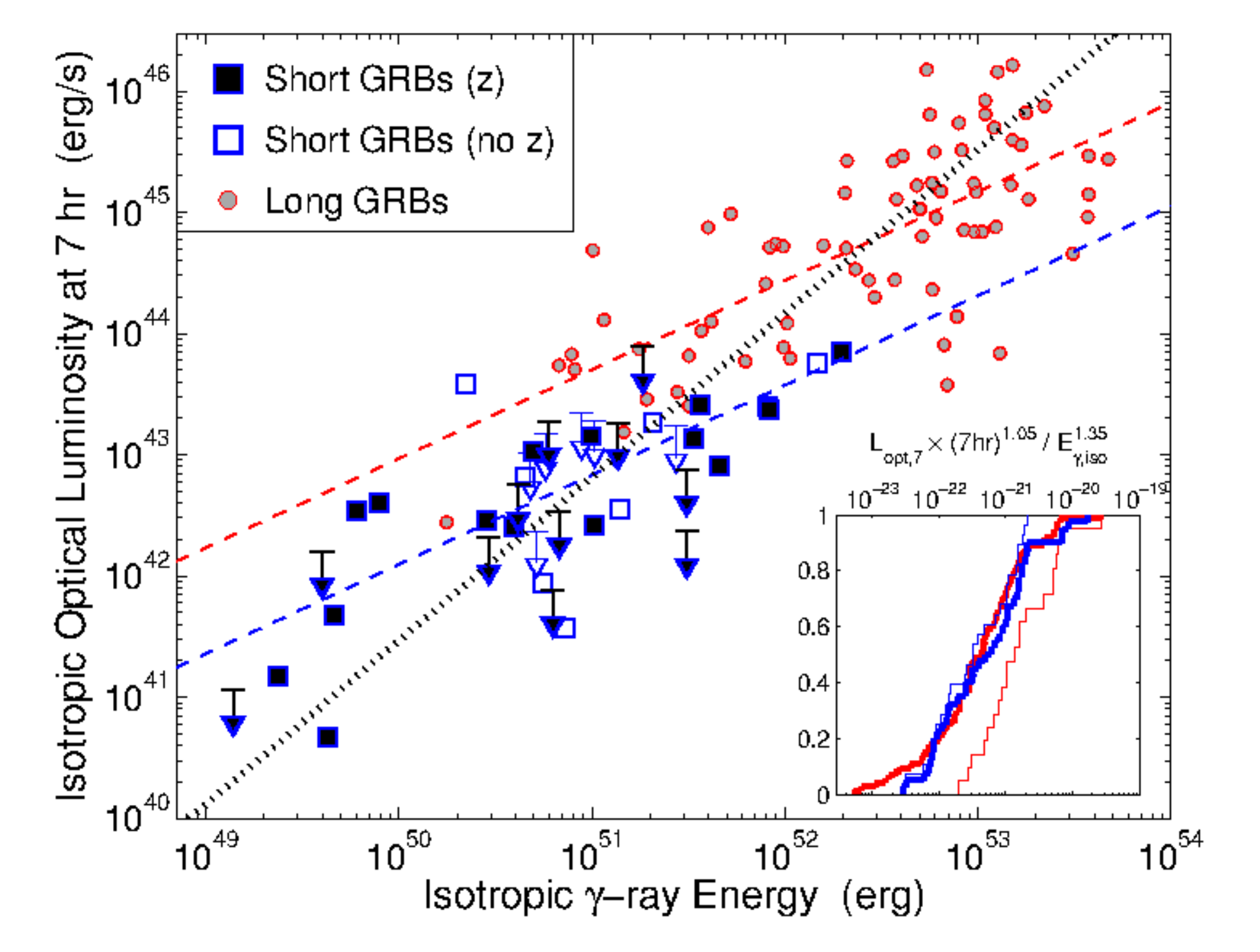}  
\caption{\footnotesize Left panel: ``$F_{O,1\rm{d}}$ in the $R$ band 
plotted against the bolometric $E_{\gamma,prompt}$ ($E_{iso,bol}$ in the picture) for all GRBs in the optically selected sample from \cite{kann10}
(except GRB 991208, which was only discovered after several days, and GRBs 060210, 060607A, 060906 and 080319C, where the follow-up 
does not extend to one day). While no tight relation is visible, there is a trend of increasing optical luminosity with increasing
prompt energy release. This is confirmed by a linear fit (in log-log space), using a Monte Carlo analysis to account for the
asymmetric errors. The dashed line shows the best fit, while the dotted line marks the 3 $\sigma$ error region. Several special 
GRBs are marked". Middle panel: ``$L_{X,11}$ vs. $E_{\gamma,prompt}$ for
  SGRBs (blue) and LGRBs (grey) from \cite{berger14}.  Open symbols for SGRBs
  indicate events without a known $z$, for which a fiducial value
  of $z=0.75$ is assumed.  The dashed blue and red lines are the
  best-fit power law relations to the trends for SGRBs and LGRBs,
  respectively, while the dotted black line is the expected
  relation based on the afterglow synchrotron model with
  $\nu_X>\nu_c$ and $p=2.4$ ($\log L_{X,11}\sim1.1\times \log E_{\gamma,prompt}$).
  The inset shows the distribution of the ratio $\log (L_{X,11} \times
  (11\,{\rm hr})^{1.3}/ E_{\gamma,prompt}^{1.1})$, for the full samples
  (thick lines) and for the region where SGRBs and LGRBs have equal $E_{\gamma,prompt}$ values (thin lines).  
  The lower level of $L_{X,11}$ relative to
  $E_{\gamma,prompt}$ for SGRBs is evident from these various
  comparisons". Right panel: same as in the middle panel, ``but for the
  isotropic-equivalent afterglow optical luminosity at a rest frame
  time of 7 hours ($L_{O,7}$), still from \cite{berger14}. The dotted black line is the
  expected relation based on the afterglow model for
  $\nu_m<\nu_{O}<\nu_c$ and $p=2.4$ ($\log L_{O,7}\sim 1.35\times \log E_{\gamma,prompt}$).  The inset 
  shows the distribution of the
  ratio $\log (L_{O,7} \times (7\,{\rm hr})^{1.05}/
  E_{\gamma,prompt}^{1.35})$, for the full samples (thick lines) and for the region where SGRBs and LGRBs 
  have equal $E_{\gamma,prompt}$ values (thin
  lines). The lower level of $L_{O,7}$ relative to
  $E_{\gamma,prompt}$ for SGRBs is evident from these various
  comparisons".}
\label{fig:berger}
\end{figure}

\cite{Liang2010} confirmed his results in the optical range using a sample of 32 Swift GRBs 
($E_{\gamma,prompt}-L_{O,peak}$ with $b=1.40 \pm 0.08$, see the right panel of Fig. 
\ref{fig:kaneko}). In addition, \cite{kann10} also confirmed his results with a sample of 76 LGRBs 
($E_{\gamma,prompt}-L_{O,1\rm{d}}$ with $b=0.36$, see the left panel of Fig. \ref{fig:berger}).

\begin{figure}[htbp]
\centering
\includegraphics[width=8.1cm,angle=0,clip]{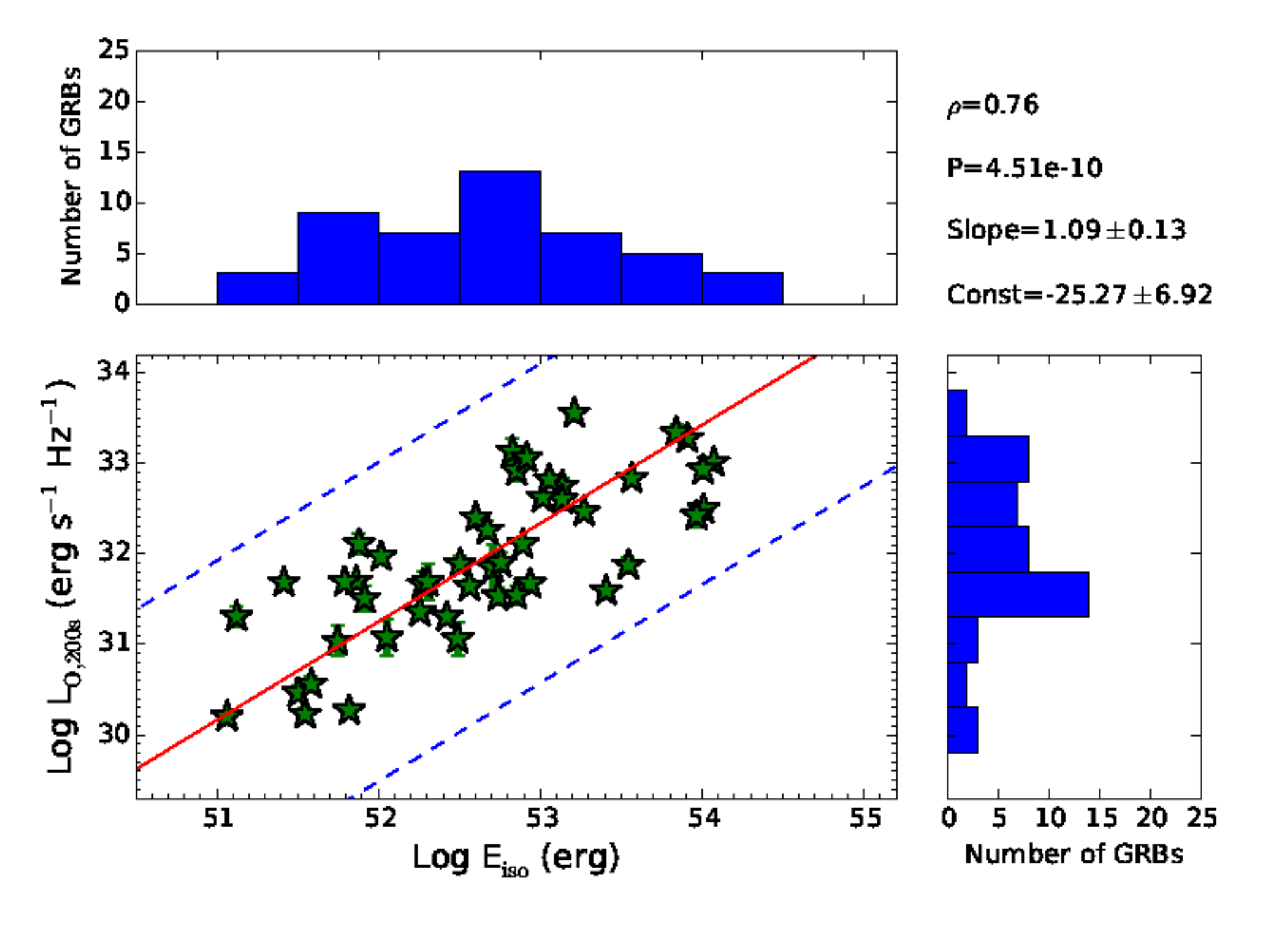}  
\includegraphics[width=8.1cm,angle=0,clip]{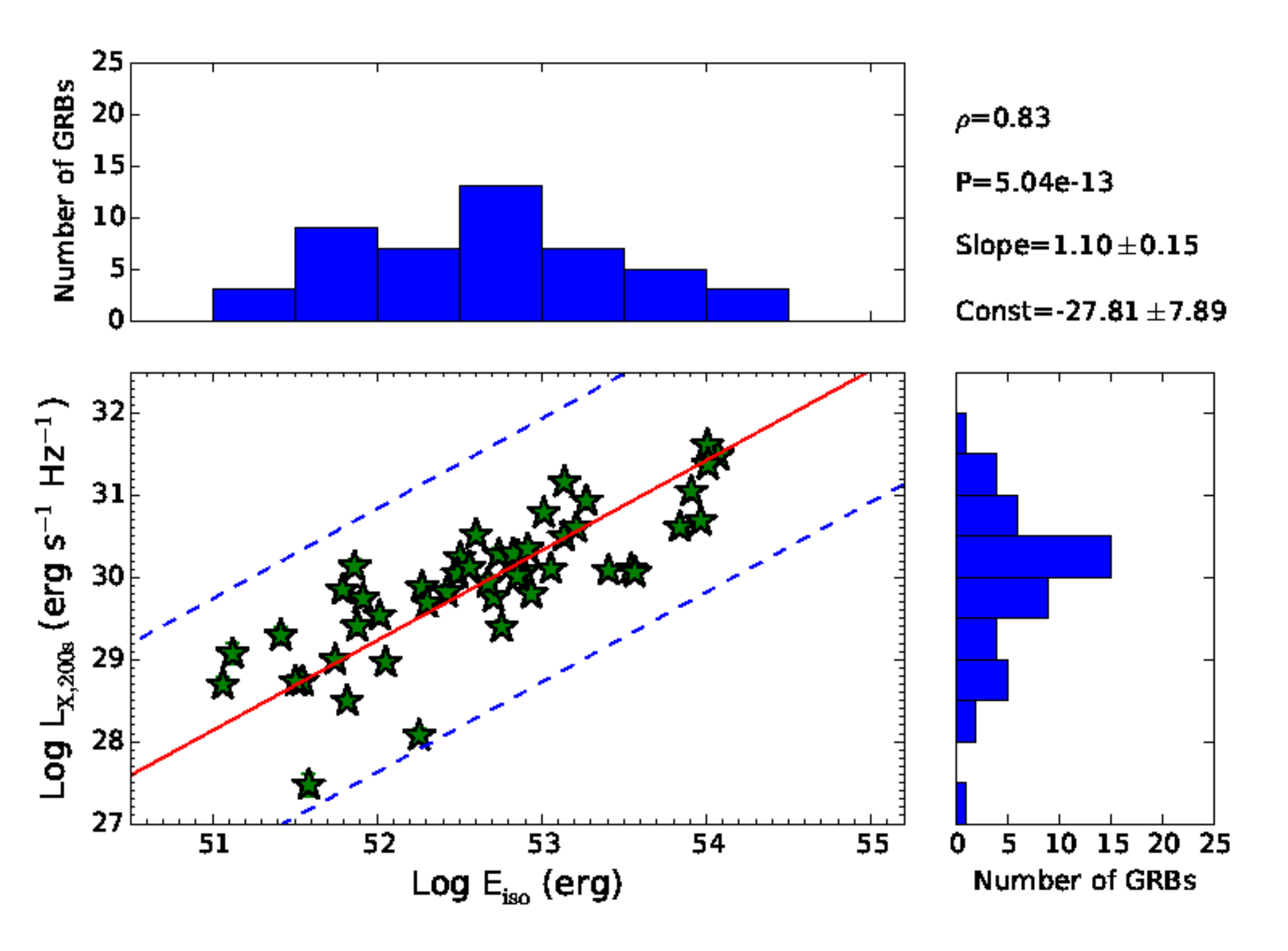}  
\caption{\footnotesize Left panel: $\log L_{O,200\rm{s}}$-$\log E_{\gamma,prompt}$ ($E_{iso}$ in the picture) relation from \cite{oates15}. Right 
  panel: $\log L_{X,200\rm{s}}$-$\log E_{\gamma,prompt}$ ($E_{iso}$ in the picture) relation from \cite{oates15}.}
\label{fig:liang2}
\end{figure}

Similarly, \cite{Dainotti11b} analyzed the relation between 
$\log L_{X,a}$ and $\log E_{\gamma,prompt}$ 
using the light curves of 66 LGRBs from the Swift BAT+XRT repository, http://www.swift.\\ac.uk/burst$_{-}$analyser/. 
Their sample has been divided into two subsamples: E4 formed of 62 LGRBs and E0095 consisting of 8 LGRBs, assuming 
$\sigma_E$ as a parameter representing the goodness of the fit.
For the E4 subsample it was found:

\begin{equation}
\log L_{X,a}=28.03^{+2.98}_{-2.97}+0.52_{-0.06}^{+0.07}\times \log E_{\gamma,prompt},
\end{equation}

with $\rho=0.43$ and $P=1.4\times10^{-5}$, while for the E0095 subsample

\begin{equation}
\log L_{X,a}=29.82^{+7.11}_{-7.82}+0.49_{-0.16}^{+0.21}\times \log E_{\gamma,prompt}, 
\end{equation}

with $\rho=0.83$ and $P=3.2 \times 10^{-2}$. 
Thus, it was concluded that the small error energy sample led to a higher relation 
and to the existence of a subset of GRBs which can yield a ``standardizable candle".
Furthermore, since $\log L_{X,a}$ and $\log T_{X,a}^*$ are strongly correlated, and the slope is roughly -1, the energy
reservoir of the plateau is roughly constant. Since $\log E_{\gamma,peak}$ and $\log E_{\gamma,prompt}$ are both linked with 
$\log L_{X,a}$,
then the $\log E_{\gamma,peak}-\log E_{\gamma,prompt}-\log E_{X,plateau}$ relation is straightforward. 
For its modification
taking into account $\log E_{\gamma,iso}$ of the whole X-ray light curves see \cite{Bernardini13}.
As further confirmations of the $L_{X,a}-E_{\gamma,prompt}$ relation, \cite{davanzo12} and \cite{margutti13} 
found a relation between $\log L_{X,a}$ and $E_{\gamma,prompt}$ with slope $b\sim1$ and $\rho\approx 0.70$, 
using 58 and 297 Swift LGRBs respectively. \\
Furthermore, \cite{berger14} studied the relation between the X-ray luminosities at 11 hours, $L_{X,11}$, and $E_{\gamma,peak}$, and the relation between 
the optical luminosity at 7 hours $L_{O,7}$ and $E_{\gamma,peak}$ for a sample of 70 SGRBs
and 73 LGRBs detected mostly by Swift. He found that the observed relations are flatter 
than the ones simulated by \cite{kann10}, see the middle and right panels of Fig. \ref{fig:berger}. \\
Regarding the relation between $E_{\gamma,prompt}$ and the optical luminosities, \cite{oates15} analyzed the relation 
between $L_{O,200\rm{s}}$ or $L_{X,200\rm{s}}$ and $\log E_{\gamma,prompt}$ with a sample of 48 LGRBs. They claimed
a strong connection between prompt and afterglow phases, see Fig. \ref{fig:liang2} and Table \ref{tbl4}. This relation permits to 
study some important spectral characteristics of GRBs, the optical and X-ray components of the radiation process and 
the standard afterglow model. In Table \ref{tbl4}, a summary of the relations described in this section is shown.\\
Regarding the physical interpretation of the $L_{X,afterglow}-E_{\gamma,prompt}$ relation, 
\cite{gehrels08} underlined that the optical
and X-ray radiation are characterized by $\beta_{OX,a}\approx0.75$. This value matches the slow cooling case, 
important at 11 hours, when the electron distribution power law index is $p=2.5$ for $\nu_m < \nu_O < \nu_X < \nu_c$.\\
\cite{oates15} pointed out that within the standard afterglow model, the $\log E_{\gamma,prompt}-(\log L_{O,200\rm{s}}, \log L_{X,200\rm{s}})$ 
relations are expected.
However, the slopes of the simulated and observed relations are inconsistent at $>$ 3 $\sigma$ due to values set for the $\eta$
parameter. If the distribution of the efficiencies is not sufficiently narrow the relation will be more disperse. Thus, the 
simulations repeated with $\eta=0.1$ and $\eta=0.9$ gave, anyway, incompatible results between the simulated and observed slopes at 
$>$ 3 $\sigma$.

\begin{table}[htbp]
\footnotesize
\begin{center}
\begin{tabular}{|c|c|c|c|c|c|c|c|}
\hline
Correlations & Author & N & Slope& Norm & Corr.coeff. & P \\
\hline
$ F_{X,1\rm{d}}- S_{\gamma,prompt}$ & Berger (2007)& 16& $1.01^{+0.09}_{-0.09}$&& 0.86 &$5.3\times10^{-5}$\\
\hline
$F_{X,11}-S_{\gamma,prompt}$ & Gehrels et al. (2008) & 111 & $0.63^{+0.04}_{-0.04}$&$2.11^{+0.21}_{-0.21}$&0.53&$4\times10^{-9}$ \\
 &Gehrels et al. (2008) & 10  &$0.36^{+0.17}_{-0.17}$&$0.06^{+1.07}_{-1.07}$&0.35&0.31\\
\hline
$ F_{O,11}-E_{\gamma,prompt}$ & Nysewander et al. (2009)&421&$\sim 1$&&& \\
$ F_{O,11}-E_{\gamma,prompt}$ & Nysewander et al. (2009)&37&$\sim 1$&&& \\
\hline
$F_{X,11}- E_{\gamma,prompt}$ & Nysewander et al. (2009)&421&$\sim 1$&&&\\
$F_{X,11}- E_{\gamma,prompt}$ & Nysewander et al. (2009)&37&$\sim 1$&&&\\
\hline
$ F_{O,a}-E_{\gamma,prompt}$ & Panaitescu\&Vestrand (2011)&37&1.67&&0.75&$10^{-7.3}$ \\
\hline
$L_{X,1\rm{d}}-E_{\gamma,prompt}$ & Berger (2007) &13& $1.13^{+0.16}_{-0.16}$&$43.43^{+0.20}_{-0.20}$& 0.94 &$3.2\times10^{-6}$\\
\hline
$ L_{O,peak}-E_{\gamma,prompt}$ & Liang et al. (2010) &32& $1.40^{+0.08}_{-0.08}$&$0.83^{+0.15}_{-0.15}$& 0.87 &$10^{-4}$\\
\hline
$L_{O,1\rm{d}}- E_{\gamma,prompt}$ & Kann et al. (2010) &76&0.36&&&\\
\hline
$ L_{X,a}- E_{\gamma,prompt}$ & Dainotti et al. (2011b) & 62& $0.52_{-0.06}^{+0.07}$&$28.03^{+2.98}_{-2.97}$&0.43&$1.4\times10^{-5}$\\
& Dainotti et al. (2011b) & 8 & $0.49_{-0.16}^{+0.21}$& $29.82^{+7.11}_{-7.82}$&0.83&$3.2 \times10^{-2}$\\
& D'Avanzo et al. (2012)& 58 & $\sim 1$& &$\approx 0.70$&\\
& Margutti et al. (2013)& 297 & $\sim 1$& &$\approx 0.70$ &\\
\hline
$L_{X,11}-E_{\gamma,prompt}$ & Berger (2014)&73&0.72&44.75&&\\
& Berger (2014)&70&0.83&43.93&&\\
\hline
$ L_{O,7}-E_{\gamma,prompt}$ & Berger (2014)&73&0.73&43.70&&\\
& Berger (2014)&70&0.74& 42.84&&\\
\hline
$ L_{X,200\rm{s}}- E_{\gamma,prompt}$ & Oates et al. (2015)&48&$1.10^{+0.15}_{-0.15}$&$-27.81^{+7.89}_{-7.89}$&0.83&$5.04\times10^{-13}$\\
\hline
$L_{O,200\rm{s}}-E_{\gamma,prompt}$ & Oates et al. (2015)&48&$1.09^{+0.13}_{-0.13}$&$-25.27^{+6.92}_{-6.92}$&0.76&$4.51\times10^{-10}$\\
\hline
\end{tabular}
\caption{\footnotesize Summary of the relations in this section. The first column represents the relation in log scale,
the second one the authors, and the third one the number of GRBs in the used sample. Afterwards, the fourth and fifth columns are
the slope and normalization of the relation and the last two columns are the correlation coefficient and the chance probability, P.}
\label{tbl4}
\end{center}
\end{table}

\subsection{The \texorpdfstring{$L_{X,a}-L_{O,a}$}{Lg} relation and its physical interpretation}
\begin{figure}[htbp]
\includegraphics[width=5.5cm,height=4.9cm]{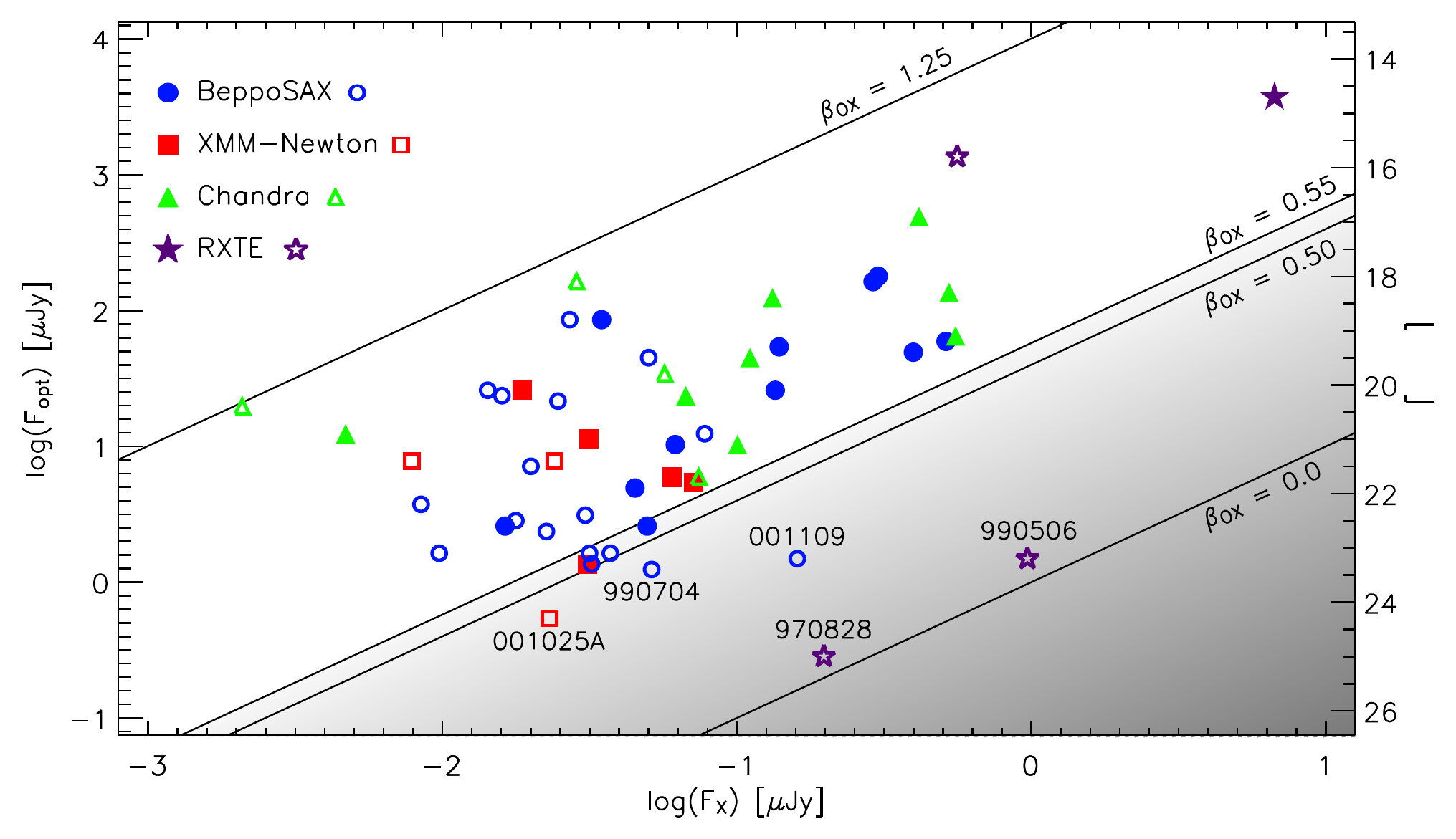}
\includegraphics[width=5.5cm,height=5.2cm]{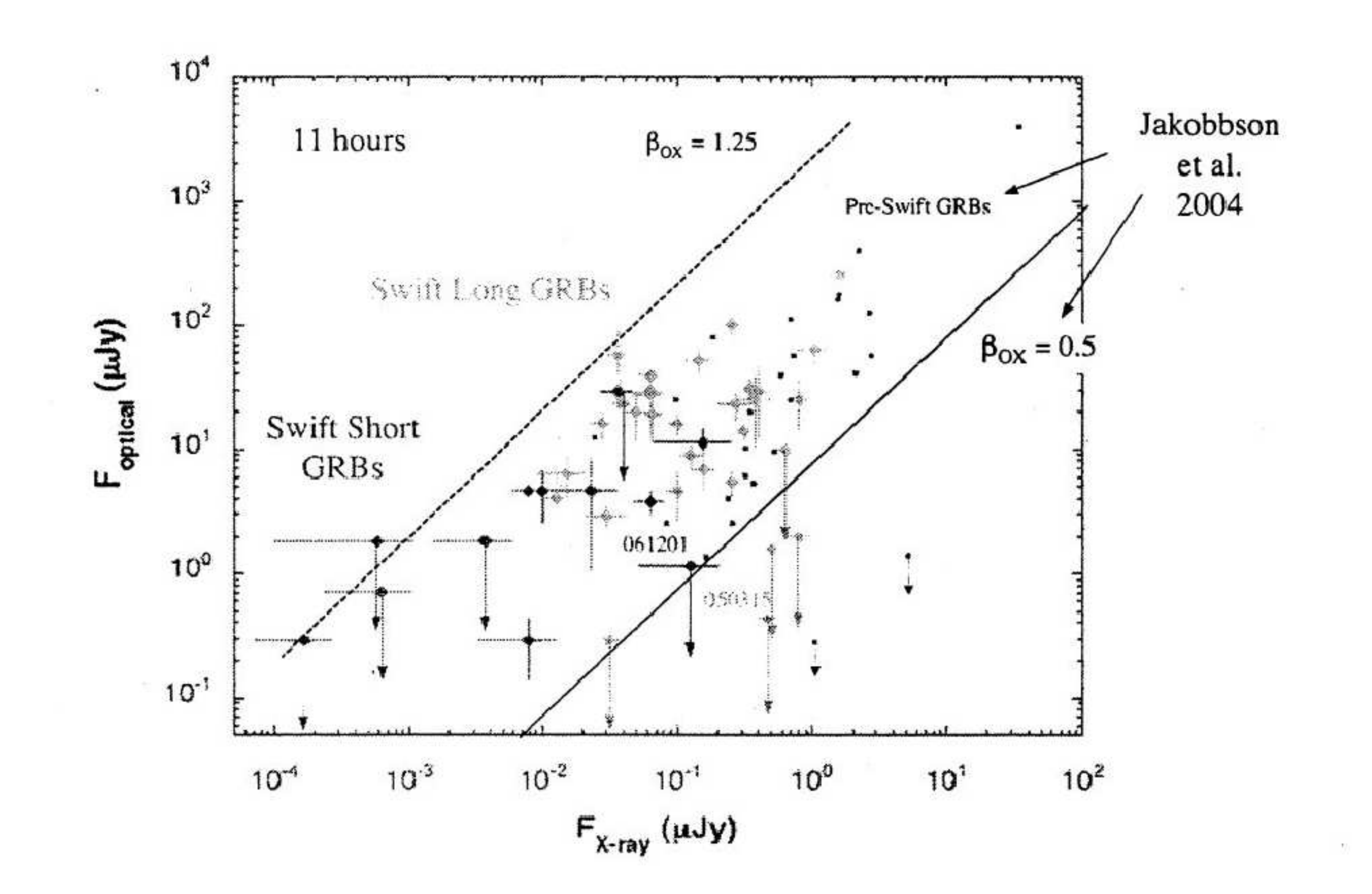}
\includegraphics[width=5.2cm,height=5.2cm,angle=0]{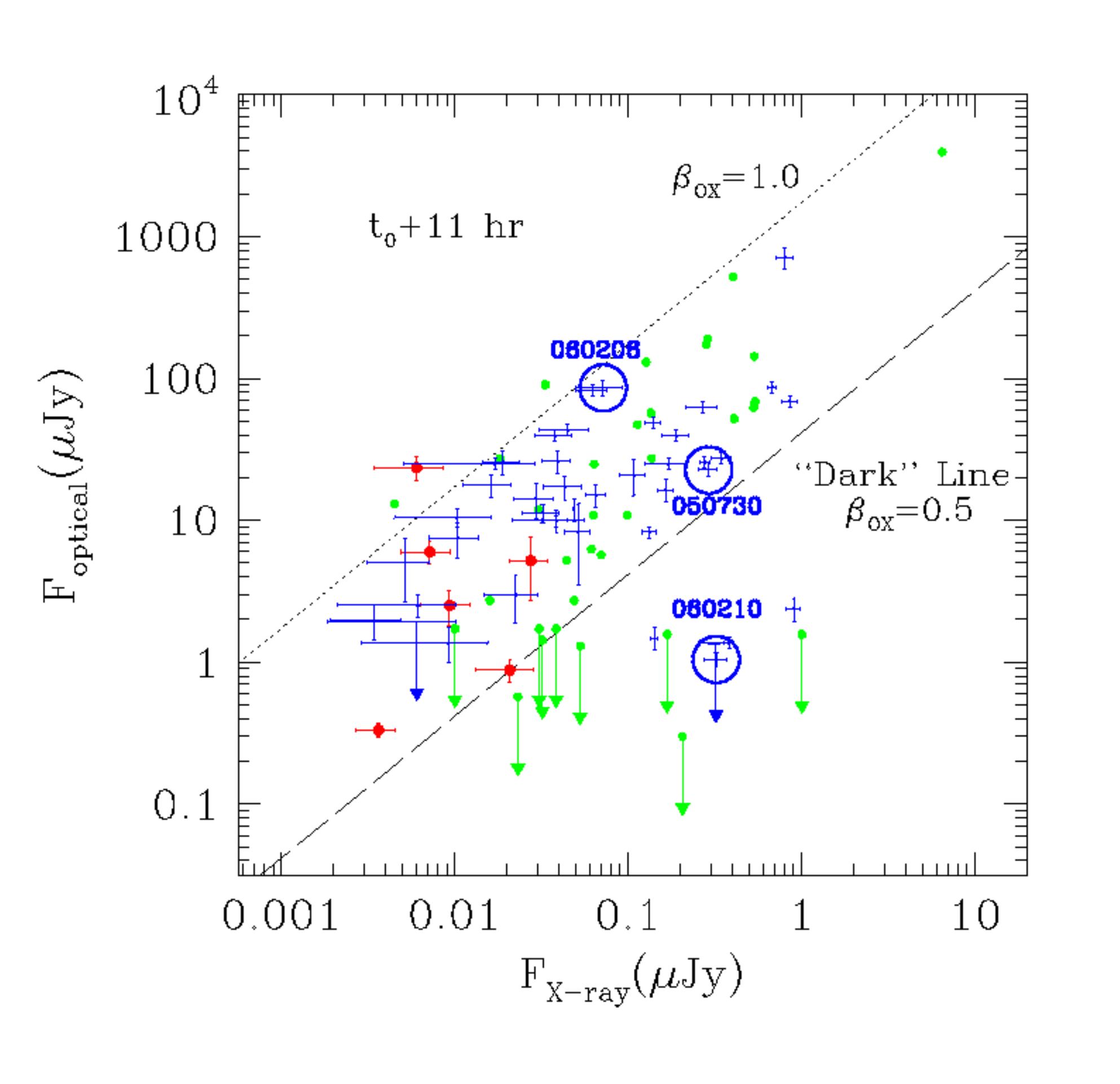}  
\caption{\footnotesize Left panel: $\log F_{O,11}$-$\log F_{X,11}$ ($F_{opt}$ and $F_X$ respectively in the
plot) distribution for the data set from \cite{jakobsson04}. 
Filled symbols show optical detections while open symbols represent upper limits. Lines of constant $\beta_{OX,a}$
are displayed with the corresponding value. Dark bursts are those which have $\beta_{OX,a}< 0.5$. 
 Middle panel: $F_{X,11}$-$F_{O,11}$ relation for Swift SGRBs and LGRBs from Gehrels (2007). Comparison is made to pre-Swift GRBs 
 and to lines of optical to X-ray spectral index from \cite{jakobsson04}. The grey points indicate LGRBs,
the black points represent SGRBs and the small black points without error bars
are the pre-Swift GRBs. Right panel: $F_{O,11}$-$F_{X,11}$ relation for Swift SGRBs (shown in red) and LGRBs (shown in blue) 
from \cite{gehrels08}.  The three circled bursts are those
for which $z>3.9$. The pre-Swift GRBs taken from \cite{jakobsson04} are presented in green.
Also the dark burst separation line 
$\beta_{OX,a} = 0.5$ \citep{jakobsson04} and a line showing $\beta_{OX,a} = 1.0$ are represented.}
 \label{OX_comp}
\end{figure}

In the observed frame, \cite{jakobsson04} studied the $\log F_{O,11}$ versus $\log F_{X,11}$ distribution, in 
the optical R band and in the $2-10$ keV band respectively, using all known GRBs with a detected X-ray 
afterglow, see the left panel of Fig. \ref{OX_comp}. 
Different from the previous definition of dark bursts (where dark bursts were simply defined as
those bursts in which the optical transient is not observed), they defined these bursts as 
GRBs where the optical-to-X-ray
spectral index, $\beta_{OX,a}$, is shallower than the X-ray spectral index minus 0.5, $\beta_{X,a}-0.5$. 
They found out 5 dark bursts among 52 observed by Beppo-SAX\footnote{Beppo-SAX, (1996-2003), was an Italian-Dutch 
satellite
capable of simultaneously observing targets over more than 3 decades of energy,
from $0.1$ to $600$ keV with relatively large area, good (for that time) energy resolution and imaging capabilities (with a spatial
resolution of $1$ arcminute between $0.1$ and $10$ keV). The instruments on board Beppo-SAX are Low Energy Concentrator Spectrometer
(LECS), Medium Energy Concentrator Spectrometer (MECS), High Pressure Gas Scintillation Proportional Counter (HPGSPC), Phoswich 
Detector System (PDS) and Wide Field Camera (WFC, from $2-30$ keV and from $100-600$ keV). The first four instruments point to the 
same direction allowing observations in the broad energy range (0.1-300 keV). With the WFC it was possible to model the afterglow 
as a simple power law, mainly due to the lack of observations during a certain period in the GRB light curve.}.
This analysis aimed at distinguishing dark GRBs through Swift.
\cite{gehrels07} and \cite{gehrels08} confirmed the results  
using a data sample of 19 SGRBs and 37 LGRBs$+$6 SGRBs respectively, see the middle and right panels of 
Fig. \ref{OX_comp}. In particular, \cite{gehrels08} obtained a slope $b=0.38 \pm 0.03$
for LGRBs and $b=0.14\pm0.45$ for SGRBs.

\begin{figure}[htbp]
\includegraphics[width=0.47\hsize]{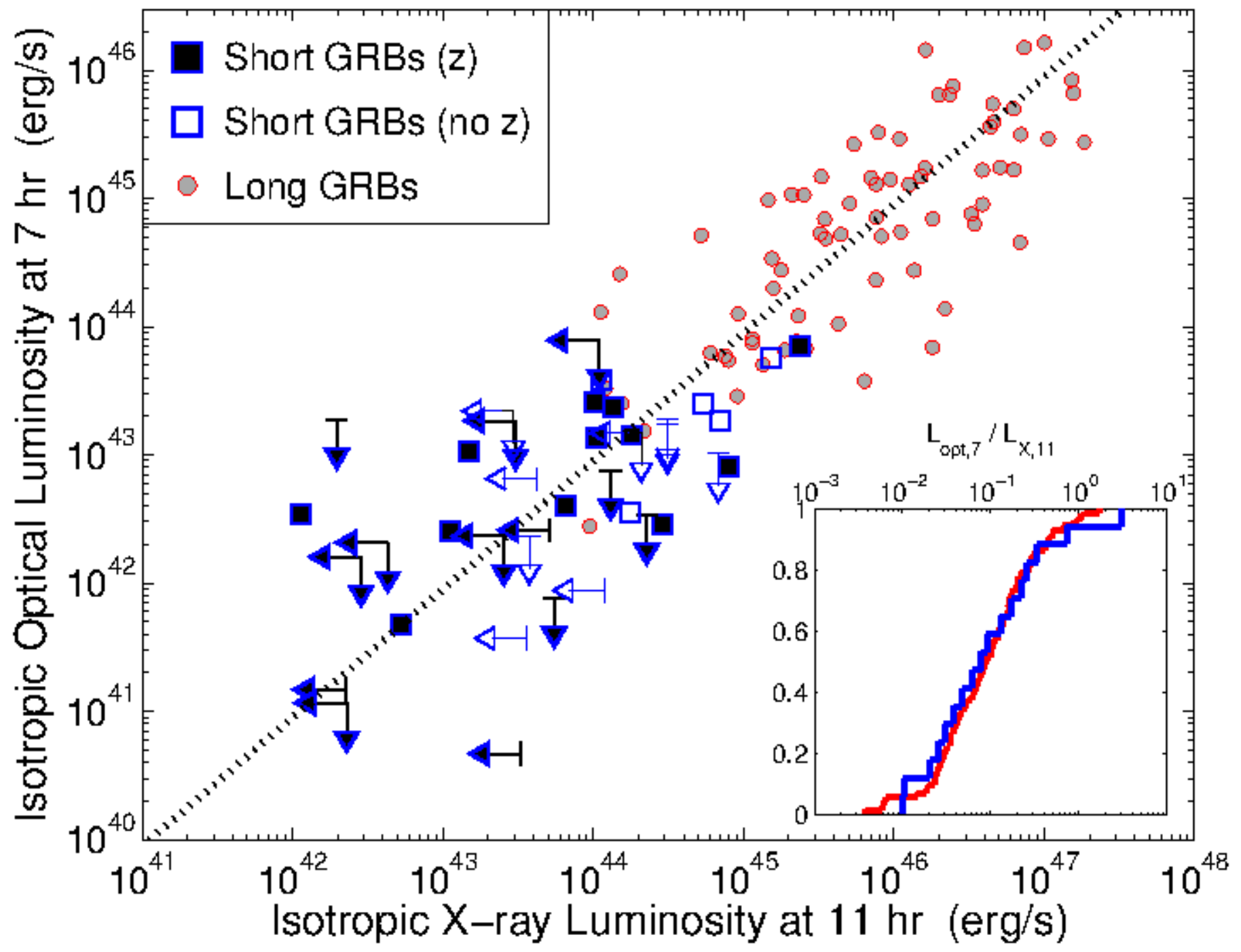}
\includegraphics[width=0.5\hsize]{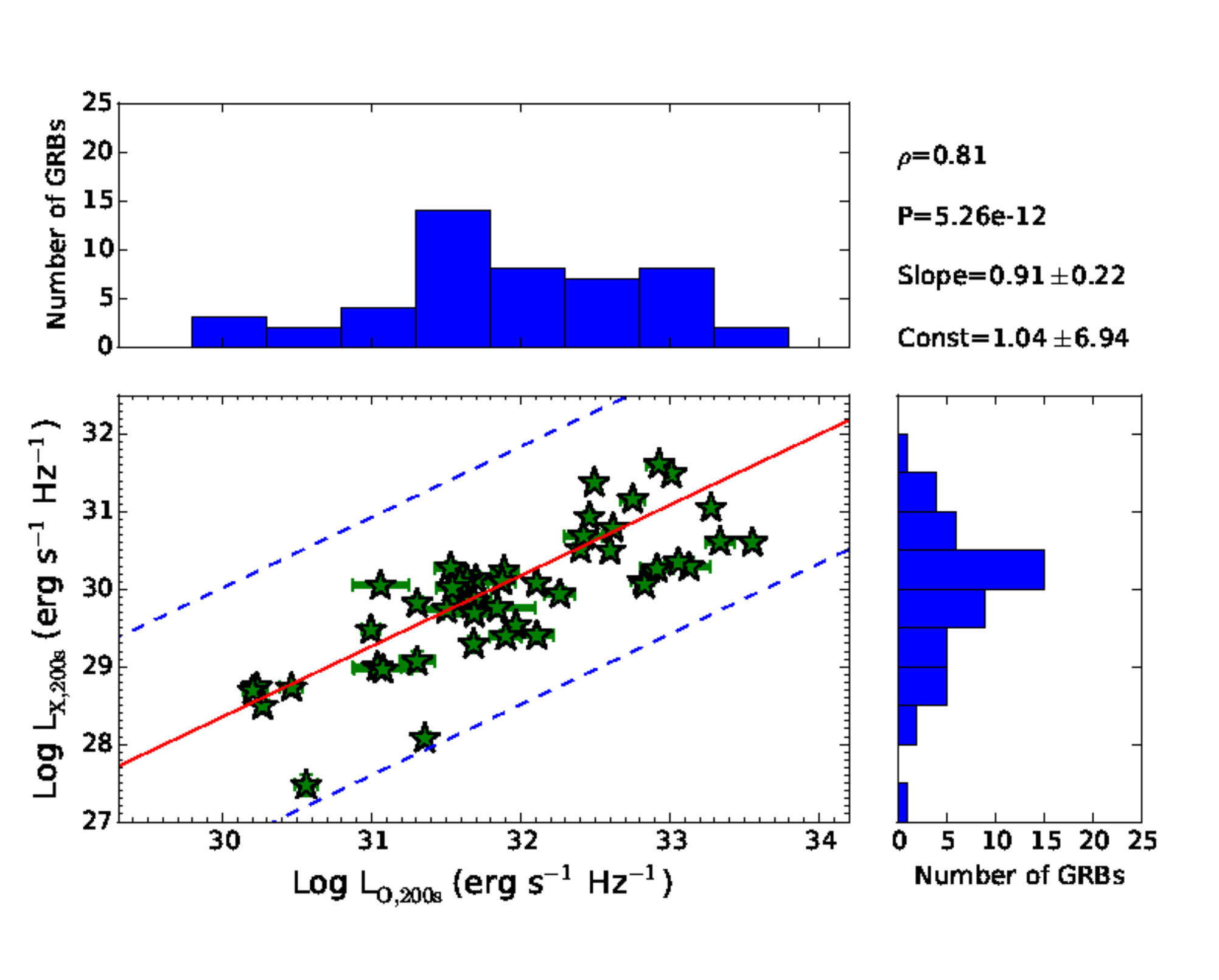}
\caption{\footnotesize Left panel: ``$L_{O,7}$ vs.
 $L_{X,11}$ from \cite{berger14}. The dotted black line marks a linear relation, expected for
  $\nu_X\sim \nu_c$.  The inset shows the distribution of the ratio
  $L_{O,7}/L_{X,11}$, indicating that both SGRBs and LGRBs
  exhibit a similar ratio, and that in general $L_{O,7}/
  L_{X,11}\sim 1$, indicative of $\nu_X\sim \nu_c$ for SGRBs". 
  Right panel: ``$\log L_{O,200\rm{s}}$ vs. $\log L_{X,200\rm{s}}$ from \cite{oates15}. 
The red solid line represents the best fit regression and the blue dashed line 
represents 3 times the RMS deviation. In the top right corner, it is given
 $\rho$ and $P$ and it is provided the best-fit slope and constant determined by linear regression".}
\label{fig:berger14}
\end{figure}

Instead in the rest-rest frame, \cite{berger14} studied the relation between $L_{O,7}$ and $L_{X,11}$ on 70 SGRBs 
and 73 LGRBs, finding some similarities between SGRBs and LGRBs and a central value 
$<L_{O,7}/L_{X,11}> \approx 0.08$, see the left panel of Fig. \ref{fig:berger14}.\\
\cite{oates15} improved their study. They analyzed a similar relation with a sample of 48 LGRBs, 
but using $L_{O,200\rm{s}}$ and $L_{X,200\rm{s}}$, see the right panel of Fig. \ref{fig:berger14}. 
The slope obtained has a value $b=0.91 \pm 0.22$.\\
This relation helps to explore the synchrotron spectrum of GRBs and to obtain some constraints on the circumburst 
medium for both LGRBs and SGRBs. In Table \ref{tblgehrels} a summary of the relations
described in this section is displayed.

\begin{table}[htbp]
\footnotesize
\begin{center}
\begin{tabular}{|c|c|c|c|c|c|c|c|}
\hline
Correlations & Author & N& Slope& Norm & Corr.coeff.& P \\
\hline
$F_{X,11}-F_{O,11}$ &Gehrels et al. (2008) & 6 &$0.14\pm0.45$&$0.72 \pm 0.94$&$0.06$&$0.68$ \\
 &&37&$0.38 \pm 0.03$&$1.62\pm0.04$&$0.44$&$0.006$\\
  \hline
$L_{X,11}-L_{O,7}$ & Berger (2014) &70&0.08&&& \\
 &&73&0.08&&&\\
\hline
$L_{X,200\rm{s}}-L_{O,200\rm{s}}$ & Oates et al. (2015)&48&$0.91 \pm 0.22$&$1.04\pm6.94$&0.81&$5.26\times10^{-12}$ \\
\hline
\end{tabular}
\caption{\footnotesize Summary of the relations in this section. The first column represents the relation in log scale,
the second one the authors, and the third one the number of GRBs in the used sample. Afterwards, the fourth and fifth columns are
the slope and normalization of the relation and the last two columns are the correlation coefficient and the chance probability, P.}
\label{tblgehrels}
\end{center}
\end{table}

Regarding the physical interpretation of the $L_{X,a}-L_{O,a}$ relation, \cite{berger14} showed that, 
in the context of the synchrotron model, the comparison of $L_{O,7}$ and $L_{X,11}$ indicated that usually $\nu_c$ is
near or higher than the X-ray band. Indeed, LGRBs have often
greater circumburst medium densities (about $50$ times greater than SGRBs) and therefore $\nu_c \sim \nu_X$.

\subsection{The \texorpdfstring{$L_X(T_a)-L_{\gamma,iso}$}{Lg} relation}\label{Dainotti2011b}
\begin{figure}[htbp]
\centering
\includegraphics[width=5.6cm,height=5.85cm,angle=0,clip]{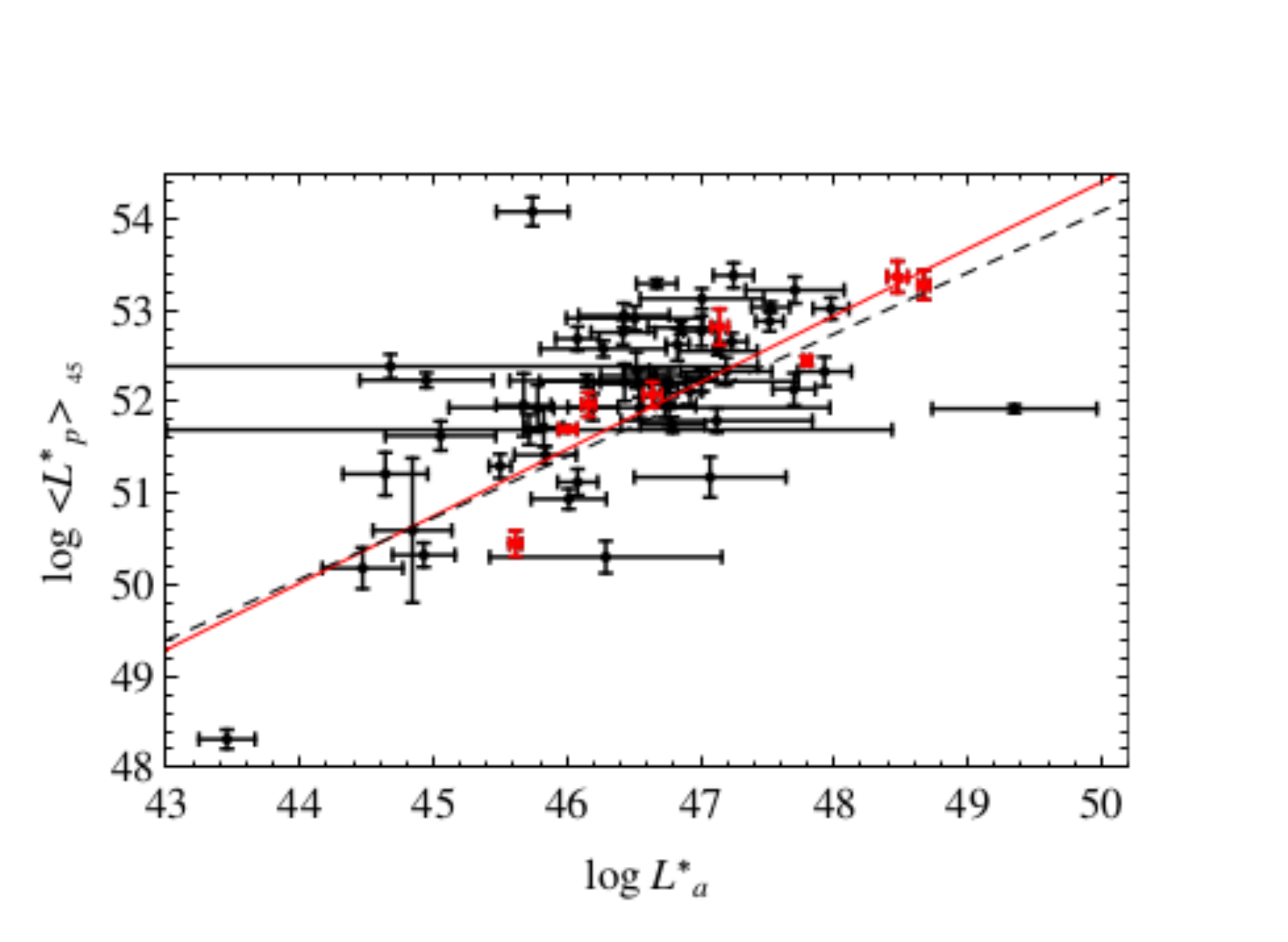}   
\includegraphics[width=5.05cm,height=4.85cm,angle=0,clip]{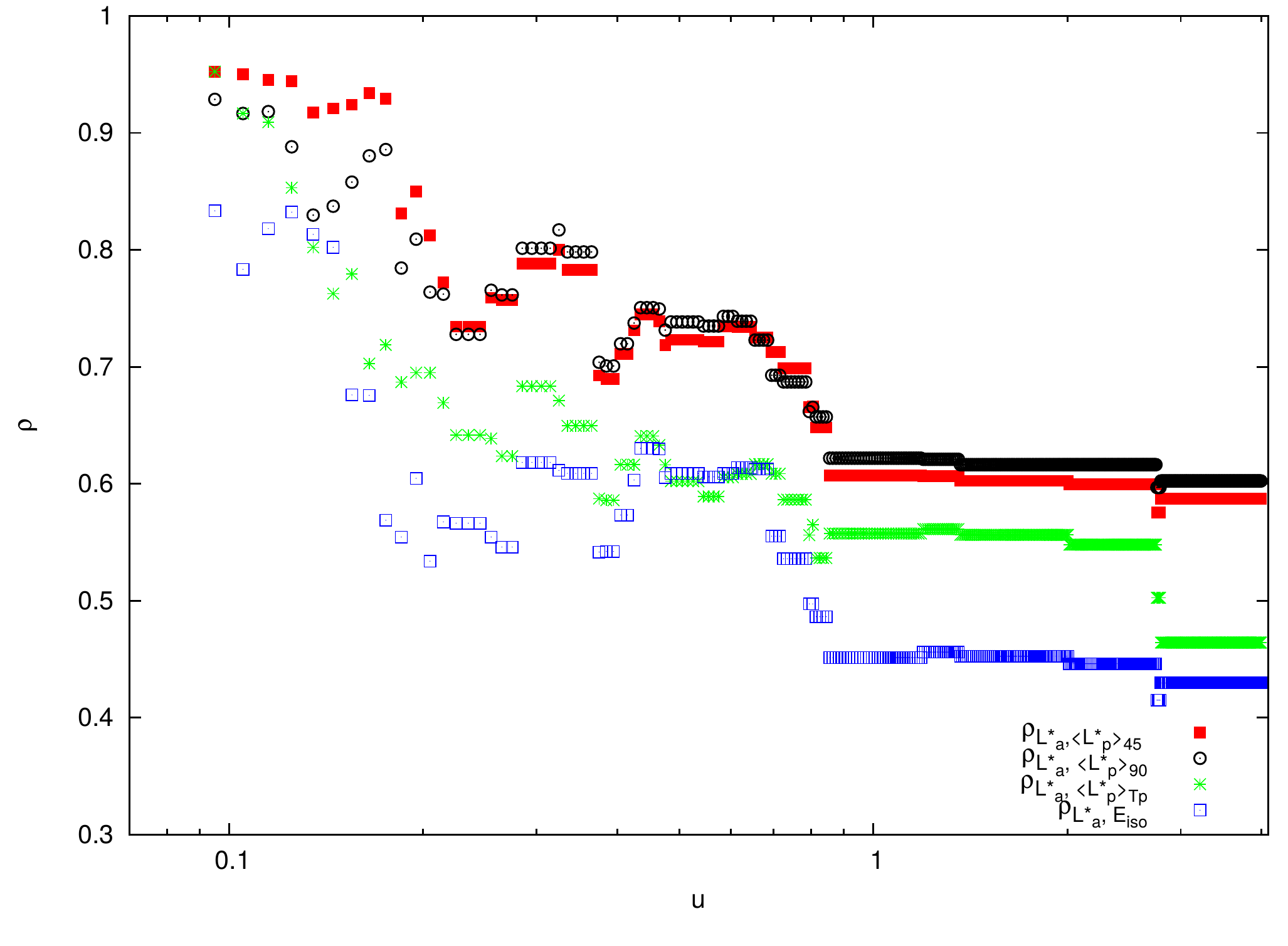} 
\includegraphics[width=5.6cm,height=5cm,angle=0,clip]{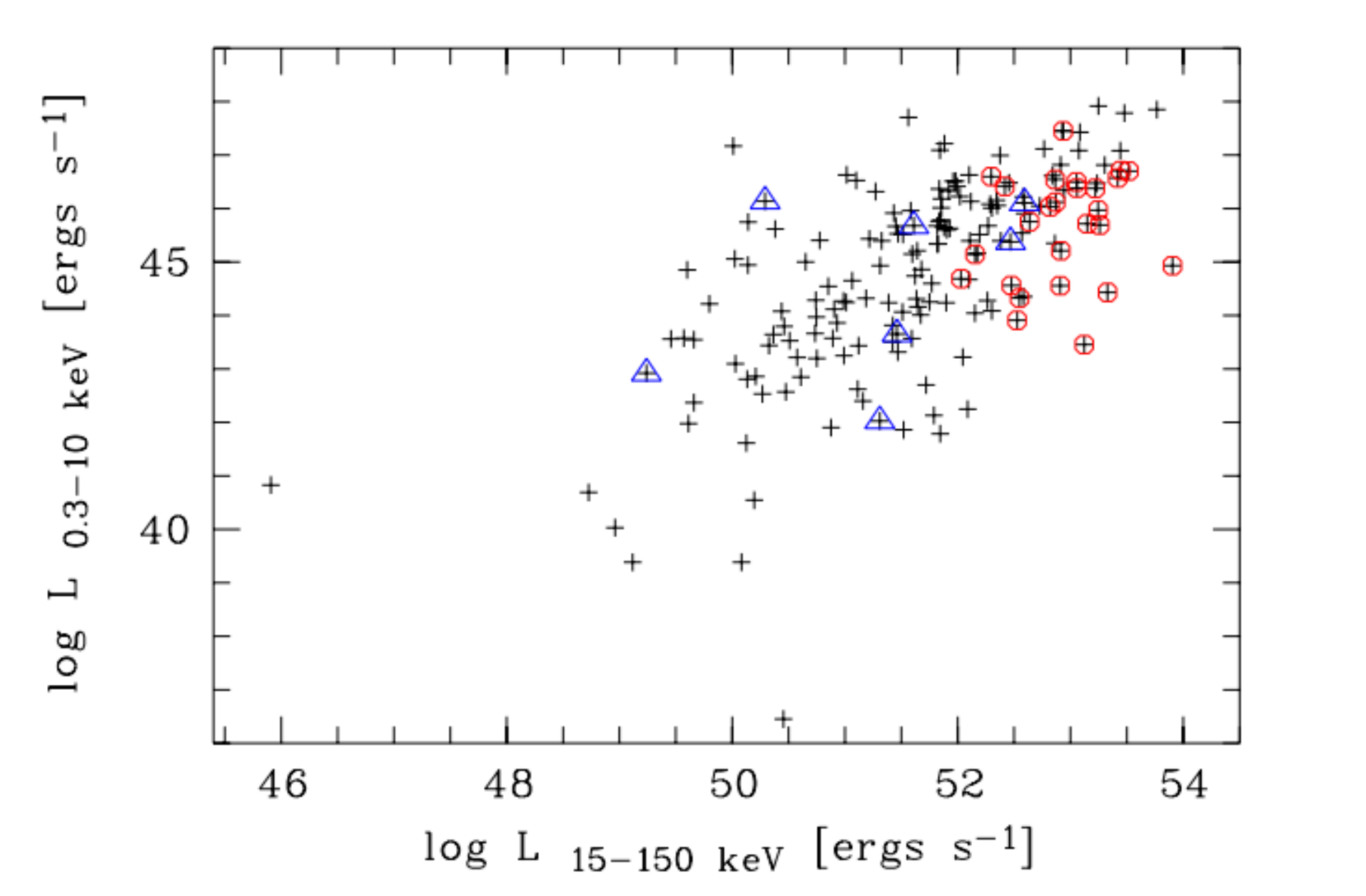} 
   \caption{\footnotesize Left panel: $\log L_{X,a}$ vs. $\log <L_{\gamma,iso}>_{45}$ relation ($\log L^*_{X,a}$ and
   $\log L^*_p$ respectively) for the E4 data set from \cite{Dainotti11b},
   with the fitted relation dashed line in black. The red line is fitted to the 8 lowest error (red) points of the 
   E0095 subset.
   Middle panel: correlation coefficients $\rho$ for $\log\ L_{X,a}-\log\ <L_{\gamma,iso}>_{45}$ (red squares), 
   $\log\ L_{X,a}-\log\ <L_{\gamma,iso}>_{90}$ (black circles), $\log\ L_{X,a}-\log\ <L_{\gamma,iso}>_{T_{X,p}}$ (green asterisks) and 
   $\log\ L_{X,a}-\log\ E_{\gamma,prompt}$
   (blue squares) relations, obtained using the LGRB subset with the maximum \textcolor{red}{u $=$ }$\sigma_E$ from \cite{Dainotti11b}. 
   Right panel: Luminosity in the 0.3-10 keV XRT band ($L_{X,a}$) vs. luminosity in the 
   15-150 keV BAT band ($L_{\gamma,iso}$) from \cite{grupe2013}. 
   Short bursts are represented by triangles and high-redshift (z $> 3.5$) bursts by circles.}
   \label{fig:16}
\end{figure}

In \cite{Dainotti11b} the connections between the physical properties of the prompt emission and $\log L_{X,a}$ were analyzed
using a sample of 62 Swift LGRBs. A relation was found between $\log L_{X,a}$ in the XRT band and the isotropic prompt luminosity,
$\log < L_{\gamma,iso}>_{45}\equiv \log (E_{\gamma,prompt}/T_{45})$, in the BAT energy band,
see the left panel of Fig. \ref{fig:16}. This relationship can be fitted with the following equation:

 \begin{equation}
\log\ L_{X,a} = 20.58^{+6.66}_{-6.73} + 0.67^{+0.14}_{-0.15} \times \log\ <L_{\gamma,iso}>_{45}, 
\end{equation}

obtaining $\rho=0.59$ and $P=7.7\times10^{-8}$.
In this paper $\log L_{X,a}$ was related to several prompt luminosities defined using different time scales, 
such as $T_{90}$, $T_{45}$ (the time in which the 
45\% between 5\%-50\% of radiation is emitted in the prompt emission), and $T_{X,p}$ (the time at 
the end of the prompt emission within the W07 model).
Furthermore, the E4 (defined in Table \ref{abbreviations}) subsample of $62$ LGRBs with known $z$ from a sample of
77 Swift LGRBs and the E0095 subsample of 8 GRBs with smooth light curves were used, see black and
red points in the left panel of Fig. \ref{fig:16}.\\
Therefore, it has been shown that the GRB subsample with the strongest correlation coefficient 
for the LT relation also implies the tightest prompt-afterglow relations. 
This subsample can be used as a standard one for astrophysical and cosmological studies.\\
In the middle panel of Fig. \ref{fig:16}, the correlation coefficients $\rho$ are shown for the following distributions: $(\log <L_{\gamma,iso}>_{45},\log <L_{\gamma,iso}>_{90},
\log <L_{\gamma,iso}>_{T_{X,p}},\log E_{\gamma,prompt})-\log L_{X,a}$, represented by different colours,
namely red, black, green and blue respectively.\\
No significant relations for the IC bursts have been found out. However, the paucity of the data does not allow a 
definitive statement.
From this analysis, it is clear that the inclusion of the IC GRB class does not strengthen the existing relations. 

\begin{table}[htbp]
\footnotesize
\begin{center}
\begin{tabular}{|c|l c|c c|}
\hline
 & \multicolumn{2}{c|}{E4} & \multicolumn{2}{c|}{E0095} \\
\hline
Correlations & \multicolumn{2}{l|}{\hspace{20pt}$\rho$ \hspace{60pt}  (b, a)} & \multicolumn{2}{l|}{\hspace{20pt}$\rho$ \hspace{60pt} (b, a)} \\
\cline{2-5} & \multicolumn{2}{l|}{\hspace{95pt}{P}} & \multicolumn{2}{l|}{\hspace{95pt}{P}} \\
\hline
$ L_{X,a}- <L_{\gamma,iso}>_{45}$ & 0.59 & $(0.67_{-0.15}^{+0.14}, 20.58 _{-6.73}^{+6.66})$  & 0.95 & $(0.73 _{-0.11}^{+0.16}, 17.90_{-6.0}^{+5.29})$  \\
& 0.62 &  $7.7 \times 10^{-8}$  & 0.90& $2.3 \times 10^{-3}$ \\
$ L_{X,a}- <L_{\gamma,iso}>_{90}$ & 0.60 & $(0.63 _{-0.16}^{+0.15}, 22.05_{-7.31}^{+7.14})$ & 0.93 &  $(0.84 _{-0.12}^{+0.11}, 11.86_{-3.44}^{+3.43})$  \\
& 0.62 & $7.7 \times 10^{-8}$ &0.94 & $2.7 \times 10^{-3}$\\
$ L_{X,a}- <L_{\gamma,iso}>_{T_{X,p}}$ & 0.46 & $(0.73_{-0.14}^{+0.09}, 16.61_{-4.35}^{+4.35})$  & 0.95 & $(0.93_{-0.23}^{+0.20},7.70_{-3.46}^{+3.47})$\\
& 0.56 & $2.21 \times 10^{-6}$ &0.90 & $2.3 \times 10^{-3}$\\
$ L_{X,a}- E_{\gamma,prompt}$ & 0.43 & $(0.52 _{-0.06}^{+0.07}, 28.03_{-2.97}^{+2.98})$  & 0.83 & $(0.49 _{-0.16}^{+0.21}, 29.82 _{-7.82}^{+7.11})$ \\
 &0.52 & $1.4 \times 10^{-5}$ & 0.75 & $3.2\times 10^{-2}$ \\
$ T^*_{X,a}- E_{\gamma,prompt}$ &  -0.19 & $(-0.49_{-0.08}^{+0.09},54.51_{-0.30}^{+0.37})$  &  -0.81 &$(-0.96_{-0.22}^{+0.21}, 54.67_{-0.69}^{+0.69})$\\
&-0.21 & $1.0 \times 10^{-1}$  & -0.69 & $5.8 \times 10^{-2}$\\
$ L_{X,a}- E_{\gamma,peak}$ & 0.54 & $(1.06_{-0.23}^{+0.53}, 43.88_{-1.00}^{+0.61})$  & $0.74$ & $(1.5_{-0.94}^{+0.65}, 43.10_{-2.26}^{+2.53})$\\
& 0.51& $2.2 \times 10^{-5}$ &0.80 & $1.7 \times 10^{-2}$\\
$ T^*_{X,a}- E_{\gamma,peak}$ &  -0.36 & $(-0.66_{-0.29}^{+0.20}, 4.96_{-0.80}^{+0.81})$ &  -0.74 &  $(-1.40_{-0.65}^{+0.66}, 7.04_{-1.77}^{+1.79})$\\
& -0.35& $5.2 \times 10^{-3}$ & -0.77 & $2.5 \times 10^{-2}$ \\
$ <L_{\gamma,iso}>_{45}- E_{\gamma,peak}$ & 0.81 & $(1.14_{-0.25}^{+0.22}, 49.27_{-0.60}^{+0.61})$ &  0.76 &   $(1.45_{-0.54}^{+0.26}, 48.48_{-1.04}^{+1.05})$\\
&0.67&  $2.6 \times 10^{-9}$ & 0.92& $1.2 \times 10^{-3}$\\
\hline
\end{tabular}
\caption{\footnotesize Correlation coefficients $\rho$, the respective relation fit line parameters (a, b), and the 
correlation coefficient $r$ with the respective random occurrence probability $P$, for the considered prompt-afterglow and 
prompt-prompt distributions in log scale from \cite{Dainotti11b}.}
\label{tbl9}
\end{center}
\end{table}

In general, this study pointed out that the plateau phase results connected to the inner engine. 
In addition, also relations between $\log L_{X,a}$ and several other prompt 
emission parameters were analyzed, including $\log E_{\gamma,peak}$ and the variability, $\log V$.
As a result, relevant relations are found between these quantities, except for the variability parameter, see 
Table \ref{tbl9}.

%%%%%
%%%

Finally, as shown in Table \ref{tbl9}, only
a very small relationship exists between $\log T^{*}_{X,a}-\log E_{\gamma,prompt}$ with $\rho=-0.19$. 
Also \cite{grupe2013} claimed the existence of relations between $\log <L_{\gamma,iso}>_{90}$ 
and $\log L_{X,a}$ (see the right panel of Fig. \ref{fig:16}) and between $\log <L_{\gamma,iso}>_{90}$ and 
$\log T^*_{X,a}$ using a sample of 232 GRBs. The latter can be derived straightforwardly from the 
$\log T^*_{X,a}-\log E_{\gamma,prompt}$ relation, 
being $\log <L_{\gamma,iso}>_{90}$ computed as $\log (E_{\gamma,prompt}/T_{90})$.

\subsection{The \texorpdfstring{$L_{X,peak}-L_X(T_a)$}{Lg} relation}\label{lpromptla}
\begin{figure}[htbp]
\includegraphics[width=0.495\hsize,angle=0]{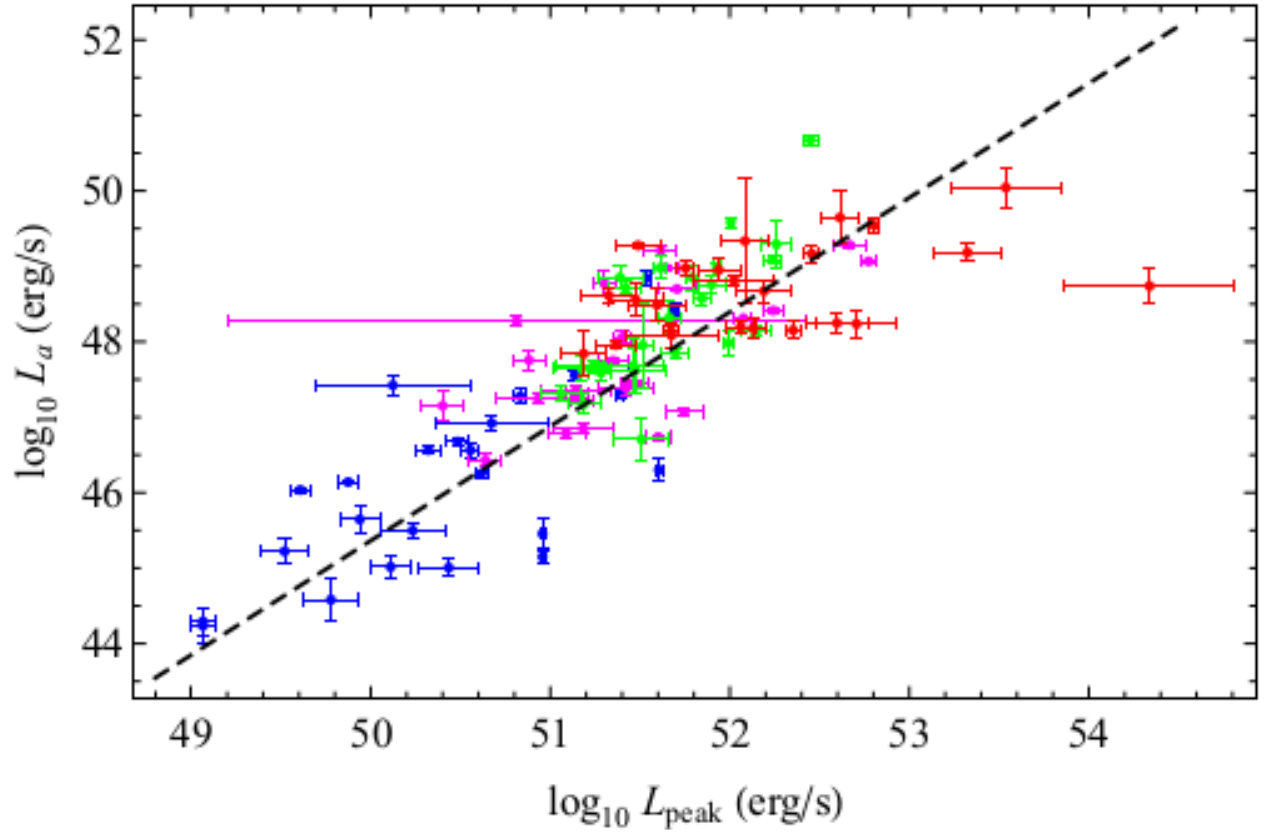}
\includegraphics[width=0.495\hsize,angle=0]{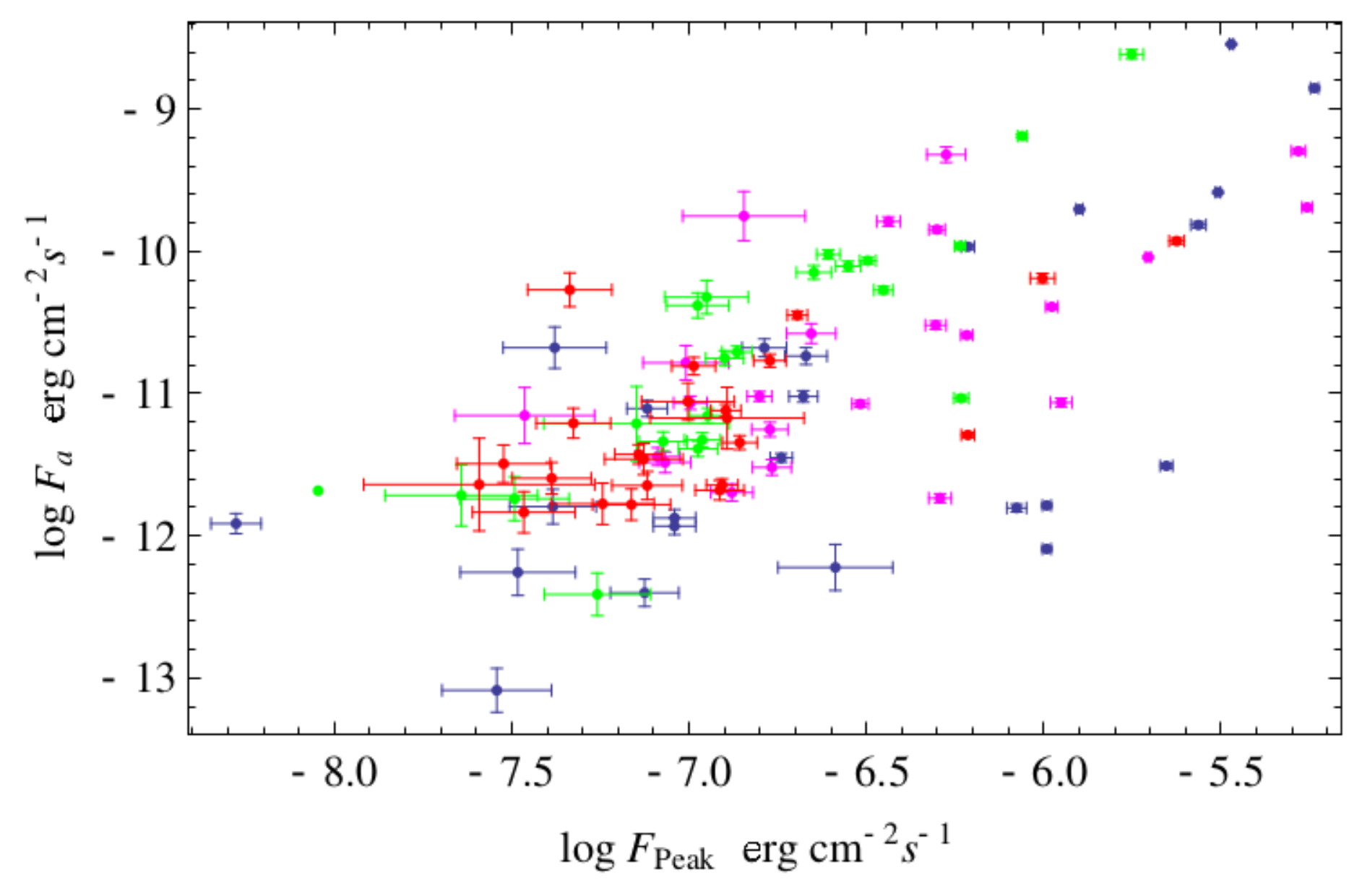}
\caption{\footnotesize Left panel: ``GRB distributions in redshift bins on the $\log L_{X,a}$-$\log L_{X,peak}$
 plane from \cite{dainotti15}, where $\log L_{X,peak}$ is computed using the approach used in the Second BAT Catalogue. The
sample is split into 4 different equipopulated redshift bins: $z \le 0.84$
(blue), $0.84 \leq z < 1.8$ (magenta), $1.8 \leq z < 2.9$ (green) and $z \geq
2.9$ (red). The dashed line is the fitting relation line". Right panel: ``GRB distributions in redshift bins on the
$\log F_{X,a}-\log F_{X,peak}$ plane from \cite{dainotti15}, where $\log F_{X,peak}$ 
is computed following the approach used in the Second BAT Catalogue. The
sample is split into 4 different equipopulated redshift bins: $z \le 0.84$
(blue), $0.84 \leq z < 1.8$ (magenta), $1.8 \leq z < 2.9$ (green) and $z \geq
2.9$ (red)".}
\label{fig:lpeakla}
\end{figure}

\cite{dainotti15} further investigated the prompt-afterglow relations presenting an updated analysis of 
123 Swift BAT+XRT light curves of LGRBs with known $z$ and afterglow plateau phase. The relation between the peak 
luminosity of the prompt phase in the X-ray, $\log L_{X,peak}$, and $\log L_{X,a}$ can be written as follows:

\begin{equation}
\log L_{X,a}=A+B\times \log L_{X,peak},
\end{equation}

with $A=-14.67 \pm 3.46$, $B=1.21^{+0.14}_{-0.13}$ and with $\rho=0.79$ and $P <0.05$, see the left panel of 
Fig. \ref{fig:lpeakla}. In the literature $L_{X,peak}$ is denoted as:

\begin{equation}
 L_{X,peak}=4\pi\times D_L(z, \Omega_M, h)^2\times F_{X,peak}\times K.
 \end{equation}

The relation $<\log L_{\gamma,iso}>-\log L_{X,a}$ \citep{Dainotti11b} 
for the same GRB sample presented a correlation coefficient, $\rho=0.60$, smaller than the one of  
the $\log L_{X,peak}-\log L_{X,a}$ relation, see sec. \ref{Dainotti2011b}. 
This implied that a better definition of the luminosity
or energy parameters improves $\rho$ by 24\%.
In the left panel of Fig. \ref{fig:lpeakla} $\log L_{X,peak}$ is calculated directly from the peak
flux in X-ray, $F_{X,peak}$, considering a broken or a simple power law as the best fit of the spectral model. 
Thus, the error propagation due to time and energy
is not involved, differently from the earlier considered luminosities. In addition, \cite{dainotti15} preferred the $\log L_{X,peak}-\log L_{X,a}$ to the 
relations presented in \cite{Dainotti11b}, namely the $(\log E_{\gamma,prompt}, \log E_{\gamma,peak})-\log L_{X,a}$,
due to the fact that $\log E_{\gamma,prompt}$ and $\log E_{\gamma,peak}$ can 
undergo double bias truncation due to high and low energy detector threshold. Instead, this problem does not appear for $\log L_{X,peak}$ \citep{Lloyd99}.
Furthermore, to show that the $\log L_{X,peak}-\log L_{X,a}$ relation is robust,
the redshift dependence induced by the distance 
luminosity was eliminated employing fluxes rather than luminosities. A relation 
between $\log F_{X,a}$ and $\log F_{X,peak}$ was obtained with $\rho=0.63$, see the right panel of 
Fig. \ref{fig:lpeakla}.

\begin{figure}[htbp]
\includegraphics[width=0.495\hsize,angle=0]{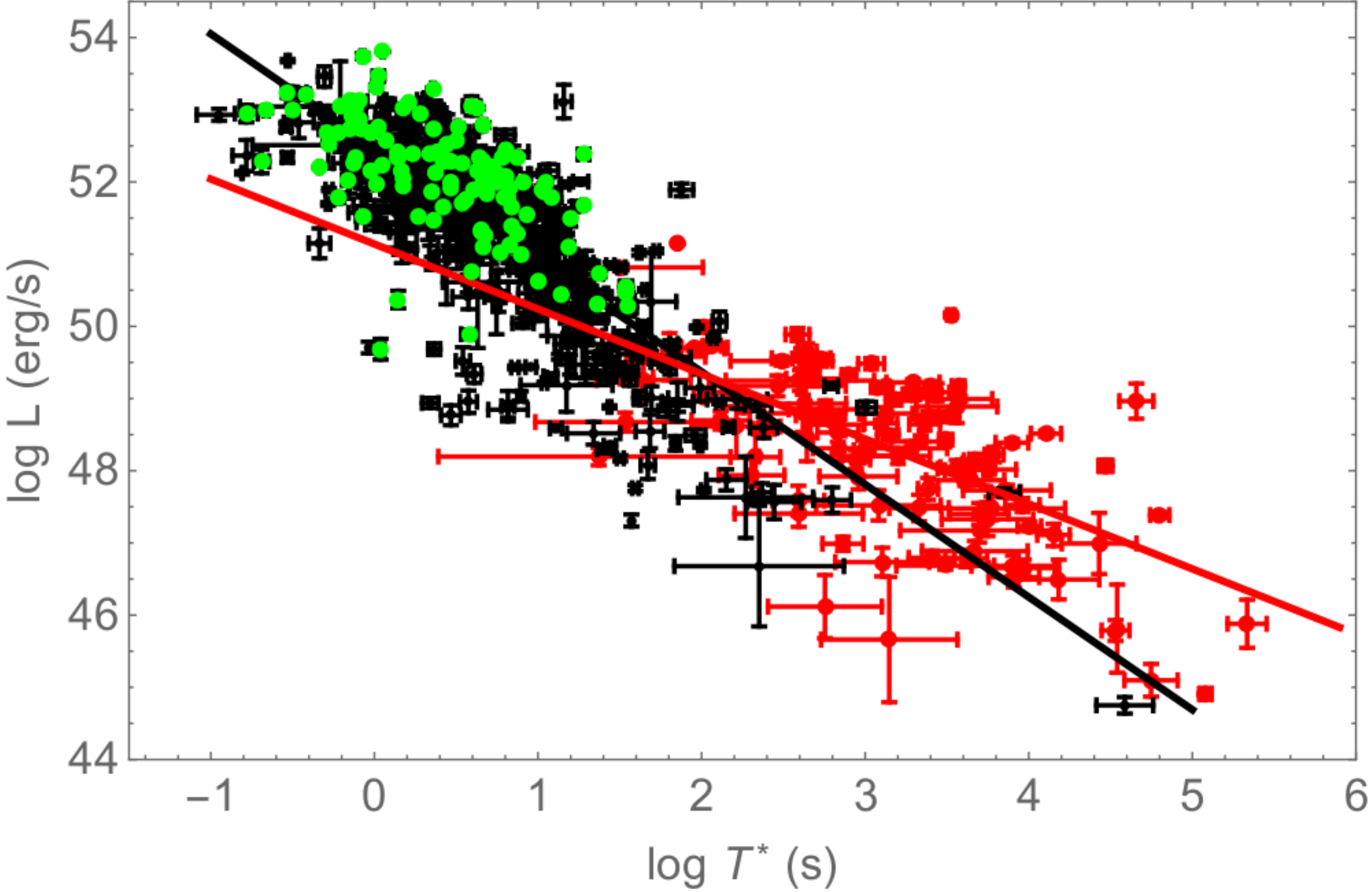}
\includegraphics[width=0.495\hsize,angle=0]{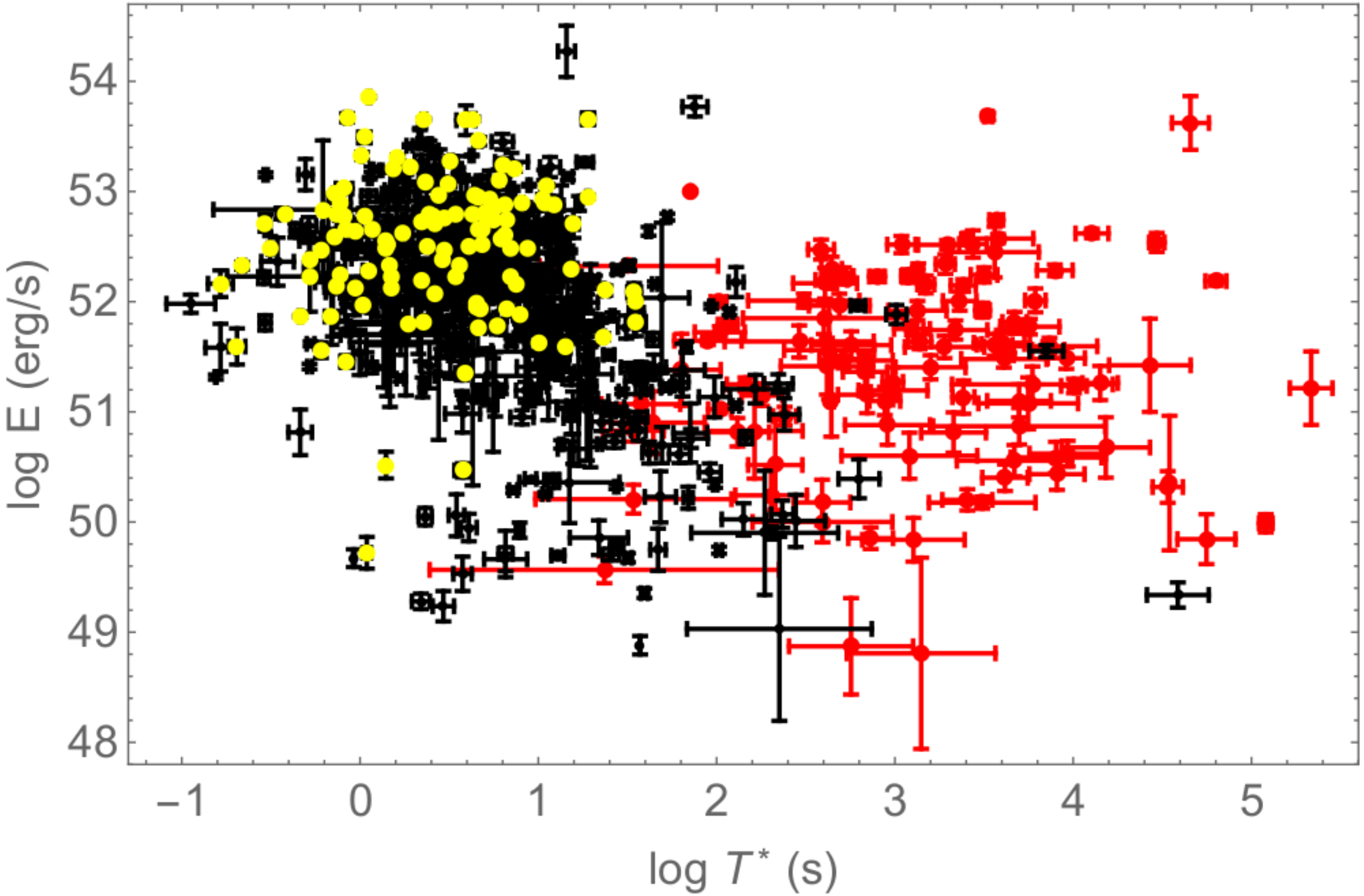}
\caption{\footnotesize Left panel: $\log L$-$\log T^*$ relation for all the pulses in the prompt (black symbols) and in the
afterglow (red symbols) emissions from \cite{dainotti15}. $\log L$ is the same as $\log L_{X,f}$ for the prompt 
emission pulses, while indicates $\log L_{X,a}$ 
for the afterglow emission pulses, and, the time $\log T^*$ indicates $\log T^*_{X,f}$ for the prompt emission pulses and $\log T^*_{X,a}$ 
for the afterglow phase. The green points show the maximum luminosity prompt emission pulses ($\log T_{Lmax}$, $\log L_{max}$). 
Right panel: $\log E$ vs. $\log T^*$ relation for all the pulses in the prompt (black symbols) and in the
afterglow (red symbols) emissions from \cite{dainotti15}. $\log E$ represents $\log E_{X,f}$
for the prompt emission pulses, while it represents $\log E_{X,plateau}$ for the afterglow emission pulses, and
the time $\log T^*$ indicates $\log T^*_{X,f}$ for the prompt emission pulses and $\log T^*_{X,a}$ for the
afterglow phase. The yellow points display the maximum energy prompt emission pulses ($\log T_{Emax}, \log E_{max}$).}
\label{fig:energylpeakla}
\end{figure}

However, for further details about a quantitative analysis of the selection effects see sec. 
\ref{Selection effects}.\\ 
Finally, \cite{dainotti15} showed that the LT relation has a different slope, 
at more than 2 $\sigma$, from the one of the prompt phase relation between the
time since ejection of the pulse and the respective luminosity, $\log L_{X,f}-\log T^{*}_{X,f}$ 
\citep{willingale2010}, see the left panel of 
Fig. \ref{fig:energylpeakla}. 
This difference also implied a discrepancy in the distributions of the energy and time, see the right panel of 
Fig. \ref{fig:energylpeakla}.
The interpretation of this discrepancy between the slopes opens a new perspective in the theoretical understanding of 
these observational facts, see the next section for details.\\
As a further step, \cite{dainotti16a} analyzed this relation
adding as a third parameter $T_{X,a}$ with a sample of 122 LGRBs (without XRFs and GRBs associated to SNe). They found
a tight relation:

\begin{equation}
 \log L_{X,a}=(15.69 \pm 3.8) + (0.67 \pm 0.07) \times \log L_{X,peak}-(0.80 \pm 0.07) \log T_{X,a},
\end{equation}

with $\rho=0.93$, $P\leq 2.2\times 10^{-16}$, and $\sigma_{int}= 0.44 \pm 0.03$.
Additionally, the scatter
could be further reduced considering the subsample of 40 LGRBs having light curves with good data coverage and flat 
plateaus:

\begin{equation}
 \log L_{X,a}=(15.75 \pm 5.3) + (0.67 \pm 0.1) \times \log L_{X,peak}-(0.77 \pm 0.1) \log T_{X,a},
\end{equation}

with $\rho=0.90$, $P=4.41 \times 10^{-15}$, and $\sigma_{int}= 0.27 \pm 0.04$.
These results may suggest the use of this plane as a ``fundamental" plane for GRBs and for 
further cosmological studies.

\subsubsection{Physical interpretation of the \texorpdfstring{$L_X(T_a)-L_{\gamma,iso}$ and the $L_{X,peak}-L_X(T_a)$}{Lg} relations}
In \cite{dainotti15}, the two distinct slopes of the luminosity-duration and
the energy-duration distributions of prompt and plateau pulses could reveal that these two are different characteristics
of the radiation: the former may be generated by internal
shocks and the latter by the external shocks. Indeed, if the plateau is produced by synchrotron radiation from the 
external shock, then all the pulses have analogous physical conditions (e.g. the power
law index of the electron distribution). In addition, the prompt-afterglow 
connections were analyzed in order to better explain the existing physical
models of GRB emission predicting the $\log L_{X,a}-\log L_{\gamma,iso}$ and the $\log L_{X,peak}-\log L_{X,a}$ 
relations together with the 
LT one in the prompt and afterglow phases. They claimed that the model better explaining these
relationships is the one by \cite{hascoet2014}.
In this work they considered two scenarios: one in the standard FS model assuming a modification 
of the microphysics parameters to decrease the efficiency at initial stages of the GRB evolution; in the 
latter the early afterglow stems from a long-lived RS in the FS scenario. In the FS 
scenario a wind external medium is assumed together with a microphysics parameter $\epsilon_e \propto n^{-\nu}$, the amount of the internal energy going into electrons 
(or positrons), where n is the density medium. In the case of $\nu \approx 1$ is possible to reproduce a flat plateau. Thus, even operating on
just one parameter can lead to the formation of a 
plateau that also reproduces the $\log L_{X,a}-\log L_{X,iso}$ and the $\log L_{X,peak}-\log L_{X,a}$ relations.
Alternatively, in the RS scenario, in order to obtain the observed prompt-afterglow relationships, the typical 
$\Gamma$ of the ejecta should rise with the burst energy.

\begin{figure}[htbp]
\centering
 \includegraphics[width=0.95\hsize,angle=0]{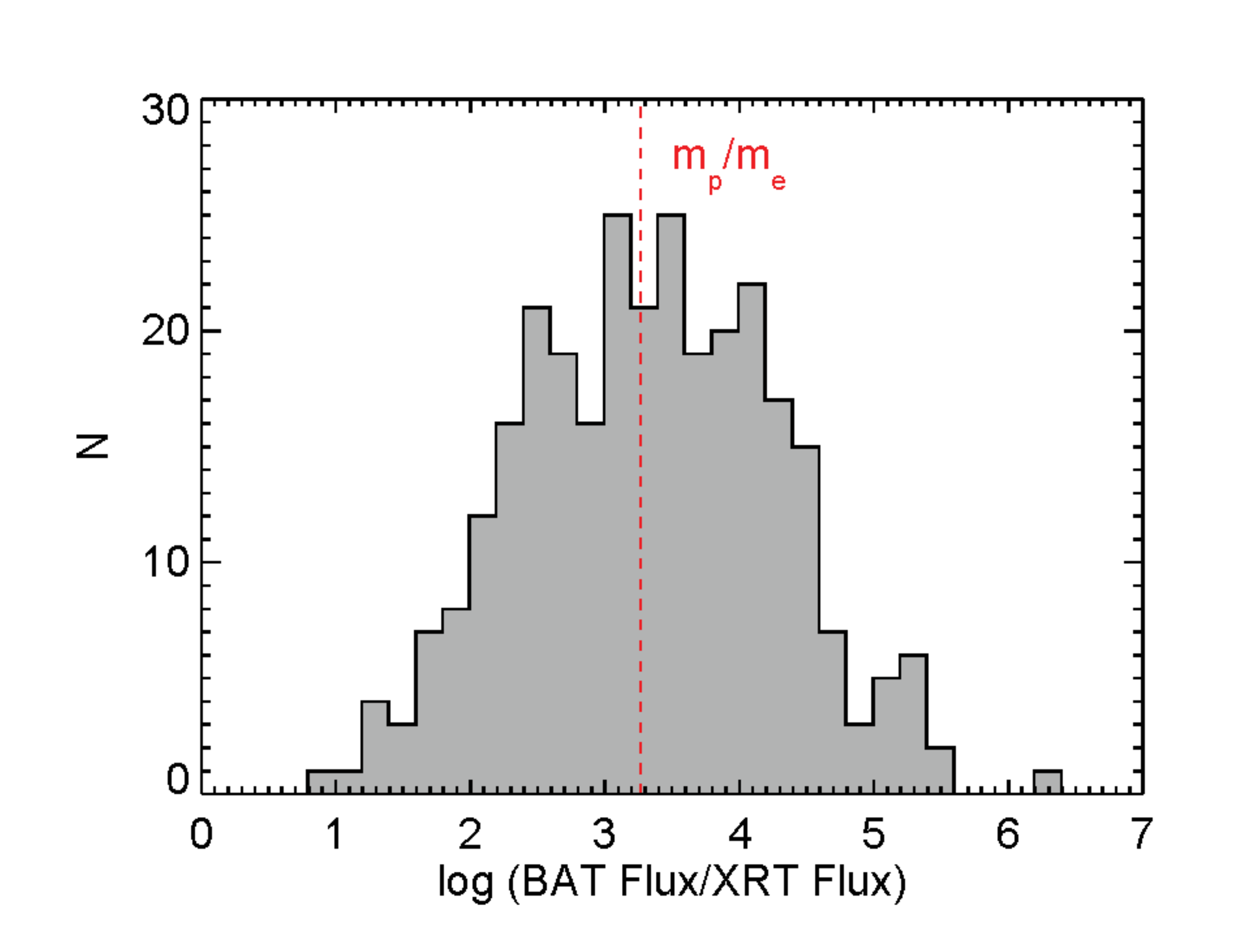}
 \caption{\footnotesize ``The histogram of the BAT to XRT flux ratio for a number of Swift GRB from \cite{kazanas15}. 
 The distribution shows clearly a
preferred value for this ratio of order $\sim 10^3-10^4$. The vertical line shows also the proton to electron mass
ratio $m_p/m_e$".}
\label{fig:kazanas15}
 \end{figure}

In addition, \cite{ruffini14} pointed out that the induced gravitational collapse paradigm
can recover the $\log L_{X,a}-\log L_{\gamma,iso}$ and the $\log L_{X,peak}-\log L_{X,a}$ relations. This model
considers the very energetic ($10^{52}-10^{54}$ erg) LGRBs for which the SNe has been seen. 
The light curves were built assuming for the external medium either a radial structure for the wind 
\citep{Guida2008,bernardini06,bernardini07,caito09} or a partition of the 
shell \citep{Dainotti2007}, therefore well matching the afterglow plateau and the prompt emission.\\
Recently, \cite{kazanas15} within the context of the Supercritical Pile GRB Model claimed that they can reproduce
the $\log L_{X,a}-\log L_{\gamma,iso}$ 
and the $\log L_{X,peak}-\log L_{X,a}$ relations, because the ratio, R, between the luminosities appears consistent with the
one between the mean prompt energy flux from BAT and the afterglow plateau fluxes detected by XRT. In particular, 
$R \approx 2000$ is close to the proton to electron mass ratio, see Fig. \ref{fig:kazanas15}.\\
Indeed, this is a new challenge for theoretical modelling that would need to consider, simultaneously, the several 
prompt-afterglow connections in order to better reproduce the phenomenology of the relations from a statistical point of view.

\subsection{The \texorpdfstring{$L^F_{O,peak}-T^{*F}_{O,peak}$}{Lg} relation and its physical interpretation}\label{lisotp}
\begin{figure}[htbp]
\centering
\includegraphics[width=0.5\hsize,angle=0]{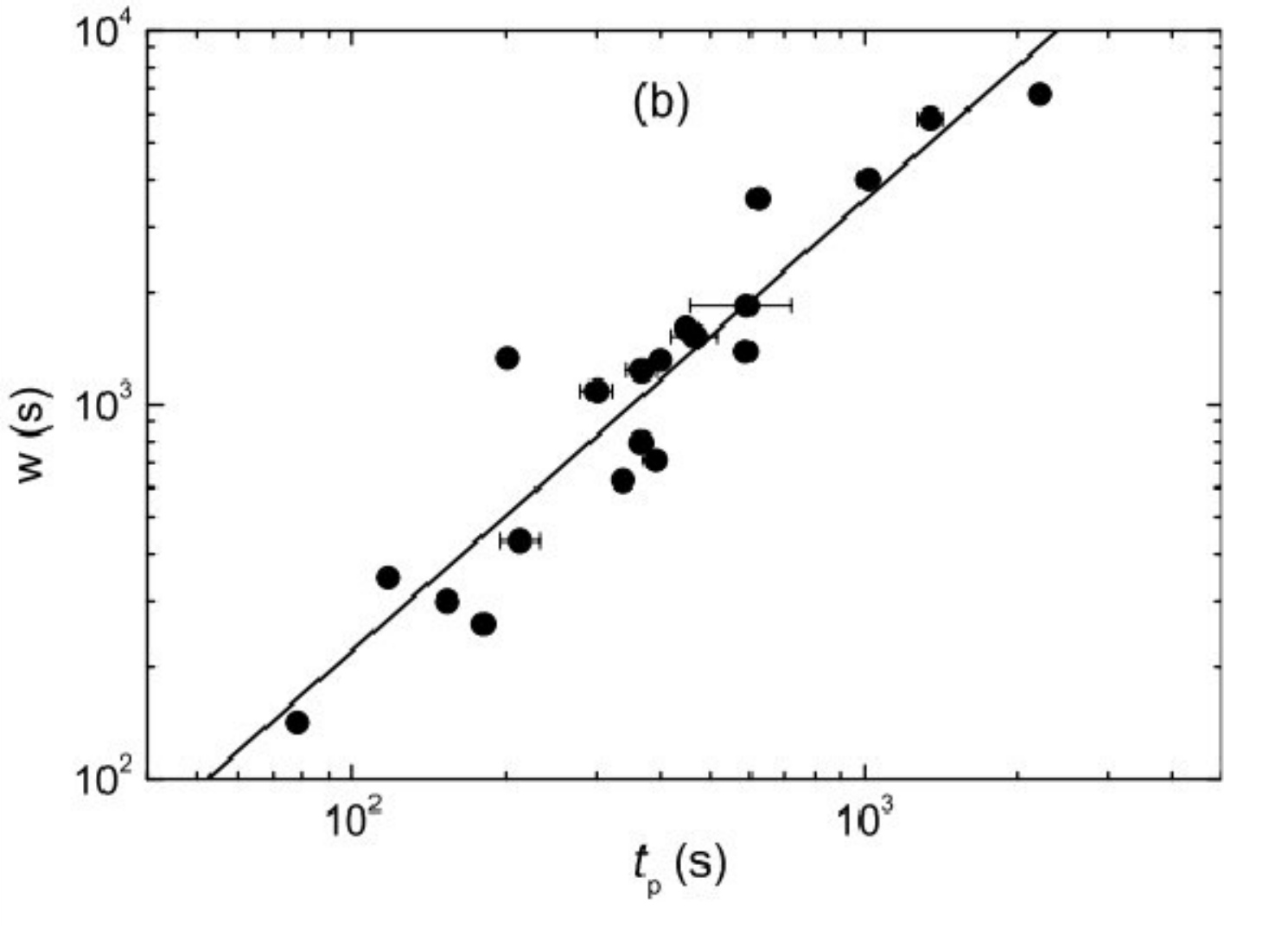}
\includegraphics[width=0.47\hsize,angle=0]{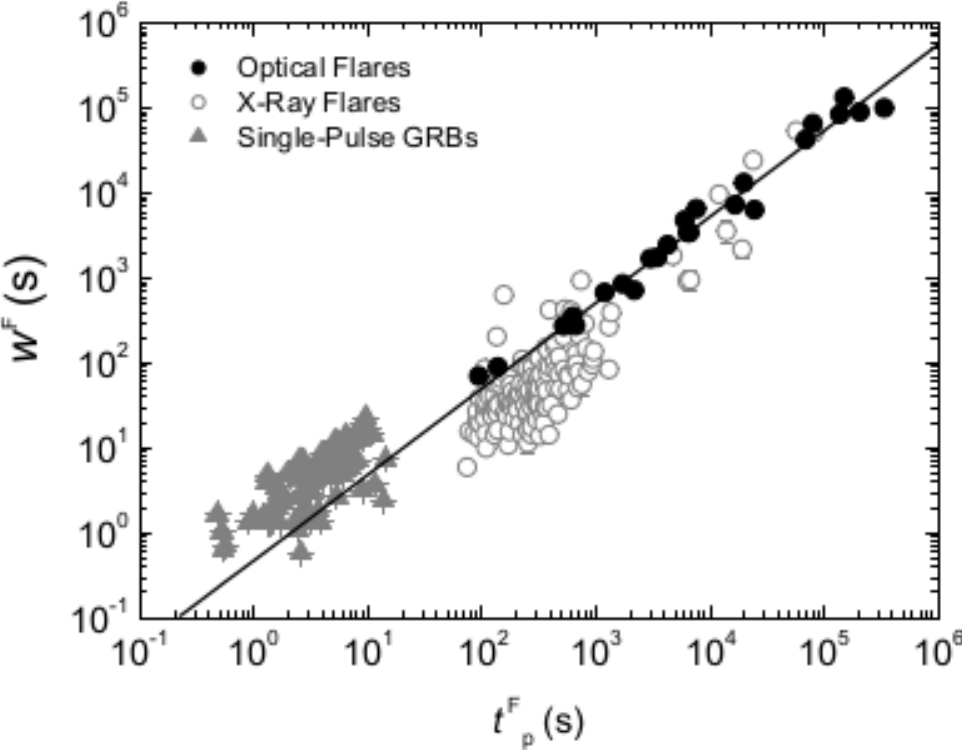}
\caption{\footnotesize Left panel: $\log w^F$-$\log T_{O,peak}^F$ distribution from \cite{Liang2010}.
Right panel: $\log w^F$-$\log T_{O,peak}^F$ distribution from \cite{Li2012}. In both panels lines represent the best fit.}
  \label{fig:liang3}
\end{figure}

\cite{Liang2010} studied the relation between the width of the light curve 
flares, $w$, and $T_{O,peak}$ of the flares, denoted with the index F, using a sample of 32 Swift GRBs, see the left panel of Fig. \ref{fig:liang3}. 
This relation reads as follows:

\begin{equation}
\log w^F = (0.05 \pm 0.27) + (1.16 \pm 0.10) \times \log T_{O,peak}^{F}, 
\end{equation}

with $\rho=0.94$.

Later, \cite{Li2012} found the same relation as \cite{Liang2010}, but with smaller values of normalization
and slope, using 24 flares from 19 single-pulse GRBs observed with CGRO/BATSE\footnote{Among the instruments of the
Compton Gamma Ray Observatory (CGRO) satellite, running from 1991 to 2001, the
Burst and Transient Source Experiment (BATSE) played a 
fundamental role in the measurements of GRB spectral features in the range from $20$ keV to $8$ MeV. Bursts were typically 
detected at rates of roughly one per day over the 9-year CGRO mission within a time interval ranging from $\sim0.1$ s up to 
about $100$ s. Therefore, this satellite enabled careful analysis of the spectral properties of the GRB prompt emission.}, 
see the right panel of Fig. \ref{fig:liang3}. 
However, for these 19 GRBs only in 14 GRBs a flare activity is distinctly visible.
The relationship was given by:

\begin{equation}
\log w^F = -0.32 + 1.01 \times \log T^F_{O,peak}.
\end{equation}

\begin{figure}[htbp]
\centering
\includegraphics[width=0.95\hsize,angle=0]{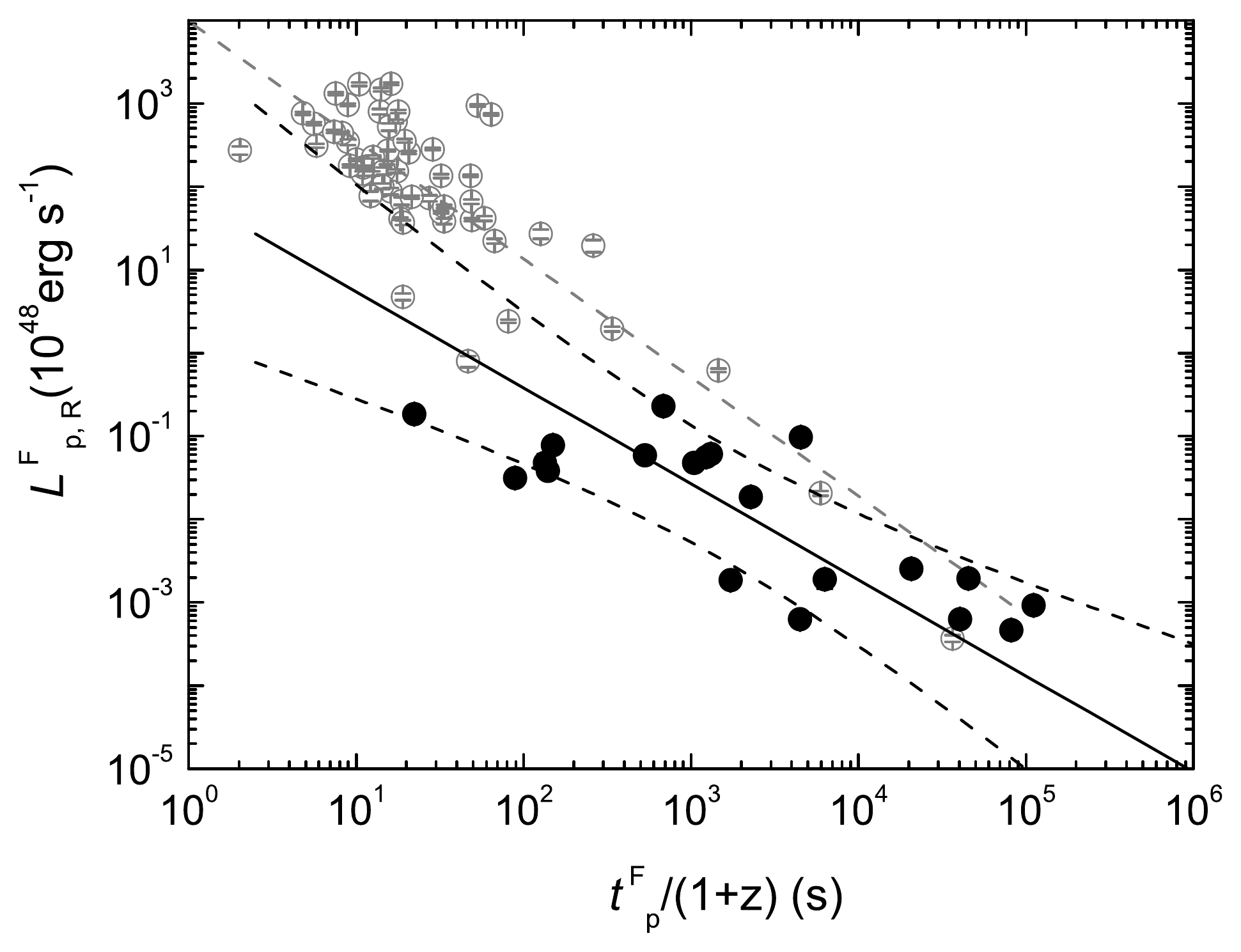}
\caption{\footnotesize $\log L^F_{O,peak}$-$\log T_{O,peak}^{*F}$ relation from \cite{Li2012}. Lines represent 
the best fits, black dots indicate optical flares, and the grey circles with errors show X-ray flares associated
with the optical flares.}
  \label{fig:li2012}
\end{figure}

They claimed that earlier flares are brighter and narrower than later ones.
They compared the $w^F-T^F_{O,peak}$ distribution for the
X-ray flares detected by Swift/XRT with the one for the
optical flares in the R band. As a conclusion, they seemed to have a similar behaviour 
\citep{chincarini07,margutti2010}, see the right panel of Fig. \ref{fig:liang3}.\\
Furthermore, in the rest frame band, they found a relation between the
$L_{O,peak}$ of the flares in the R energy in units of $10^{48}$ erg s$^{-1}$ and $T^*_{O,peak}$ of the flares
using 19 GRBs, see Fig. \ref{fig:li2012}.
Both prompt pulses and X-ray and optical flares are correlated and present a visible temporal evolution, as seen in 
Fig. \ref{fig:li2012}. This relation is given by:

\begin{equation}
\log L^{F}_{O,peak}=(1.89\pm 0.52)-(1.15\pm0.15)\times \log T^{*F}_{O,peak},
\end{equation}

with $\rho=0.85$ and $P<10^{-4}$. 
$T^{F}_{O,peak}$ spans from $\sim$ tens of seconds to $\sim 10^6$ seconds, instead the $L^{F}_{O,peak}$ varies from $10^{43}$ to 
$10^{49}$ erg s$^{-1}$, with an average value of $10^{46}$ erg s$^{-1}$.
In addition, considering only the most luminous GRBs, they found that $T^{*F}_{O,peak}$ was strongly anti-correlated to 
$E_{\gamma,prompt}$ in the $1-10^4$ keV energy band:

\begin{equation}
\log T^{*F}_{O,peak}=(5.38\pm 0.30)-(0.78\pm 0.09)\times \log E_{\gamma,prompt}/10^{50},
\end{equation}

with $\rho=0.92$. 
These outcomes revealed that the GRB flares in the optical wavelength with higher $E_{\gamma,prompt}$ peak earlier and 
are much more luminous. In Table \ref{tbl2} a summary of the relations described in this section is displayed.

\begin{table}[htbp]
\footnotesize
\begin{center}
\begin{tabular}{|c|c|c|c|c|c|c|}
\hline
Correlations & Author & N & Slope& Norm & Corr.coeff. & P \\
\hline
$w^F-T^F_{O,peak}$ & Liang et al. (2010) & 32 & $1.16^{+0.10}_{-0.10}$&$0.05^{+0.27}_{-0.27}$&0.94& $<10^{-4}$\\
 &Li et al. (2012)&19&1.01&-0.32&&\\
\hline
$L^{F}_{O,peak}-T^{*F}_{O,peak}$ & Li et al. (2012)&19&$-1.15^{+0.15}_{-0.15}$& $1.89^{+0.52}_{-0.52}$&0.85&$<10^{-4}$ \\
\hline
$T^{*F}_{O,peak}-E_{\gamma,prompt}$ &Li et al. (2012)&19&$-0.78^{+0.09}_{-0.09}$&$5.38^{+0.30}_{-0.30}$&0.92&$<10^{-4}$ \\
\hline
\end{tabular}
\caption{\footnotesize Summary of the relations in this section. The first column represents the relation in log scale,
the second one the authors, the third one the number of GRBs in the used sample, and the fourth and the fifth columns are the slope and 
normalization of the relation. The last two columns are the correlation coefficient and the chance 
probability, P.}
\label{tbl2}
\end{center}
\end{table}

As regards the physical interpretation of the $L^F_{O,peak}-T^{*F}_{O,peak}$ relationship, 
\cite{Li2012} found that the flares are separated
components superimposed to the afterglow phase. The coupling between $L^{F}_{O,peak}$ and $T^{*F}_{O,peak}$ suggested that the prompt $\gamma$-ray and 
late optical flare emission may arise from the same mechanism, namely from a central
engine that can periodically eject a number of shells during the emission. Impacts of these shells could
create internal shocks or magnetic turbulent reconnections,
which would emerge from the variability \citep{kobayashi97,zhang2011}. \cite{fenimore95} revealed 
no relevant pattern in the width and intensity distributions using gamma ray data only. 
In addition, the usual tendency of the $w^F-T^F_{O,peak}$ relation
cannot be due to hydrodynamical diffusion of the shells emitted at recent times, but it is necessary that the central engine radiates
thicker and fainter shells at late stages \citep{maxham09}. This could be explained as flares generated by clumps, 
such that the diffusion during the accretion mechanism would extend the accretion duration onto the BH \citep{perna06,proga06}.

\section{Selection Effects} \label{Selection effects}
Selection effects are distortions or biases that usually occur when the sample observed is not representative of the
``true'' population itself. This kind of biases usually affects GRB relations. 
\cite{Efron1992}, \cite{Lloyd99}, \cite{Dainotti2013a,Dainotti2015b} and \cite{petrosian14} emphasized that when 
dealing with a multivariate data set, it is imperative to determine first the true 
relations among the variables, not those introduced by the observational selection effects, 
before obtaining the individual distributions of the variables themselves. 
This study is needed for claiming the existence of the intrinsic relations. A relation can be called 
intrinsic only if it is carefully tested and corrected for these biases.\\
The selection effects present in the relations discussed above are mostly due to the dependence of 
the parameters on the redshift,
like in the case of the time and the luminosity evolution, or due to the threshold of the detector used.\\
In this section, we describe several different methods to deal with selection biases.\\
In paragraph \ref{redind}, we discuss the redshift induced relation
through a qualitative method, while in \ref{efronpetmethod} we present a more quantitative approach through the EP method.
In \ref{intrinsicorrelation}, we describe how to obtain the intrinsic relations corrected by selection biases, and 
in \ref{ghisellinisel} we report the selection effects for the optical and X-ray luminosities. Lastly, in \ref{montecarlosimulation}
we show the evaluation of the intrinsic relation through Monte Carlo simulations.

\subsection{Redshift induced relations}\label{redind}
\begin{figure}[htbp]
\centering
\includegraphics[width=0.495\hsize,angle=0]{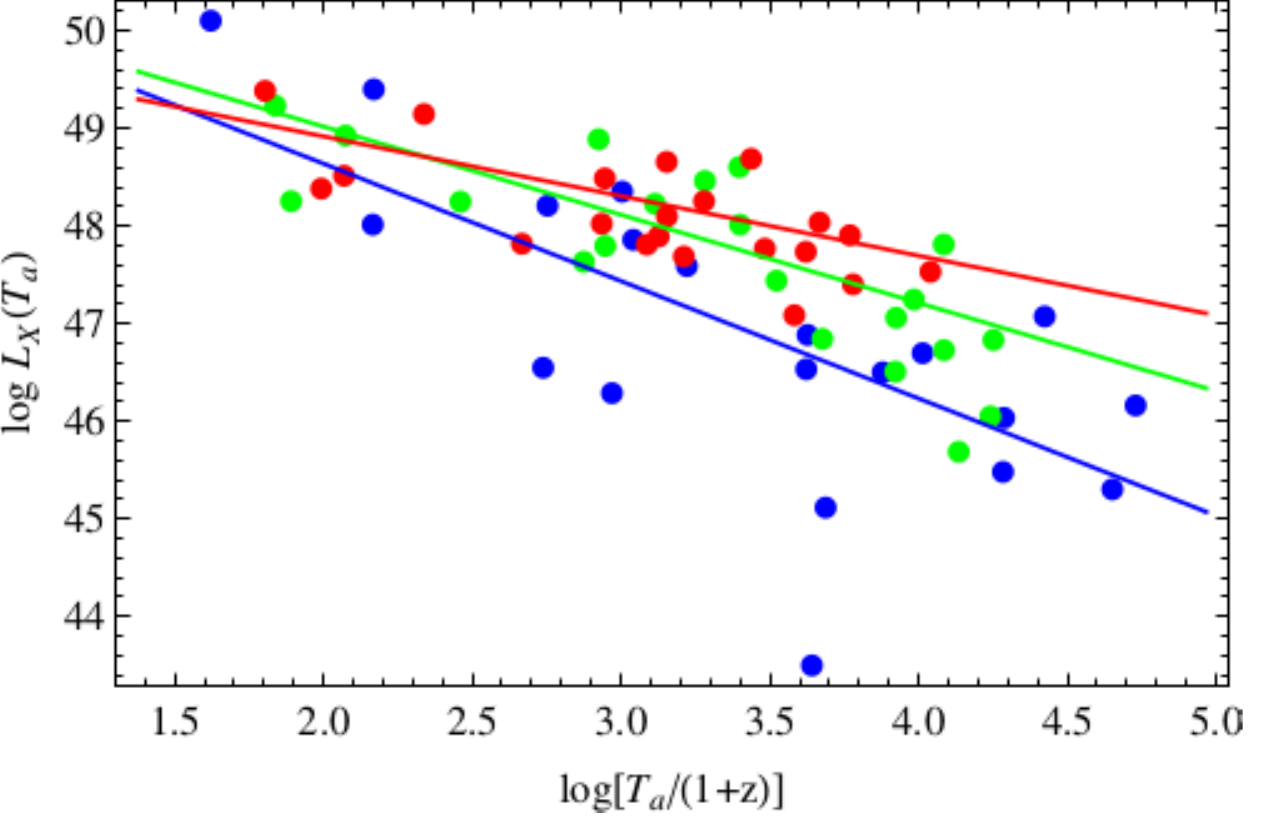}  
\includegraphics[width=0.495\hsize,angle=0]{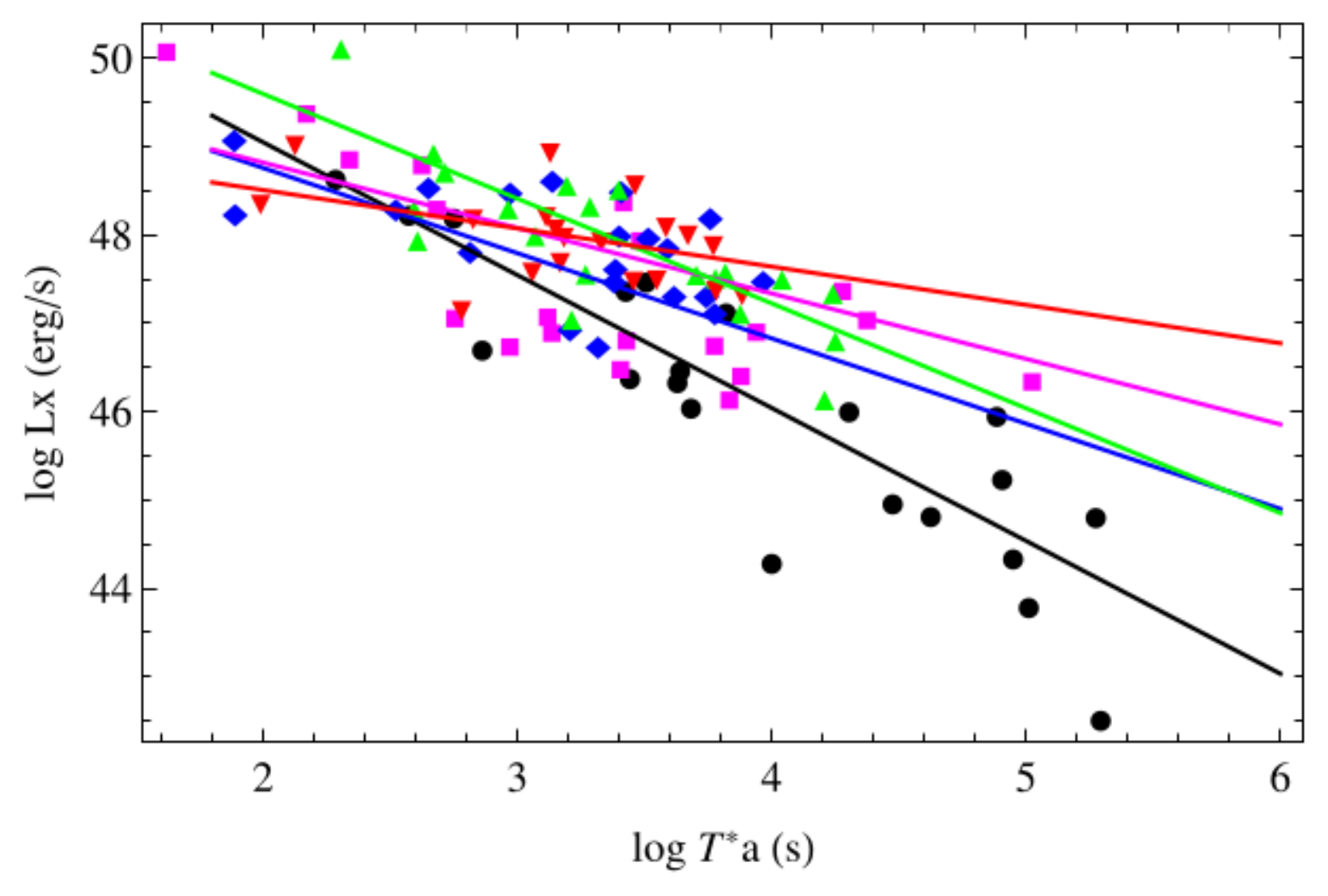}   
   \caption{\footnotesize Left panel: ``$\log L_{X,a}-\log T^*_{X,a}$ relation divided in the three redshift 
   bins $Z1=(0.08, 1.56)$, $Z2=(1.71, 3.08)$ and $Z3=(3.21, 8.26)$ from \cite{dainotti11a}. With the blue points it is 
   represented the $Z1$ sample, with the green ones the $Z2$ sample and with the red points the $Z3$ sample. The respective fitted 
   lines are
   in the same colours". Right panel: ``$\log L_{X,a}-\log T^*_{X,a}$ distribution from \cite{Dainotti2013a} 
   for the sample of 101 GRB afterglows divided in $5$ equipopulated redshift bins shown by different colours: black for 
   $z < 0.89$, magenta for $0.89 \leq z \leq 1.68$, 
   blue for $1.68 < z \leq 2.45$, green $2.45 < z \leq 3.45$, red for $ z \geq 3.45$. Solid
   lines show the fitted  relations".}
   \label{fig:redshiftbin}
\end{figure}

An important source of possible selection effects is the dependence of the variables on the redshift. To this
end, \cite{dainotti11a} investigated the redshift evolution of the parameters of the LT relation, because a change of
the relation slope has been observed when comparing several analyses \citep{Dainotti2008,Dainotti2010}. Namely,
in the first paper, it was found $b=-0.74^{+0.20}_{-0.19}$ and in the latter $b=-1.06^{+0.27}_{-0.28}$. Therefore, it became crucial to 
understand the reason of this change, even if the two slopes are still comparable at the 1-$\sigma$ level.
The distribution of the 62 LGRBs in the sample is not uniform within the range $(z_{min}, z_{max}) = (0.08, 8.26)$ with 
few data points at large redshifts. Even if this sample is sparse, it was important to investigate whether the calibration
coefficients $(a, b, \sigma_{int})$ were in agreement within the error bars over this large redshift interval, see the left panel
of Fig. \ref{fig:redshiftbin}.

For this reason, the data set was separated in three redshift bins with the same number of elements, 
$Z1=(0.08, 1.56)$, $Z2=(1.71, 3.08)$ and $Z3=(3.21, 8.26)$ presented as blue, green and red points respectively in the left panel 
of Fig. \ref{fig:redshiftbin}. The results are presented in Table \ref{tbl8}.

\begin{table}[htbp]
\footnotesize
\begin{center}{
\begin{tabular}{|c|c|c|c|c|}
\hline
Id & $\rho$ & $(b, a, \sigma_{int})_{bf}$ & $b_{median}$ & $(\sigma_{int})_{median}$ \\
\hline
Z1 & -0.69 & (-1.20, 51.04, 0.98) & $-1.08_{-0.30}^{+0.27}$ & $1.01_{-0.16}^{+0.20}$ \\
Z2 & -0.83 & (-0.90, 50.82, 0.43) & $-0.86_{-0.16}^{+0.18}$ & $0.45_{-0.08}^{+0.09}$ \\
Z3 & -0.63 & (-0.61, 50.14, 0.26) & $-0.58_{-0.15}^{+0.14}$ & $0.26_{-0.06}^{+0.07}$ \\
\hline
\end{tabular}}
\caption{\footnotesize Results of the calibration procedure for GRBs divided in three 
equally populated redshift bins with $(z_{min}, z_{max}) = (0.08, 1.56)$, 
$(1.71, 3.08)$, $(3.21, 8.26)$ for bins Z1, Z2, Z3 from \cite{dainotti11a}. The subscript $bf$ displays
the best fit values, while the $median$ subscript shows the median values.}
\label{tbl8}
\end{center}
\end{table}       

\begin{figure}[htbp]
\includegraphics[width=0.495\hsize,angle=0]{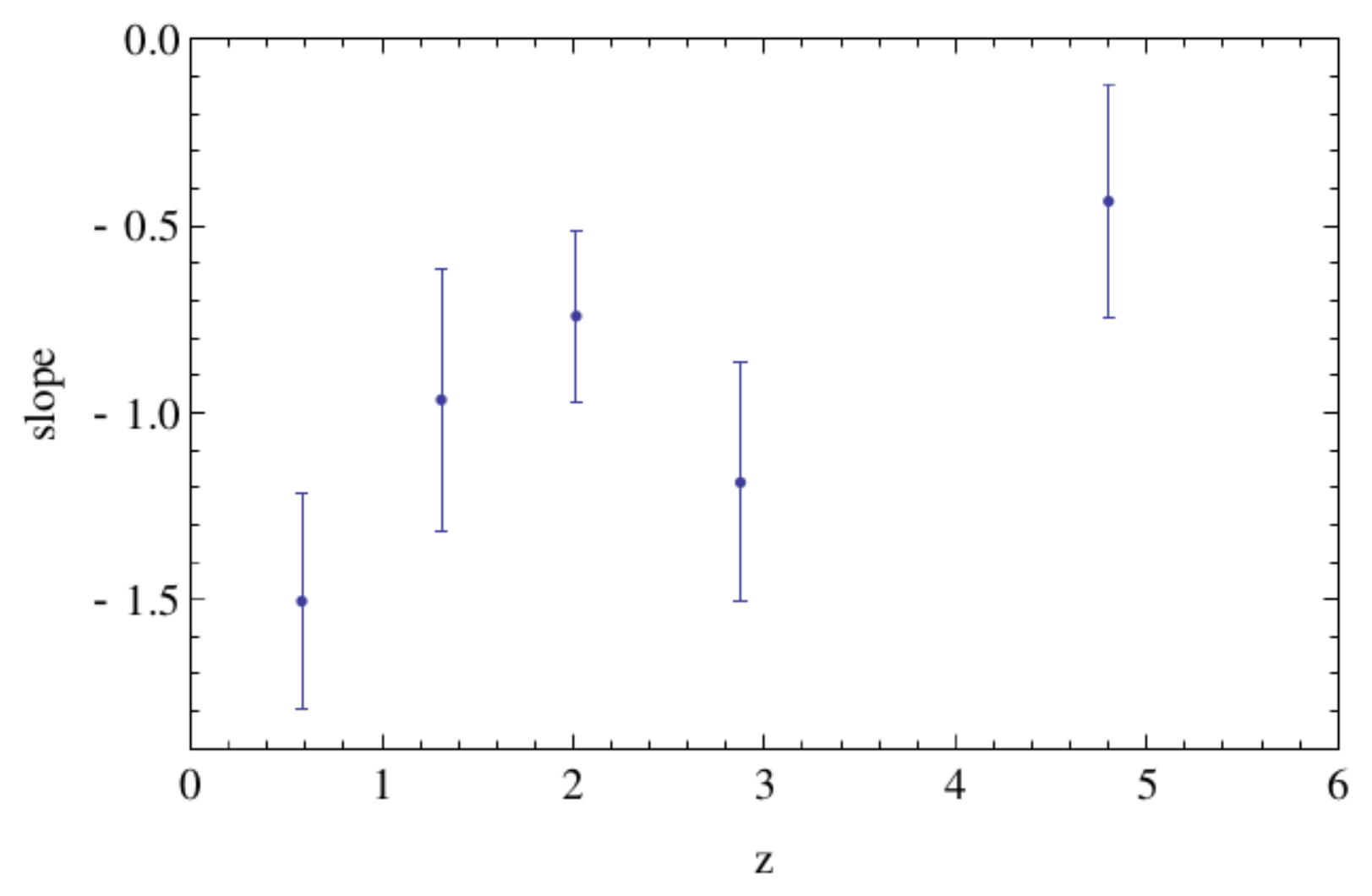} 
\includegraphics[width=0.485\hsize,angle=0]{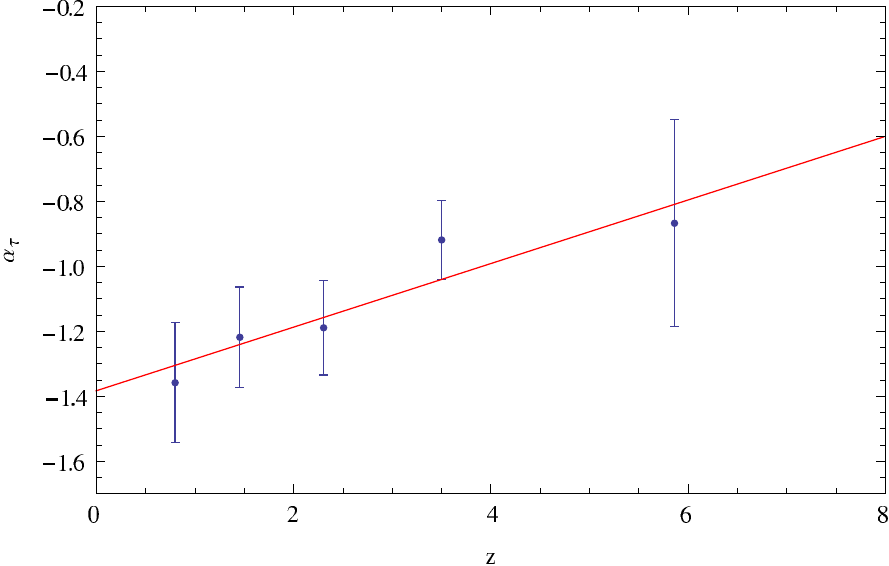}
\caption{\footnotesize Left panel: ``the variation of $b$ (and its error range) with the mean value of the
   redshift bins from \cite{Dainotti2013a}". Right panel: ``$\alpha_{\tau}$, which is equivalent to the slope $b$, 
   vs. z using a linear function $\alpha_{\tau}=0.10 z-1.38$ from \cite{Dainotti2015b}".}
	 \label{fig:slopevsz}
\end{figure}

The correlation coefficient $\rho$ was found quite high in each redshift bins, supporting the 
independence of the LT relation on $z$. The slopes $b$ for 
bins $Z1$ and $Z2$ are comparable within the $68\%$ CL, while the slopes in bins $Z1$ and $Z3$ only
within the $95\%$ CL, see Table \ref{tbl8}. On the contrary, the normalization $a$ is comparable in all the bins. 
From this analysis, it is not possible to confirm that the LT relation is shallower for larger
$z$ GRBs, due to the low number of data points and the presence of high $\sigma_E$ GRBs. Finally, 
bigger samples with small $\sigma_E$ values and a more uniform $z$ binning are required to overcome this problem.\\
For this reason, \cite{Dainotti2013a} performed a similar analysis, but with a larger sample consisting of $101$ GRBs. 
Specifically, this updated sample was split in $5$ redshift ranges with the same number of elements, 
thus having $20$ GRBs in each subgroup, represented in the right panel of Fig. \ref{fig:redshiftbin} by different 
colours: black for $z < 0.89$, magenta for $0.89 \leq z \leq 1.68$, blue for $1.68 < z \leq 2.45$, green for 
$2.45 < z \leq 3.45$ and red for $ z \geq 3.45$. The fitted lines for each
redshift bin are also shown in the same colour code. The distribution of the subsamples presented different power law
slopes when 
the whole sample was divided into bins. The objects in the different bins exhibited some separation into different regions 
of the LT plane.
Moreover, the slope of the relation for each redshift bin versus the averaged redshift range has also been presented, 
see the left panel of Fig. \ref{fig:slopevsz}.\\
In addition, in \cite{Dainotti2015b}, the updated sample of $176$ GRBs was divided into $5$ redshift bins consisting
of about 35 GRBs for each group, as shown in the right panel of Fig. \ref{fig:slopevsz}. A small evolution in $z$ has
been confirmed 
with the following linear function $b(z)=0.10z-1.38$.

Regarding the $\log L_{X,peak}-\log L_{X,a}$ relation, 
\cite{dainotti15} showed that it is not produced by the dependence on the redshift of its variables. 
To estimate the redshift
evolution, the sample was separated into $4$ redshift bins as shown in the left panel of Fig. \ref{fig:lpeakla}. 
The GRB distribution in each bin is not grouped
or constrained within a specific region, therefore indicating no strong redshift evolution. 
For $\log L_{X,a}$ it was found that there was negligible
redshift evolution of the afterglow X-ray luminosity \citep{Dainotti2013a}, while for $\log L_{X,peak}$ has been 
demonstrated that there is significant 
redshift evolution \citep{yonetoku04,petrosian14,dainotti15}.
For more details, see sec. \ref{luminosityev} and \ref{timeev}.

\subsection{Redshift induced relations through Efron and Petrosian method}\label{efronpetmethod}
For a quantitative study of the redshift evolution, which is the dependence of the variables on the redshift, we here refer to the
EP method which is specifically designed to overcome the biases resulting from incomplete data. 
The Efron \& Petrosian technique, applied to GRBs \citep{petrosian09,Lloyd99,Lloyd2000}, allows to compute the intrinsic 
slope of the relation by creating new bias-free observables, called local variables and denoted with the symbol ${'}$.
For these quantities, 
the redshift evolution and the selection effects due to instrumental thresholds are removed.
The EP method uses a modification of the Kendall tau test\footnote{The Kendall $\tau$ is a non-parametric statistical test used to 
measure the association between two measured quantities. It is a measure of rank relation: the similarity of the
orderings of the data when ranked by each of the quantities.} $\tau$ to compute the best fit values of the parameters which represent
the luminosity and time evolutionary functions. For details about the definition of $\tau$ see \cite{Efron1992}.

\subsubsection{Luminosity evolution}\label{luminosityev}
\begin{figure}[htbp]
\centering
   \includegraphics[width=0.49\hsize,angle=0,clip]{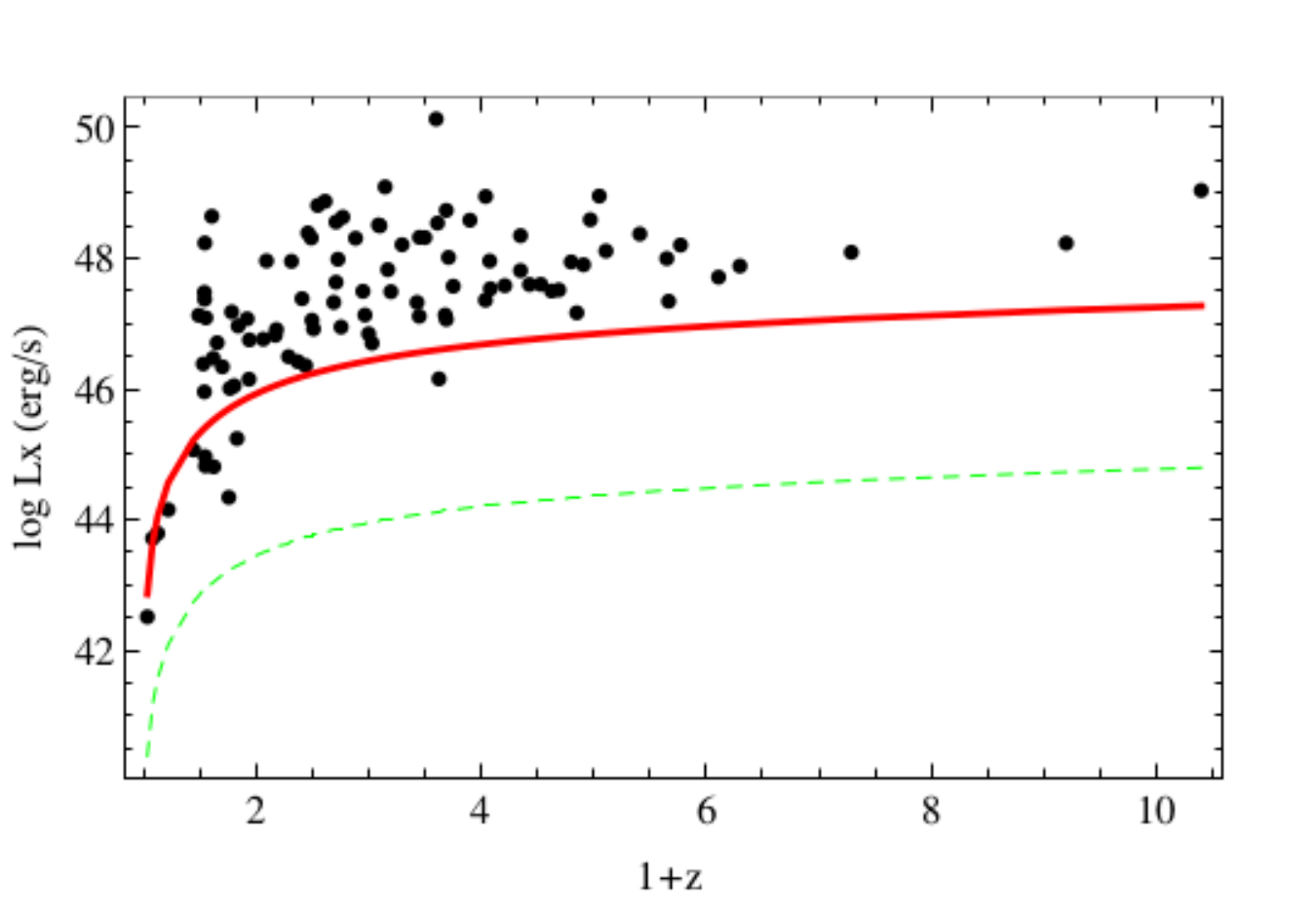}
   \includegraphics[width=0.48\hsize,angle=0,clip]{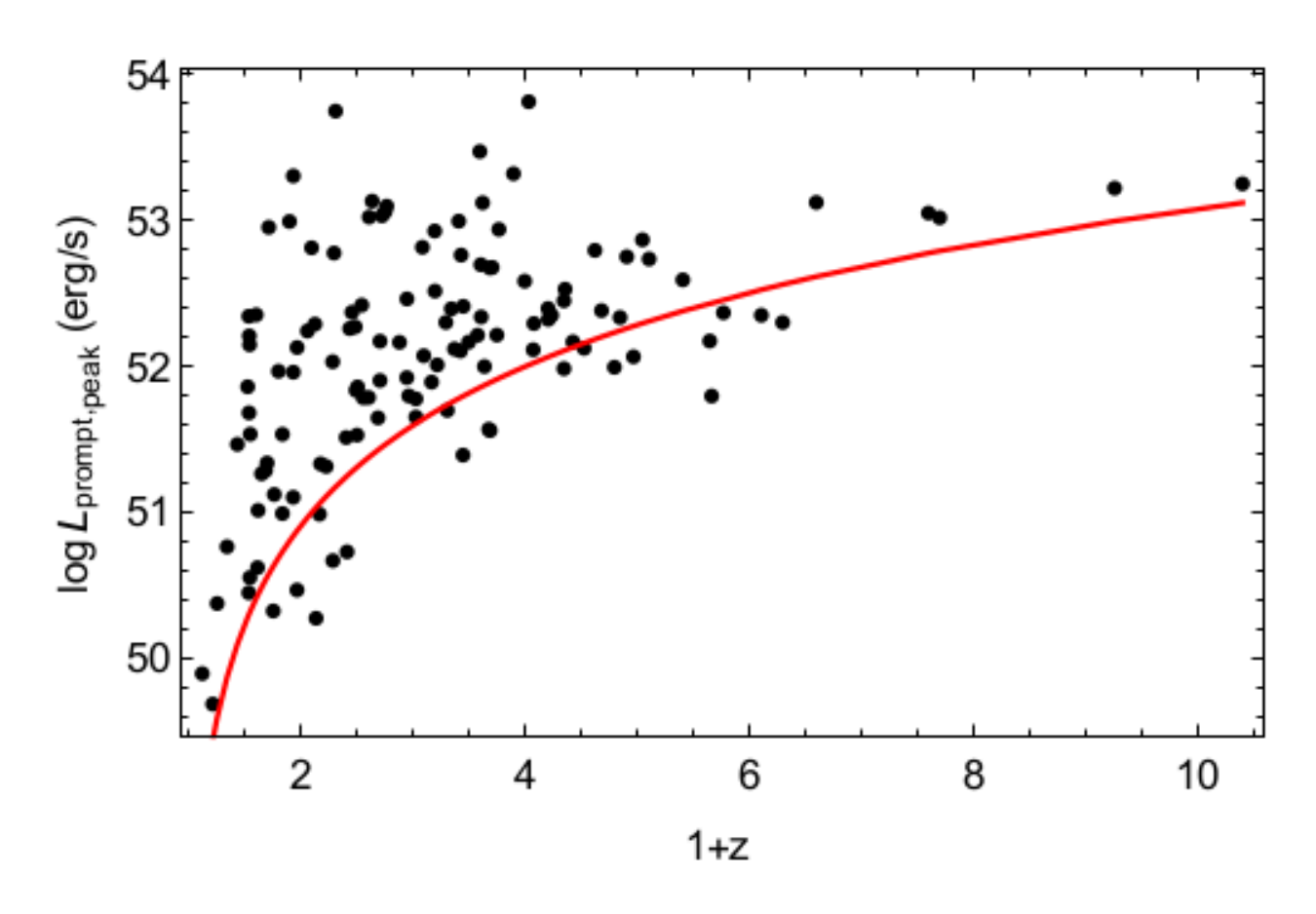}
   \caption{\footnotesize Left panel: ``the bivariate distribution of $\log L_{X,a}$ and $z$ with two different flux limits 
   from \cite{Dainotti2013a}. The instrumental XRT flux limit, $1.0\times10^{-14}$ erg cm$^{-2}$ s$^{-1}$ (dashed 
   green line), is too low to be representative of the flux limit, $1.5 \times 10^{-12}$ erg cm$^{-2}$ s$^{-1}$ (solid 
   red line) represents better the limit of the sample". Right panel: ``the bivariate distribution of $\log L_{X,peak}$ and $z$ with the 
   flux limit assuming the K correction $K = 1$ from \cite{dainotti15}. The BAT flux limit, $4.0 \times 10^{-8}$ erg cm$^{-2}$ s$^{-1}$ 
   (solid red line), better represents the limit of the sample".}
   \label{fig:19}
\end{figure}

The relation between luminosity and $z$ is called luminosity evolution.
We discuss the luminosity evolution for both prompt and plateau phases. Before applying the EP method to the 
plateau phase, the limiting plateau flux, $F_{\rm{lim}}$, which gives the minimum observed luminosity for a given
$z$ needs to be parameterized. 
The XRT sensitivity, $F_{lim,XRT}=10^{-14}$ erg cm$^{-2}$ s$^{-1}$, is not high enough to represent 
the truncation of the data set.
Hence, as claimed by \cite{cannizzo2011}, a better choice for the flux threshold is $10^{-12}$ erg cm$^{-2}$ s$^{-1}$. 
Several threshold fluxes were analyzed \citep{Dainotti2013a}, finally  
$F_{lim,XRT} = 1.5 \times $10$^{-12}$ erg cm$^{-2}$ s$^{-1}$, which leaves 90 out of 101 GRBs, was selected (see the left panel of 
Fig. \ref{fig:19}).
Regarding instead the prompt limiting flux, \cite{dainotti15} chose a BAT flux limit 
$F_{lim,BAT}=4\times10^{-8}$ erg cm$^{-2}$ s$^{-1}$, which also allows 90\% of GRBs in the sample, see the right panel of Fig. \ref{fig:19}.

In \cite{Dainotti2013a}, the relation function, g(z), is defined when determining the evolution of $L_{X,a}$ 
so that the local variable $L'_{X,a} \equiv L_{X,a}/g(z)$ is not dependent anymore from $z$. 
The evolutionary function is parameterized by a simple relation function:

\begin{equation}
g(z)=(1+z)^{k_{L_{X,a}}}.
\label{lxev}
\end{equation}
More complex evolution functions lead to comparable results, see Dainotti et al. (2013a, 2015b).

\begin{figure}[htbp]
\centering
   \includegraphics[width=5.05cm,height=4.9cm,angle=0,clip]{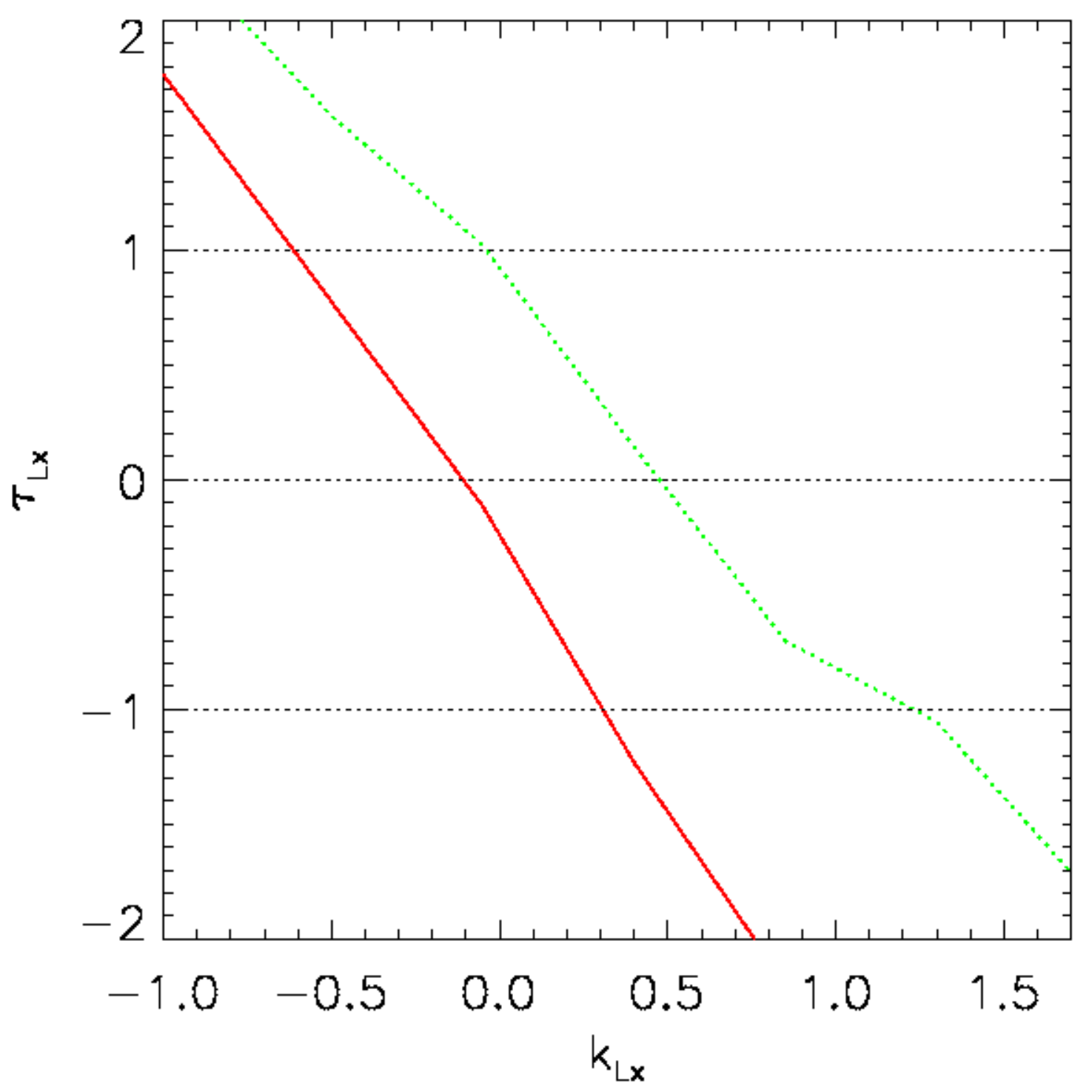}
   \includegraphics[width=5.6cm,height=5.2cm,angle=0,clip]{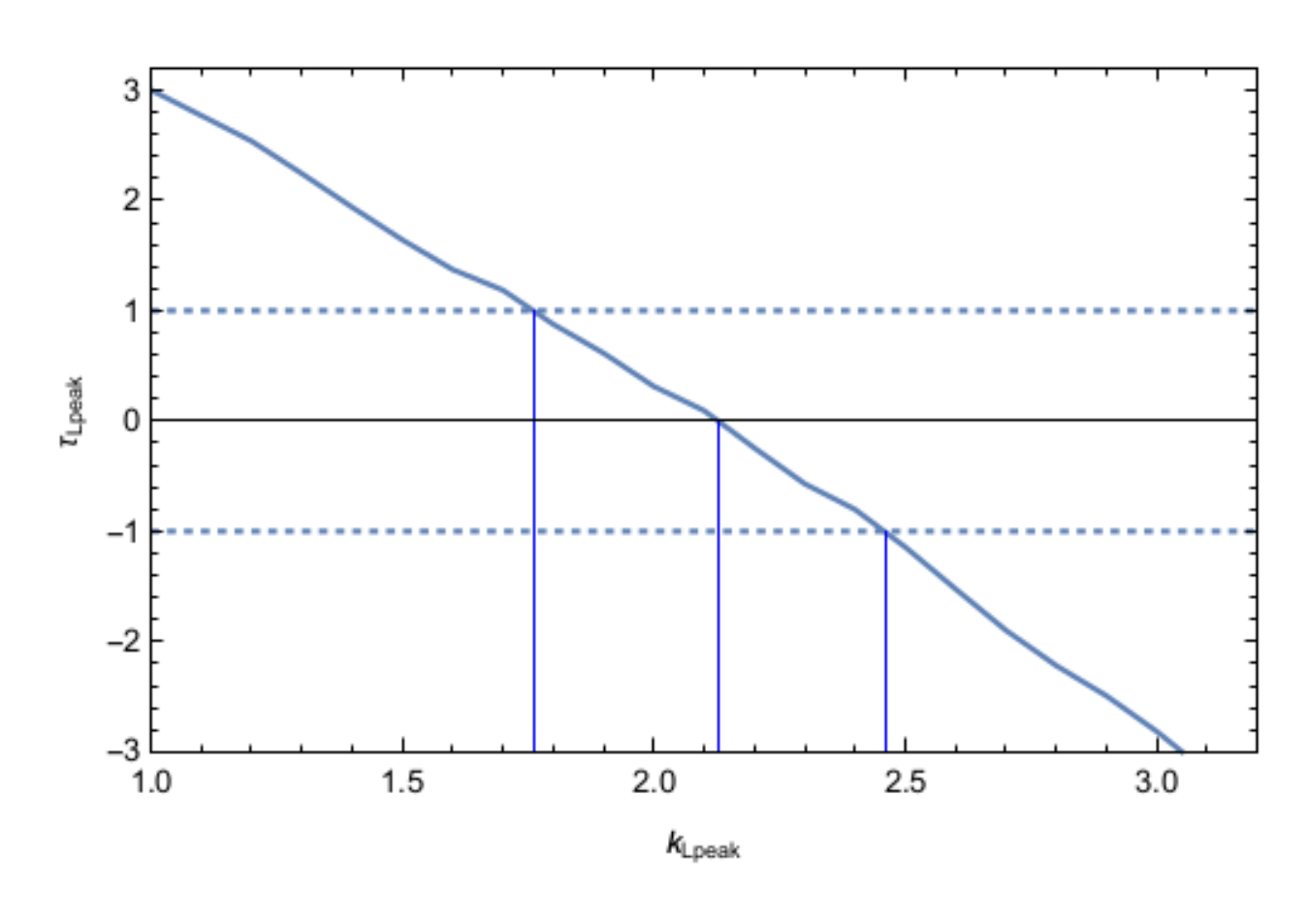}
    \includegraphics[width=5.6cm,height=5.3cm,angle=0,clip]{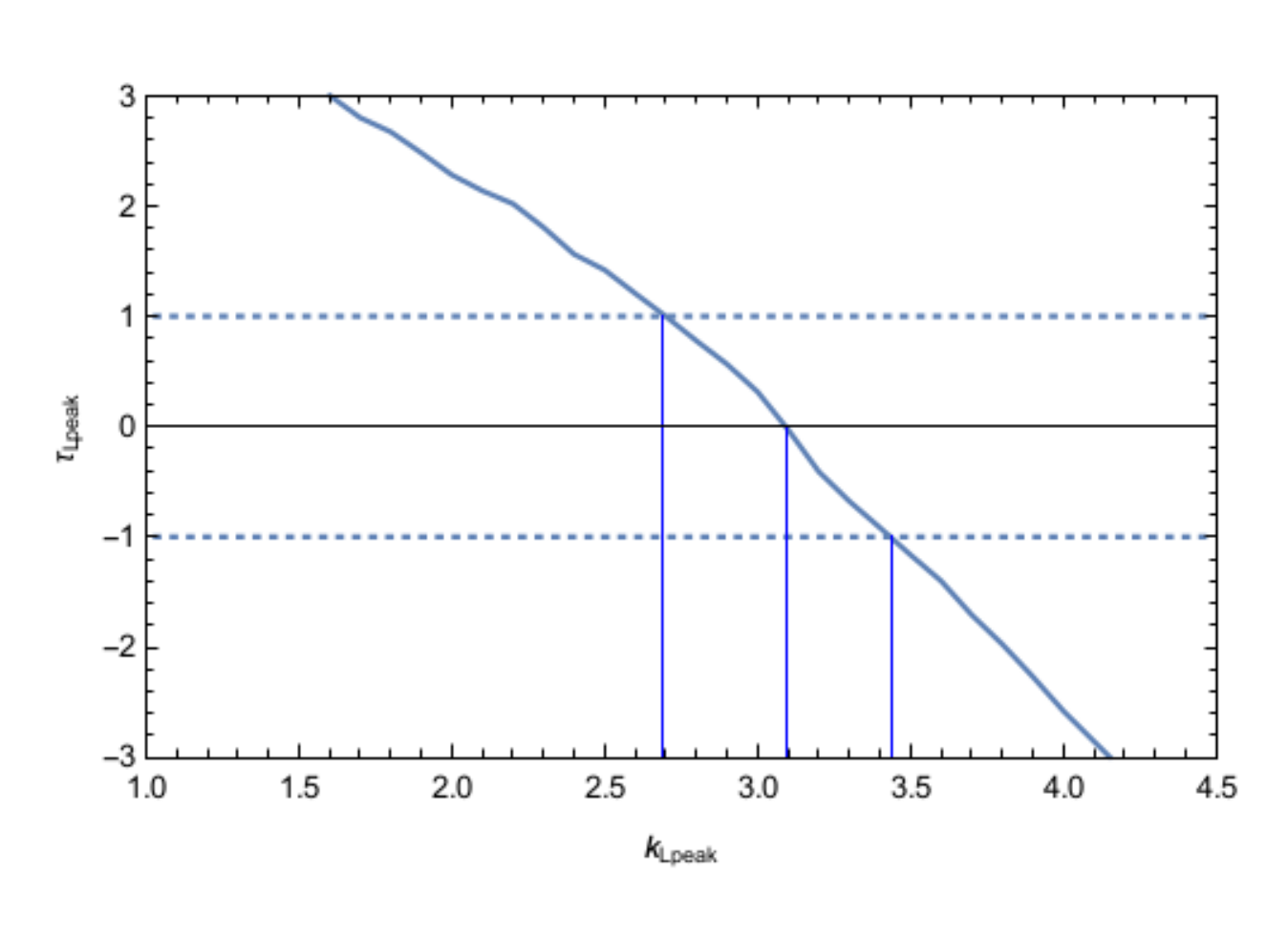}
     \caption{\footnotesize Left panel: $\tau$ vs. $k_{L_{X,a}}$ from \cite{Dainotti2013a}.
   The red line indicates the full sample, while the green dotted line indicates the sample of 47 GRBs 
   in common with the 77 LGRBs in \cite{dainotti11a}. Middle panel: $\tau$ vs. $k_{L_{X,peak}}$, using the 
   eq. \ref{lxev}, from \cite{dainotti15}. Right panel: $\tau$ vs. $k_{L_{X,peak}}$, using the eq. \ref{compl1}, from 
   \cite{dainotti15}.}
   \label{fig:18bis}
\end{figure}

With this modified version of $\tau$, the value of $k_{L_{X,a}}$ for which 
$\tau_{L_{X,a}} = 0$ is the one that best represents the luminosity evolution at the 1 $\sigma$ level.
$k_{L_{X,a}}=-0.05_{-0.55}^{+0.35}$ means that this evolution is negligible, see the left panel of Fig. \ref{fig:18bis}.
In the same panel, this distribution is also plotted
for a smaller sample of 47 GRBs (green dotted line) in common with the previous one of 77 LGRBs presented in \cite{dainotti11a}.\\
The results of the afterglow luminosity evolution among the two samples are compatible at 2 $\sigma$.
Instead, regarding the study of the evolution of $L_{X,peak}$, the simple relation function 
(see eq. \ref{lxev}) was compared to a more complex function \citep{dainotti15} given by:

\begin{equation} 
  g(z)=\frac{Z^{k_L}(1+Z_{cr}^{k_L})}{Z^{k_L}+Z_{cr}^{k_L}},
  \label{compl1}
     \end{equation}
where $Z = 1 + z$ and $Z_{cr} = 3.5$. A relevant luminosity evolution was obtained in the prompt, 
$k_{L_{X,peak}} = 2.13^{+0.33}_{-0.37}$, using the simple relation, while $k_{L_{X,peak}} = 3.09^{+0.40}_{-0.35}$
for the more complex function, see the middle and right panels of Fig. \ref{fig:18bis} respectively.
The results of the prompt luminosity evolution among the two different functions are compatible at 2 $\sigma$.

\subsubsection{Time Evolution}\label{timeev}
\begin{figure}[htbp]
\centering
   \includegraphics[width=0.495\hsize,angle=0,clip]{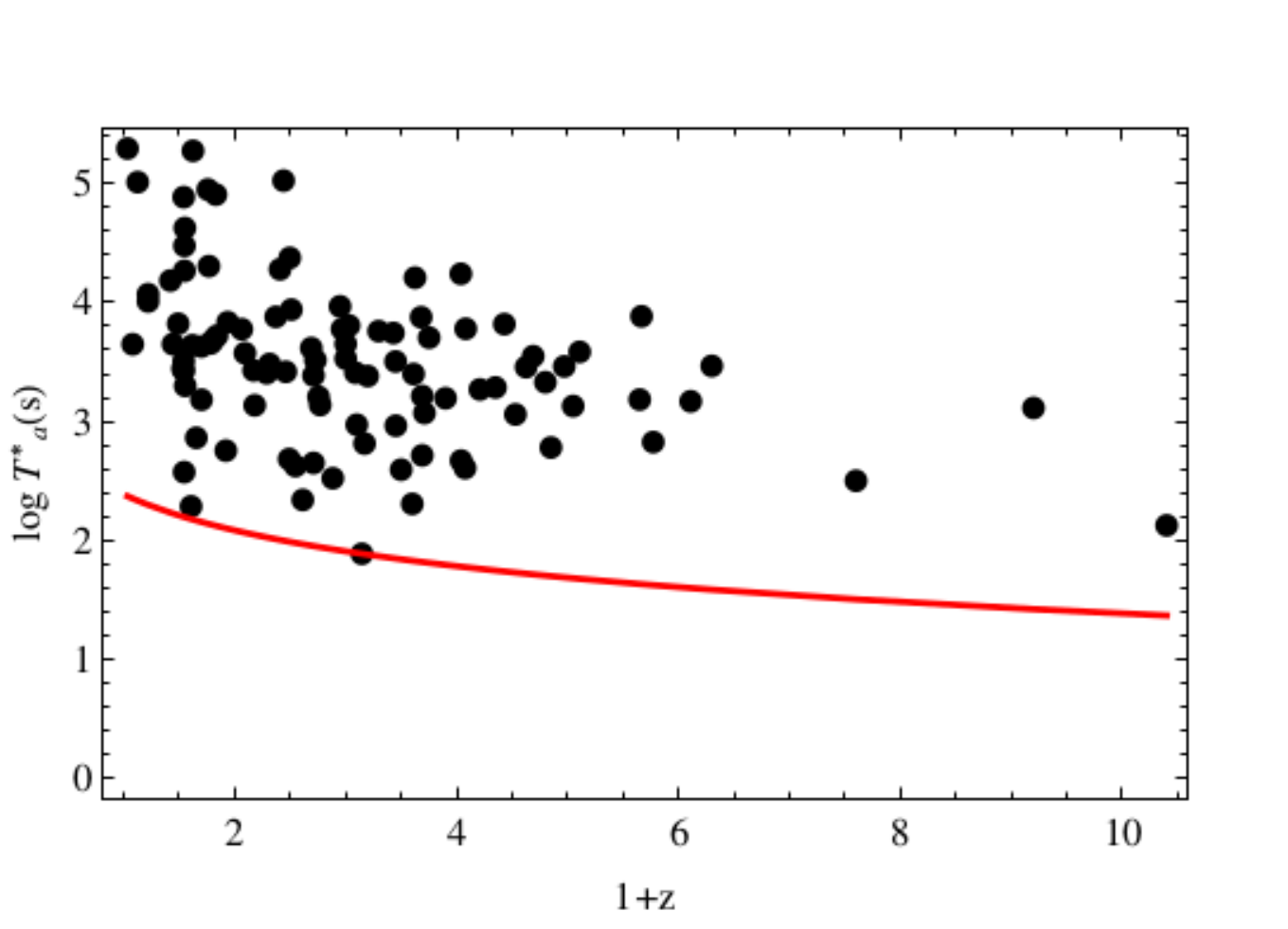}
   \includegraphics[width=0.495\hsize,angle=0,clip]{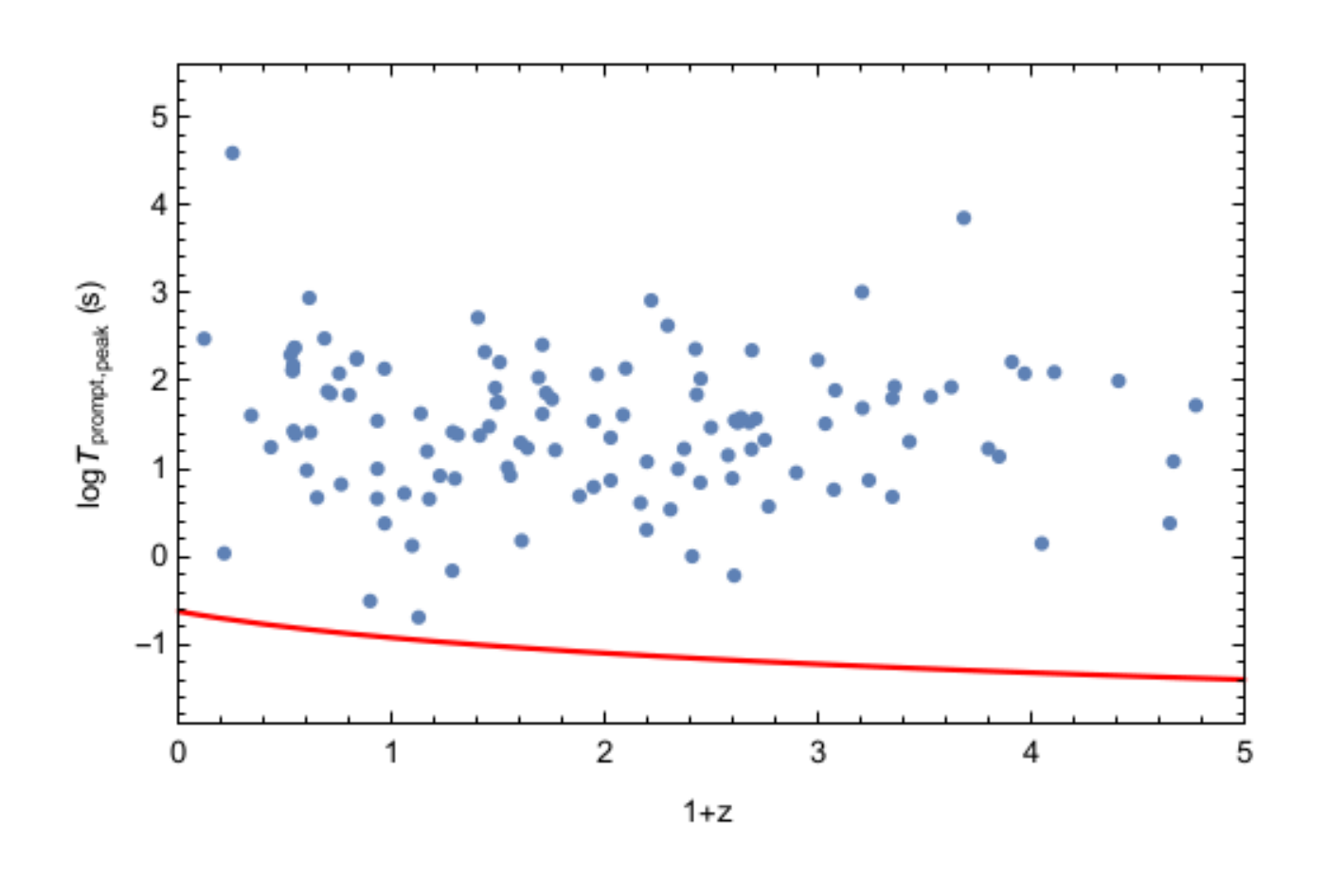}
   \caption{\footnotesize Left panel: ``the bivariate distribution of the rest frame time $\log T^{*}_{X,a}$ and $z$ from
   \cite{Dainotti2013a}. The red line is the limiting rest frame time, $\log (T_{X,a,lim}/(1 + z))$ where the chosen limiting value of the
   observed end-time of the plateau in the sample is $T_{X,a,lim} = 242$ s". Right panel: ``the bivariate distribution of the rest frame
   time $\log T^*_{X,prompt}$ and $z$ from \cite{dainotti15}, where with $\log T^*_{X,prompt}$ they denoted the sum of the 
   peak pulses width of each single
   pulse in each GRB. The chosen limiting value of the observed pulse width in the sample is $\log T_{X,prompt,lim} = 0.24$ s. 
   The red line is the limiting rest frame time, $\log (T_{X,prompt,lim}/(1 + z))$".}
   \label{fig:19bis}
\end{figure}

\begin{figure}
 \centering
   \includegraphics[width=1\hsize,angle=0,clip]{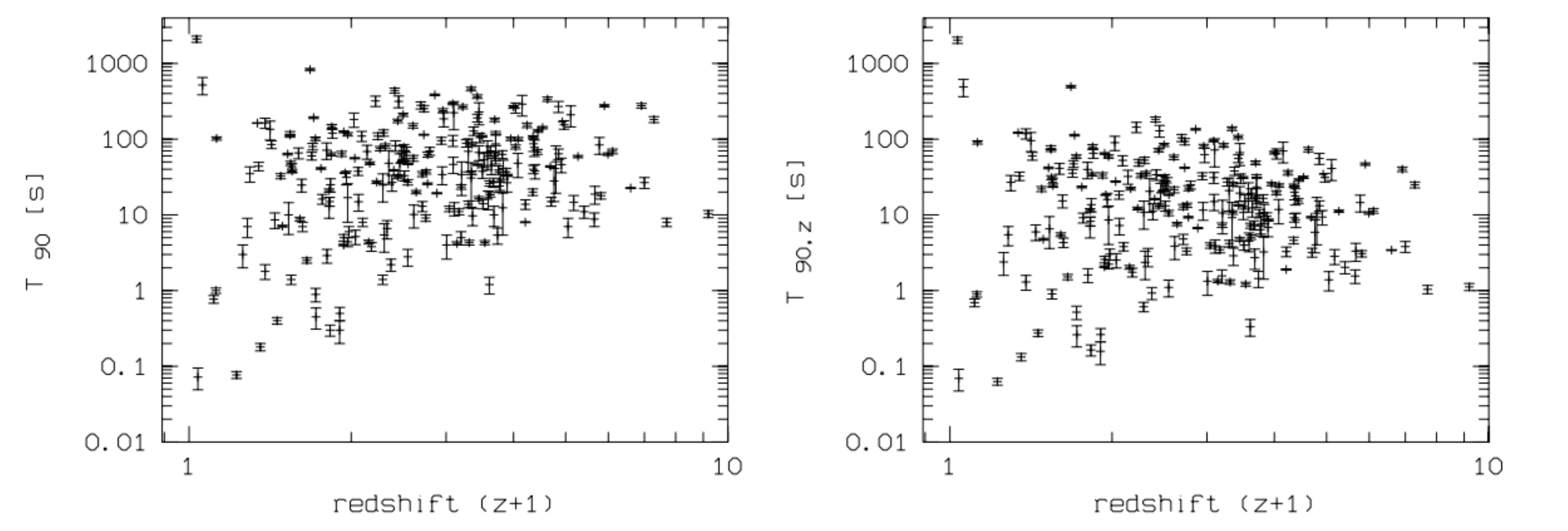}
   \caption{\footnotesize Distributions between redshift and the observed (left panel) and rest-frame (right panel) 
   $T_{90}$ in the BAT energy range from \cite{grupe2013}.}
   \label{fig:grupetime}
\end{figure}

Similarly to the treatment of the luminosity evolution, one has also to determine the limit of the plateau end time, 
$T^{*}_{X,a,{lim}}= 242/(1+z)$ s \citep{Dainotti2013a}, and of the prompt peak time 
$T^*_{X,prompt,lim} = 1.74/(1+z)$ s \citep{dainotti15}, see the left and right panels of Fig. \ref{fig:19bis} 
and Fig. \ref{fig:grupetime} respectively.

\begin{figure}[htbp]
\centering
   \includegraphics[width=5.05cm,height=4.75cm,angle=0,clip]{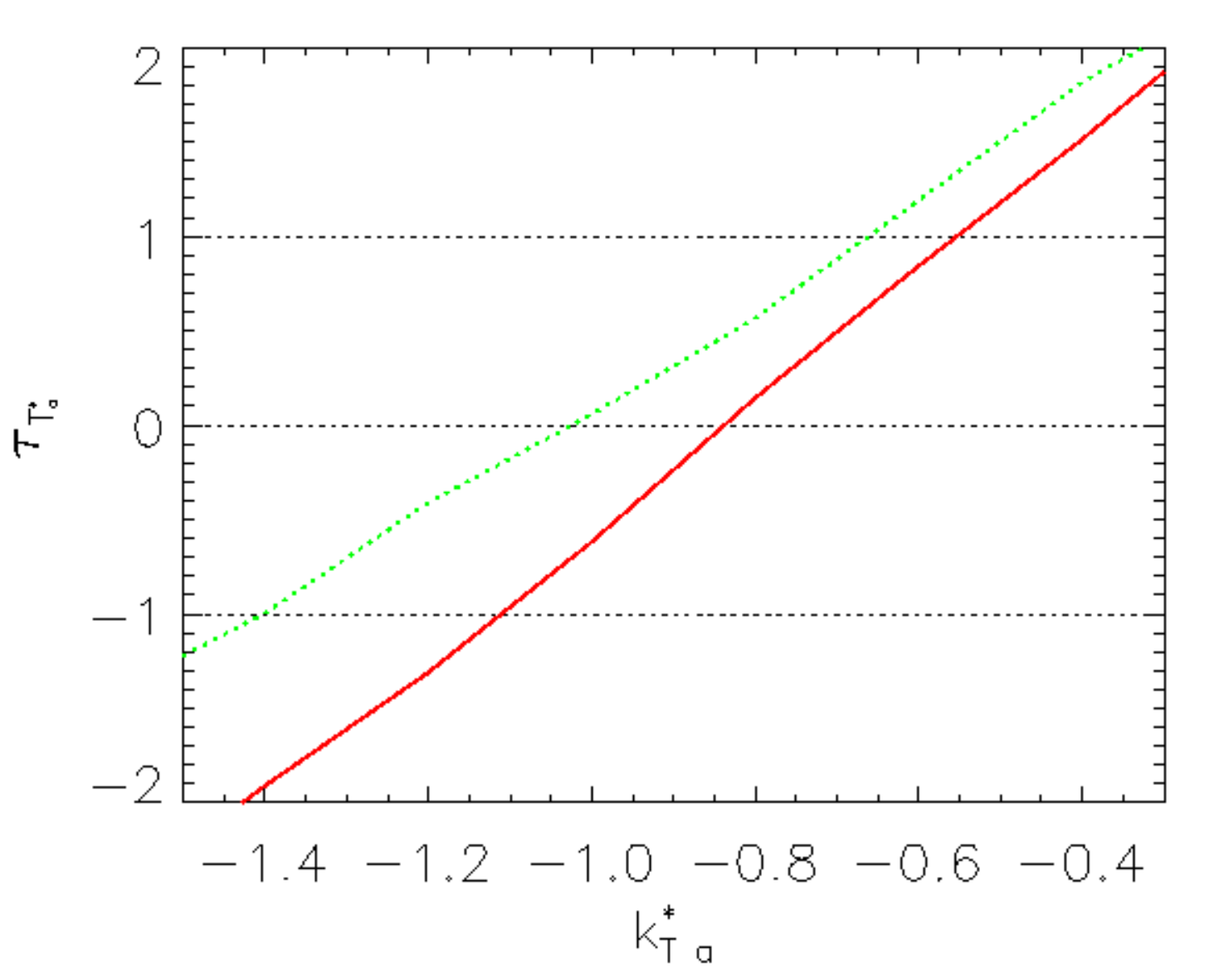}
   \includegraphics[width=5.6cm,height=4.65cm,angle=0,clip]{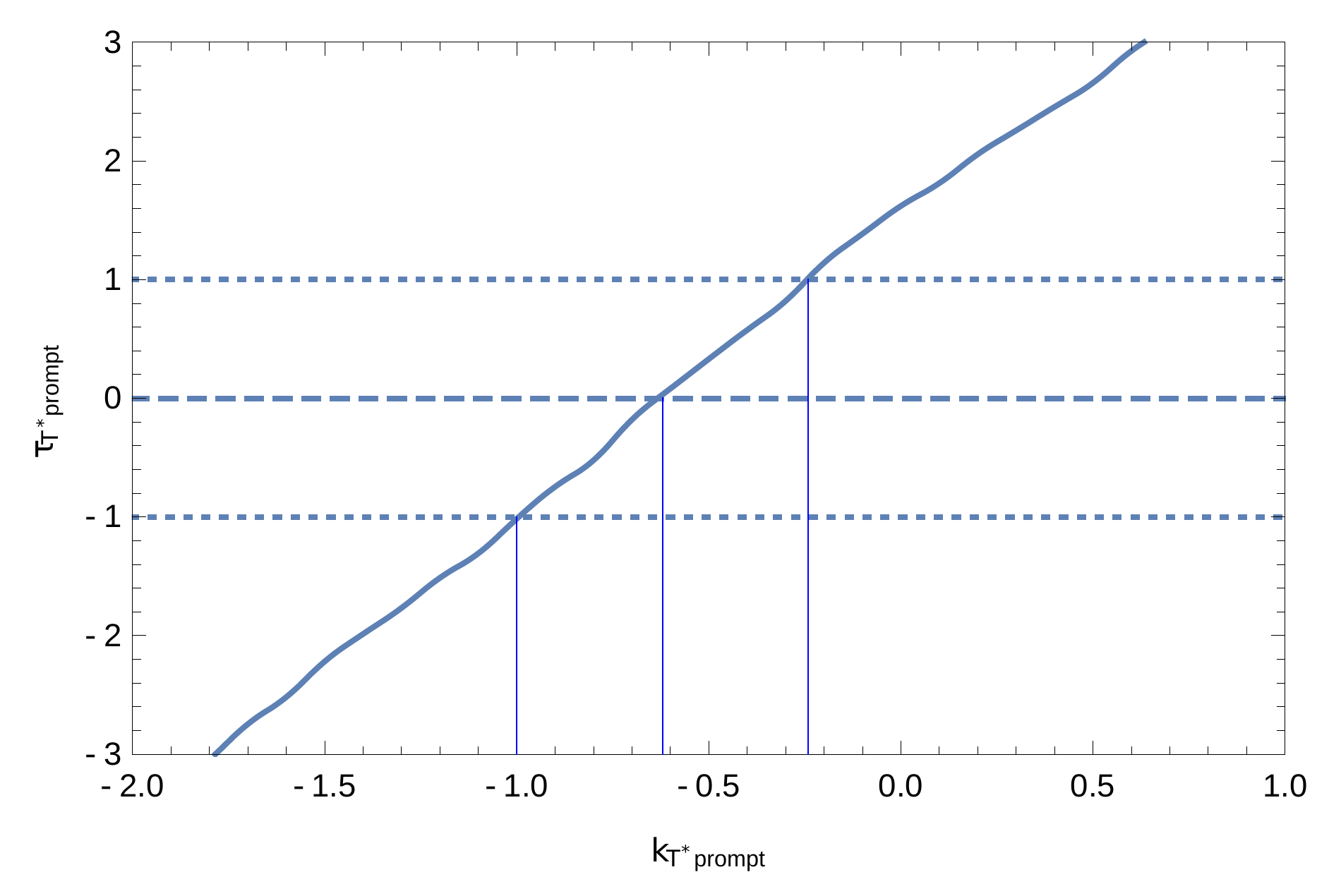}
   \includegraphics[width=5.6cm,height=4.65cm,angle=0,clip]{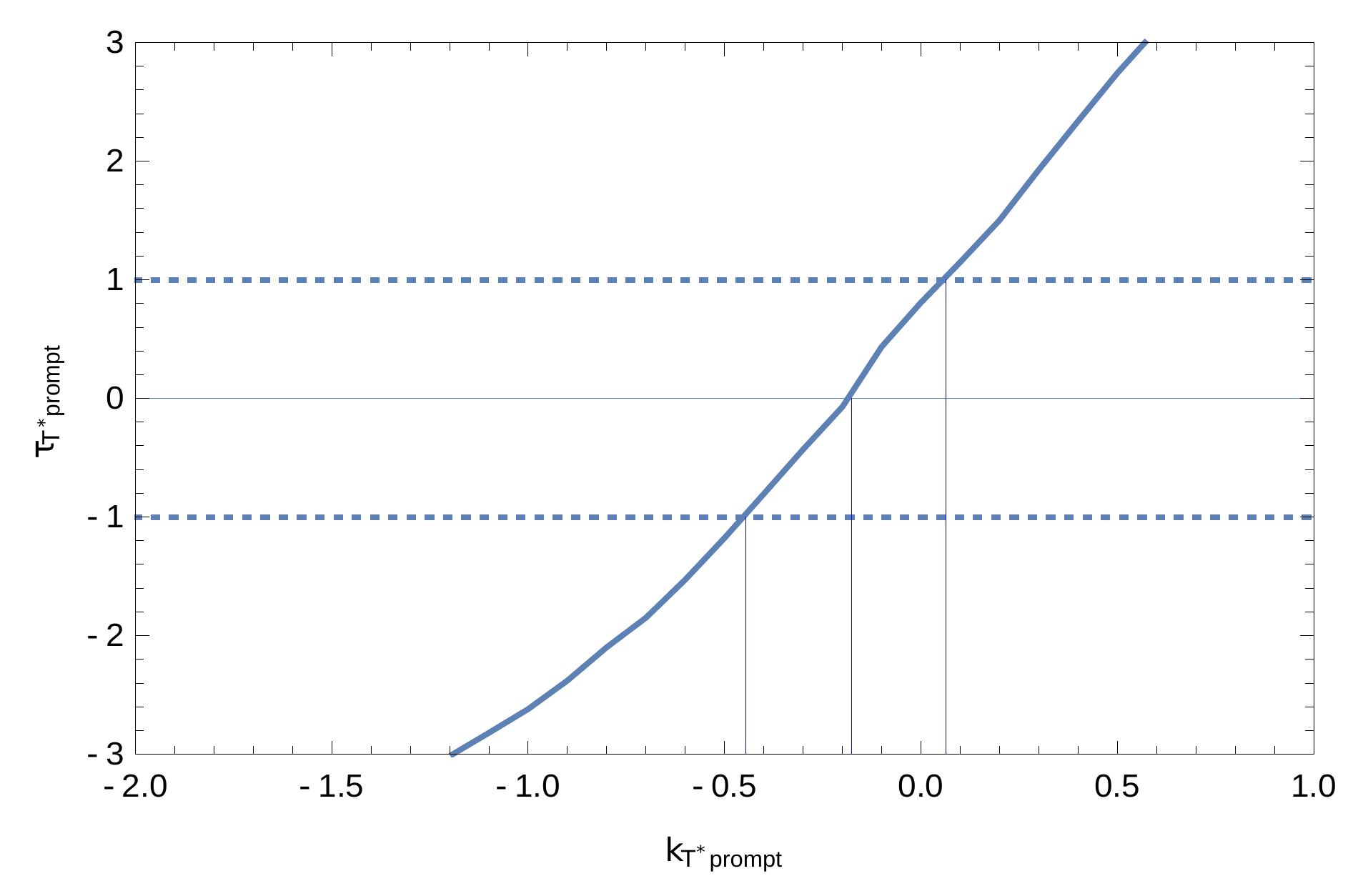}
     \caption{\footnotesize Left panel: $\tau$ vs. $k_{T^{*}_{X,a}}$ from \cite{Dainotti2013a}. The red line indicates the full sample, while the 
green dotted line indicates the 47 GRBs in common with the sample presented in \cite{dainotti11a}. Middle panel: $\tau$ 
vs. $k_{T^*_{X,prompt}}$, using the eq. \ref{taev}, from \cite{dainotti15}. Right panel: $\tau$ vs. $k_{T^*_{X,prompt}}$, using the eq. \ref{compl2}, from \cite{dainotti15}.}
   \label{fig:18}
\end{figure}

To determine the evolution of $T^*_{X,a}$, so that the de-evolved variable $T'_{X,a} \equiv T^{*}_{X,a}/f(z)$ is not correlated 
with z, the relation function $f(z)$ \citep{Dainotti2013a} was specified:

\begin{equation}
f(z)=(1+z)^{k_{T^{*}_{X,a}}}.
\label{taev}
\end{equation}

The values of $k_{T^{*}_{X,a}}$ for which $\tau_{T^{*}_{X,a}} = 0$ is the one that best matches the plateau end time evolution 
at the 1 $\sigma$ uncertainty. $\tau_{T^{*}_{X,a}}$ versus $k_{T^{*}_{X,a}}$ distribution
shows a consistent evolution in $T^{*}_{X,a}$, as seen in the left panel of Fig. \ref{fig:18}, namely 
$k_{T^{*}_{X,a}}=-0.85_{-0.30}^{+0.30}$. In the same panel this distribution is also displayed for a smaller 
sample of 47 GRBs (green dotted line) in common with the previous one of 77 GRBs presented 
in \cite{dainotti11a}.
The results of the afterglow time evolution among the two samples are compatible at 1.5 $\sigma$.\\
Regarding the prompt time evolution,
a more complex function was also used in addition to the simple relation function \citep{dainotti15}:

\begin{equation} 
  f(z)=\frac{Z^{k^*_T}(1+Z_{cr}^{k^*_T})}{Z^{k^*_T}+Z_{cr}^{k^*_T}},
  \label{compl2}
    \end{equation}

where $Z = 1 + z$ and $Z_{cr} = 3.5$.

As a conclusion, a not relevant time evolution in the prompt was found for both the simple function, 
$k_{T^*_{X,prompt}}= -0.62^{+0.38}_{-0.38}$, and for the more complex one $k_{T^*_{X,prompt}}=-0.17^{+0.24}_{-0.27}$,
see the middle and right panels of Fig. \ref{fig:18} respectively.
The results of the prompt time evolution among the two different functions are compatible at 1 $\sigma$.

\subsection{Evaluation of the intrinsic slope}\label{intrinsicorrelation}
\begin{figure}[htbp]
\centering
      \includegraphics[width=0.43\hsize,angle=0,clip]{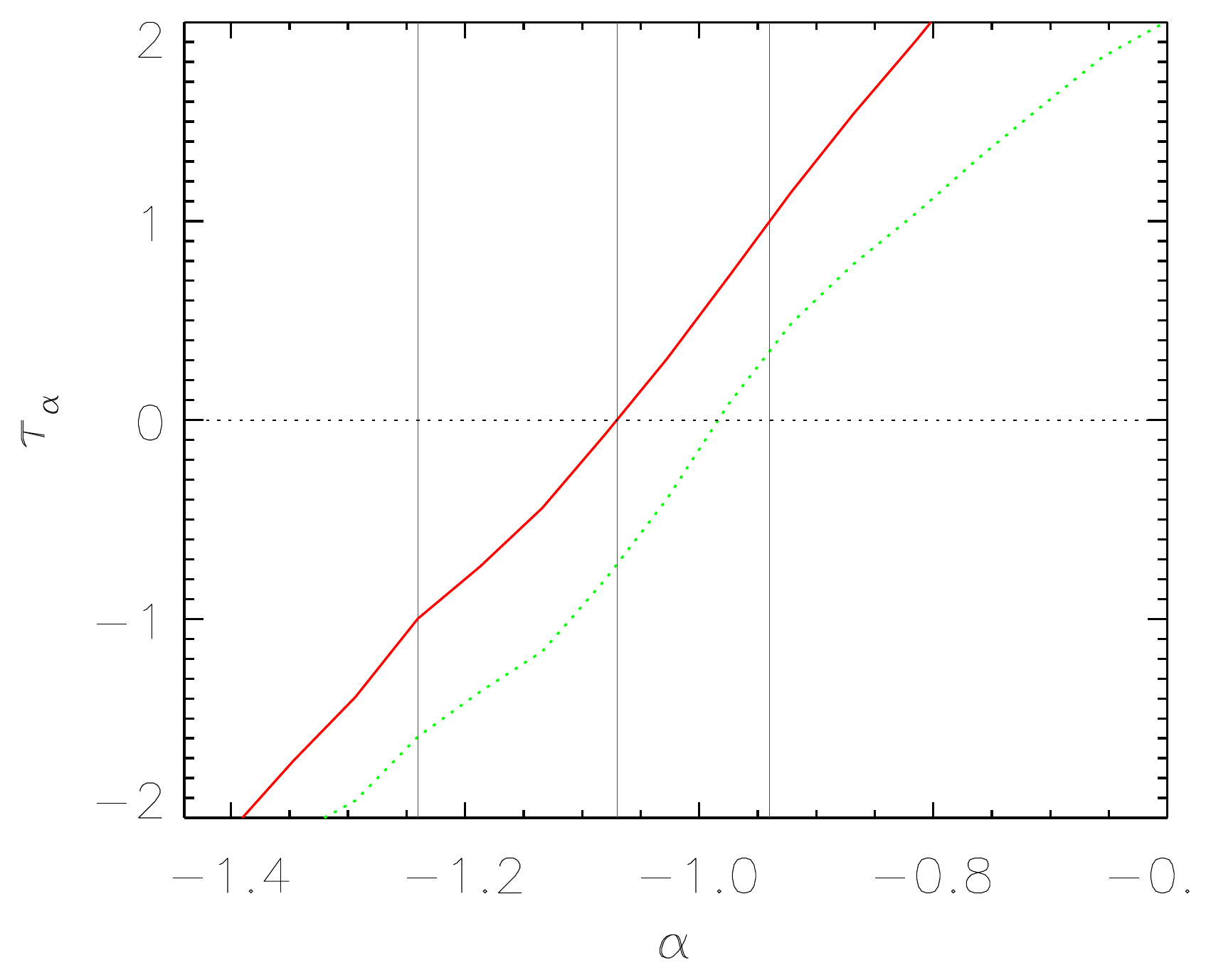}
      \includegraphics[width=0.54\hsize,angle=0,clip]{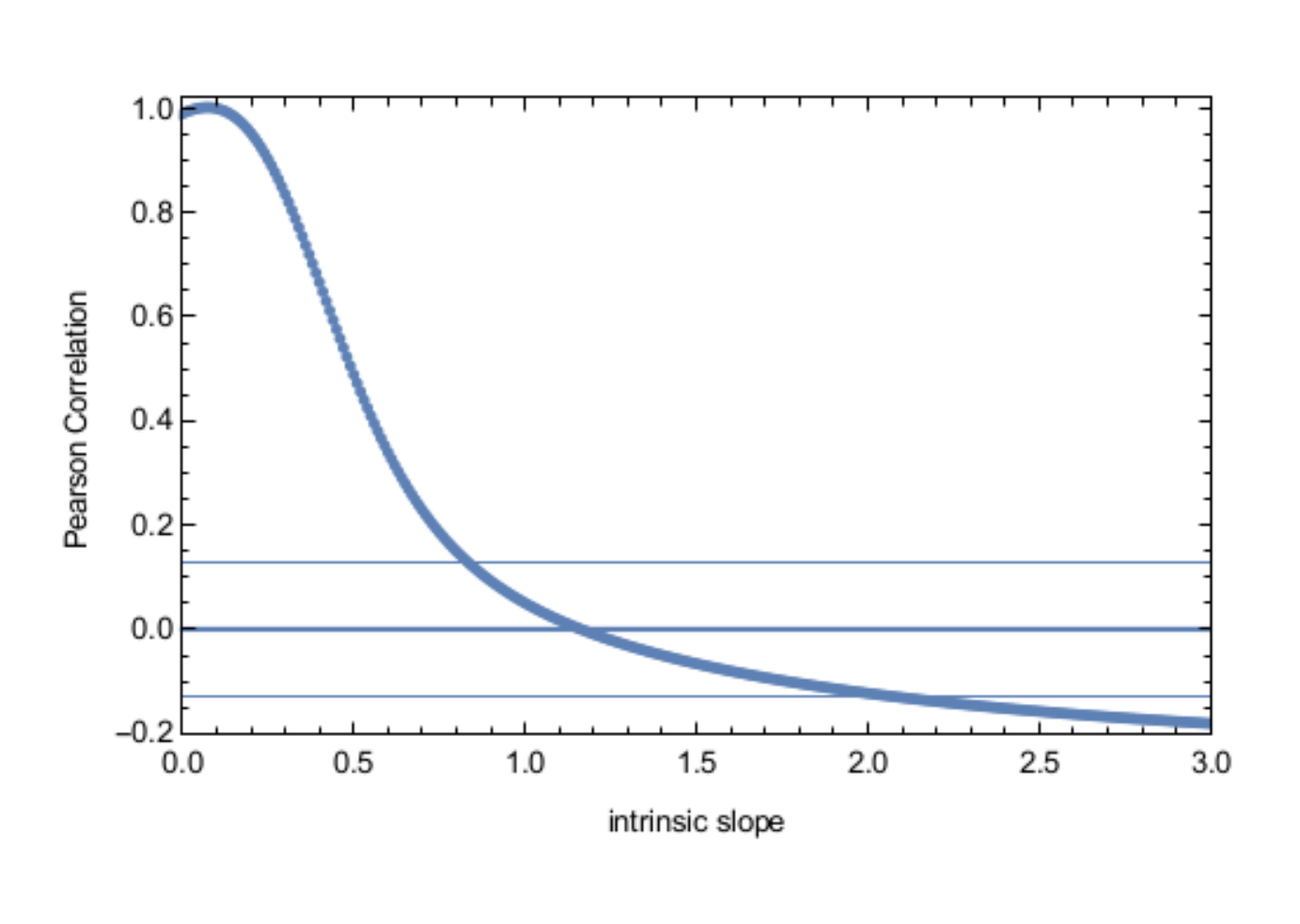}
   \caption{\footnotesize Left panel: $\tau$ vs. $b_{int}$ (indicated with $\alpha$ in the picture) from \cite{Dainotti2013a}.
   Right panel: r vs. $b_{int}$ from \cite{dainotti15} with the 
   best value where $\log L_{X,peak}$ and $\log L_{X,a}$ are strongly correlated (the central thick line). The two thinner 
   lines indicate the $0.05\%$ probability that the sample is drawn by chance.}
   \label{fig:17}
\end{figure}

The last step to determine if a relation is intrinsic is to evaluate its ``true" slope.
To this end, the EP method was used in the local time ($T'_{X,a}$) and luminosity ($L'_{X,a}$) space 
obtaining an intrinsic slope for the LT relation $b_{int}=1/\alpha=-1.07^{+0.09}_{-0.14}$. 
The significance of this relation is at 12 $\sigma$ level. It
can be derived directly from the left panel of Fig. \ref{fig:17} \citep{Dainotti2013a}, because 
if there was no relation it would have been that $\tau=0$ for $b_{int}=0$ at 1 $\sigma$.\\
Instead, regarding the evaluation of the intrinsic slope in the $\log L_{X,peak}-\log L_{X,a}$ relation,
\cite{dainotti15} used a different method, namely the partial correlation coefficient.
This is the degree of association between two random variables calculated as a function of $b_{int}$ in 
the following way:

\begin{equation}
r_{L^{'}_{X,peak} L^{'}_{X,a}, D_L}=\frac{r_{L^{'}_{X,peak},L^{'}_{X,a}}-r_{L^{'}_{X,peak},D_L}*r_{L^{'}_{X,a},D_L}}{(1-r^{2}_{L^{'}_{X,peak},D_L})*{(1-r^{2}_{L^{'}_{X,a},D_L})}},
\end{equation} 

where $\log L^{'}_{X,a}=L^{'}_{X,a}$ and $\log L^{'}_{X,peak}=L^{'}_{X,peak}$.

As displayed in the right panel of Fig. \ref{fig:17}, the relation is
highly significant when $b_{int}= 1.14^{+0.83}_{-0.32}$, which is at 1 $\sigma$ of the observed slope.\\
In addition, following an analysis similar to the one of \cite{Butler2010}, \cite{Dainotti2015b} simulated a sample for 
which biases on both time and luminosity are considered. Particularly, they assumed the biases to be roughly the same whichever 
monotonic efficiency function for the luminosity detection is taken. This method presented how an unknown 
efficiency function could affect the slope
of any relation and the GRB density rate. Then, biases in slope or normalization
caused by the truncations were analyzed. This gave distinct fit values that allow for studying the scatter of the relation and its
selection effects. This analysis has shown, together with the one in \cite{Dainotti2013a}, that the LT
relation can be corrected by selection effects and therefore can be used in principle as redshift estimator
(see sec. \ref{redshiftestimator}) and as a valuable cosmological tool (see sec. \ref{cosmology}).
As regards other relations, \cite{davanzo12} for the $L_{X,a}-E_{\gamma,prompt}$ relation, 
\cite{oates15} for the $L_{O,200\rm{s}}-\alpha_{O,>200\rm{s}}$ relation,
and \cite{racusin16} for the $L_{X,200\rm{s}}$\,-\,$\alpha_{X,>200\rm{s}}$ relation, also used the partial correlation 
coefficient method to show that the redshift dependence does not induce these relations.

\subsection{Selection effects for the optical and X-ray luminosities}\label{ghisellinisel}
In this section we discuss the selection effects due to the limiting optical and X-ray luminosities
relevant for the relations mentioned above.
\cite{nardini08a} investigated 
if the observed luminosity distribution can be the result of selection effects by studying the optically dark afterglows.
By simulating the $\log L_{O,12}$, $z$, host galaxy dust
absorption, $A^{host}_V$, and telescope limiting magnitude for each of the 30000 GRBs, the observed optical luminosity 
distribution was contrasted to 
the simulated one.
From this simulated distribution regarding the intrinsic one, it is necessary 
to take only GRBs with a flux which is larger than the threshold flux of the associated detector. This corresponds 
with a lower luminosity truncation, which is around $\log \ L_{O,12} \approx 31.2$ (erg s$^{-1}$ Hz$^{-1}$). Therefore, the 
fact that we do not
observe GRBs with such a luminosity puts a limit to the luminosity function.\\
They also checked statistically the presence of a low luminosity category of events which are at $3.6$
$\sigma$ off the central value of the distribution. 
They pointed out that if the absorption is chromatic, the observed luminosity distribution does not match with any unimodal
one. If many GRBs are absorbed by
``grey" achromatic dust, then a unimodal luminosity distribution
can be obtained. 
In summary, dark bursts could belong to an optically subluminous group or to a category of bursts for which
a high achromatic absorption is present.\\
As regards the evaluation of the selection effects of $L_{O,peak}$, the biases
in the detection of $F_{O,peak}$ need to be considered.
As found from \cite{panaitescu08}, for a typical optical afterglow spectrum ($F_{O,a}\propto T_{O,a}^{-1}$), 
variations in 
the observer offset angle induce a $\log F_{O,peak}-\log T_{O,peak}$ anti-relation that is flatter than
what is measured. In fact, an observational selection
effect could steepen the slope of the anti-relation between $\log F_{O,peak}$ and $\log T_{O,peak}$.\\
In addition, SGRBs observed by Swift seem to be fluence-limited, while LGRBs detected with the same telescope 
are flux-limited \citep{gehrels08} due to the instrument trigger.\\
\cite{nysewander09} pointed out that the ratio $F_{O,11}/F_{X,11}$ may be influenced 
by absorption of photons in the host galaxy. Furthermore, they showed that $F_{X,11}$ should 
be precise, because the LGRBs observed in the XRT passband do not present X-ray column absorptions, differently 
from the majority of LGRBs.
The computed optical absorption of LGRB afterglows indicates smaller column densities ($N_H$) 
than in the X-ray, with optical absorptions ($A_V$) about one-tenth to one magnitude
\citep{shady07,cenko09}. Regarding the SGRBs, they have more luminous optical emission relative to the X-ray
than what is assumed by the standard model.
Later, \cite{kann10} claimed that the grouping of the optical luminosity at the time of 1 day, 
$L_{O,1\rm{d}}$, is less remarkable than the one described by \cite{liang06} and \cite{nardini06a} 
for GRBs observed by Swift.
This suggested that the grouping pointed out in pre-Swift data can
be due to selection effects only. Finally, \cite{berger14} claimed that the optical afterglow detection
can influence the luminosity distribution towards places with larger densities medium.

\subsection{Selection effects in the \texorpdfstring{$L_{O,200\rm{s}}-\alpha_{O,>200\rm{s}}$}{Lg} relation}\label{montecarlosimulation}

\cite{oates2012} ensured that a high S/N light curve, covering both early and
late times, can be constructed from the UVOT multi-filter observations using the criteria from \cite{oates2009}.
If the faintest optical/UV afterglows decay more slowly than the brightest ones,
then at late time the luminosity distribution is less dispersed and the correlation coefficient of the 
$\log L_{O,200\rm{s}}-\alpha_{O,>200\rm{s}}$ relation must become smaller and/or
negligible. Indeed, both of these effects were observed in their sample.
Furthermore, the $\log L_{O,200\rm{s}}-\alpha_{O,>200\rm{s}}$ relation may arise, by chance, from the way in which the 
sample is chosen. Thus, to verify if this is not the case, they computed Monte Carlo simulations. 
Among the $10^6$ trials, 34 have a correlation 
coefficient indicating a more significant relation than the original one. This points out that, at 4.2 $\sigma$ confidence,
the $\log L_{O,200\rm{s}}-\alpha_{O,>200\rm{s}}$ relation is not caused by the selection criteria nor does it happens by chance, and 
thus it is intrinsic.

\section{Redshift Estimator}\label{redshiftestimator}
As we have pointed out in the introduction, the study of GRBs as possible distance estimators is relevant, 
since for many of them
$z$ is unknown. Therefore, having a relation which is able to infer the distance from known quantities observed
independently of $z$ would allow a better investigation of the GRB population.
Moreover, in the cases in which $z$ is uncertain, the estimator can give hints on the 
upper and lower limits of the distance at which the GRB is placed. Some examples of redshift estimators for the 
prompt relations \citep{atteia03,yonetoku04,tsutsui13} have been reported. In these papers, a method 
is developed for inverting GRB luminosity relations in respect to the redshift to have an expression of the distance 
as a function of z. 
The methodology used for the prompt emission relations can be then applied also to the afterglow or prompt-afterglow 
phase relations.

In this respect, \cite{dainotti11a} investigated the LT relation as a redshift estimator.
From this relation, the best fit parameters of the slope and normalization are derived, while parameters such as 
$\log F_{X,a}$, $\log T_{X,a}$ and $\beta_{X,a}$ are known, because they are measured. 
Therefore, the LT relation can be inverted to obtain
an estimate of $z$ as it has been done for the prompt relations by \cite{yonetoku04}.
With this intention, let us return to the eq. \ref{eq: lx} and write it in another form:

\begin{eqnarray}
\log{L_{X,a}} & = & \log{(4 \pi F_{X,a})} + 2 \log{D_L(z, \Omega_M, h)} - (1 - \beta_{X,a}) \log{(1 + z)} \nonumber \\
~ & = & \log{(4 \pi F_{X,a})} + (1 + \beta_{X,a}) \log{(1 + z)} + 2 \log{r(z)} + 2 \log{(c/H_0)} \nonumber \\
~ & = & a \log{\left ( \frac{T_{X,a}}{1 + z} \right )} + b
\end{eqnarray}

where $r(z)=D_L(z, \Omega_M, h)\times(H_0/c)$. Solving respect to $z$, it was obtained:

\begin{equation}
(1 + \beta_{X,a} + a) \log{(1 + z)} + 2 \log{r(z)} = a \log T_{X,a} + b - \log{(4 \pi F_{X,a})} - 2 \log{(c/H_0)}.
\label{eq: zedest}
\end{equation}

\begin{figure}[htbp]
\centering
   \includegraphics[width=1\hsize,angle=0]{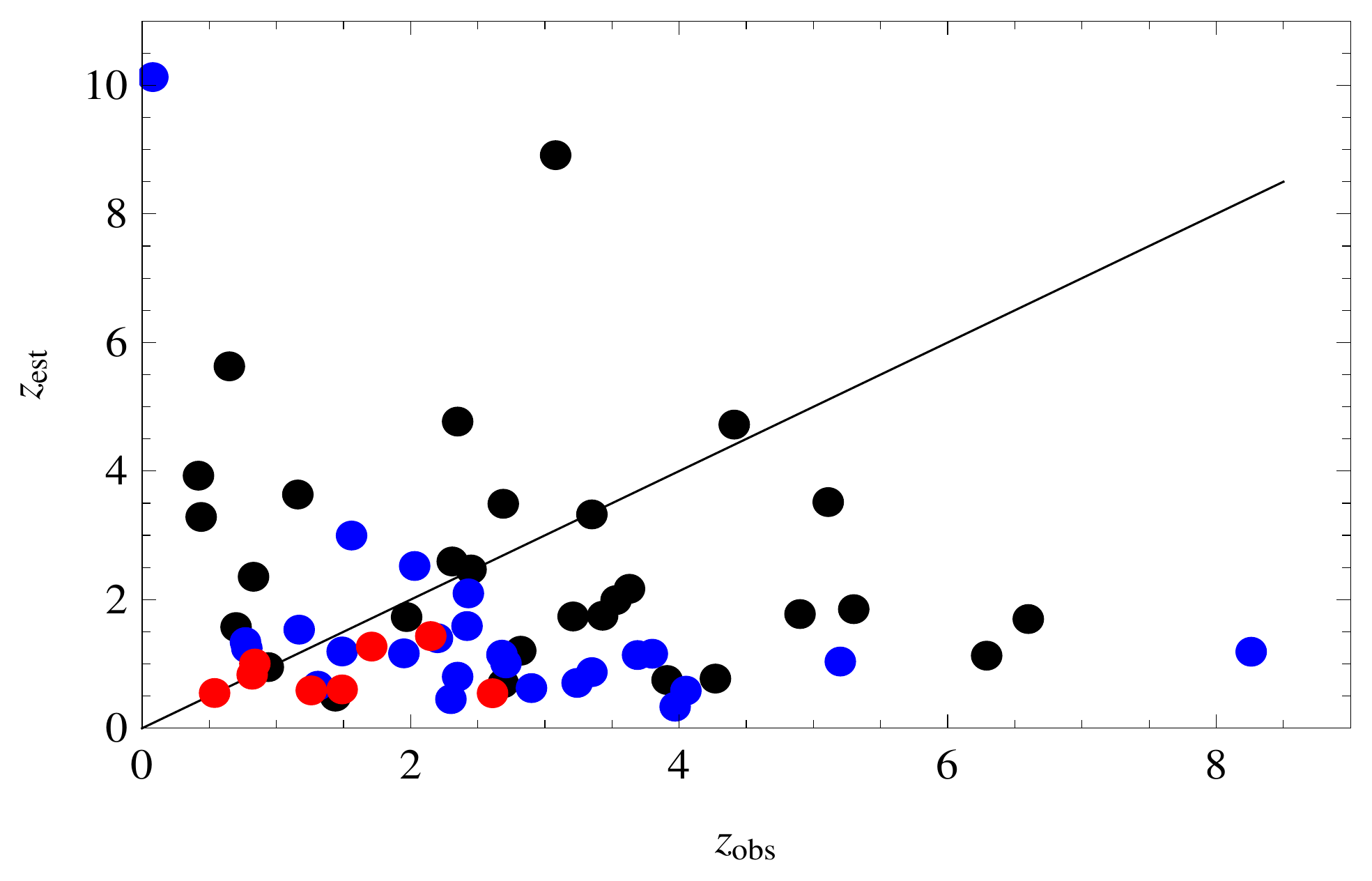}
   \caption{\footnotesize $z_{obs}$-$z_{est}$ distribution for the 62 LGRBs divided in three $\sigma_E$ ranges from 
   \cite{dainotti11a}: $\sigma_E \le 0.095$ is represented by red points, $0.095 \le \sigma_E \le 0.3$ is represented 
   by blue points, $0.3 \le \sigma_E \le 4$ is represented by black points.}
   \label{fig:theorobservredshift}
\end{figure}

The numerical solution of this equation may encounter some problems that must be taken into account:  
$(\log T_{X,a}, \log \\F_{X,a}, \beta_{X,a})$ and the LT calibration parameters $(a, b)$ are influenced by their 
own errors. Furthermore, the errors on $(a, b)$ are not symmetric and $\sigma_{int}$ is summed to the total 
error in a nonlinear way. For details about possible solutions on how to consider the errors see \cite{dainotti11a}.
The above solution was employed for the E4 and the E0095 samples, pointing out that the LT 
relation can still not be considered as a precise redshift estimator, see Fig. \ref{fig:theorobservredshift}. 

Assuming $\Delta z = z_{obs} - z_{est}$, where $z_{obs}$ and $z_{est}$ are the observed and the estimated redshifts respectively, 
it has been shown that $\sim 20\%$ of GRBs in the E4 sample (black, $0.3 \le \sigma_E \le 4$, and blue,
$0.095 \le \sigma_E \le 0.3$, points in Fig. \ref{fig:theorobservredshift}) has
$|\Delta z/\sigma(z_{est})| \le 1$. While for the E0095 subsample $28\%$ has $|\Delta z/\sigma(z_{est})| \le 1$, 
red dots in Fig. \ref{fig:theorobservredshift}.
The percentage of successful solutions rises 
to $\sim 53\%$ ($\sim 57\%$) for the E4 (E0095) sample if $|\Delta z/\sigma(z_{est})| \le 3$ is considered.
The comparison of the results for both the E4 and 
E0095 samples is proof that $\sigma_E$ has no strong influence on the redshift estimate.
The reason why the redshift indicator has not yet given successful results depends on the intrinsic scatter of the LT 
relation. 
Thus, it is useful to check whether better results can be achieved by increasing the
data sample size. For this reason, an E0095 subsample was simulated creating $(\log T_{X,a}, \beta_{X,a}, z)$ values from a distribution 
similar to the observed one for the E4 sample. 
Then, $\log L_{X,a}$ was selected from a Gaussian distribution with mean 
value obtained by the LT relation and with $\sigma_{Gauss}=\sigma_{int}$. 
These values were employed to compute $\log F_{X,a}$ and to reproduce the noise for all the quantities so that the relative errors 
resembled the observations. Then, using Markov chains as input to the
redshift estimate formula, it is concluded that only enlarging the sample is not an appropriate methodology to increase the
success of the LT relation as a redshift estimator.\\
In fact, with ${\cal{N}} \simeq 50$, the number of GRBs with 
$|\Delta z/\sigma(z_{est})| \le 1$ first rises to $\sim 34\%$ and then
diminish to $\sim 20\%$ for ${\cal{N}} \simeq 200$, while $\langle \Delta z/z_{obs} \rangle \simeq -17\%$ for both 
${\cal{N}} \simeq 50$ and ${\cal{N}} \simeq 200$.
The fact that enlarging the sample does not improve the result could be expected. Indeed, 
a bigger sample conducts
to tighter constraints on the $(a, b, \sigma_{int})$ values, but does not affect $\sigma_{int}$ which is the principal cause of
inconsistencies between the observed and the estimated $z$.\\
Therefore, an alternative way was explored: $\sigma_{int}$ was decreased and the best fit $(a, b)$ parameters of the E0095 subsample 
were chosen. In fact, fixing $\sigma_{int} = 0.10$ gives  $f(|\Delta z/z_{obs}| \le 1) \simeq 66\%$. 
These outcomes suggested that the LT relation could be employed as a redshift estimator only in the case that a subsample of 
GRBs could be determined with $\sigma_{int} = 0.10-0.20$. If such a sample is achievable 
is not clear yet due to the paucity of the E0095 subsample. In fact, it is difficult to find out
some useful indicators that can help to define GRBs close to the best fit line of the LT relation. To obtain $\sim 50$ GRBs to calibrate 
the LT relation with $\sigma_{int} \sim 0.20$ it has been estimated that a sample of $\sim 600$ GRBs with 
observed $(\log T_{X,a}, \log F_{X,a}, \beta_{X,a}, z)$ values is needed. However, even if this is a challenging goal, it may be possible to find out 
properties of GRB afterglows which enable us to reduce the $\sigma_{int}$ of the LT relation with a much smaller sample. 
Finally, an interesting feature would be to correct for the selection effects all the physical quantities of 
the relations mentioned above. 
In this manner, it would be possible to average them in order to create a more precise redshift 
estimator.

\section{Cosmology}\label{cosmology}
The study of the Hubble Diagram (HD), namely the distribution of the distance modulus $\mu(z)$\footnote{The difference between 
the apparent magnitude m, ideally corrected from the effects of interstellar
absorption, and the absolute magnitude M of an astronomical object.} versus $z$ of SNe Ia, opened the way to the investigation of 
the nature of DE.
As it is known from the literature, $\mu(z)$ is proportional to the logarithm of the luminosity distance 
$D_L(z, \Omega_M, h)$ through the following equation:

\begin{equation}
 \mu(z)  =  25 + 5 \times \log D_L(z, \Omega_M, h).
\end{equation}

In addition, $D_L(z, \Omega_M, h)$ is related to different DE EoSs.

\subsection{The problem of the calibration}
One of the most important issues presented in the use of GRB relations for cosmological studies is the so-called 
circularity problem. Namely, a cosmological model needs to be assumed to compute 
$D_L(z, \Omega_M, h)$. This is due to the fact that local GRBs are not available apart from the case of GRB 980425.
Indeed, this kind of GRBs would be observed at $z < 0.01$ and their measure would be independent of a particular 
cosmological setting.
This issue could be overcome in three ways: a) through the calibration of these relations by several low $z$ GRBs 
(in fact, at $z\le 0.1$ the luminosity distance is not sensitive to the balance of $\Omega_M$ and $\Omega_{\Lambda}$
for a given $H_0$, where $H_0$ is between 65 and 72); b)
through a solid theoretical model in order to explain the observed 2D relations. Namely, this would fix their slopes and 
normalizations independently of cosmology, but this task still has to be achieved; c) through the calibration of the standard 
candles using GRBs in a narrow redshift range ($\Delta z$) near a fiducial redshift, $z_c$. We here describe some examples on how to overcome the problem of circularity using prompt 
relations.
%0.6\hsize
\begin{figure}[htbp]
\centering 
 \includegraphics[width=16.5cm, height=8cm,angle=0]{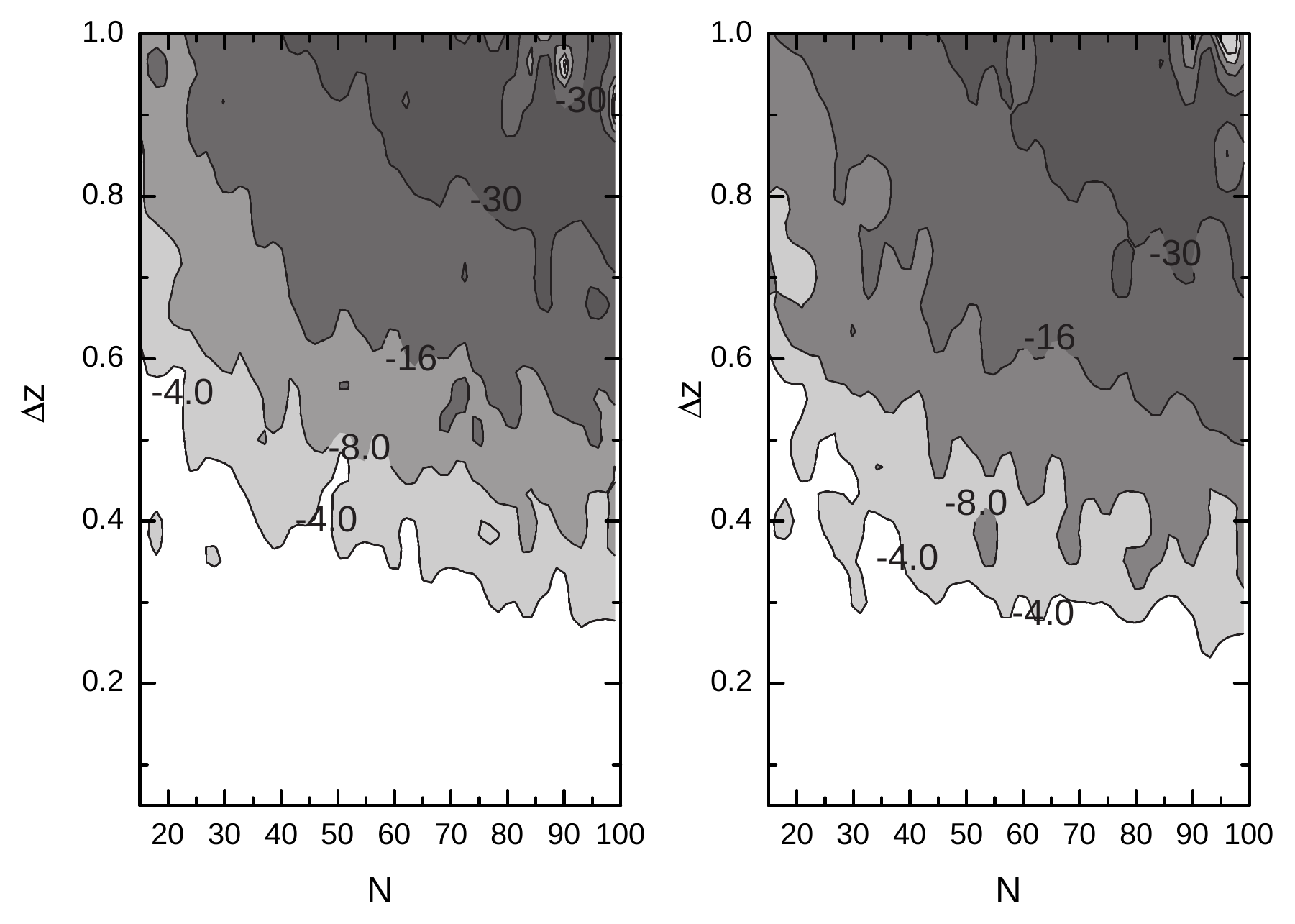}
 \caption{\footnotesize ``Distribution of $\log P$ in the (N, $\Delta z$) plane from \cite{liang06}. The grey contours
mark the areas where the dependencies of $b_1$ and $b_2$ on $\Omega_M$ are statistically significant (P $< 10^{-4}$). The white region is suitable for the calibration purpose".}
 \label{fig:liang06}
\end{figure}

The treatment of this problem will be the same once we consider afterglow or prompt-afterglow relations. 
\cite{liang06} suggested a new GRB luminosity indicator, $E_{\gamma,iso} = aE^{b_1}_{\gamma,peak}T^{b_2}_{O,a}$, 
different from the previous GRB luminosity indicators that are generally
written in the form of $L=a\prod x^{b_i}_i$, where a is the normalization, $x_i$
is the i-th observable, and $b_i$ is its corresponding power law index. It was demonstrated
that while $a$ relies on the cosmological parameters, this is not the case for $b_1$ and $b_2$ until $\Delta z$ is sufficiently 
little, see Fig. \ref{fig:liang06}. The choice of $\Delta z$ for a given GRB sample could be evaluated 
depending on its dimension and the errors on the variables. 
The most suitable approach would be to assemble GRBs within a small redshift range
around a central $z_c$ ($z_c \sim 1$ or $z_c \sim 2$), because the GRB $z$ distribution peaks in this interval
\textcolor{red}{(see also \citealt{wang2011} and \citealt{wang15})}.\\
In addition, also \cite{ghirlanda06} defined the 
luminosity indicator $E_{\gamma,peak}= a \times E^b_{\gamma,cor}$ using the $\log E_{\gamma,peak}-\log E_{\gamma,cor}$ relation \citep{Ghirlanda2004},
where

\begin{equation}
E_{\gamma,cor}=(1-\cos\theta_{jet})\times 4\pi \times D_L^2(z, \Omega_M, h)\times S_{\gamma,prompt}/(1+z)^2 
\end{equation}

is the energy corrected for the beaming factor and $\theta_{jet}$ is the opening angle of the jet.
They calculated the minimum number of GRBs (N), within $\Delta z$ around a certain $z_c$, needed to 
calibrate the relation, considering a sample of 19 GRBs detected mostly by Beppo-SAX and Swift.
Particularly, they fitted the relation for each value of $\Omega_M$ and $\Omega_{\Lambda}$ using
a set of N GRBs distributed in the interval $\Delta z$ (centered around $z_c$). If the variation
of the slope, b, is less than $1\%$ the relation is assumed calibrated.
$N$, $\Delta z$ and $z_c$ are free parameters. They checked several $z_c$ and distinct $z$ dispersions 
$\Delta z \in (0.05, 0.5)$ by Monte Carlo simulations.
At every $z$ the smaller the N the bigger the variation of the slope, $\Delta b$ (for the same $\Delta z$), 
because the relation is more scattered. On the other hand, for greater $z_c$ a
tinier $\Delta z$ is necessary to maintain $\Delta b$ in its little state. Finally, they found that 12 GRBs with
$z \in (0.9, 1.1)$ can be sufficient to calibrate the slope of the $\log E_{\gamma,peak}-\log E_{\gamma,cor}$ 
relation.
Instead, at $z_c = 2$ a narrower redshift bin is needed, for example $z \in (1.95, 2.05)$.\\
However, this method might becomes unsuccessful, because the sample size of the observed GRBs is not sufficiently big. 
Another method for a model-independent calibration may be obtained employing SNe Ia as distance 
indicators. This method is based on
the assumption that a GRB at redshift z must have the same distance modulus $\mu(z)$ of a SNe Ia at the same redshift.
In this way, GRBs should be considered as complementary to SNe Ia at very high z, thus allowing for the construction of
a very long distance
ladder. Therefore, interpolating the SNe Ia HD provides the value of $\mu(z)$ for a subsample
of GRBs with $z \le 1.4$, which can be employed for the calibration of the 2D relations \citep{kodama2008,liang08,wei09}. 
This value is given by the formula:

\begin{eqnarray}
\mu(z) & = & 25 + (5/2) (\log \ y - k) \nonumber \\ ~ & = & 25 + (5/2) (a + b \log\ x -k),
\label{modulus}
\end{eqnarray}

where $y = k D_L^2(z, \Omega_M, h)$ is a given quantity with $k$ a redshift independent constant, 
and $a$ and $b$ are the relation parameters. Presuming that this calibration is redshift independent, 
the HD at higher $z$ can be constructed using the calibrated relations for the 
other GRBs in the data set.\\
Finally, \cite{li2014} analyzed the light curves of 8 LGRBs associated with SNe finding a relation
between the peak magnitude and the decline rate at 5, 10 and 15 days as in SNe Ia. However, from the 
comparison with the well-known relation for SNe Ia \citep{phillips93}, it was pointed out that these two objects have 
two different progenitors. More importantly, this discovery allowed 
to use GRBs associated with SNe as possible standard candles. In addition, \cite{cano2014} investigated the optical 
light curves of 8 LGRBs associated with SNe discovering evidence of a relation between their luminosity 
and the width of the GRB light curves relative to the template of the well-known SN 1998bw. This result also confirmed 
the possibility of using GRBs associated with SNe as standard candles.

\subsection{Applications of GRB afterglow relations}
In this section, we describe some applications to cosmology only for the LT relation, because this is the only 
afterglow relation that has been used so far as a cosmological probe. However, the method is very general and it can 
be employed for all the other relations presented in the review. 
The idea to use afterglow GRBs phase as cosmological rulers was proposed for the first time in 2009, when the LT relation
was used to derive a new HD \citep{cardone09,cardone10}.\\ 
More specifically, \cite{cardone09} revised the data set  
used in \cite{schaefer2007} appending the LT relation. They used a Bayesian fitting method, similar to that used
in \cite{firmani06} for the $\log E_{\gamma,peak}$-$\log E_{\gamma,cor}$ relation, to calibrate the 
different GRB relations known at that time assuming a fiducial $\Lambda$CDM model compatible with the data provided 
by the Wilkinson Microwave Anisotropy Probe, WMAP5.\\
A new HD including $83$ objects was obtained (69 from
\cite{schaefer2007} plus 14 new GRBs obtained by the LT relation) computing the mean performed over six relations 
($\log E_{\gamma,cor}-\log E_{\gamma,peak}$, $\log L_{\gamma,iso}-\log V$, with $V$ the variability which measures
the difference between the observed light curve and a smoothed version of that light curve, 
$\log L_{X,a}-\log T_{X,a}^*$, 
$\log L_{\gamma,iso}-\log \tau_{lag}$, with $\tau_{lag}$ the difference in arrival time to the 
observer of the high energy photons and low energy photons, $\log L_{\gamma,iso}-\log \tau_{RT}$, 
with $\tau_{RT}$ the shortest time over which the light curve increases by the $50\%$ of the 
peak flux of the pulse, and $\log L_{\gamma,iso}-\log E_{\gamma,peak}$).\\
To elude the circularity problem, local regression was run to calculate $\mu(z)$ from
the newest SNe Ia sample containing 307 SNe Ia in the range
$0.015 \le z \le 1.55$. Indeed, 
the GRB relations mentioned before were calibrated while considering only GRBs with $z \le 1.4$ in order to cover the same
redshift range spanned by the SNe Ia data. This SNe Ia sample is the input for the local
regression estimate of $\mu(z)$.\\
The basic idea of the local regression analysis consists of several stages described in \cite{cardone09}.
To find out which are the optimal parameters of this procedure, a large sample of simulations was carried out. 
They set the value of the model parameters 
$(\Omega_M, w_0, w_a, h)$, with $w_0$ and $w_a$ given by the coefficient of the DE EoS $w(z)=w_0+w_a z(1+z)^{-1}$ 
\citep{schaefer2007}, in the ranges $0.15 \le \Omega_M \le 0.45$, $-1.5 \le w_0 \le -0.5$, $-2.0 \le w_a \le 2.0$ 
and $0.60 \le h \le 0.80$. For each $z$ value, $\mu(z_i)$ was selected from a Gaussian distribution centered on 
the predicted value and with $\sigma_{int}= 0.15$, consistent with the
$\sigma_{int}$ of the SNe Ia absolute magnitude. This way, a mock 
catalogue with the same $z$ and error distribution of the SNe sample was built. Each 
$\mu(z)$ value derived from this procedure is compared to the input one.
The local regression method correctly produces the underlying $\mu(z)$ at each $z$ from the SNe Ia sample, whichever 
is the cosmological model.\\
Furthermore, comparing their HD to the one derived by
\cite{schaefer2007}, referred as the Schaefer HD, they have updated the Schaefer HD in three ways, namely
updating the $\Lambda$CDM model parameters, using a Bayesian fitting procedure and adding the LT relation. 
To analyze the influence of these changes, the sample of 69 GRBs adopted by 
\cite{schaefer2007} was also used and the distance moduli were computed with the new calibration,
but without considering the LT relation. It was found that $\mu_{new}/\mu_{old}$ is close to 1 within 5\%. 
Thus, this calibration procedure has not modified the results.\\
In conclusion, it was pointed out that the $\mu(z)$ for each of the GRBs in common to \cite{schaefer2007}
and \cite{Dainotti2008} samples is compatible with the one computed using the set of \cite{schaefer2007} relations. 
Therefore, no systematic bias is added by also considering the LT relation.
On the other hand, the addition of the LT relation to the pre-existing ones not only decreases the errors on 
$\mu(z)$ by $\sim 14$\%, but also expands the data set from $69$ to $83$ GRBs.\\
While \cite{cardone09} added the LT relation to a set of other $5$ known prompt emission relations, \cite{cardone10} 
used instead the LT relation alone (66 LGRBs) or in combination with other cosmological tools in order to find 
some constraints on the cosmological parameters at large $z$. 
The GRBs were divided in E0095 and E4 samples, indicating that the introduction of 
the LT relation alone also provides constraints compatible with previous outcomes, since the HD spans over a 
large redshift range $(0.033, 8.2)$.\\
Furthermore, considering three different cosmological models, namely the $\Lambda$CDM, the CPL \citep{Chevallier2001} 
and the quintessence (QCDM), it was discovered that the $\Lambda$CDM model is preferred. To better show the 
impact of GRBs, the fit was repeated only with other probes, such as SNe Ia or Baryon acoustic Oscillations, 
excluding the GRBs. The addition of GRBs does not significantly narrow the parameters
confidence ranges, but GRBs drive the constraints on $w_a$ to $0$. This result indicates
that the consideration of a big sample of E0095 GRBs may lead to a constant
EoS DE model.\\
In addition, we may note that, different from what was done in the literature at the time of their publication, the HD for the 
E0095 and E4 samples is the only GRB HD built with a single relation in the afterglow containing a statistically significant
sample.\\
Furthermore, the LT relation does not request the mix of several relations to rise the number of
GRBs with a known $\mu(z)$. In fact, each relation is influenced by its own biases and intrinsic scatter; therefore, 
using all of them in the same HD can affect the evaluation of the cosmological parameters.
The $\sigma_{int}$ of the LT relation may be considerably decreased if only the E0095 subsample
is analyzed. However, considering the whole sample of 66 LGRBs, \cite{cardone10} 
constrained $\Omega_M$ and $H_0$ obtaining values compatible with the ones presented in the literature.\\ 
This analysis clearly claimed that the LT relation can be considered for building a GRB HD without adding 
any bias in the study of the cosmological parameters. 
Equivalent findings were achieved considering E0095 GRBs even if they are just $12\%$ of the whole sample. 
Therefore, a further
investigation of E0095 GRBs can boost their use as standard sample for studying the DE mystery.\\
As a further development, \cite{Dainotti2013b} pointed out
to what extent a separation of 5 $\sigma$ above and below the intrinsic value, $b_{int}=-1.07_{-0.14}^{+0.09}$, of 
the slope of the LT relation can influence the cosmological results.\\
For this study, a simulated data set of 101 GRBs obtained through a Monte Carlo simulation was 
collected assuming 
$b=-1.52$, $\sigma_{int}=0.93$ (larger than the scatter computed from the original data set, namely $\sigma_{int}=0.66$),
and the fiducial $\Lambda$CDM flat cosmological model with $\Omega_M = 0.291$ and $H_0 = 71$ Km s$^{-1}$ Mpc$^{-1}$. 
They investigated how much the scatter in the cosmological parameters can be diminished if, instead of the total 
sample (hereafter Full), a highly luminous subsample (hereafter High Luminosity)
is considered, constrained by the condition that $\log L_{X,a} \geq 48.7$.
The choice of this selection cut at a given luminosity is explained in \cite{Dainotti2013a}, who
showed that the local luminosity function is similar to the observed luminosity one
for $\log L_{X,a} \geq 48$.\\
The methodology is similar to what has been done by \cite{Amati2008} for the $\log E_{\gamma,peak}-\log E_{\gamma,iso}$ relation, namely 
the fit has been performed varying simultaneously both the calibration parameters, $p_{GRB}=(a, b, \sigma_{int})$, and the 
cosmological parameters, $p_c=(\Omega_M , \Omega_{\Lambda}, w_0, w_a, h)$, each time for a 
given model in order to correctly take this issue into account.\\
In order to have stronger limits on the cosmological parameters two samples were added to the data set, the $H(z)$
sample ($H(z)=H_0 \times \sqrt{\Omega_M(1+z)^3+\Omega_k(1+z)^2+\Omega_{\Lambda}}$) over the 
redshift range $0.10 \le z \le 1.75$ \citep{stern2010} and the Union $2.1$ SNe Ia
sample containing 580 objects over the redshift range $0.015 \le z \le 1.414$ \citep{suzuki2012}.\\
A Markov Chain Monte Carlo (MCMC) method was used, running three parallel chains and applying
the Gelman-Rubin test\footnote{The Gelman-Rubin diagnostics relies on parallel chains 
to test whether they all converge to the same posterior distribution.} in order to analyze the convergence for an assumed 
cosmological model characterized by a given set of cosmological parameters $p_c$ to be determined.\\
From this statistical analysis results regarding the Full GRB sample, $b$, $a$ and $\sigma_{int}$ of the LT
relation are independent of the chosen cosmological model and the presence of the
SNe Ia and $H(z)$ data in the sample. 
In addition, even if a 5 $\sigma$ scatter in $b_{int}$ is assumed, the results for the Full sample
are in agreement with earlier outcomes \citep{Dainotti2008,dainotti11a} where exclusively flat models were assumed.\\ 
On the other hand, due to the wide errors on the simulated data, 
the cosmological parameters are not emerging in the calibration procedure. However, the signature of the cosmology 
will appear considering a greater data set with low errors on $(\log \ T_{X,a}^*, \log \ L_{X,a})$.\\
Furthermore, for the Full sample, it was studied how much the deviation from the $b_{int}$ of the LT relation influences 
the cosmological parameters. To analyze this 
problem, a model parameterized in terms of the present day values of $\Omega_M$, $\Omega_{\Lambda}$ and $H_0$ 
was considered.

\begin{figure}[htbp]
\includegraphics[width=0.95\hsize]{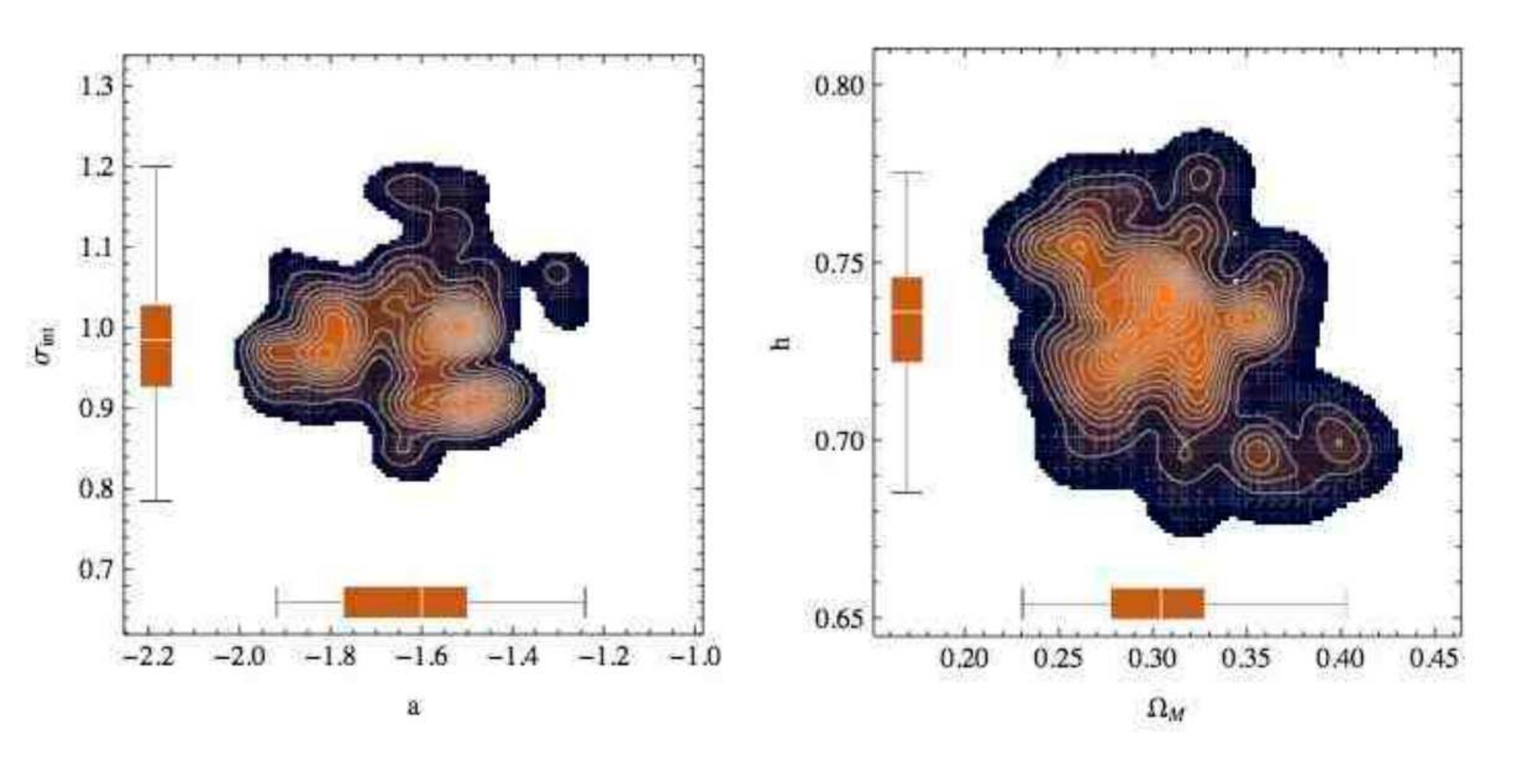}
\includegraphics[width=0.95\hsize]{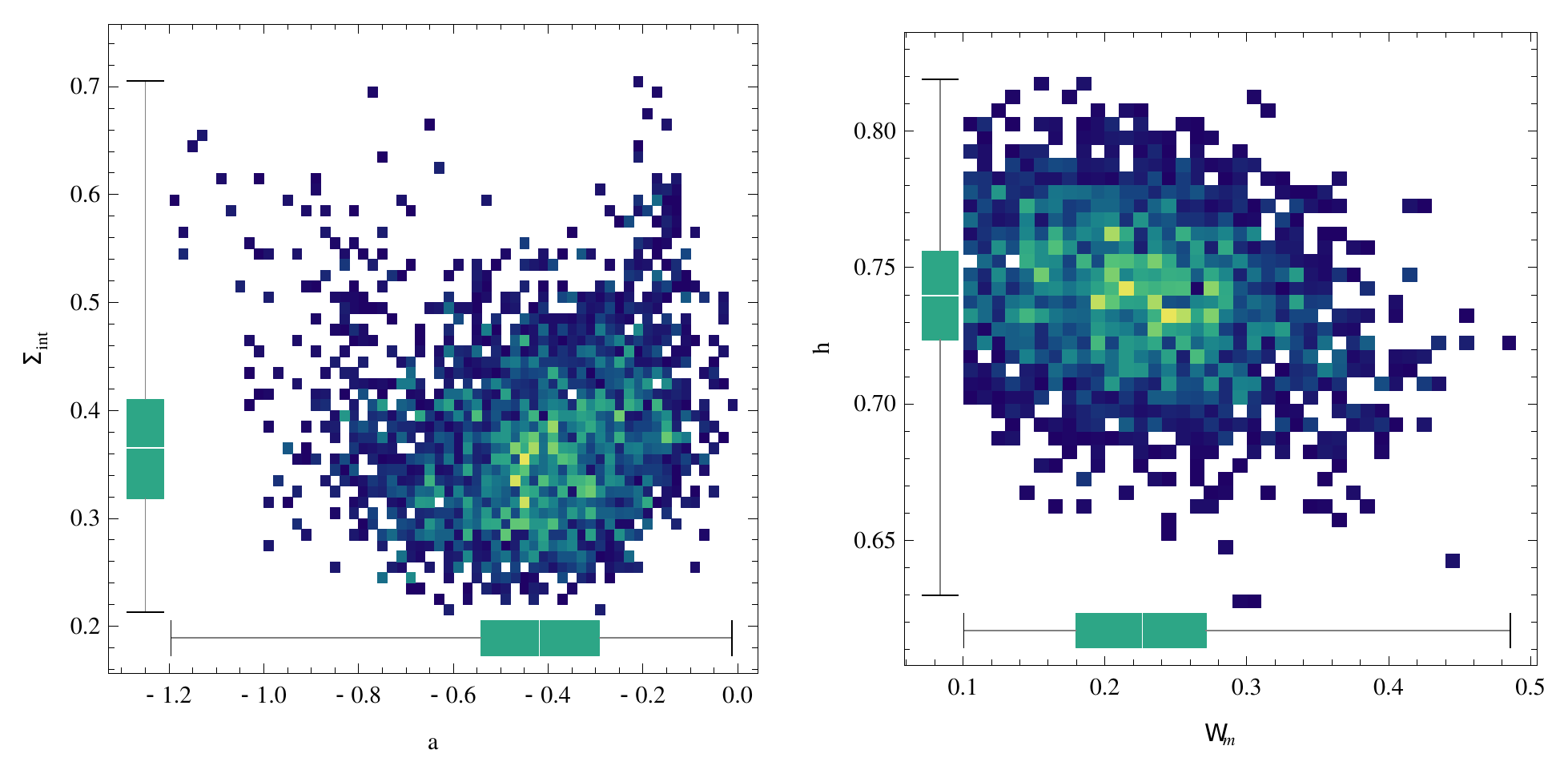}
\caption{\footnotesize Upper left panel: ``regions of confidence for the marginalized likelihood function $ {\mathcal L}(b,\sigma)$ from
\cite{Dainotti2013b}, obtained
marginalizing over $a$ and the cosmological parameters using the Full sample. The bright brown regions indicate the 
1 $\sigma$ (full zone) and 2 $\sigma$ (bright grey) regions of confidence respectively. On the axes are plotted the box-and-whisker 
diagrams relatively to the $b$ and
$\sigma_{int}$ parameters: the bottom and top of the diagrams are the 25th and 75th percentile (the lower and upper quartiles, 
respectively), and the band near the middle of the box is the 50th percentile (the median)". Upper right panel: ``regions of confidence
for the marginalized likelihood function ${ \mathcal L}(\Omega_M, h)$, obtained using the Full sample, from \cite{Dainotti2013b}". 
Bottom left panel: ``regions of 
confidence for the marginalized likelihood function $ {\mathcal L}(b,\sigma)$ from \cite{Dainotti2013b}, 
obtained marginalizing over $a$ and the cosmological parameters for the High Luminosity sample. The bright brown regions indicate the
1 $\sigma$ (full zone) and 2 $\sigma$ (bright grey) regions of confidence respectively. On the axes are plotted the box-and-whisker 
diagrams relatively to the $b$ and
$\sigma_{int}$ parameters: the bottom and top of the diagrams are  the 25th and 75th percentile (the lower and upper quartiles, 
respectively), and the band near the middle of the box is  the 50th percentile (the median)". Bottom right panel: ``regions of confidence
for the marginalized likelihood function ${ \mathcal L}(\Omega_M, h)$, obtained using the High Luminosity sample, from 
\cite{Dainotti2013b}".}
\label{conreg}
\end{figure}

\begin{figure}[htbp]
\centering
   \includegraphics[width=8.1cm,height=5.4cm,angle=0]{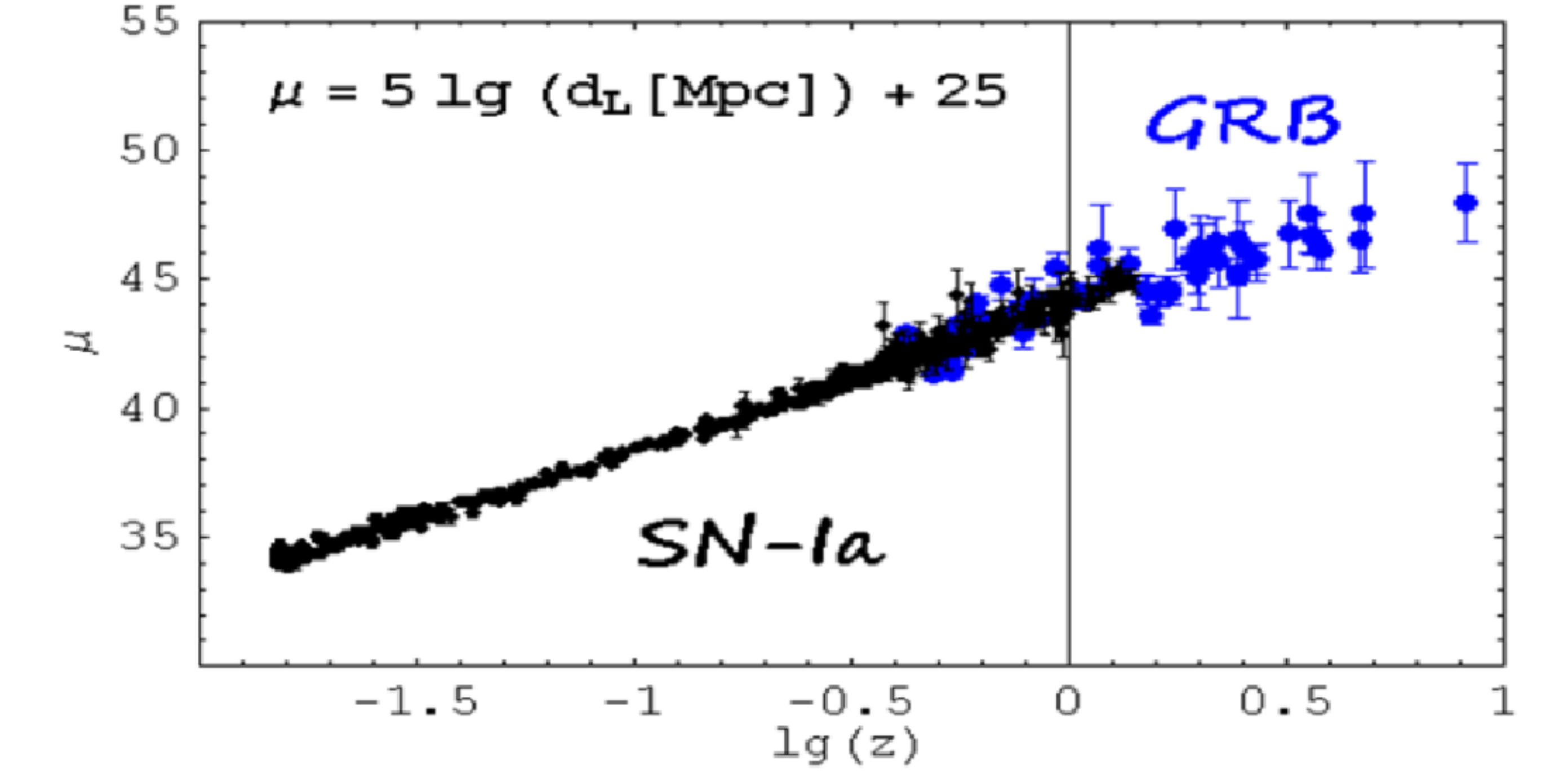}
   \includegraphics[width=8.1cm,height=5.5cm]{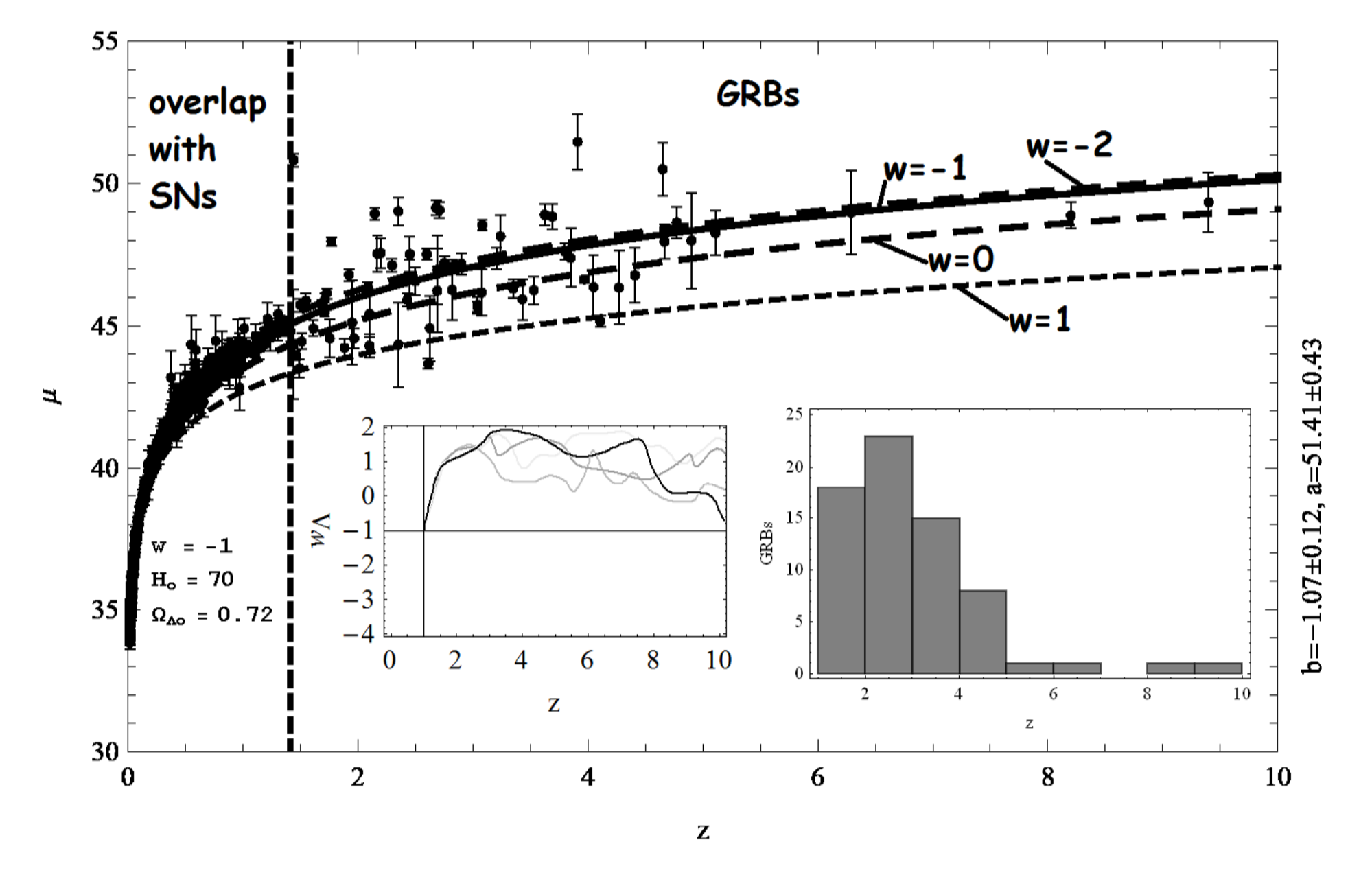}
   \caption{\footnotesize Left panel: ``($z_j$, $\mu_j \pm \Delta \mu_j$) for SNe Ia from \cite{postnikov14}. GRBs are inferred 
   from the relation assuming a flat $w = -1$ cosmology and stand out only from their larger error bars, no
discontinuity is evident, implying a first order consistency of a $w = -1$ model out to very high $z$.
The SNe Ia data were taken from the Union 2.1 compendium \citep{suzuki2012}". Right panel: ``distance ladder from 
\cite{postnikov14}. GRBs in the SNe Ia overlap redshift range, where cosmology is well 
constrained, are used to calculate the GRB 
intrinsic correlation coefficient. This relation is then used to calculate $D_L(z, \Omega_M, h)$ for high $z$ GRBs from 
their 
X-ray afterglow luminosity curves. Standard constant $w$ solutions are shown for reference. Vertical dashed line marks farthest SNe Ia
event. Inset to the right shows a histogram of the GRB sample distribution in $z$. Inset to the left shows resulting most probable
EoS, together with a small sample of models probed, confidence intervals are so large, that only extreme variations with respect to
$w=-1$ can be excluded".}
\label{fig:hubblediagram}
\end{figure}

Although $h$ is comparable with the values from both the local distance estimators \citep{riess2009} and CMBR
data \citep{komatsu2011}, the median values for
$(\Omega_M, \Omega_{\Lambda})$ are broader if compared to a fiducial $\Omega_M \sim 0.27$ recovered in earlier works
\citep{davis07}. 
For this reason, considering for the Full sample, a distinct $b_{int}$ will lead to a disagreement of $13\%$ with the
best value of the $\Omega_M$ parameter (see the upper panels of Fig. \ref{conreg}).
Even if the median values of the fit for the sample that also has SNe Ia and $H(z)$ data do not conduct towards 
flat models, a spatially flat Universe accords with, for example, 
the WMAP7 cosmological parameters within $95\%$ giving $\Omega_k = -0.080_{-0.093}^{+0.071}$.
This difference can be deduced, because in this case it is not possible to distinguish among flat and not 
flat models and this distinction is still not possible when SNe Ia data are present in the 
fit. Thus, constraining the model to be spatially flat, but shaping the DE EoS with $w(z)$,
leads to a couple $(w_0, w_a)$ completely different irrespective of whether SNe Ia and $H(z)$ data are considered or not in the 
sample. Regarding instead the High Luminosity subsample, the limits on the calibration parameters mostly do not
depend on either the used cosmological model or if SNe Ia and $H(z)$ data are considered in the sample. 
Furthermore, for the High Luminosity subsample it is shown that adding the SNe Ia and 
$H(z)$ data does not ameliorate the constraints on the calibration parameters.\\
Finally, the Full sample outcomes are comparable to those of the flat cosmological model for the SNe Ia sample, while the 
High Luminosity subsample diverges by $5\%$ in the value of $H_0$ as computed in \cite{Petersen2010}, and the scatter 
in $\Omega_M$ is underestimated by $13\%$, see the bottom panels of Fig. \ref{conreg}. 
In conclusion, an optimal procedure is to consider a High luminosity subsample provided by a cut exactly at 
$\log L_{X,a}=48$; otherwise, the 
luminosity and time evolutions should be added in the computation of the cosmological parameters.\\
Later, another application of GRBs to cosmology is presented in \cite{postnikov14} where the DE EoS was analyzed as a 
function of $z$ without assuming any a priori $w(z)$ functional form.\\ 
To build a GRB $(\mu, z)$ diagram, 580 SNe Ia from the Union 2.1 compendium \citep{suzuki2012} were used together with
54 LGRBs in the overlapping
redshift ($z \leq 1.4$ see the left panel of Fig. \ref{fig:hubblediagram})
region between GRBs and SNe Ia. In addition, a standard $w=-1$ cosmological model was assumed.

\begin{figure}[htbp]
\centering
   \includegraphics[width=1\hsize]{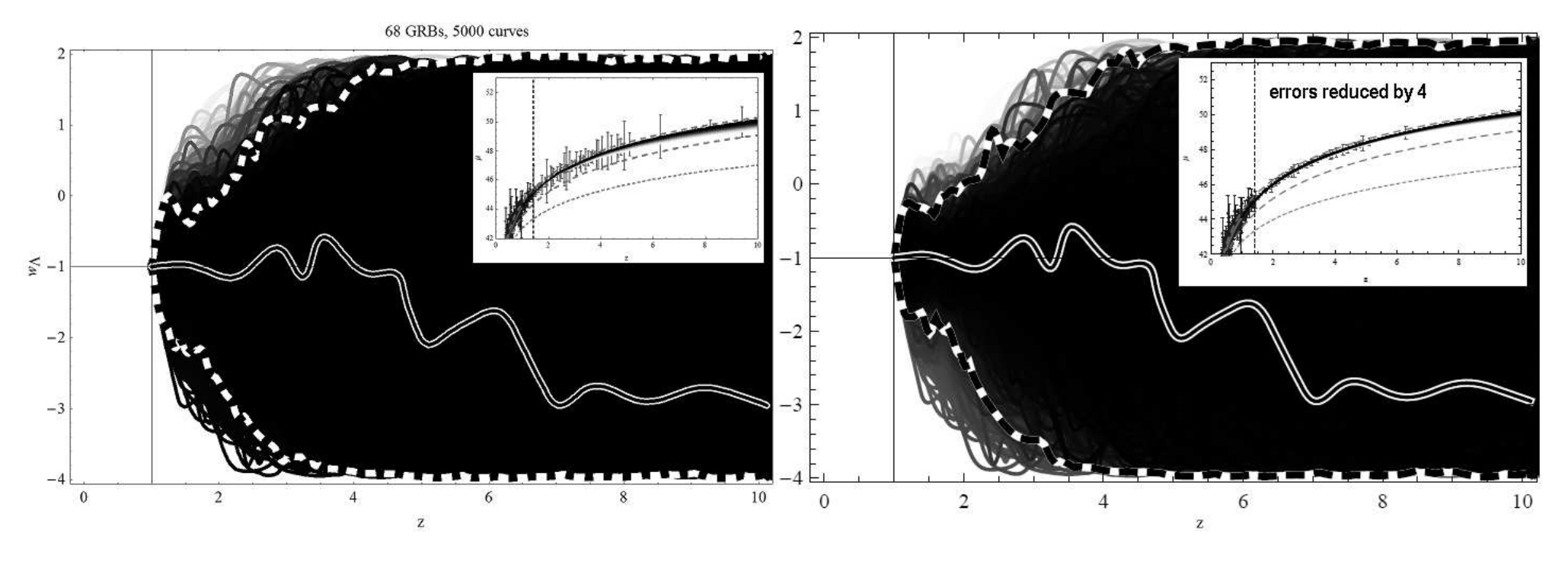}
   \caption{\footnotesize ``Tree of $w(z>1)$ curves inferred from synthetic GRB samples constructed for $w(z)=-1$ cosmologies
in \cite{postnikov14}, showing to what extent correlated GRB errors constrain EoS at high $z$ ($z>1$). 
In the left side plot GRB errors taken from
actual data are used, while in the right side plot GRB errors reduced by a factor of $4$ are considered".}
\label{fig:hubblediagram2}
\end{figure}

One order of magnitude expansion
in redshift interval is supplied by the GRB data set considering the correlation coefficients obtained for the SNe Ia.
This detail allows for the enlargement of the cosmological model out to $z=8.2$.
In fact, a relation was found given by:

\begin{equation}
\log L_{X,a}=53.27_{-0.48}^{+0.54}-1.51^{+0.26}_{-0.27} \times \log T^*_{X,a},
\end{equation}

with $\rho=-0.74$ and $P=10^{-18}$.\\
\cite{postnikov14} used a Bayesian statistical analysis, similarly to 
\cite{firmani06} and \cite{cardone10}, in which the hypothesis is related to a particular $w(z)$ function with the 
selection of $H_0$ and the present DE density parameter, $\Omega_{\Lambda0}$.
The assumption of isotropy for the 
cosmological model, reliable limits on the EoS and also a
fixed value for $w(z)$ in the $z \leq 0.01$ redshift interval were employed.
In addition, a huge number of randomly chosen $w(z)$ models were used.\\
To test the procedure, their pattern is verified through the simulated data sets obtained from several input cosmological 
models with relative errors and $z$ distribution equal to the real data. Through this procedure, employing the LT
relation, a data set of GRBs detected by the
Swift satellite, with $z$ from $0.033$ to $9.44$, was adopted (see inset in the right panel of 
Fig. \ref{fig:hubblediagram}). Thus, it is possible to investigate the history of the Universe out to $z \approx 10$.
(However, an additional analysis would be beneficial if we would consider the sample without the GRB at
$z=9.4$. We note that indeed in \cite{cardone10} a sample of canonical GRBs was used in which this burst has not
been included).\\
In order to do that, they simulated $2000$ constant EoSs uniformly spaced between $-4 \leq w_{\Lambda} \leq 2$, 
with $w_{\Lambda}$ the DE EoS. Beginning from SNe Ia data sample, a precise solution was found to be 
in agreement with the cosmological constant and a small confidence interval, $w=-0.99 \pm 0.2$, see the right panel of Fig. \ref{fig:hubblediagram}.
Furthermore, it is shown that assuming also that the BAO limits do not differ from the solution of the EoS, but it 
considerably decreases the 
confidence interval ($w=-0.99 \pm 0.06$). In fact, the insertion of the BAO notably constrained the confidence
region of the solutions, especially for the present DE density parameter, giving $\Omega_{\Lambda0}=0.723 \pm 0.025$.\\
As a further step, the $w(z)$ model which leads to the best evaluation of $D_L(z, \Omega_M, h)$, $z$ of
the SNe Ia sample and the BAO constraints needs to be selected.
The confidence region of the allowed $w(z)$ curves is significantly constrained taking into account 
also the BAO data.\\
Afterwards, also considering that GRB data should constrain the cosmological parameters, apart from obtaining 
one order of magnitude expansion in the redshift range, it was extremely difficult to 
constrain the high $z$ $w(z)$ functional form, considered the paucity of points over a broad 
redshift interval and the error bars related to these data.
This is visible in the left panel of Fig. \ref{fig:hubblediagram2}, where a simulated GRB data set having the same $z$
distribution and error bars as the real data, but with assumed $w=-1$ Universe, is provided.
It is noted that only strong $w(z)$ fluctuations are not allowed. Then, decreasing the errors by a 
factor of $4$ led to more intriguing high $z$ DE constraints, see the right panel of Fig. \ref{fig:hubblediagram2}.\\
In addition, the small number of elements in the SNe Ia overlapping region indicated broad error bars on the GRB correlation 
coefficients. Meanwhile, the broad error bars for high $z$ GRBs generated a very flat probability
distribution (represented by the uniform black shading area in the left panel of Fig. \ref{fig:hubblediagram2}) 
for the several EoSs checked. Therefore, there will be great interest for the $1<z<4$ region of the GRB HD as soon as 
the GRB data set is enlarged and the quality of data is upgraded.

\section{Summary and discussion}\label{discussion}
\textcolor{red}{From the analysis of the relations mentioned in previous sections, it is visible that:}

\begin{enumerate}
 \item \textcolor{red}{The accretion model \citep{Cannizzo2009,cannizzo2011} and the magnetar model 
 \citep{usov92,dallosso2011,rowlinson12} seem to give the best explanation of the Dainotti relation 
 (giving best fit slopes -3/2 and -1 respectively). The magnetar model seems to be favoured compared to the accretion 
 one, because the intrinsic slope computed in \cite{Dainotti2013a} is exactly $-1.07_{-0.9}^{+0.14}$.} 
 \item \textcolor{red}{A more complex jet structure is needed for interpreting the $\log L_{O,200\rm{s}}$\,-\,$\alpha_{O,>200\rm{s}}$ 
 relation \citep{oates2012}. Indeed, \cite{oates2012} showed that the standard afterglow model cannot explain this 
 relation, especially taking into account the closure relations \citep{Sari98}, which relate temporal decay and spectral 
 indices. Therefore, in order to interpret their results, they claimed either the presence of some features of the central engine
 which dominate the energy release or that the observations were made by observers at different angular distances from the source's axis.
 \cite{Dainotti2013a} pointed out a similarity between the $\log L_{O,200\rm{s}}$\,-\,$\alpha_{O,>200\rm{s}}$ 
 relation and the $L_X-T^{*}_a$ relation, making worthy of investigating the possibility of a single physical mechanism inducing both of them.}
\item \textcolor{red}{In the external shock model the $L_X(T_a)-L_{\gamma,iso}$ and the $L_{X,peak}-L_X(T_a)$ relations
cannot lead to a net distinction among constant or wind type density media, but they are able to 
exclude so far the thin shell models and to favour the thick shell ones. Among the models that very well 
describe the $L_X(T_a)-L_{\gamma,iso}$ and the $L_{X,peak}-L_X(T_a)$ relations there is the one 
by \cite{hascoet2014}. They investigated the standard FS model with 
a wind external medium and a microphysics parameter $\epsilon_e \propto n^{-\nu}$, and they found out 
that for values $\nu \approx 1$ is possible to reproduce a flat plateau phase, and consequently the relations 
mentioned above.
This shows how important the study of correlations especially with the aim of discriminating among models.}
\item \textcolor{red}{In regard to the prompt-afterglow relations, mentioned in section \ref{promptaftcor}, 
involving the energies and the luminosities for the prompt and the afterglow phases, it is pointed out that they help 
to interpret the connection between these two GRB phases. For example, \cite{racusin11} pointed out that
the fraction of kinetic energy transferred from the prompt phase to the afterglow one, for BAT-detected GRBs,
is around 10\%, in agreement with the analysis by \cite{zhang07b}. 
However, from the investigation of these relations, the synchrotron radiation process seems to not explain completely the observations, and also the scatter present in these 
relations is significant. Therefore, further analysis will be useful.}
\item \textcolor{red}{The study of the $L^F_{O,peak}-T^{*F}_{O,peak}$ relation sheds light on the nature of the flares in
the GRB light curves. From the analysis carried on by \cite{Li2012}, it was found out that the flares are additional and 
distinct components of the afterglow phase. They also claimed that a periodically-emitting energy central engine can explain the optical 
and $\gamma$-ray flares in the afterglow phase.}
\item \textcolor{red}{One of the greatest issues that may undermine the GRB relations as model discriminators and as cosmological tools are selection bias and the evolution with the redshift of the physical quantities involved in these relations. 
An example of selection biases is given by \cite{Dainotti2013a}, who used the \cite{Efron1992} method to deal with
the redshift evolution of the X-ray luminosity and the time, to evaluate the intrinsic $L_X-T^{*}_a$ relation. 
Furthermore, \cite{Dainotti2015b} assumed an unknown efficiency function for the detector and investigated the biases due 
to the detector's threshold and how they affect the X-ray luminosity and the time measurements. The methods described
can be also useful to deal with the selection effects for the optical luminosity and in the $\log L_{O,200\rm{s}}$\,-\,$\alpha_{O,>200\rm{s}}$ relation and any other relation.}
\item \textcolor{red}{Regarding the use of correlations as cosmological tools, we still have to further reduce the scatter of the GRB 
measurements and the dispersion of the relations themselves to allow GRBs to be complementary with the measurement 
of SNe Ia. Indeed, the redshift evolution effect and the threshold of the detector can generate
relevant selection biases on the physical quantities which however we know how to treat analytically with robust 
statistical techniques as we have shown in several sections. Nevertheless, more precise calibration methods, with the help of
other cosmological objects, and more space missions dedicated to detect faint GRBs and GRBs at high redshift 
(for example the future SVOM mission) can shed new light on the use of GRBs as cosmological tools.
Lastly, other open questions are concerned with how much cosmological parameters can reduce their degeneracy adding 
GRBs into the set of cosmological standard candles. For example, different results of the value of $w$ can lead to 
scenarios which can be compatible with a non-flat cosmological model.}
\end{enumerate}

\section{Conclusions}\label{conclusion} 
In this work, we have summarized the bivariate relations among the GRB afterglow parameters and their characteristics in order
to discuss 
their intrinsic nature and the possibility to use them as standardizable candles. 
It has been shown with different methodologies
that some of the relations presented are intrinsic. However, the intrinsic slope has been determined 
only for a few relations.
For the other relations, we are not aware of their intrinsic slopes and
consequently how far the use of the observed relations can influence the evaluation of the theoretical models and the ``best" 
cosmological settings \citep{Dainotti2013b}.
Therefore, the estimate of the intrinsic relations is crucial for the determination of the most plausible model that can explain 
the plateau phase and the afterglow emission.\\
In fact, though there are several theoretical interpretations describing each relation, as we 
have shown, in many cases, more than one is viable. This result indicates 
that the emission processes that rule the GRBs still have to be further investigated.
To this end, it is necessary to use the intrinsic relations and not the observed ones affected by
selection biases to test the theoretical models. Moreover, the pure afterglow relations have the advantage
of not presenting the double truncation in the flux limit, thus facilitating the correction for selection 
effects and their use as redshift estimators and cosmological tools.\\ 
A very challenging future step would be to use the corrected relations as a reliable redshift estimator and to determine a 
further estimate of $H_0$, $\Omega_{\Lambda}$ and $w$. In particular, it is advisable to
use all the afterglow relations which are not yet employed for cosmological studies as new probes, after they 
are corrected for selection biases, in order to reduce the 
intrinsic scatter as it has been done in \cite{schaefer2007} for the prompt relations.

\section{Acknowledgments}
This work made use of data supplied by the UK Swift Science Data Centre at the University of Leicester. We thank S. Capozziello
for fruitful comments. M.G.D is grateful to the Marie Curie Program, because the research leading to these results has received funding from the 
European Union Seventh FrameWork Program (FP7-2007/2013) under grant agreement N 626267. R.D.V. is grateful to the 
Polish National Science Centre through the grant DEC-2012/04/A/ST9/00083. 

\bibliography{biblioReview4}

\end{document}